\documentclass[11pt]{article}
\usepackage{makeidx}
\usepackage{amsfonts}
\usepackage{amssymb}
\usepackage{graphicx}
\usepackage{amsmath}
\usepackage{geometry}
\usepackage{appendix}

\let\dprod\prod
\DeclareMathSizes{10.95}{10}{7}{6}
\input{tcilatex}
\begin{document}

\title{Statistical Field Theory and Neural Structures Dynamics I: Action
Functionals, Background States and External Perturbations}
\author{Pierre Gosselin\thanks{%
Pierre Gosselin: Institut Fourier, UMR 5582 CNRS-UGA, Universit\'{e}
Grenoble Alpes, BP 74, 38402 Saint Martin d'H\`{e}res, France\ E-Mail:
Pierre.Gosselin@univ-grenoble-alpes.fr} \and A\"{\i}leen Lotz\thanks{%
A\"{\i}leen Lotz: Cerca Trova, BP 114, 38001 Grenoble Cedex 1, France.\
E-mail: a.lotz@cercatrova.eu}}
\maketitle

\begin{abstract}
This series of papers models the dynamics of a large set of interacting
neurons within the framework of statistical field theory. The system is
described using a two-field model. The first field represents the neuronal
activity, while the second field accounts for the interconnections between
cells. This model is derived by translating a probabilistic model involving
a large number of interacting cells into a field formalism. The current
paper focuses on deriving the background fields of the system, which
describe the potential equilibria in terms of interconnected groups.
Dynamically, we explore the perturbation of these background fields, leading
to processes such as activation, association, and reactivation of groups of
cells.
\end{abstract}

\section{Introduction}

Bridging the micro and macroscopic behaviors remains largely problematic
when dealing with systems characterized by a large number of degrees of
freedom. When investigating neural activity, one typically either begins
directly with a macroscopic description of the system or initiates a study
with a microscopic description that is subsequently treated numerically. In
a previous study (\cite{GL}), we introduced a statistical field-theoretic
approach to establish a connection between the micro and macro levels. For a
dynamic system consisting of a large number of interacting spiking neurons
distributed within a defined spatial region, also referred to as the
"thread," we can associate a field-theoretic framework that encompasses the
fundamental microscopic characteristics of the system.

This field-theoretic framework enables the determination of the system's
effective action, along with the associated background field, which
corresponds to the minimum of the effective action. This background field
characterizes the collective state of the system. The field framework
facilitates the computation of neurons firing frequencies, i.e., neural
activity, at each point within the system in a specific background state.
Furthermore, we can derive the propagation of perturbations in neural
activity from one point to another. We showed the presence of persistent
nonlinear traveling waves along the thread.

Nonetheless, (\cite{GL}) considered the connectivity functions between
various points in the thread as endogenous functions of the neural
activities. We introduced an extension that involved dynamic equations for
these functions driven by the activities dynamics. However, even in this
extended case, the connectivities between the points in the thread were not
treated as a dynamically interacting system that could be addressed as a
field-theoretic system in its own right.

The current series of papers addresses this issue and takes a significant
step towards the field-theoretic representation of a system of interacting
neurons. We introduce a two-fields model that characterizes both the
dynamics of neural activity and the connectivity between cells within the
thread. Furthermore, we investigate the implications of this model in the
context of neural and connectivity dynamics.

This field theory results from a two-step process and is grounded in a
method initially developed by \cite{Kl} and subsequently adapted to complex
interacting systems in \cite{GL1}\cite{GL2}\cite{GL3}\cite{GL4}. In the
first step, the conventional formalism governing the dynamic equations of a
large assembly of interacting neurons, as outlined in (\cite{V11}), is
extended to encompass a dynamic system that accounts for the dynamic nature
of neuronal connectivity. We use the formalism for connectivity functions as
presented in (\cite{IFR}), rewritten in a format suitable to translation
into field theory. In the second step, these two sets of dynamic equations
are transformed into a second-quantized Euclidean field theory, as
elucidated in \cite{GL1}\cite{GL2}\cite{GL3}. The action functional of this
field theory relies on two fields. The first field, akin to the one
introduced in (\cite{GL}), characterizes the assembly of neurons and their
activity. Meanwhile, the second field delineates the dynamics of
connectivity between cells. Both fields exhibit self-interaction, portraying
interactions across the network, and interact with each other, encapsulating
the interdependencies between neural activities and connectivities. This
field-based description encompasses both collective and individual aspects
of the system. The system with these two fields is described by a field
action functional that comprehensively records the interactions at the
microscopic level. This action functional encapsulates the dynamics of the
entire system.

Our description enables the derivation of background fields for both neural
interactions and connectivities, which minimize the action functional. These
background fields depict the collective configuration of the system and
determine the potential static equilibria for neural activities and
connectivities. These equilibria serve as the foundational framework for the
system, organizing fluctuations and signal propagation within it. The
background fields are contingent upon internal system parameters and
external stimuli, and thus, the entire system may undergo transitions in
response to variations in these parameters and stimuli.

Throughout the four articles of this series, we investigate the implications
of our formalism. We explore the conditions and the possible forms for
collective interconnected structures, their interactions, and the mechanisms
governing their merger. This study progressively reorients the relative
importance of our prior research between neural activity and connectivity
dynamics. The structures under examination are defined by sets of
interconnected cells, and the activity within the system is determined by
the specific configuration of these sets.

We accomplish this by deriving an effective action for the connectivity
field alone, with neural activity becoming an endogenous variable dependent
on connectivity. Ultimately, based on the findings presented in these
papers, the fourth section of this work will introduce an expanded effective
model designed to address a substantial or even infinite number of potential
collective states. The system will be characterized by fields representing
collective states, each of these fields depending on a large set of
parameters. These sets of parameters describe potential interconnected
structures with various amplituds and frequencies of activity.

From a field theory perspective, this approach is tantamount to engaging in
a second-quantized formulation of the initial formalism, that is, to
second-quantizing a previously second-quantized formalism. Starting directly
with a field formalism for collective states allows us to model how these
structures interact, activate or deactivate, and experience transitions in
terms of activities amplitudes and activity frequencies.

In this first paper, we expound upon the development of the field model that
characterizes the system, encompassing both neural activity and connectivity
functions. We present the derivation of an effective action for the
connectivities and subsequently compute its quasi-static background field.
This background field delineates the average connectivities among elements
within the thread, as well as the average activities between these elements.
Subsequently, we outline the dynamic implications of these findings by
examining the effects of external perturbations that alter the activities
between individual cells.

Taking into account that the timescale for connectivities is slower than
that of individual cells, we demonstrate how repeated activations at certain
points can propagate throughout the network and progressively alter the
connectivity functions. In the presence of oscillatory perturbations, the
oscillatory response may exhibit interference phenomena. At points where
constructive interference occurs, the background state for connectivities
and the average connectivities experience modification. These long-term
alterations manifest as emerging states characterized by enhanced
connectivities between specific points.

These states reflect the impact of external activations and can be regarded
as a record of these activations. They exhibit a gradual fading over time
but remain capable of reactivation by external perturbations. Furthermore,
the association of such emerging states is possible when their activation
occurs at closed times. The resulting state is a composite of two distinct
states, which can be described as a modification of the initial background
state at various points. Activating one of these two states leads to the
reactivation of their combined effect. Thus, regardless of the cause of
their activation, these states of enhanced connectivity present
characteristics akin to interacting partial neuronal assemblies.

The second application focuses on the system of connectivities as an
independent self-interacting object. By replacing the individual cells'
field as an effective quantity dependent on connectivities, we can derive
the effective dynamics for the connectivity fields. This approach engenders
internal dynamics that can induce shifts in the static background
state,particularly at certain points within the system. Self-interactions
triggered by perturbations may initiate internal patterns of connections
among specific cells. Depending on internal parameters, we observe the
potential for enduring shifts in connectivity background states within
certain regions of the network, while other areas remain unaffected. This
effective theory can also be applied to study the mechanisms underpinning
connectivity reinforcement among multiple cells.

This paper is organised as follows: Section 2 provides a literature review.
Part I of the paper develops a theoretical model for a neural system.
Section 3 introduces a general method for translating a system with a large
number of agents into a field framework. In Section 4, we elucidate the
individual dynamics of interacting neurons and its probabilistic
interpretation. Subsequently, we transcribe this framework in terms of
fields in Section 5.

The second section of the paper applies this model to explore structural
aspects of the neural system. We investigate the background states of the
system along with the corresponding cellular activities and connectivities.
In Section 6, we establish the saddle point equation for the neural field
and derive the general form of the equation for activities. Section 7
provides the saddle point equation for the connectivity field. In Section 8,
we deduce solutions to these equations, which represent the background
fields for connectivities. Additionally, we derive equations for the average
values of connectivities within these states.

Section 9 involves the computation of static equilibrium, yielding the
possible average connectivities and activities within the background field.
Multiple solutions arise, each corresponding to different potential
configurations of interconnected states. Section 10 presents a model
extension in which $n$ different types of cells interact. The background
states for such a model are derived, alongside the connectivities and
activities within these background states.

The third part of this work delves into dynamic aspects of the system,
investigating how external perturbations can induce transitions between
connected equilibrium states. In Section 11, we revisit the dynamics of the
neural field and elucidate the dynamic aspects of the neurons'background
field. Section 12 subsequently derives the dynamic component of cellular
activity.

Within this dynamic framework, stable oscillations in activity emerge, some
of which may be driven by external signals. These oscillations, stemming
from source terms, are expounded upon in Section 13. Building on these
results, Section 14 examines how oscillations in activities can modify the
background states of connectivities and the groups of connected states.

The mechanism of interferences between waves of activity, isolating and
synchronizing groups of interacting cells, is explored in Section 15.
Sections 16, 17, 18 are dedicated to the study of group reactivation by
subsequent signals, group associations through signals, the reactivation of
associated groups by subsequent signals, and the impact of sequences of
signals on state associations, respectively. Section 19 concludes the paper.

\section{Litterature review}

Several branches of the literature are related to our work. First, at the
macroscopic scale, and at the modeling level we are considering, our
approach shares common goals with the literature on mean fields or neural
fields. Neural fields model large populations of neurons as homogeneous
structures, with individual neurons indexed by spatial coordinates. These
models are employed to describe various patterns of brain activity.
Following the work of Wilson, Cowan, and Amari (\cite{R1}\cite{R2}\cite{R3}%
\cite{R4}\cite{R5}\cite{R7}\cite{R8}\cite{R9}\cite{R10}), neural field
dynamics are typically investigated in the continuum limit, with neural
activity represented by a macroscopic variable---the population-averaged
firing rate. Mean field theory has been extended in various ways and has
found a wide range of applications.

It permits the existence of traveling wave solutions (see \cite{Wv1}\cite%
{Wv2}, and related literature). Incorporating stochastic effects in firing
rates allows to model perturbations and diffusion patterns in pulse wave
dynamics as well as noisy transitions between various mean field regimes
(see \cite{Tl} for instance). Besides, mean field approaches can be expanded
to investigate the influence of neural network topology on spatial
arrangements of neural activity, with relevant work found in \cite{N1},
further developments in \cite{N2}, and related sources.

Nevertheless, the mean field approach is an effective theoretical framework
in which the degrees of freedom of certain underlying processes are
aggregated. Despite the convenience and practical applications of mean field
formalism, it relies on simplifications to represent the microscopic level,
such as the neglect of interaction delays or variations in neuronal
connectivity. Furthermore, owing to its aggregated nature, this framework is
unable to capture emerging behaviors.

Compared to this approach, our statistical field theory model maintains a
detailed account of the individual dynamics and connectivities of
interacting neurons while retaining certain features and objectives of
neural field dynamics. For example, we assign spatial coordinates to neurons
in order to derive continuous dynamic equations for the entire system.
However, unlike Mean Field Theory and its extensions, our fields do not
directly represent neural activity. Similar to Statistical Field Theory (as
seen in \cite{Kl}), they are rather abstract, complex-valued functionals
that encapsulate microscopic information on a larger collective scale. It is
only after translating the microscopic model into the language of fields
that we reconstruct specific quantities to describe neural activities.

Our approach incorporates features that have been explored in certain
extensions of mean field theory. Our results inherently involve stochastic
elements: the field accounts for the interactions of neurons subject to
dynamic uncertainty. We are able to recover certain patterns of traveling
waves. However, beyond that, our formalism offers several advantages. It
sheds light on the influence of internal variables on firing rate dynamics.
It can provide a direct approach to phenomena related to phase transitions,
such as the impact of collective patterns on individual ones, by examining
the system's effective action. Additionally, it allows for a wide range of
extensions, as demonstrated in this series of paper, which incorporates the
dynamics of connectivity functions within the framework of field theory.

Please note that the Mean Field approach has already been extended using the
tools of statistical field theory, albeit in a manner different from ours
(see \cite{P1}\cite{P2}\cite{P3}\cite{P4}\cite{P5}\cite{P6}\cite{P7}). In
these extensions, statistical fields represent neural activity, or spike
counts, at each point within the network. An effective action is formulated
for neural activity. Because these extensions account for covariances
between neural activities at different network points, the perturbation
expansion of the effective action goes beyond the mean field approximation.
However, these models are constructed based on the mean field model and
incorporate some deviations from it, ultimately remaining at the collective
level rather than emerging from the network's microscopic features. Closer
to our approach are studies such as those in (\cite{FC1}) and (\cite{FC2}),
which utilize partition functions for the entire neural system, or (\cite%
{FC3}), which uses an effective action. Nevertheless, these approaches
either make simplified micro-level assumptions or impose \textit{a priori}
constraints on the effective action.

At the same scale as the neural field, a specific portion of the literature
has focused on the role of connectivities and plasticity in systems of
interacting neurons. This literature is situated within the context of
neural networks (see (\cite{NN}) and its references for network-related
studies) or neural field modeling. In (\cite{Jrs}), the author investigated
the stability of the resting-state activity within a neural field concerning
variations in connectivity with respect to a homogeneous connectivity
matrix. This concept was further extended in (\cite{Jrsb}) in the context of
a generic network differential equation (see (\cite{Jrf}) and (\cite{Spg})
for a comprehensive account) to explore the impact of non-homogeneous
connectivities on network properties and neural dynamics. The authors
revealed that symmetry breaking in network connectivity, or inhomogeneous
connectivities, leads to the emergence of an attractive functional subspace.
The states of symmetry breaking share some similarities with the background
states considered in this paper, by encompassing the decomposition into
background and fluctuations. Nevertheless, in their context, the backgrounds
are postulated, and parameters must be fine-tuned to produce stable states.

A more detailed examination of connectivity dynamics and their
characteristics is provided in (\cite{Rbn}) (see (\cite{Rbt}) and (\cite{Rbh}%
) for applications). Neural field theory is employed to define connectivity
tensors in terms of bare and dressed propagators, and a diagrammatic
analysis, akin to a Feynman graph expansion, is conducted. This approach to
connectivities enables to surpass the typical phenomenological approach,
allowing for the characterization and exploration of patterns in brain
connectivity and activity. The utilization of graph expansion is similar to
our approach in the sense that connectivities are comprehended through
propagators and series expansion. However, this approach is tailored within
the framework of neural fields and lacks the emergence of global or
background states. Connectivities are not regarded as a dynamic system on
their own.

At the microscopic scale, another branch of literature closely related to
our work lies at the intersection of dynamical systems, complex systems, and
neural networks. This body of research is primarily concerned with the
dynamics and interactions of individual neurons (see \cite{P8}\cite{P9}\cite%
{P10}, and the references therein).

In literature strands such as cognitive neurodynamics or computational
neuroscience, neural processes arise from the interactions of assemblies of
individual neurons. This finer-grained approach enables a more detailed
account of the interplay between neurons' connectivity and firing rates
compared to that of neural fields. Typically, it does not assume spatial
indices; neurons are not arranged within a spatial structure, and the
model's resolution relies on numerical studies. This approach accommodates
neurons' cyclical dynamics, variations in oscillation regimes (for further
details, see \cite{P8} and the references therein), and, of greater
relevance to our work, the emergence of local connectivity and higher-scale
phenomena. These phenomena include the binding problem or polychronization
(see \cite{V1}\cite{V2}\cite{V3}\cite{V4}\cite{V5}\cite{V6}\cite{V7}\cite{V8}%
\cite{V9}\cite{V10}\cite{V11}).

However, unlike mean fields, these models lack an analytical treatment of
collective effects. Our method aims to bridge the gap between macro-scale
modeling in neural field theory and the assembly of interacting neurons.
Nonetheless, they provide the foundational elements for developing a
microscopic basis for a field-theoretic description. Our work is based on
one of these frameworks described in (\cite{IFR}). Initially designed to
address polychronization, we employ it to describe the dynamics of
connectivity functions and investigate the emergence of connected patterns
across the thread of individual cells.

A third branch of the literature is of relevance to our work. Our research
is closely tied to the ongoing debate regarding the existence of specific
sets of connected cells responsible for storing information, particularly
related to memory, often referred to as 'engrams.' Recent empirical and
theoretical studies provide increasing evidence for the mechanisms
underlying engram formation and memory linkage. The articles \cite{Qfh} (and
its references), \cite{Rbs}, \cite{Dhr} (and its references), \cite{Brw},
among others, support the engram hypothesis and its significance in memory
storage. Engrams exhibit persistence and can be subject to reactivation.
They underscore the non-local aspects of engrams, which may span different
regions and types of neurons, as well as the importance of interactions
among multiple engram ensembles that result in enhanced memory recall
compared to the reactivation of a single engram ensemble: multiple engram
ensembles are conferred a greater level of memory recall than reactivation
of a single engram ensemble.

In (\cite{Shn}) engram allocation is linked to the excitability of neurons,
offering insights into engram interactions. Engrams are associated with
dynamic interactions among connectivity states through co-allocation and
overlapping. Some characteristics of these interactions are further examined
in studies such as (\cite{Ssm}), (\cite{Lcr}), and (\cite{Tkc}).

These studies provide strong support for the notion that changes in the
strength of neuronal connections are stored in the brain. They emphasize the
growing consensus regarding the role of interactions between sets of neurons
in the formation of recurrent memories and the composition of complex
behaviors. This confirms the idea that changes in connectivities and the
emergence of associated patterns correspond to the formation of engrams. In
other words, the formation of engrams is directly linked to the dynamic
aspects of connectivity functions, including their modifications and
interactions. Engram states can thus be viewed as dynamic interacting states
of connectivities, akin to those studied in our work. Furthermore, the
mechanism of engram interaction proposed in (\cite{Sht}), which involves a
competitive process that integrates memories of events occurring closely in
time (co-allocating overlapping populations of neurons to both engrams) and
separates memories of events occurring at distant times, bears similarity to
our mechanism of associating connectivity states. In our model, connectivity
states interact and may associate to produce more stable states, with the
timing of activation being crucial for the formation of such associated
states.

Last but not least, several studies converge on the significance of
regulatory mechanisms involving connectivities and interactions between
connectivities and neuronal activity. In (\cite{Shr}), the authors review
recent research on the mechanisms that shape engrams and regulate memory
functions. They speculate that countervailing forces within local
microcircuits contribute to the generation and maintenance of engrams. On
the other hand, some studies emphasize the role of homeostatic processes
that stabilize neuronal activity, as seen in (\cite{Shk}), (\cite{Frd}), (%
\cite{Gln}), (\cite{Sp1}), (\cite{Sp2}), (\cite{Sp3}), (\cite{Sp4}), (\cite%
{Sp5}). The interaction of Hebbian homosynaptic plasticity with rapid
non-Hebbian heterosynaptic plasticity, when complemented with slower
homeostatic changes and consolidation, is sufficient for the formation of
assemblies. This confirms both the role of interactions between
connectivities and neural activity in the formation of connectivity states,
as well as the importance of certain homeostatic processes that we aim to
capture through the concept of activity and connectivity potential.

\part*{Part I: Field formalism for large set of neurons activities and
connectivity functions}

In this initial section, we present the method to transform a dynamic system
with a large number of agents into a field-theoretic model. In general, this
translation entails the introduction of one field for each type of dynamic
variable. Subsequently, we delve into neural dynamics, where we elucidate
the conventional dynamic equations governing neuronal activities and
connectivities (also referred to as transfer functions in (\cite{GL})).
These equations are then translated into a formalism based on two fields.

\section{Translation of dynamical systems with large number of degrees of
freedom into fields models: General method}

The formalism we propose transforms a dynamic system with large number of
agents into a statistical field model. In the present work, the term agent
will refer to individual neurons or the connectivity between cells that are
themselves dynamic variables.\ In classical models, each agent's dynamics is
described by an optimal path\ for\ some vector variable, say $A_{i}\left(
t\right) $, from an initial to a final point, up to some fluctuations.

But this system of agents could also be seen as probabilistic: each agent
could be described by a \emph{probability density }centered around the
classical optimal path, up to some idiosyncratic uncertainties\footnote{%
Because the number of possible paths is infinite, the probability of each
individual path is null.\ We, therefore, use the word "probability density"
rather than "probability".} \footnote{%
See Gosselin, Lotz and Wambst (2017, 2020, 2021).}. In this probabilistic
approach, each possible trajectory of the whole set of $N$ agents has a
specific probability. The classical model is therefore described by the set
of trajectories of the group of $N$ agents, each one being endowed with its
own probability, its statistical weight. The statistical weight is therefore
a function that associates a probability with each trajectory of the group.

This probabilistic approach can be translated into a more compact \emph{%
field formalism}\footnote{%
Ibid.} that preserves the essential information encoded in the model but
implements a change in perspective. A field model is a structure governed by
its own intrinsic rules that encapsulate the dynamic model chosen.\ This
field model contains all possible realizations that could arise from the
initial economic model, i.e. all the possible global outcomes, or collective
state, permitted by the economic model.\ So that, once constructed, the
field model provides a unique advantage over a standard dynamic model: it
allows to compute the probabilities of each of the possible outcomes for
each collective state of the model. These probabilities are computed
indirectly through the \emph{action functional} of the model, a function
that assigns a specific value to each realization of the field. Technically,
the random $N$\ agents' trajectories $\left\{ \mathbf{A}_{i}\left( t\right)
\right\} $ are replaced by a field, a random variable whose realizations are
complex-valued functions $\Psi $ of the variables $\mathbf{A}$\textbf{,} and
the statistical weight of the $N$\ agents' trajectories $\left\{ \mathbf{A}%
_{i}\left( t\right) \right\} $ in the probabilistic approach is translated
into a statistical weight for each realization $\Psi $. They encapsulate the
collective states of the system.

Once the probabilities of each collective state computed, the most probable
collective state among all other collective states, can be found. In other
words, a field model allows to consider the true global outcome induced by
any standard economic model. This is what we will call the \emph{expression}
of the field model, more usually called the \emph{background field} of the
model.

This most probable realization of the field, the expression or background
field of the model, should not be seen as a final outcome resulting from a
trajectory, but rather as its most recurring realization. Actually, the
probability of the realizations of the model is peaked around the expression
of the field.\ This expression, which is characteristic of the system, will
determine the nature of individual trajectories within the structure, in the
same way as a biased dice would increase the probability of one event.

\subsection{Statistical weight and minimization functions for a classical
system}

In a dynamic system with a large number of agents, each agent is
characterized by one or more stochastic dynamic equations. Some of these
equations result from the optimization of one or several objective
functions. Deriving the statistical weight from these equations is
straightforward: it associates, to each trajectory of the group of agents $%
\left\{ T_{i}\right\} $, a probability that is peaked around the set of
optimal trajectories of the system, of the form:%
\begin{equation}
W\left( s\left( \left\{ T_{i}\right\} \right) \right) =\exp \left( -s\left(
\left\{ T_{i}\right\} \right) \right)  \label{wdt}
\end{equation}%
where $s\left( \left\{ T_{i}\right\} \right) $ measures the distance between
the trajectories $\left\{ T_{i}\right\} $ and the optimal ones.

As explained above, this paper studies two types of agents: cells and
connectivities between cells. To remain at a general level in this section,
we rather consider two arbitrary types of agents characterized by
vector-variables $\left\{ \mathbf{A}_{i}\left( t\right) \right\}
_{i=1,...N}, $ and $\left\{ \mathbf{\hat{A}}_{l}\left( t\right) \right\}
_{i=1,...\hat{N}} $ respectively, where $N$ and $\hat{N}$ are the number of
agents of each type, with vectors $\mathbf{A}_{i}\left( t\right) $\ and $%
\mathbf{\hat{A}}_{l}\left( t\right) $ of arbitrary dimension. For such a
system, the statistical weight writes:%
\begin{equation}
W\left( \left\{ \mathbf{A}_{i}\left( t\right) \right\} ,\left\{ \mathbf{\hat{%
A}}_{l}\left( t\right) \right\} \right) =\exp \left( -s\left( \left\{ 
\mathbf{A}_{i}\left( t\right) \right\} ,\left\{ \mathbf{\hat{A}}_{l}\left(
t\right) \right\} \right) \right)  \label{wdh}
\end{equation}

The optimal paths for the system are assumed to be described by the sets of
equations:%
\begin{equation}
\frac{d\mathbf{A}_{i}\left( t\right) }{dt}-\sum_{j,k,l...}f\left( \mathbf{A}%
_{i}\left( t\right) ,\mathbf{A}_{j}\left( t\right) ,\mathbf{A}_{k}\left(
t\right) ,\mathbf{\hat{A}}_{l}\left( t\right) ,\mathbf{\hat{A}}_{m}\left(
t\right) ...\right) =\epsilon _{i}\text{, }i=1...N  \label{gauche}
\end{equation}%
\begin{equation}
\frac{d\mathbf{\hat{A}}_{l}\left( t\right) }{dt}-\sum_{i,j,k...}\hat{f}%
\left( \mathbf{A}_{i}\left( t\right) ,\mathbf{A}_{j}\left( t\right) ,\mathbf{%
A}_{k}\left( t\right) ,\mathbf{\hat{A}}_{l}\left( t\right) ,\mathbf{\hat{A}}%
_{m}\left( t\right) ...\right) =\hat{\epsilon}_{l}\text{, }i=1...\hat{N}
\label{dnw}
\end{equation}%
where the $\epsilon _{i}$ and $\hat{\epsilon}_{i}$ are idiosynchratic random
shocks.\ These equations describe the general dynamics of the two types
agents, including their interactions with other agents. They may\ encompass
the dynamics of optimizing agents where interactions act as externalities so
that this set of equations is the full description of a system of
interacting agents\footnote{%
Expectations of agents could be included by replacing $\frac{d\mathbf{A}%
_{i}\left( t\right) }{dt}$ with $E\frac{d\mathbf{A}_{i}\left( t\right) }{dt}$%
, where $E$ is the expectation operator. This would amount to double some
variables by distinguishing "real variables" and expectations. However, for
our purpose, in the context of a large number of agents, at least in this
work, we discard as much as possible this possibility.}\footnote{%
A generalisation of equations (\ref{gauche}) and (\ref{dnw}), in which
agents interact at different times, and its translation in term of field is
presented in appendix 1.}\textbf{. }

For equations (\ref{gauche}) and (\ref{dnw}), the quadratic deviation at
time $t$ of any trajectory with respect to the optimal one for each type of
agent are:%
\begin{equation}
\left( \frac{d\mathbf{A}_{i}\left( t\right) }{dt}-\sum_{j,k,l...}f\left( 
\mathbf{A}_{i}\left( t\right) ,\mathbf{A}_{j}\left( t\right) ,\mathbf{A}%
_{k}\left( t\right) ,\mathbf{\hat{A}}_{l}\left( t\right) ,\mathbf{\hat{A}}%
_{m}\left( t\right) ...\right) \right) ^{2}  \label{pst}
\end{equation}%
and:%
\begin{equation}
\left( \frac{d\mathbf{\hat{A}}_{l}\left( t\right) }{dt}-\sum_{i,j,k...}\hat{f%
}\left( \mathbf{A}_{i}\left( t\right) ,\mathbf{A}_{j}\left( t\right) ,%
\mathbf{A}_{k}\left( t\right) ,\mathbf{\hat{A}}_{l}\left( t\right) ,\mathbf{%
\hat{A}}_{m}\left( t\right) ...\right) \right) ^{2}  \label{psh}
\end{equation}%
Since the function (\ref{wdh}) involves the deviations for all agents over
all trajectories, the function: 
\begin{equation*}
s\left( \left\{ \mathbf{A}_{i}\left( t\right) \right\} ,\left\{ \mathbf{\hat{%
A}}_{l}\left( t\right) \right\} \right)
\end{equation*}%
is obtained by summing (\ref{pst}) and (\ref{psh}) over all agents, and
integrate over $t$. We thus find:%
\begin{eqnarray}
s\left( \left\{ \mathbf{A}_{i}\left( t\right) \right\} ,\left\{ \mathbf{\hat{%
A}}_{l}\left( t\right) \right\} \right) &=&\int dt\sum_{i}\left( \frac{d%
\mathbf{A}_{i}\left( t\right) }{dt}-\sum_{j,k,l...}f\left( \mathbf{A}%
_{i}\left( t\right) ,\mathbf{A}_{j}\left( t\right) ,\mathbf{A}_{k}\left(
t\right) ,\mathbf{\hat{A}}_{l}\left( t\right) ,\mathbf{\hat{A}}_{m}\left(
t\right) ...\right) \right) ^{2}  \label{prw} \\
&&+\int dt\sum_{l}\left( \frac{d\mathbf{\hat{A}}_{l}\left( t\right) }{dt}%
-\sum_{i,j,k...}\hat{f}\left( \mathbf{A}_{i}\left( t\right) ,\mathbf{A}%
_{j}\left( t\right) ,\mathbf{A}_{k}\left( t\right) ,\mathbf{\hat{A}}%
_{l}\left( t\right) ,\mathbf{\hat{A}}_{m}\left( t\right) ...\right) \right)
^{2}  \notag
\end{eqnarray}%
There is an alternate, more general, form to (\ref{prw}). We can assume that
the dynamical system is originally defined by some equations of type (\ref%
{gauche}) and (\ref{dnw}), plus some objective functions for agents $i$ and $%
l$, and that these agents aim at minimizing respectively:%
\begin{equation}
\sum_{j,k,l...}g\left( \mathbf{A}_{i}\left( t\right) ,\mathbf{A}_{j}\left(
t\right) ,\mathbf{A}_{k}\left( t\right) ,\mathbf{\hat{A}}_{l}\left( t\right)
,\mathbf{\hat{A}}_{m}\left( t\right) ...\right)  \label{glf}
\end{equation}%
and:%
\begin{equation}
\sum_{i,j,k..}\hat{g}\left( \mathbf{A}_{i}\left( t\right) ,\mathbf{A}%
_{j}\left( t\right) ,\mathbf{A}_{k}\left( t\right) ,\mathbf{\hat{A}}%
_{l}\left( t\right) ,\mathbf{\hat{A}}_{m}\left( t\right) ...\right)
\label{gln}
\end{equation}%
In the above equations, the objective functions depend on other agents'
actions seen as externalities\footnote{%
We may also assume intertemporal objectives, see (\cite{GL1}).{}}. The
functions (\ref{glf}) and (\ref{gln}) could themselves be considered as a
measure of the deviation of a trajectory from the optimum. Actually, the
higher the distance, the higher (\ref{glf}) and (\ref{gln}).

Thus, rather than describing the systm by a full system of dynamic
equations, we can consider some ad-hoc equations of type (\ref{gauche}) and (%
\ref{dnw}) and some objective functions (\ref{glf}) and (\ref{gln}) to write
the alternate form of (\ref{prw}) as:%
\begin{eqnarray}
&&s\left( \left\{ \mathbf{A}_{i}\left( t\right) \right\} ,\left\{ \mathbf{%
\hat{A}}_{l}\left( t\right) \right\} \right)  \label{mNZ} \\
&=&\int dt\sum_{i}\left( \frac{d\mathbf{A}_{i}\left( t\right) }{dt}%
-\sum_{j,k,l...}f\left( \mathbf{A}_{i}\left( t\right) ,\mathbf{A}_{j}\left(
t\right) ,\mathbf{A}_{k}\left( t\right) ,\mathbf{\hat{A}}_{l}\left( t\right)
,\mathbf{\hat{A}}_{m}\left( t\right) ...\right) \right) ^{2}  \notag \\
&&+\int dt\sum_{l}\left( \frac{d\mathbf{\hat{A}}_{l}\left( t\right) }{dt}%
-\sum_{i,j,k...}\hat{f}\left( \mathbf{A}_{i}\left( t\right) ,\mathbf{A}%
_{j}\left( t\right) ,\mathbf{A}_{k}\left( t\right) ,\mathbf{\hat{A}}%
_{l}\left( t\right) ,\mathbf{\hat{A}}_{m}\left( t\right) ...\right) \right)
^{2}  \notag \\
&&+\int dt\sum_{i,j,k,l...}\left( g\left( \mathbf{A}_{i}\left( t\right) ,%
\mathbf{A}_{j}\left( t\right) ,\mathbf{A}_{k}\left( t\right) ,\mathbf{\hat{A}%
}_{l}\left( t\right) ,\mathbf{\hat{A}}_{m}\left( t\right) ...\right) +\hat{g}%
\left( \mathbf{A}_{i}\left( t\right) ,\mathbf{A}_{j}\left( t\right) ,\mathbf{%
A}_{k}\left( t\right) ,\mathbf{\hat{A}}_{l}\left( t\right) ,\mathbf{\hat{A}}%
_{m}\left( t\right) ...\right) \right)  \notag
\end{eqnarray}

In the sequel, we will refer to the various terms arising in equation (\ref%
{mNZ}) as the "minimization functions",\textbf{\ i.e}. the functions whose
minimization yield the dynamics equations of the system\footnote{%
A generalisation of equation (\ref{mNZ}), in which agents interact at
different times, and its translation in term of field is presented in
appendix 1.{}}.

We have shown in \cite{GL1}\cite{GL2}\cite{GL3} that the probabilistic
description of the system (\ref{mNZ}) is equivalent to a statistical field
formalism. In such a formalism, the system is collectively described by a
field that is an element of the Hilbert space of complex functions. The
arguments of\ these functions are the same as those describing an individual
neuron and the connectivity function between two cells. A shortcut of the
translation of systems similar to (\ref{mNZ}) in terms of field, is given in 
\cite{GL4} . The next paragraph gives an account of this method.

\subsection{Translation techniques}

Once the statistical weight $W\left( s\left( \left\{ T_{i}\right\} \right)
\right) $ defined in (\ref{wdt}) iscomputed, it can be translated in terms
of field. To do so, and for each type $\alpha $ of agent, the sets of
trajectories $\left\{ \mathbf{A}_{\alpha i}\left( t\right) \right\} $ are
replaced by a field $\Psi _{\alpha }\left( \mathbf{A}_{\alpha }\right) $, a
random variable whose realizations are complex-valued functions $\Psi $ of
the variables $\mathbf{A}_{\alpha }$\footnote{%
In the following, we will use indifferently the term "field" and the
notation $\Psi $ for the random variable or any of its realization $\Psi $.}%
. The statistical weight for the whole set of fields $\left\{ \Psi _{\alpha
}\right\} $ has the form $\exp \left( -S\left( \left\{ \Psi _{\alpha
}\right\} \right) \right) $. The function $S\left( \left\{ \Psi _{\alpha
}\right\} \right) $\ is called the \emph{fields action functional}. It
represents the interactions among different types of agents. Ultimately, the
expression $\exp \left( -S\left( \left\{ \Psi _{\alpha }\right\} \right)
\right) $ is the statistical weight for the field\footnote{%
In general, one must consider the integral of $\exp \left( -S\left( \left\{
\Psi _{\alpha }\right\} \right) \right) $\ over the configurations $\left\{
\Psi _{\alpha }\right\} $. This integral is the partition function of the
system.} that computes the probability of any realization $\left\{ \Psi
_{\alpha }\right\} $\ of the field.

The form of $S\left( \left\{ \Psi _{\alpha }\right\} \right) $\ is obtained
directly from the classical description of our model. For two types of
agents, we start with expression (\ref{mNZ}). The various minimizations
functions involved in the definition of $s\left( \left\{ \mathbf{A}%
_{i}\left( t\right) \right\} ,\left\{ \mathbf{\hat{A}}_{l}\left( t\right)
\right\} \right) $ will be translated in terms of field and the sum of these
translations will produce finally the action functional $S\left( \left\{
\Psi _{\alpha }\right\} \right) $. The translation method can itself be
divided into two relatively simple processes, but varies slightly depending
on the type of terms that appear in the various minimization functions.

\subsubsection{Terms without temporal derivative}

In equation (\ref{mNZ}), the terms that involve indexed variables but no
temporal derivative terms are the easiest to translate.\ They are of the
form:%
\begin{equation*}
\sum_{i}\sum_{j,k,l,m...}g\left( \mathbf{A}_{i}\left( t\right) ,\mathbf{A}%
_{j}\left( t\right) ,\mathbf{A}_{k}\left( t\right) ,\mathbf{\hat{A}}%
_{l}\left( t\right) ,\mathbf{\hat{A}}_{m}\left( t\right) ...\right)
\end{equation*}%
These terms describe the whole set of interactions both among and between
two groups of agents. Here, agents are characterized by their variables $%
\mathbf{A}_{i}\left( t\right) ,\mathbf{A}_{j}\left( t\right) ,\mathbf{A}%
_{k}\left( t\right) $... and $\mathbf{\hat{A}}_{l}\left( t\right) ,\mathbf{%
\hat{A}}_{m}\left( t\right) $... respectively, for instance in our model
firms and investors.

In the field translation, agents of type $\mathbf{A}_{i}\left( t\right) $
and $\mathbf{\hat{A}}_{l}\left( t\right) $ are described by a field $\Psi
\left( \mathbf{A}\right) $ and $\hat{\Psi}\left( \mathbf{\hat{A}}\right) $,
respectively.

In a first step, the variables indexed $i$ such as $\mathbf{A}_{i}\left(
t\right) $ are replaced by variables $\mathbf{A}$ in the expression of $g$.
The variables indexed $j$, $k$, $l$, $m$..., such as $\mathbf{A}_{j}\left(
t\right) $, $\mathbf{A}_{k}\left( t\right) $, $\mathbf{\hat{A}}_{l}\left(
t\right) ,\mathbf{\hat{A}}_{m}\left( t\right) $... are replaced by $\mathbf{A%
}^{\prime },\mathbf{A}^{\prime \prime }$, $\mathbf{\hat{A}}$, $\mathbf{\hat{A%
}}^{\prime }$ , and so on for all the indices in the function. This yields
the expression:

\begin{equation*}
\sum_{i}\sum_{j,k,l,m...}g\left( \mathbf{A},\mathbf{A}^{\prime },\mathbf{A}%
^{\prime \prime },\mathbf{\hat{A},\hat{A}}^{\prime }...\right)
\end{equation*}%
In a second step, each sum is replaced by a weighted integration symbol: 
\begin{eqnarray*}
\sum_{i} &\rightarrow &\int \left\vert \Psi \left( \mathbf{A}\right)
\right\vert ^{2}d\mathbf{A}\text{, }\sum_{j}\rightarrow \int \left\vert \Psi
\left( \mathbf{A}^{\prime }\right) \right\vert ^{2}d\mathbf{A}^{\prime }%
\text{, }\sum_{k}\rightarrow \int \left\vert \Psi \left( \mathbf{A}^{\prime
\prime }\right) \right\vert ^{2}d\mathbf{A}^{\prime \prime } \\
\sum_{l} &\rightarrow &\int \left\vert \hat{\Psi}\left( \mathbf{\hat{A}}%
\right) \right\vert ^{2}d\mathbf{\hat{A}}\text{, }\sum_{m}\rightarrow \int
\left\vert \hat{\Psi}\left( \mathbf{\hat{A}}^{\prime }\right) \right\vert
^{2}d\mathbf{\hat{A}}^{\prime }
\end{eqnarray*}%
which leads to the translation:%
\begin{eqnarray}
&&\sum_{i}\sum_{j}\sum_{j,k...}g\left( \mathbf{A}_{i}\left( t\right) ,%
\mathbf{A}_{j}\left( t\right) ,\mathbf{A}_{k}\left( t\right) ,\mathbf{\hat{A}%
}_{l}\left( t\right) ,\mathbf{\hat{A}}_{m}\left( t\right) ...\right)  \notag
\\
&\rightarrow &\int g\left( \mathbf{A},\mathbf{A}^{\prime },\mathbf{A}%
^{\prime \prime },\mathbf{\hat{A},\hat{A}}^{\prime }...\right) \left\vert
\Psi \left( \mathbf{A}\right) \right\vert ^{2}\left\vert \Psi \left( \mathbf{%
A}^{\prime }\right) \right\vert ^{2}\left\vert \Psi \left( \mathbf{A}%
^{\prime \prime }\right) \right\vert ^{2}\times ...d\mathbf{A}d\mathbf{A}%
^{\prime }d\mathbf{A}^{\prime \prime }...  \label{tln} \\
&&\times \left\vert \hat{\Psi}\left( \mathbf{\hat{A}}\right) \right\vert
^{2}\left\vert \hat{\Psi}\left( \mathbf{\hat{A}}^{\prime }\right)
\right\vert ^{2}\times ...d\mathbf{\hat{A}}d\mathbf{\hat{A}}^{\prime }... 
\notag
\end{eqnarray}%
where the dots stand for the products of square fields and integration
symbols needed.

\subsubsection{Terms with temporal derivative}

In equation (\ref{mNZ}), the terms that involve a variable temporal
derivative are of the form:%
\begin{equation}
\sum_{i}\left( \frac{d\mathbf{A}_{i}^{\left( \alpha \right) }\left( t\right) 
}{dt}-\sum_{j,k,l,m...}f^{\left( \alpha \right) }\left( \mathbf{A}_{i}\left(
t\right) ,\mathbf{A}_{j}\left( t\right) ,\mathbf{A}_{k}\left( t\right) ,%
\mathbf{\hat{A}}_{l}\left( t\right) ,\mathbf{\hat{A}}_{m}\left( t\right)
...\right) \right) ^{2}  \label{edr}
\end{equation}%
This particular form represents the dynamics of the $\alpha $-th coordinate
of a variable $\mathbf{A}_{i}\left( t\right) $ as a function of the other
agents.

The method of translation is similar to the above, but the time derivative
adds an additional operation.

In a first step, we translate the terms without derivative inside the
parenthesis:%
\begin{equation}
\sum_{j,k,l,m...}f^{\left( \alpha \right) }\left( \mathbf{A}_{i}\left(
t\right) ,\mathbf{A}_{j}\left( t\right) ,\mathbf{A}_{k}\left( t\right) ,%
\mathbf{\hat{A}}_{l}\left( t\right) ,\mathbf{\hat{A}}_{m}\left( t\right)
...\right)  \label{ntr}
\end{equation}%
This type of term has already been translated in the previous paragraph, but
since there is no sum over $i$ in equation (\ref{ntr}), there should be no
integral over $\mathbf{A}$\textbf{,} nor factor $\left\vert \Psi \left( 
\mathbf{A}\right) \right\vert ^{2}$.

The translation of equation (\ref{ntr}) is therefore, as before:%
\begin{equation}
\int f^{\left( \alpha \right) }\left( \mathbf{A},\mathbf{A}^{\prime },%
\mathbf{A}^{\prime \prime },\mathbf{\hat{A},\hat{A}}^{\prime }...\right)
\left\vert \Psi \left( \mathbf{A}^{\prime }\right) \right\vert
^{2}\left\vert \Psi \left( \mathbf{A}^{\prime \prime }\right) \right\vert
^{2}d\mathbf{A}^{\prime }d\mathbf{A}^{\prime \prime }\left\vert \hat{\Psi}%
\left( \mathbf{\hat{A}}\right) \right\vert ^{2}\left\vert \hat{\Psi}\left( 
\mathbf{\hat{A}}^{\prime }\right) \right\vert ^{2}d\mathbf{\hat{A}}d\mathbf{%
\hat{A}}^{\prime }  \label{trn}
\end{equation}%
A free variable $\mathbf{A}$ remains, which will be integrated later, when
we account for the external sum $\sum_{i}$. We will call $\Lambda (\mathbf{A}%
)$ the expression obtained:%
\begin{equation}
\Lambda (\mathbf{A})=\int f^{\left( \alpha \right) }\left( \mathbf{A},%
\mathbf{A}^{\prime },\mathbf{A}^{\prime \prime },\mathbf{\hat{A},\hat{A}}%
^{\prime }...\right) \left\vert \Psi \left( \mathbf{A}^{\prime }\right)
\right\vert ^{2}\left\vert \Psi \left( \mathbf{A}^{\prime \prime }\right)
\right\vert ^{2}d\mathbf{A}^{\prime }d\mathbf{A}^{\prime \prime }\left\vert 
\hat{\Psi}\left( \mathbf{\hat{A}}\right) \right\vert ^{2}\left\vert \hat{\Psi%
}\left( \mathbf{\hat{A}}^{\prime }\right) \right\vert ^{2}d\mathbf{\hat{A}}d%
\mathbf{\hat{A}}^{\prime }  \label{bdt}
\end{equation}%
In a second step, we account for the derivative in time by using field
gradients. To do so, and as a rule, we replace :%
\begin{equation}
\sum_{i}\left( \frac{d\mathbf{A}_{i}^{\left( \alpha \right) }\left( t\right) 
}{dt}-\sum_{j}\sum_{j,k...}f^{\left( \alpha \right) }\left( \mathbf{A}%
_{i}\left( t\right) ,\mathbf{A}_{j}\left( t\right) ,\mathbf{A}_{k}\left(
t\right) ,\mathbf{\hat{A}}_{l}\left( t\right) ,\mathbf{\hat{A}}_{m}\left(
t\right) ...\right) \right) ^{2}  \label{inco}
\end{equation}%
by:%
\begin{equation}
\int \Psi ^{\dag }\left( \mathbf{A}\right) \left( -\nabla _{\mathbf{A}%
^{\left( \alpha \right) }}\left( \frac{\sigma _{\mathbf{A}^{\left( \alpha
\right) }}^{2}}{2}\nabla _{\mathbf{A}^{\left( \alpha \right) }}-\Lambda (%
\mathbf{A})\right) \right) \Psi \left( \mathbf{A}\right) d\mathbf{A}
\label{Trll}
\end{equation}%
The variance $\sigma _{\mathbf{A}^{\left( \alpha \right) }}^{2}$ reflects
the probabilistic nature of the model which is hidden behind the field
formalism. This variance represents the characteristic level of uncertainty
of the system's dynamics. It is a parameter of the model. Note also that in (%
\ref{Trll}), the integral over $\mathbf{A}$ reappears at the end, along with
the square of the field $\left\vert \Psi \left( \mathbf{A}\right)
\right\vert ^{2}$.\ This square is split into two terms, $\Psi ^{\dag
}\left( \mathbf{A}\right) $ and $\Psi \left( \mathbf{A}\right) $, with a
gradient operator inserted in between.

\subsection{Action functional}

The field description is ultimately obtained by summing all the terms
translated above and introducing a time dependency. This sum is called the
action functional. It is the sum of terms of the form (\ref{tln}) and (\ref%
{Trll}), and is denoted $S\left( \Psi ,\Psi ^{\dag }\right) $.

For example, in a system with two types of agents described by two fields $%
\Psi \left( \mathbf{A}\right) $and $\hat{\Psi}\left( \mathbf{\hat{A}}\right) 
$, the action functional has the form:%
\begin{eqnarray}
S\left( \Psi ,\Psi ^{\dag }\right) &=&\int \Psi ^{\dag }\left( \mathbf{A}%
\right) \left( -\nabla _{\mathbf{A}^{\left( \alpha \right) }}\left( \frac{%
\sigma _{\mathbf{A}^{\left( \alpha \right) }}^{2}}{2}\nabla _{\mathbf{A}%
^{\left( \alpha \right) }}-\Lambda _{1}(\mathbf{A})\right) \right) \Psi
\left( \mathbf{A}\right) d\mathbf{A}  \label{notime} \\
&&\mathbf{+}\int \hat{\Psi}^{\dag }\left( \mathbf{\hat{A}}\right) \left(
-\nabla _{\mathbf{\hat{A}}^{\left( \alpha \right) }}\left( \frac{\sigma _{%
\mathbf{\hat{A}}^{\left( \alpha \right) }}^{2}}{2}\nabla _{\mathbf{\hat{A}}%
^{\left( \alpha \right) }}-\Lambda _{2}(\mathbf{\hat{A}})\right) \right) 
\hat{\Psi}\left( \mathbf{\hat{A}}\right) d\mathbf{\hat{A}}  \notag \\
&&+\sum_{m}\int g_{m}\left( \mathbf{A},\mathbf{A}^{\prime },\mathbf{A}%
^{\prime \prime },\mathbf{\hat{A},\hat{A}}^{\prime }...\right) \left\vert
\Psi \left( \mathbf{A}\right) \right\vert ^{2}\left\vert \Psi \left( \mathbf{%
A}^{\prime }\right) \right\vert ^{2}\left\vert \Psi \left( \mathbf{A}%
^{\prime \prime }\right) \right\vert ^{2}\times ...d\mathbf{A}d\mathbf{A}%
^{\prime }d\mathbf{A}^{\prime \prime }...  \notag \\
&&\times \left\vert \hat{\Psi}\left( \mathbf{\hat{A}}\right) \right\vert
^{2}\left\vert \hat{\Psi}\left( \mathbf{\hat{A}}^{\prime }\right)
\right\vert ^{2}\times ...d\mathbf{\hat{A}}d\mathbf{\hat{A}}^{\prime }... 
\notag
\end{eqnarray}%
where the sequence of functions $g_{m}$\ describes the various types of
interactions in the system.

\section{Probabilistic description of large set of cells and connectivity
functions}

We describe a dynamic system of a large number of neurons ($N>>1$) and their
connectivity functions. We define their individual equations. Then, we write
a probability density for the configurations of the whole system over time.

\subsection{Cells Individual dynamics}

We follow the description of \cite{V11} for coupled quadratic
integrate-and-fire (QIF) neurons, but use the additional hypothesis that
each neuron is characterized by its position in some spatial range.

Each neuron's potential $X_{i}\left( t\right) $ satisfies the differential
equation:%
\begin{equation}
\dot{X}_{i}\left( t\right) =\gamma X_{i}^{2}\left( t\right) +J_{i}\left(
t\right)  \label{ptn}
\end{equation}%
for $X_{i}\left( t\right) <X_{p}$, where $X_{p}$ denotes the potential level
of a spike. When $X=X_{p}$, the potential is reset to its resting value $%
X_{i}\left( t\right) =X_{r}<X_{p}$. For the sake of simplicity, following (%
\cite{V11}) we have chosen the squared form $\gamma X_{i}^{2}\left( t\right) 
$ in (\ref{ptn}). However any form $f\left( X_{i}\left( t\right) \right) $\
could be used. The current of signals reaching cell $i$ at time $t$ is
written $J_{i}\left( t\right) $.

Our purpose is to find the system dynamics in terms of the spikes'
frequencies, that is neural activities. First, we consider the time for the $%
n$-th spike of cell $i$, $\theta _{n}^{\left( i\right) }$. This is written
as a function of $n$, $\theta ^{\left( i\right) }\left( n\right) $. Then, a
continuous approximation $n\rightarrow t$ allows to write the spike time
variable as $\theta ^{\left( i\right) }\left( t\right) $. We thus have
replaced:

\begin{equation*}
\theta _{n}^{\left( i\right) }\rightarrow \theta ^{\left( i\right) }\left(
n\right) \rightarrow \theta ^{\left( i\right) }\left( t\right)
\end{equation*}%
The continuous approximation could be removed, but is convenient and
simplifies the notations and computations. We assume now that the timespans
between two spikes are relatively small. The time between two spikes for
cell $i$ is obtained by writing (\ref{ptn}) as:%
\begin{equation*}
\frac{dX_{i}\left( t\right) }{dt}=\gamma X_{i}^{2}\left( t\right)
+J_{i}\left( t\right)
\end{equation*}%
and by inverting this relation to write:%
\begin{equation*}
dt=\frac{dX_{i}}{\gamma X_{i}^{2}+J^{\left( i\right) }\left( \theta ^{\left(
i\right) }\left( n-1\right) \right) }
\end{equation*}%
Integrating the potential between two spikes thus yields:%
\begin{equation*}
\theta ^{\left( i\right) }\left( n\right) -\theta ^{\left( i\right) }\left(
n-1\right) \simeq \int_{X_{r}}^{X_{p}}\frac{dX}{\gamma X^{2}+J^{\left(
i\right) }\left( \theta ^{\left( i\right) }\left( n-1\right) \right) }
\end{equation*}%
Replacing $J^{\left( i\right) }\left( \theta ^{\left( i\right) }\left(
n-1\right) \right) $ by its average value during the small time period $%
\theta ^{\left( i\right) }\left( n\right) -\theta ^{\left( i\right) }\left(
n-1\right) $, we can consider $J^{\left( i\right) }\left( \theta ^{\left(
i\right) }\left( n-1\right) \right) $ as constant in first approximation, so
that:

\begin{eqnarray}
\theta ^{\left( i\right) }\left( n\right) -\theta ^{\left( i\right) }\left(
n-1\right) &\simeq &\frac{\left[ \arctan \left( \sqrt{\frac{\gamma }{%
J^{\left( i\right) }\left( \theta ^{\left( i\right) }\left( n-1\right)
\right) }}X\right) \right] _{X_{r}}^{X_{p}}}{\sqrt{\gamma J^{\left( i\right)
}\left( \theta ^{\left( i\right) }\left( n-1\right) \right) }}  \notag \\
&=&\frac{\arctan \left( \frac{\left( \frac{1}{X_{r}}-\frac{1}{X_{p}}\right) 
\sqrt{\frac{J^{\left( i\right) }\left( \theta ^{\left( i\right) }\left(
n-1\right) \right) }{\gamma }}}{1+\frac{J^{\left( n\right) }\left( \theta
^{\left( n-1\right) }\right) }{\gamma X_{r}X_{p}}}\right) }{\sqrt{\gamma
J^{\left( i\right) }\left( \theta ^{\left( i\right) }\left( n-1\right)
\right) }}  \label{FG}
\end{eqnarray}%
To work at the highest level of generality when possible, we write:%
\begin{equation*}
\theta ^{\left( i\right) }\left( n\right) -\theta ^{\left( i\right) }\left(
n-1\right) \equiv G\left( \theta ^{\left( i\right) }\left( n-1\right) \right)
\end{equation*}%
understood that for computations and numerical approximations we will use
formula (\ref{FG}) for $G$.

For $\gamma <<1$, (\ref{FG}) yields:%
\begin{equation*}
\theta ^{\left( i\right) }\left( n\right) -\theta ^{\left( i\right) }\left(
n-1\right) \equiv G\left( \theta ^{\left( i\right) }\left( n-1\right)
\right) =\frac{X_{p}-X_{r}}{J^{\left( i\right) }\left( \theta ^{\left(
i\right) }\left( n-1\right) \right) }
\end{equation*}%
For $\gamma =O\left( 1\right) $ and for $\gamma $ normalized to $1$ and $%
\frac{J^{\left( n\right) }\left( \theta ^{\left( n-1\right) }\right) }{%
X_{r}X_{p}}<<1$, this is: 
\begin{equation}
\theta ^{\left( i\right) }\left( n\right) -\theta ^{\left( i\right) }\left(
n-1\right) \equiv G\left( \theta ^{\left( i\right) }\left( n-1\right)
\right) =\frac{\arctan \left( \left( \frac{1}{X_{r}}-\frac{1}{X_{p}}\right) 
\sqrt{J^{\left( i\right) }\left( \theta ^{\left( i\right) }\left( n-1\right)
\right) }\right) }{\sqrt{J^{\left( i\right) }\left( \theta ^{\left( i\right)
}\left( n-1\right) \right) }}  \label{spt}
\end{equation}%
The activity or firing rate at $t$, $\omega _{i}\left( t\right) $, is
defined by the inverse time span (\ref{spt}) between two spikes:%
\begin{eqnarray*}
\omega _{i}\left( t\right) &=&\frac{1}{G\left( \theta ^{\left( i\right)
}\left( n-1\right) \right) } \\
&\equiv &F\left( \theta ^{\left( i\right) }\left( n-1\right) \right) =\frac{%
\sqrt{J^{\left( i\right) }\left( \theta ^{\left( i\right) }\left( n-1\right)
\right) }}{\arctan \left( \left( \frac{1}{X_{r}}-\frac{1}{X_{p}}\right) 
\sqrt{J^{\left( i\right) }\left( \theta ^{\left( i\right) }\left( n-1\right)
\right) }\right) }
\end{eqnarray*}%
Since we consider small time intervals between two spikes, we can write:%
\begin{equation}
\theta ^{\left( i\right) }\left( n\right) -\theta ^{\left( i\right) }\left(
n-1\right) \simeq \frac{d}{dt}\theta ^{\left( i\right) }\left( t\right)
-\omega _{i}^{-1}\left( t\right) =\varepsilon _{i}\left( t\right)
\label{dnm}
\end{equation}%
where the white noise perturbation $\varepsilon _{i}\left( t\right) $ for
each period was added to account for any internal uncertainty in the time
span $\theta ^{\left( i\right) }\left( n\right) -\theta ^{\left( i\right)
}\left( n-1\right) $. This white noise is independent from the instantaneous
inverse activity $\omega _{i}^{-1}\left( t\right) $. We assume these $%
\varepsilon _{i}\left( t\right) $ to have variance $\sigma ^{2}$, so that
equation (\ref{dnm}) writes: 
\begin{equation}
\frac{d}{dt}\theta ^{\left( i\right) }\left( t\right) -G\left( \theta
^{\left( i\right) }\left( t\right) ,J^{\left( i\right) }\left( \theta
^{\left( i\right) }\left( t\right) \right) \right) =\varepsilon _{i}\left(
t\right)  \label{dnq}
\end{equation}%
The $\omega _{i}\left( t\right) $ are computed by considering the overall
current which, using the discrete time notation, is given by:%
\begin{equation}
\hat{J}^{\left( i\right) }\left( \left( n-1\right) \right) =J^{\left(
i\right) }\left( \left( n-1\right) \right) +\frac{\kappa }{N}\sum_{j,m}\frac{%
\omega _{j}\left( m\right) }{\omega _{i}\left( n-1\right) }\delta \left(
\theta ^{\left( i\right) }\left( n-1\right) -\theta ^{\left( j\right)
}\left( m\right) -\frac{\left\vert Z_{i}-Z_{j}\right\vert }{c}\right)
T_{ij}\left( \left( n-1,Z_{i}\right) ,\left( m,Z_{j}\right) \right)
\label{crt}
\end{equation}%
The quantity $J^{\left( i\right) }\left( \left( n-1\right) \right) $ denotes
an external current. The term inside the sum is the average current sent to $%
i$ by neuron $j$ during the short time span $\theta ^{\left( i\right)
}\left( n\right) -\theta ^{\left( i\right) }\left( n-1\right) $. The
function $T_{ij}\left( \left( n-1,Z_{i}\right) ,\left( m,Z_{j}\right)
\right) $ is the connectivity function between cells $j$ and $i$. It
measures the level of connectivity between $i$ and $j$. If we consider%
\textbf{\ }$T_{ij}\left( \left( n-1,Z_{i}\right) ,\left( m,Z_{j}\right)
\right) $\textbf{\ }as exogenous, we may assume that (see \cite{NN1}): 
\begin{equation*}
T_{ij}\left( \left( n-1,Z_{i}\right) ,\left( m,Z_{j}\right) \right) =T\left(
\left( n-1,Z_{i}\right) ,\left( m,Z_{j}\right) \right)
\end{equation*}%
so that the connectivity function of $Z_{j}$ on $Z_{i}$ only depends on
positions and times. It models the connectivity function as an average
connectivity between local zones of the thread. this transfer function is
typically considered as gaussian or decreasing exponentially with the
distance between neurons, so that the closer the cells, the more connected
they are.

However, in this paper, the connectivity function is a dynamical object
whose dynamic equations are described in the next paragraph.

We can justify the other terms arising in (\ref{crt}): given the distance $%
\left\vert Z_{i}-Z_{j}\right\vert $ between the two cells and the signals'
velocity $c$, signals arrive with a delay $\frac{\left\vert
Z_{i}-Z_{j}\right\vert }{c}$. The spike emitted by cell $j$ at time $\theta
^{\left( j\right) }\left( m\right) $ has thus to satisfy: 
\begin{equation*}
\theta ^{\left( i\right) }\left( n-1\right) <\theta ^{\left( j\right)
}\left( m\right) +\frac{\left\vert Z_{i}-Z_{j}\right\vert }{c}<\theta
^{\left( i\right) }\left( n\right)
\end{equation*}%
to reach cell $i$ during the timespan $\left[ \theta ^{\left( i\right)
}\left( n-1\right) ,\theta ^{\left( i\right) }\left( n\right) \right] $.
This relation must be represented by a step function in the current formula.
However given our approximations, this can be replaced by: 
\begin{equation*}
\delta \left( \theta ^{\left( i\right) }\left( n-1\right) -\theta ^{\left(
j\right) }\left( m\right) -\frac{\left\vert Z_{i}-Z_{j}\right\vert }{c}%
\right)
\end{equation*}%
as in (\ref{crt}). However, this Dirac function must be weighted by the
number of spikes emitted during the rise of the potential. This number is
the ratio $\frac{\omega _{j}\left( m\right) }{\omega _{i}\left( n-1\right) }$
that counts the number of spikes emitted by neuron $j$ towards neuron $i$
between the spikes $n-1$ and $n$ of neuron $i$. Again, this is valid for
relatively small timespans between two spikes. For larger timespans, the
firing rates should be replaced by their average over this period of time.

The sum over $m$ and $i$ is the overall contribution to the current from any
possible spike of the thread, provided it arrives at $i$ during the interval 
$\theta ^{\left( i\right) }\left( n\right) -\theta ^{\left( i\right) }\left(
n-1\right) $ considered. Note that the current (\ref{crt}) is partly an
endogenous variable. It depends on signals external to $i$, but depends also
on $i$ through $\omega _{i}\left( n-1\right) $. This is a consequence of the
intrication between the system's elements.

In the sequel, we will work in the continuous approximation, so that formula
(\ref{crt}) is replaced by:%
\begin{equation}
\hat{J}^{\left( i\right) }\left( t\right) =J^{\left( i\right) }\left(
t\right) +\frac{\kappa }{N}\int \sum_{j}\frac{\omega _{j}\left( s\right) }{%
\omega _{i}\left( t\right) }\delta \left( \theta ^{\left( i\right) }\left(
t\right) -\theta ^{\left( j\right) }\left( s\right) -\frac{\left\vert
Z_{i}-Z_{j}\right\vert }{c}\right) T_{ij}\left( \left( t,Z_{i}\right)
,\left( s,Z_{j}\right) \right) ds  \label{crT}
\end{equation}

Formula (\ref{crT}) shows that the dynamic equation (\ref{dnm}) has to be
coupled with the activity equation:%
\begin{eqnarray}
\omega _{i}\left( t\right) &=&G\left( \theta ^{\left( i\right) }\left(
t\right) ,\hat{J}\left( \theta ^{\left( i\right) }\left( t\right) \right)
\right) +\upsilon _{i}\left( t\right)  \label{cstrt} \\
&=&\frac{\sqrt{\hat{J}^{\left( i\right) }\left( t\right) }}{\arctan \left(
\left( \frac{1}{X_{r}}-\frac{1}{X_{p}}\right) \sqrt{\hat{J}^{\left( i\right)
}\left( t\right) }\right) }+\upsilon _{i}\left( t\right)  \notag
\end{eqnarray}%
and $J^{\left( i\right) }\left( t\right) $ is defined by (\ref{crT}). A
white noise $\upsilon _{i}\left( t\right) $ accounts for the possible
deviations from this relation, due to some internal or external causes for
each cell. We assume that the variances of $\upsilon _{i}\left( t\right) $
are constant, and equal to $\eta ^{2}$, such that $\eta ^{2}<<\sigma ^{2}$.

\subsection{Connectivity functions dynamics}

We describe the dynamics for the connectivity functions $T_{ij}\left( \left(
n-1,Z_{i}\right) ,\left( m,Z_{j}\right) \right) $ between cells. To do so we
adapt the description of (\cite{IFR}) to our context. In this work, the
connectivity functions depend on some intermediate variables and do not
present any space index. The connectivity between neurons $i$ and $j$
satisifies a differential equation:%
\begin{equation}
\frac{dT_{ij}}{dt}=-\frac{T_{ij}\left( t\right) }{\tau }+\lambda \hat{T}%
_{ij}\left( t\right) \sum_{l}\delta \left( t-\Delta t_{ij}-t_{j}^{l}\right)
\label{spn}
\end{equation}%
where $\hat{T}_{ij}\left( t\right) $ represents the variation in
connectivity, due to the synaptic interactions between the two neurons. The
delay $\Delta t_{ij}$ is the time of arrival at neuton $i$ for a spike of
neuron $j$. The time $t_{j}^{l}$ accounts for time of neuron $j$ spikes. The
sum:%
\begin{equation*}
\sum_{l}\delta \left( t-\Delta t_{ij}-t_{j}^{l}\right)
\end{equation*}%
counts the number of spikes emitted by neuron $j$ and arriving at time $t$
at neuron $i$.

The variation in connectivity satisfies itself an equation:%
\begin{equation}
\frac{d\hat{T}_{ij}}{dt}=\rho \left( 1-\hat{T}_{ij}\left( t\right) \right)
C_{ij}\left( t\right) \sum_{k}\delta \left( t-t_{i}^{k}\right) -\hat{T}%
_{ij}\left( t\right) D_{i}\left( t\right) \sum_{l}\delta \left( t-\Delta
t_{ij}-t_{j}^{l}\right)  \label{SPt}
\end{equation}%
where $C_{ij}\left( t\right) $ and $D_{i}\left( t\right) $\ measure the
cumulated postsynaptic and presynaptic activity. The sum:%
\begin{equation*}
\sum_{k}\delta \left( t-t_{i}^{k}\right)
\end{equation*}%
counts the number of spikes emitted at time $t$. Quantities $C_{ij}\left(
t\right) $ and $D_{i}\left( t\right) $ follow the dynamics:%
\begin{equation}
\frac{dC_{ij}}{dt}=-\frac{C_{ij}\left( t\right) }{\tau _{C}}+\alpha
_{C}\left( 1-C_{ij}\left( t\right) \right) \sum_{l}\delta \left( t-\Delta
t_{ij}-t_{j}^{l}\right)  \label{spr}
\end{equation}%
\begin{equation}
\frac{dD_{i}}{dt}=-\frac{D_{i}\left( t\right) }{\tau _{D}}+\alpha _{C}\left(
1-D_{i}\left( t\right) \right) \sum_{k}\delta \left( t-t_{i}^{k}\right)
\label{spf}
\end{equation}

To translate these equations in our set up, we have to consider connectivity
functions of the form:%
\begin{equation*}
T_{ij}\left( \left( n_{i},Z_{i}\right) ,\left( n_{j},Z_{j}\right) \right)
\end{equation*}%
that include the positions of neurons $i$ and $j$ and the parameter $n_{i}$
and $n_{j}$ which are our counting variables of neurons spikes. However,
equations (\ref{spn}), (\ref{SPt}), (\ref{spr}), (\ref{spf}) include a time
variable.

In our formalism, the time variable $\theta ^{\left( i\right) }\left(
n_{i}\right) $ is the time at which neuron $i$ produces its $n_{i}$-th
spike. We should write classical equations depending on these variables.

Moreover, the number of spikes $\sum_{l}\delta \left( t-\Delta
t_{ij}-t_{j}^{l}\right) $ emitted by cell $j$ at time $t_{j}^{l}$ and the
number of spikes $\sum_{k}\delta \left( t-t_{i}^{k}\right) $ emitted by cell 
$i$ at time $t$ are proportional to $\delta \left( \theta ^{\left( j\right)
}\left( n_{j}\right) -\left( t-\Delta t_{ij}\right) \right) \omega
_{j}\left( n_{j}\right) $ and $\delta \left( \theta ^{\left( i\right)
}\left( n_{i}\right) -t\right) \omega _{j}\left( n_{i}\right) $
respectively. Given the introduction of a spatial indices, we have the
relation:%
\begin{equation*}
\Delta t_{ij}=\frac{\left\vert Z_{i}-Z_{j}\right\vert }{c}
\end{equation*}%
and the first $\delta $ function writes:%
\begin{equation*}
\delta \left( \theta ^{\left( j\right) }\left( n_{j}\right) -\left( t-\Delta
t_{ij}\right) \right) =\delta \left( \theta ^{\left( j\right) }\left(
n_{j}\right) -\left( \theta ^{\left( i\right) }\left( n_{i}\right) -\frac{%
\left\vert Z_{i}-Z_{j}\right\vert }{c}\right) \right) \delta \left( \theta
^{\left( i\right) }\left( n_{i}\right) -t\right)
\end{equation*}

As a consequence, we will write first the connectivity functions from $i$ to 
$j$ as:%
\begin{equation*}
T\left( \left( Z_{i},\theta ^{\left( i\right) }\left( n_{i}\right) ,\omega
_{i}\left( n_{i}\right) \right) ,\left( Z_{j},\theta ^{\left( j\right)
}\left( n_{j}\right) ,\omega _{j}\left( n_{j}\right) \right) \right)
\end{equation*}%
This function, together with the variation in connectivity:%
\begin{equation*}
\hat{T}\left( \left( Z_{i},\theta ^{\left( i\right) }\left( n_{i}\right)
,\omega _{i}\left( n_{i}\right) \right) ,\left( Z_{j},\theta ^{\left(
j\right) }\left( n_{j}\right) ,\omega _{j}\left( n_{j}\right) \right) \right)
\end{equation*}%
along with the auxiliary variables:%
\begin{equation*}
C\left( \left( Z_{i},\theta ^{\left( i\right) }\left( n_{i}\right) ,\omega
_{i}\left( n_{i}\right) \right) ,\left( Z_{j},\theta ^{\left( j\right)
}\left( n_{j}\right) ,\omega _{j}\left( n_{j}\right) \right) \right)
\end{equation*}%
and:%
\begin{equation*}
D\left( \left( Z_{i},\theta ^{\left( i\right) }\left( n_{i}\right) ,\omega
_{i}\left( n_{i}\right) \right) \right)
\end{equation*}%
satisfy the following translations of equations (\ref{spn}), (\ref{SPt}), (%
\ref{spr}), (\ref{spf}): 
\begin{eqnarray}
&&\nabla _{\theta ^{\left( i\right) }\left( n_{i}\right) }T\left( \left(
Z_{i},\theta ^{\left( i\right) }\left( n_{i}\right) ,\omega _{i}\left(
n_{i}\right) \right) ,\left( Z_{j},\theta ^{\left( j\right) }\left(
n_{j}\right) ,\omega _{j}\left( n_{j}\right) \right) \right) \\
&=&-\frac{1}{\tau }T\left( \left( Z_{i},\theta ^{\left( i\right) }\left(
n_{i}\right) ,\omega _{i}\left( n_{i}\right) \right) ,\left( Z_{j},\theta
^{\left( j\right) }\left( n_{j}\right) ,\omega _{j}\left( n_{j}\right)
\right) \right)  \notag \\
&&+\lambda \left( \hat{T}\left( \left( Z_{i},\theta ^{\left( i\right)
}\left( n_{i}\right) ,\omega _{i}\left( n_{i}\right) \right) ,\left(
Z_{j},\theta ^{\left( j\right) }\left( n_{j}\right) ,\omega _{j}\left(
n_{j}\right) \right) \right) \right) \delta \left( \theta ^{\left( i\right)
}\left( n_{i}\right) -\theta ^{\left( j\right) }\left( n_{j}\right) -\frac{%
\left\vert Z_{i}-Z_{j}\right\vert }{c}\right)  \notag
\end{eqnarray}%
where $\hat{T}$ measures the variation of $T$ due to the signals send from $%
j $ to $i$ and the signals emitted by $i$. It satisfies the following
equation: 
\begin{eqnarray}
&&\nabla _{\theta ^{\left( i\right) }\left( n_{i}\right) }\hat{T}\left(
\left( Z_{i},\theta ^{\left( i\right) }\left( n_{i}\right) ,\omega
_{i}\left( n_{i}\right) \right) ,\left( Z_{j},\theta ^{\left( j\right)
}\left( n_{j}\right) ,\omega _{j}\left( n_{j}\right) \right) \right)
\label{VL} \\
&=&\rho \delta \left( \theta ^{\left( i\right) }\left( n_{i}\right) -\theta
^{\left( j\right) }\left( n_{j}\right) -\frac{\left\vert
Z_{i}-Z_{j}\right\vert }{c}\right)  \notag \\
&&\times \left\{ \left( h\left( Z,Z_{1}\right) -\hat{T}\left( \left(
Z_{i},\theta ^{\left( i\right) }\left( n_{i}\right) ,\omega _{i}\left(
n_{i}\right) \right) ,\left( Z_{j},\theta ^{\left( j\right) }\left(
n_{j}\right) ,\omega _{j}\left( n_{j}\right) \right) \right) \right) C\left(
\theta ^{\left( i\right) }\left( n\right) \right) h_{C}\left( \omega
_{i}\left( n_{i}\right) \right) \right.  \notag \\
&&\left. -D\left( \theta ^{\left( i\right) }\left( n\right) \right) \hat{T}%
\left( \left( Z_{i},\theta ^{\left( i\right) }\left( n_{i}\right) ,\omega
_{i}\left( n_{i}\right) \right) ,\left( Z_{j},\theta ^{\left( j\right)
}\left( n_{j}\right) ,\omega _{j}\left( n_{j}\right) \right) \right)
h_{D}\left( \omega _{j}\left( n_{j}\right) \right) \right\}  \notag
\end{eqnarray}%
where $h_{C}$ and $h_{D}$\ are increasing functions. In the set of equations
(\ref{spn}), (\ref{SPt}), (\ref{spr}), (\ref{spf}): 
\begin{eqnarray*}
h_{C}\left( \omega _{i}\left( n_{i}\right) \right) &=&\omega _{i}\left(
n_{i}\right) \\
h_{D}\left( \omega _{j}\left( n_{j}\right) \right) &=&\omega _{j}\left(
n_{j}\right)
\end{eqnarray*}

We depart slightly from (\cite{IFR}) by the introduction of the function $%
h\left( Z,Z_{1}\right) $ (they choose $h\left( Z,Z_{1}\right) =1$), to
implement some loss due to the distance. We may choose for example:%
\begin{equation*}
h\left( Z,Z_{1}\right) =\exp \left( -\frac{\left\vert Z_{i}-Z_{j}\right\vert 
}{\nu c}\right)
\end{equation*}%
where $\nu $ is a parameter. Equation (\ref{VL}) involves two dynamic
factors $C\left( \theta ^{\left( i\right) }\left( n-1\right) \right) $ and $%
D\left( \theta _{i}\left( n-1\right) \right) $. The factor $C\left( \theta
^{\left( i\right) }\left( n-1\right) \right) $ describes the accumulation of
input spikes. It is solution of the differential equation:%
\begin{eqnarray}
\nabla _{\theta ^{\left( i\right) }\left( n-1\right) }C\left( \theta
^{\left( i\right) }\left( n-1\right) \right) &=&-\frac{C\left( \theta
^{\left( i\right) }\left( n-1\right) \right) }{\tau _{C}} \\
&&+\alpha _{C}\left( 1-C\left( \theta ^{\left( i\right) }\left( n-1\right)
\right) \right) \omega _{j}\left( \theta ^{\left( i\right) }\left(
n-1\right) -\frac{\left\vert Z_{i}-Z_{j}\right\vert }{c}\right)  \notag
\end{eqnarray}%
The term $D\left( \theta _{i}\left( n-1\right) \right) $ is proportional to
the accumulation of output spikes and is solution of:%
\begin{equation}
\nabla _{\theta ^{\left( i\right) }\left( n-1\right) }D\left( \theta
^{\left( i\right) }\left( n-1\right) \right) =-\frac{D\left( \theta ^{\left(
i\right) }\left( n-1\right) \right) }{\tau _{D}}+\alpha _{D}\left( 1-D\left(
\theta ^{\left( i\right) }\left( n-1\right) \right) \right) \omega
_{i}\left( n_{i}\right)
\end{equation}%
For the purpose of field translation, we have to change the variables in the
derivatives by the counting variable $n_{i}$ and replace $\nabla _{\theta
^{\left( i\right) }\left( n_{i}\right) }\simeq \omega _{i}\left(
n_{i}\right) \nabla _{n_{i}}$ in the previous dynamics equations. We thus
rewrite the dynamic equations in the following form:

For the connectivity $T$:

\begin{eqnarray}
&&\nabla _{n_{i}}T\left( \left( Z_{i},\theta ^{\left( i\right) }\left(
n_{i}\right) ,\omega _{i}\left( n_{i}\right) \right) ,\left( Z_{j},\theta
^{\left( j\right) }\left( n_{j}\right) ,\omega _{j}\left( n_{j}\right)
\right) \right)  \label{nqp} \\
&=&-\frac{1}{\tau \omega _{i}\left( n_{i}\right) }T\left( \left(
Z_{i},\theta ^{\left( i\right) }\left( n_{i}\right) ,\omega _{i}\left(
n_{i}\right) \right) ,\left( Z_{j},\theta ^{\left( j\right) }\left(
n_{j}\right) ,\omega _{j}\left( n_{j}\right) \right) \right)  \notag \\
&&+\frac{\lambda }{\omega _{i}\left( n_{i}\right) }\left( \hat{T}\left(
\left( Z_{i},\theta ^{\left( i\right) }\left( n_{i}\right) ,\omega
_{i}\left( n_{i}\right) \right) ,\left( Z_{j},\theta ^{\left( j\right)
}\left( n_{j}\right) ,\omega _{j}\left( n_{j}\right) \right) \right) \right)
\delta \left( \theta ^{\left( i\right) }\left( n_{i}\right) -\theta ^{\left(
j\right) }\left( n_{j}\right) -\frac{\left\vert Z_{i}-Z_{j}\right\vert }{c}%
\right)  \notag
\end{eqnarray}%
For the variation in connectivity $\hat{T}$:%
\begin{eqnarray}
&&\nabla _{n_{i}}\hat{T}\left( \left( Z_{i},\theta ^{\left( i\right) }\left(
n_{i}\right) ,\omega _{i}\left( n_{i}\right) \right) ,\left( Z_{j},\theta
^{\left( j\right) }\left( n_{j}\right) ,\omega _{j}\left( n_{j}\right)
\right) \right)  \label{nqd} \\
&=&\frac{\rho }{\omega _{i}\left( n_{i}\right) }\delta \left( \theta
^{\left( i\right) }\left( n_{i}\right) -\theta ^{\left( j\right) }\left(
n_{j}\right) -\frac{\left\vert Z_{i}-Z_{j}\right\vert }{c}\right)  \notag \\
&&\times \left\{ \left( h\left( Z,Z_{1}\right) -\hat{T}\left( \left(
Z_{i},\theta ^{\left( i\right) }\left( n_{i}\right) ,\omega _{i}\left(
n_{i}\right) \right) ,\left( Z_{j},\theta ^{\left( j\right) }\left(
n_{j}\right) ,\omega _{j}\left( n_{j}\right) \right) \right) \right) C\left(
\theta ^{\left( i\right) }\left( n-1\right) \right) h_{C}\left( \omega
_{i}\left( n_{i}\right) \right) \right.  \notag \\
&&\left. -D\left( \theta ^{\left( i\right) }\left( n-1\right) \right) \hat{T}%
\left( \left( Z_{i},\theta ^{\left( i\right) }\left( n_{i}\right) ,\omega
_{i}\left( n_{i}\right) \right) ,\left( Z_{j},\theta ^{\left( j\right)
}\left( n_{j}\right) ,\omega _{j}\left( n_{j}\right) \right) \right)
h_{D}\left( \omega _{j}\left( n_{j}\right) \right) \right\}  \notag
\end{eqnarray}%
and for the auxiliary variables $C$ and $D$:%
\begin{eqnarray}
\nabla _{n_{i}}C\left( \theta ^{\left( i\right) }\left( n-1\right) \right)
&=&-\frac{C\left( \theta ^{\left( i\right) }\left( n-1\right) \right) }{\tau
_{C}\omega _{i}\left( n_{i}\right) }  \label{nqt} \\
&&+\alpha _{C}\left( 1-C\left( \theta ^{\left( i\right) }\left( n-1\right)
\right) \right) \frac{\omega _{j}\left( Z_{j},\theta ^{\left( i\right)
}\left( n-1\right) -\frac{\left\vert Z_{i}-Z_{j}\right\vert }{c}\right) }{%
\omega _{i}\left( n_{i}\right) }  \notag
\end{eqnarray}%
\begin{equation}
\nabla _{n_{i}}D\left( \theta ^{\left( i\right) }\left( n-1\right) \right) =-%
\frac{D\left( \theta ^{\left( i\right) }\left( n-1\right) \right) }{\tau
_{D}\omega _{i}\left( n_{i}\right) }+\alpha _{D}\left( 1-D\left( \theta
^{\left( i\right) }\left( n-1\right) \right) \right)  \label{nqQ}
\end{equation}

Then, to describe the connectivity by a field, we have to describe the
connectivity as a set of vectors depending of a set of double indices $kl$
(replacing $ij$) and interacting with the activities, seen as independent
variables indexed by $i,j...$

We thus describe connectivity by a set of matrices:%
\begin{equation*}
\left( T_{kl}\left( n_{kl}\right) ,\hat{T}_{kl}\left( n_{kl}\right) ,\left(
Z_{kl}\left( n_{kl}\right) =\left( Z_{k,},Z_{l}\right) \right) ,\theta
^{\left( kl\right) }\left( n_{kl}\right) ,\omega _{k}\left( n_{kl}\right)
,\omega _{l}^{\prime }\left( n_{kl}\right) ,C_{kl}\left( n_{kl}\right)
,D_{k}\left( n_{kl}\right) \right)
\end{equation*}%
where $n_{kl}$ is an internal parameter given by the average counting
variable for cells or synapses firing simultaneously at point $Z_{k,}$.

Then, we replace the description (\ref{nqp}), (\ref{nqd}), (\ref{nqt}), (\ref%
{nqQ}) by a set of equations in which connectivities $T_{kl}\left(
n_{kl}\right) $ interact with all pairs of neurons at points $Z_{k,}$ and $%
Z_{l}$ whose average firing rates at time $\theta ^{\left( kl\right) }\left(
n_{kl}\right) $ and $\theta ^{\left( kl\right) }\left( n_{kl}\right) -\frac{%
\left\vert Z_{k}-Z_{l}\right\vert }{c}$ are given by $\omega _{k}\left(
n_{kl}\right) ,\omega _{l}^{\prime }\left( n_{kl}\right) $ respectively. As
a consequence, we replace the notion of connectivity $T_{ij}\left( \left(
n-1,Z_{i}\right) ,\left( m,Z_{j}\right) \right) $ between two specific
neurons $i$ and $j$ by the average connectivity between the two sets of
neurons with identical activities at each extremity of the segment $\left(
Z_{i},Z_{j}\right) $ \ This approximation is justified if we consider that
neurons located at the same place and firing at the same rate can be
considered as closely connected and in average identical. Alternatively,
this can also be justified if we consider one neuron per spatial location
and assume each neuron as a complex entity sending several signals
simultaneously. Under this hypothesis, the average considered are taken over
the multiple activities of the same neuron\footnote{%
See section 5.2.1 for more details about these alternative interpretations.}.

Stated mathematically, the variable $n_{kl}$ is replacd by an average $%
n_{kl}=\bar{n}_{i}$ at a given time $\theta ^{\left( kl\right) }$ and we
assume that in average, connectivity variable $T_{kl}\left( n_{kl}\right) $
interacts with all neurons pairs located at $\left( Z_{k,},Z_{l}\right) $ at
times $\theta ^{\left( i\right) }\left( n_{i}\right) =\theta ^{\left(
kl\right) }\left( n_{kl}\right) $. Writing $\bar{\omega}\left(
Z_{i},n_{i}\right) $ for the average activity, we impose $\bar{\omega}\left(
Z_{i},n_{i}\right) =\omega _{k}\left( n_{kl}\right) $ and $\bar{\omega}%
\left( Z_{j},n_{j}\right) =\omega _{l}^{\prime }\left( n_{kl}\right) $ and $%
\theta ^{\left( j\right) }\left( n_{j}\right) =\theta ^{\left( kl\right)
}\left( n_{kl}\right) -\frac{\left\vert Z_{k}-Z_{l}\right\vert }{c}$
respectively. The densities $T_{kl}\left( n_{kl}\right) $ are thus the set
of all connections between points $Z_{k,}$ and $Z_{l}$ between sets of
synchronized neurons at $Z_{k}$ and synchronized neurons at $Z_{l}$, i.e.
between set of neurons at this points or alternatively between multiple
synapses for one or a few number of cells. In this point of view, we replace 
$\nabla _{\theta ^{\left( i\right) }\left( n_{i}\right) }\simeq \omega
_{i}\left( n_{i}\right) \nabla _{n_{i}}$ by:%
\begin{equation*}
\nabla _{\theta ^{\left( kl\right) }\left( n_{kl}\right) }\simeq \frac{%
\partial n_{kl}}{\partial \theta ^{\left( kl\right) }\left( n_{kl}\right) }%
\nabla _{n_{kl}}=\bar{\omega}\left( Z_{i},n_{i}\right) \nabla _{n_{kl}}
\end{equation*}%
As a consequence, the dynamic equations (\ref{nqp}), (\ref{nqd}), (\ref{nqt}%
), (\ref{nqQ}) are replaced by:%
\begin{eqnarray}
&\nabla _{n_{kl}}T_{kl}\left( n_{kl}\right) =&\left( -\sum_{i,n_{i}}\frac{1}{%
\tau \bar{\omega}\left( Z_{i},n_{i}\right) }T_{kl}\left( n_{kl}\right) +%
\frac{\lambda }{\bar{\omega}\left( Z_{i},n_{i}\right) }\hat{T}_{kl}\left(
n_{kl}\right) \right) \\
&&\times \delta \left( \theta ^{\left( i\right) }\left( n_{i}\right) -\theta
^{\left( kl\right) }\left( n_{kl}\right) \right) \delta \left(
Z_{k}-Z_{i}\right) \delta \left( \omega _{k}\left( n_{kl}\right) -\bar{\omega%
}\left( Z_{i},n_{i}\right) \right)  \notag
\end{eqnarray}%
\begin{eqnarray}
&&\nabla _{n_{kl}}\hat{T}\left( n_{kl}\right) \\
&=&\left( \sum_{i,n_{i}}\left( h\left( Z_{k},Z_{l}\right) -\hat{T}\left(
n_{kl}\right) \right) C_{kl}\left( n_{kl}\right) h_{C}\left( \omega
_{i}\left( n_{i}\right) \right) -\sum_{j,n_{j}}D_{k}\left( n_{kl}\right) 
\hat{T}\left( n_{kl}\right) h_{D}\left( \omega _{j}\left( n_{j}\right)
\right) \right)  \notag \\
&&\times \frac{\rho }{\bar{\omega}\left( Z_{i},n_{i}\right) }\delta \left(
\theta ^{\left( i\right) }\left( n_{i}\right) -\theta ^{\left( j\right)
}\left( n_{j}\right) -\frac{\left\vert Z_{i}-Z_{j}\right\vert }{c}\right)
\delta \left( \theta ^{\left( i\right) }\left( n_{i}\right) -\theta ^{\left(
kl\right) }\left( n_{kl}\right) \right)  \notag \\
&&\times \delta \left( \left( Z_{k,},Z_{l}\right) -\left(
Z_{i,},Z_{j}\right) \right) \delta \left( \omega _{k}\left( n_{kl}\right) -%
\bar{\omega}\left( Z_{i},n_{i}\right) \right) \delta \left( \omega
_{l}\left( n_{kl}\right) -\bar{\omega}\left( Z_{j},n_{j}\right) \right) 
\notag
\end{eqnarray}%
\begin{eqnarray}
\nabla _{n_{kl}}C\left( n_{kl}\right) &=&\left( -\frac{C\left( n_{kl}\right) 
}{\tau _{C}\bar{\omega}\left( Z_{i},n_{i}\right) }+\sum_{j,n_{j}}\alpha
_{C}\left( 1-C_{kl}\left( n_{kl}\right) \right) \frac{\omega _{j}\left(
n_{j}\right) }{\bar{\omega}\left( Z_{i},n_{i}\right) }\right) \\
&&\times \delta \left( \theta ^{\left( i\right) }\left( n_{i}\right) -\theta
^{\left( j\right) }\left( n_{j}\right) -\frac{\left\vert
Z_{i}-Z_{j}\right\vert }{c}\right) \delta \left( \theta ^{\left( i\right)
}\left( n_{i}\right) -\theta ^{\left( kl\right) }\left( n_{kl}\right)
\right) \delta \left( \left( Z_{k,},Z_{l}\right) -\left( Z_{i,},Z_{j}\right)
\right)  \notag \\
&&\times \delta \left( \omega _{k}\left( n_{kl}\right) -\bar{\omega}\left(
Z_{i},n_{i}\right) \right) \delta \left( \omega _{l}\left( n_{kl}\right) -%
\bar{\omega}\left( Z_{j},n_{j}\right) \right)  \notag
\end{eqnarray}%
\begin{eqnarray}
\nabla _{n_{kl}}D_{k}\left( n_{kl}\right) &=&\left( -\frac{D_{k}\left(
n_{kl}\right) }{\tau _{D}\bar{\omega}\left( Z_{i},n_{i}\right) }+\frac{1}{%
\bar{\omega}\left( Z_{i},n_{i}\right) }\sum_{i,n_{i}}\alpha _{D}\left(
1-D_{k}\left( n_{kl}\right) \right) \omega _{i}\left( n_{i}\right) \right) \\
&&\times \delta \left( \theta ^{\left( i\right) }\left( n_{i}\right) -\theta
^{\left( kl\right) }\left( n_{kl}\right) \right) \delta \left(
Z_{k}-Z_{i}\right) \delta \left( \omega _{k}\left( n_{kl}\right) -\bar{\omega%
}\left( Z_{i},n_{i}\right) \right) \delta \left( \omega _{l}\left(
n_{kl}\right) -\bar{\omega}\left( Z_{j},n_{j}\right) \right)  \notag
\end{eqnarray}%
Similarly, note that we can also rewrite the currents equation (\ref{crt})
as:%
\begin{equation*}
\hat{J}^{\left( i\right) }\left( \left( n-1\right) \right) =J^{\left(
i\right) }\left( \left( n-1\right) \right) +\frac{\kappa }{N}\sum_{j,m}\frac{%
\omega _{j}\left( m\right) }{\omega _{i}\left( n-1\right) }\delta \left(
\theta ^{\left( i\right) }\left( n-1\right) -\theta ^{\left( j\right)
}\left( m\right) -\frac{\left\vert Z_{i}-Z_{j}\right\vert }{c}\right)
T_{ij}\left( \left( n-1,Z_{i}\right) ,\left( m,Z_{j}\right) \right)
\end{equation*}%
with:%
\begin{equation}
T_{ij}\left( \left( n_{i},Z_{i}\right) ,\left( m_{j},Z_{j}\right) \right)
=\sum_{kl}T_{kl}\left( n_{kl}\right) \delta \left( \theta ^{\left( i\right)
}\left( n_{i}\right) -\theta ^{\left( kl\right) }\left( n_{kl}\right)
\right) \delta \left( \omega _{k}\left( n_{kl}\right) -\bar{\omega}\left(
Z_{i},n_{i}\right) \right) \delta \left( \omega _{l}\left( n_{kl}\right) -%
\bar{\omega}\left( Z_{j},n_{j}\right) \right)  \label{tkl}
\end{equation}

\subsection{Probability density for the system}

\subsubsection{Individual neurons}

Due to the stochastic nature of equations (\ref{dnq}) and (\ref{cstrt}), the
dynamics of a single neuron can be described by the probability density $%
P\left( \theta ^{\left( i\right) }\left( t\right) ,\omega _{i}^{-1}\left(
t\right) \right) $ for a path $\left( \theta ^{\left( i\right) }\left(
t\right) ,\omega _{i}^{-1}\left( t\right) \right) $ which is given by, up to
a normalization factor:

\begin{equation}
P\left( \theta ^{\left( i\right) }\left( t\right) ,\omega _{i}^{-1}\left(
t\right) \right) =\exp \left( -A_{i}\right)  \label{dnmcs}
\end{equation}%
where:%
\begin{equation}
A_{i}=\frac{1}{\sigma ^{2}}\int \left( \frac{d}{dt}\theta ^{\left( i\right)
}\left( t\right) -\omega _{i}^{-1}\left( t\right) \right) ^{2}dt+\int \frac{%
\left( \omega _{i}^{-1}\left( t\right) -G\left( \theta ^{\left( i\right)
}\left( t\right) ,\hat{J}\left( \theta ^{\left( i\right) }\left( t\right)
\right) \right) \right) ^{2}}{\eta ^{2}}dt  \label{dnmcszz}
\end{equation}%
(see \cite{GL1} and \cite{GL2}). The integral is taken over a time period
that depends on the time scale of the interactions. Actually, the
minimization of (\ref{dnmcszz})\ imposes both (\ref{dnm}) and (\ref{cstrt}),
so that the probability density is, as expected, centered around these two
conditions, i.e. (\ref{dnm}) and (\ref{cstrt}) are satisfied in mean. A
probability density for the whole system of neurons is obtained by summing $%
S_{i}$ over all agents. We thus define the statistical weight for the cells:%
\begin{equation}
P\left( \left( \theta ^{\left( i\right) }\left( t\right) ,\omega
_{i}^{-1}\left( t\right) ,Z_{i}\right) _{i=1...N}\right) =\exp \left(
-A\right)  \label{Prdn}
\end{equation}%
with:%
\begin{equation}
A=\sum_{i}A_{i}=\sum_{i}\frac{1}{\sigma ^{2}}\int \left( \frac{d}{dt}\theta
^{\left( i\right) }\left( t\right) -\omega _{i}^{-1}\left( t\right) \right)
^{2}dt+\int \frac{\left( \omega _{i}^{-1}\left( t\right) -G\left( \theta
^{\left( i\right) }\left( t\right) ,\hat{J}\left( \theta ^{\left( i\right)
}\left( t\right) \right) \right) \right) ^{2}}{\eta ^{2}}dt  \label{ctnt}
\end{equation}%
and (using (\ref{tkl})):%
\begin{eqnarray*}
\hat{J}^{\left( i\right) }\left( \left( n-1\right) \right) &=&J^{\left(
i\right) }\left( \left( n-1\right) \right) +\frac{\kappa }{N}\sum_{j,m}\frac{%
\omega _{j}\left( m\right) }{\omega _{i}\left( n-1\right) }\delta \left(
\theta ^{\left( i\right) }\left( n-1\right) -\theta ^{\left( j\right)
}\left( m\right) -\frac{\left\vert Z_{i}-Z_{j}\right\vert }{c}\right)
T_{ij}\left( \left( n-1,Z_{i}\right) ,\left( m,Z_{j}\right) \right) \\
&&\times \sum_{kl}T_{kl}\left( n_{kl}\right) \delta \left( \theta ^{\left(
i\right) }\left( n_{i}\right) -\theta ^{\left( kl\right) }\left(
n_{kl}\right) \right) \delta \left( \omega _{k}\left( n_{kl}\right) -\omega
_{i}\left( n_{i}\right) \right) \delta \left( \omega _{l}\left(
n_{kl}\right) -\omega _{j}\left( m\right) \right)
\end{eqnarray*}

\subsubsection{Connectivity functions}

The statistical exponent associated to the connectivity functions is
obtained as in the previous paragraph. We obtain the statistical weight:%
\begin{equation*}
\prod\limits_{k,l}P\left( T_{kl}\left( n_{kl}\right) ,\hat{T}_{kl}\left(
n_{kl}\right) ,\left( Z_{kl}\left( n_{kl}\right) =\left( Z_{k,},Z_{l}\right)
\right) ,\theta ^{\left( kl\right) }\left( n_{kl}\right) ,\omega _{k}\left(
n_{kl}\right) ,\omega _{l}^{\prime }\left( n_{kl}\right) ,C_{kl}\left(
n_{kl}\right) ,D_{k}\left( n_{kl}\right) \right) =\exp \left( -B\right)
\end{equation*}%
where:%
\begin{eqnarray}
B &=&\sum_{kl}\left( \nabla _{n_{kl}}T_{kl}\left( n_{kl}\right)
-B_{kl}^{\left( 1\right) }\right) ^{2}+\left( \nabla _{n_{kl}}\hat{T}\left(
n_{kl}\right) -B_{kl}^{\left( 2\right) }\nabla _{n_{kl}}\right) ^{2}
\label{wgp} \\
&&+\left( C\left( n_{kl}\right) -B_{kl}^{\left( 3\right) }\right)
^{2}+\left( \nabla _{n_{kl}}D_{k}\left( n_{kl}\right) -B_{k}\right) ^{2} 
\notag
\end{eqnarray}%
and:%
\begin{eqnarray}
&&B_{kl}^{\left( 1\right) }=\left( -\sum_{i,n_{i}}\frac{1}{\tau \bar{\omega}%
\left( Z_{i},n_{i}\right) }T_{kl}\left( n_{kl}\right) +\frac{\lambda }{\bar{%
\omega}\left( Z_{i},n_{i}\right) }\hat{T}_{kl}\left( n_{kl}\right) \right)
\label{wgd} \\
&&\times \delta \left( \theta ^{\left( i\right) }\left( n_{i}\right) -\theta
^{\left( kl\right) }\left( n_{kl}\right) \right) \delta \left(
Z_{k}-Z_{i}\right) \delta \left( \omega _{k}\left( n_{kl}\right) -\bar{\omega%
}\left( Z_{i},n_{i}\right) \right)  \notag
\end{eqnarray}%
\begin{eqnarray}
&&B_{kl}^{\left( 2\right) }=\left( \sum_{i,n_{i}}\left( h\left(
Z_{k},Z_{l}\right) -\hat{T}\left( n_{kl}\right) \right) C_{kl}\left(
n_{kl}\right) h_{C}\left( \omega _{i}\left( n_{i}\right) \right)
-\sum_{j,n_{j}}D_{k}\left( n_{kl}\right) \hat{T}\left( n_{kl}\right)
h_{D}\left( \omega _{j}\left( n_{j}\right) \right) \right)  \label{wgt} \\
&&\times \frac{\rho }{\bar{\omega}\left( Z_{i},n_{i}\right) }\delta \left(
\theta ^{\left( i\right) }\left( n_{i}\right) -\theta ^{\left( j\right)
}\left( n_{j}\right) -\frac{\left\vert Z_{i}-Z_{j}\right\vert }{c}\right)
\delta \left( \theta ^{\left( i\right) }\left( n_{i}\right) -\theta ^{\left(
kl\right) }\left( n_{kl}\right) \right)  \notag \\
&&\times \delta \left( \left( Z_{k,},Z_{l}\right) -\left(
Z_{i,},Z_{j}\right) \right) \delta \left( \omega _{k}\left( n_{kl}\right) -%
\bar{\omega}\left( Z_{i},n_{i}\right) \right) \delta \left( \omega
_{l}\left( n_{kl}\right) -\bar{\omega}\left( Z_{j},n_{j}\right) \right) 
\notag
\end{eqnarray}%
\begin{eqnarray}
B_{kl}^{\left( 3\right) } &=&\left( -\frac{C\left( n_{kl}\right) }{\tau _{C}%
\bar{\omega}\left( Z_{i},n_{i}\right) }+\sum_{j,n_{j}}\alpha _{C}\left(
1-C_{kl}\left( n_{kl}\right) \right) \frac{\omega _{j}\left( n_{j}\right) }{%
\bar{\omega}\left( Z_{i},n_{i}\right) }\right)  \label{wgq} \\
&&\times \delta \left( \theta ^{\left( i\right) }\left( n_{i}\right) -\theta
^{\left( j\right) }\left( n_{j}\right) -\frac{\left\vert
Z_{i}-Z_{j}\right\vert }{c}\right) \delta \left( \theta ^{\left( i\right)
}\left( n_{i}\right) -\theta ^{\left( kl\right) }\left( n_{kl}\right)
\right) \delta \left( \left( Z_{k,},Z_{l}\right) -\left( Z_{i,},Z_{j}\right)
\right)  \notag \\
&&\times \delta \left( \omega _{k}\left( n_{kl}\right) -\bar{\omega}\left(
Z_{i},n_{i}\right) \right) \delta \left( \omega _{l}\left( n_{kl}\right) -%
\bar{\omega}\left( Z_{j},n_{j}\right) \right)  \notag
\end{eqnarray}%
\begin{eqnarray}
B_{k} &=&\left( -\frac{D_{k}\left( n_{kl}\right) }{\tau _{D}\bar{\omega}%
\left( Z_{i},n_{i}\right) }+\frac{1}{\bar{\omega}\left( Z_{i},n_{i}\right) }%
\sum_{i,n_{i}}\alpha _{D}\left( 1-D_{k}\left( n_{kl}\right) \right) \omega
_{i}\left( n_{i}\right) \right)  \label{wgc} \\
&&\times \delta \left( \theta ^{\left( i\right) }\left( n_{i}\right) -\theta
^{\left( kl\right) }\left( n_{kl}\right) \right) \delta \left(
Z_{k}-Z_{i}\right) \delta \left( \omega _{k}\left( n_{kl}\right) -\bar{\omega%
}\left( Z_{i},n_{i}\right) \right) \delta \left( \omega _{l}\left(
n_{kl}\right) -\bar{\omega}\left( Z_{j},n_{j}\right) \right)  \notag
\end{eqnarray}

\subsubsection{Probability density for the full system}

The probability for the full system is obtained by the product:%
\begin{eqnarray}
&&\prod\limits_{k,l}P\left( T_{kl}\left( n_{kl}\right) ,\hat{T}_{kl}\left(
n_{kl}\right) ,\left( Z_{kl}\left( n_{kl}\right) =\left( Z_{k,},Z_{l}\right)
\right) ,\theta ^{\left( kl\right) }\left( n_{kl}\right) ,\omega _{k}\left(
n_{kl}\right) ,\omega _{l}^{\prime }\left( n_{kl}\right) ,C_{kl}\left(
n_{kl}\right) ,D_{k}\left( n_{kl}\right) \right)  \label{SSW} \\
&&\times P\left( \left( \theta ^{\left( i\right) }\left( t\right) ,\omega
_{i}^{-1}\left( t\right) ,Z_{i}\right) _{i=1...N}\right)  \notag \\
&=&\exp \left( -B\right) \exp \left( -A\right)  \notag
\end{eqnarray}

\section{Field theoretic description of the system}

\subsection{Translation of formula (\protect\ref{SSW}) in terms of field
theory}

In our context two fields are necessary. The field representing the set of
neurons depends on the three variables $\left( \theta ,Z,\omega \right) $,\
and is denoted $\Psi \left( \theta ,Z,\omega \right) $. The connectivity
functions are characterized by the set of variables $\left( T,\hat{T},\omega
,\omega ^{\prime },\theta ,Z,Z^{\prime },C,D\right) $ and represented by the
field $\Gamma \left( T,\hat{T},\omega ,\omega ^{\prime },\theta ,Z,Z^{\prime
},C,D\right) $.We provide an interpretation of the various fields at the end
of this paragraph.

\subsubsection{Translation of (\protect\ref{ctnt})}

The dynamics of neurons is described by an action functional for the field $%
\Psi \left( \theta ,Z,\omega \right) $ and its associated partition
function. This partition function captures both collective and individual
aspects of the system, enabling the retrieval of correlation functions for
number of neurons.

The field theoretic version of (\ref{dnmcszz}) is obtained using (\ref{ctnt}%
). The correspondence detailed in \cite{GL1}\cite{GL2}) yields an action $%
S\left( \Psi \right) $ for a field $\Psi \left( \theta ,Z,\omega \right) $
and a statistical weight $\exp \left( -\left( S\left( \Psi \right) \right)
\right) $ for each configuration $\Psi \left( \theta ,Z,\omega \right) $ of
this field. The functional $S\left( \Psi \right) $ is decomposed in two
parts corresponding to the two contributions in (\ref{ctnt}).

The first term of (\ref{ctnt}):%
\begin{equation}
\frac{1}{\sigma ^{2}}\int \left( \frac{d}{dt}\theta ^{\left( i\right)
}\left( t\right) -\omega _{i}^{-1}\left( t\right) \right) ^{2}dt  \label{fsT}
\end{equation}%
is a term with temporal derivative. Its form is simple since the function $%
f^{\left( \alpha \right) }$ in (\ref{inco}) depends only on the variable $%
\mathbf{X}_{i}\left( t\right) =\left( \theta ^{\left( i\right) }\left(
t\right) ,\omega _{i}^{-1}\left( t\right) ,Z_{i}\right) $. Actually $%
f^{\left( \theta \right) }\left( \mathbf{X}_{i}\left( t\right) \right)
=\omega _{i}^{-1}\left( t\right) $. Using (\ref{Trll}), the term (\ref{fsT})
is thus replaced by the corresponding quadratic functional in field theory:%
\begin{equation}
-\frac{1}{2}\Psi ^{\dagger }\left( \theta ,Z,\omega \right) \nabla \left( 
\frac{\sigma ^{2}}{2}\nabla -\omega ^{-1}\right) \Psi \left( \theta
,Z,\omega \right)  \label{thtdnmcs}
\end{equation}%
where $\sigma ^{2}$ is the variance of the errors $\varepsilon _{i}$.

The field functional that corresponds to the second term of (\ref{dnmcszz}):%
\begin{equation*}
V=\int \frac{\left( \omega _{i}^{-1}\left( t\right) -G\left( \theta ^{\left(
i\right) }\left( t\right) ,\hat{J}\left( \theta ^{\left( i\right) }\left(
t\right) \right) \right) \right) ^{2}}{\eta ^{2}}dt
\end{equation*}
is obtained by expanding the formula (\ref{crT}) for the current induced by
all $j$:

\begin{eqnarray}
V &=&\frac{1}{2\eta ^{2}}\int dt\sum_{i}\left( \omega _{i}^{-1}\left(
t\right) \right.  \label{Cpnt} \\
&&-\left. G\left( J\left( \theta ^{\left( i\right) }\left( t\right)
,Z_{i}\right) +\frac{\kappa }{N}\int ds\sum_{j}\frac{\omega _{j}\left(
s\right) T_{ij}\left( \left( t,Z_{i}\right) ,s,Z_{j}\right) }{\omega
_{i}\left( t\right) }\delta \left( \theta ^{\left( i\right) }\left( t\right)
-\theta ^{\left( j\right) }\left( s\right) -\frac{\left\vert
Z_{i}-Z_{j}\right\vert }{c}\right) \right) \right) ^{2}  \notag
\end{eqnarray}%
with $\eta <<1$, which is the constraint (\ref{cstrt}) imposed
stochastically. Its corresponding potential in field theory is obtained
straightforwardly by using the translation (\ref{tln}):%
\begin{equation}
\frac{1}{2\eta ^{2}}\int \left\vert \Psi \left( \theta ,Z,\omega \right)
\right\vert ^{2}\left( \omega ^{-1}-G\left( J\left( \theta ,Z\right) +\int 
\frac{\kappa }{N}\frac{\omega _{1}T\left( Z,\theta ,Z_{1},\theta -\frac{%
\left\vert Z-Z_{1}\right\vert }{c}\right) }{\omega }\left\vert \Psi \left(
\theta -\frac{\left\vert Z-Z_{1}\right\vert }{c},Z_{1},\omega _{1}\right)
\right\vert ^{2}dZ_{1}d\omega _{1}\right) \right) ^{2}  \label{ptntl}
\end{equation}%
and $T\left( Z,\theta ,Z_{1},\theta -\frac{\left\vert Z-Z_{1}\right\vert }{c}%
\right) $ is obtained by the translation of the term (\ref{tkl}):%
\begin{eqnarray*}
&&\sum_{kl}T_{kl}\left( n_{kl}\right) \delta \left( \theta ^{\left( i\right)
}\left( n_{i}\right) -\theta ^{\left( kl\right) }\left( n_{kl}\right)
\right) \delta \left( \omega _{k}\left( n_{kl}\right) -\omega _{i}\left(
n_{i}\right) \right) \delta \left( \omega _{l}\left( n_{kl}\right) -\omega
_{j}\left( m\right) \right) \\
&\rightarrow &\int T\left\vert \Gamma \left( T,\hat{T},\hat{\omega},\hat{%
\omega}^{\prime },\hat{\theta},\hat{Z},\hat{Z}^{\prime },C,D\right)
\right\vert ^{2}\delta \left( \theta -\hat{\theta}\right) \delta \left( \hat{%
\omega}-\omega \right) \delta \left( \hat{\omega}-\omega _{1}\right) \delta
\left( \left( \hat{Z},\hat{Z}^{\prime }\right) -\left( Z,Z_{1}\right) \right)
\\
&=&\int T\left\vert \Gamma \left( T,\hat{T},\omega ,\omega _{1},\theta
,Z,Z_{1},C,D\right) \right\vert ^{2}dTd\hat{T}dCdD\equiv T\left( Z,\theta
,Z_{1},\theta -\frac{\left\vert Z-Z_{1}\right\vert }{c}\right)
\end{eqnarray*}%
To simplify, we will write in the sequel:%
\begin{equation}
T\left( Z,\theta ,Z_{1},\theta -\frac{\left\vert Z-Z_{1}\right\vert }{c}%
\right) =\int T\left\vert \Gamma \left( T,\hat{T},\omega ,\omega _{1},\theta
,Z,Z_{1},C,D\right) \right\vert ^{2}dTd\hat{T}dCdD\equiv T\left( Z,\theta
,Z_{1}\right)  \label{tpr}
\end{equation}%
which represents the average connectivity between points $Z$ and $Z_{1}$ in
state $\Gamma \left( T,\hat{T},\omega ,\omega _{1},\theta
,Z,Z_{1},C,D\right) $.

The field action is then the sum of (\ref{thtdnmcs}) and (\ref{ptntl}):%
\begin{eqnarray}
S &=&-\frac{1}{2}\Psi ^{\dagger }\left( \theta ,Z,\omega \right) \nabla
\left( \frac{\sigma _{\theta }^{2}}{2}\nabla -\omega ^{-1}\right) \Psi
\left( \theta ,Z,\omega \right)  \label{lfS} \\
&&+\frac{1}{2\eta ^{2}}\int \left\vert \Psi \left( \theta ,Z,\omega \right)
\right\vert ^{2}\left( \omega ^{-1}-G\left( J\left( \theta ,Z\right) +\int 
\frac{\kappa }{N}\frac{\omega _{1}}{\omega }\left\vert \Psi \left( \theta -%
\frac{\left\vert Z-Z_{1}\right\vert }{c},Z_{1},\omega _{1}\right)
\right\vert ^{2}T\left( Z,\theta ,Z_{1}\right) dZ_{1}d\omega _{1}\right)
\right) ^{2}  \notag
\end{eqnarray}%
$\allowbreak $

\subsubsection{Translation for connectivity dynamics (\protect\ref{wgp})}

The translation of the four action terms describing the connectivity
dynamics (\ref{wgd}), (\ref{wgt}), (\ref{wgq}) and (\ref{wgc}) in (\ref{wgp}%
) is straightforward. We obtain four contributions:%
\begin{equation}
S_{\Gamma }^{\left( 1\right) }=\int \Gamma ^{\dag }\left( T,\hat{T},\omega
_{\Gamma },\omega _{\Gamma }^{\prime },\theta ,Z,Z^{\prime },C,D\right)
\nabla _{T}\left( \frac{\sigma _{T}^{2}}{2}\nabla _{T}+O_{T}^{\omega
}\right) \Gamma \left( T,\hat{T},\omega _{\Gamma },\omega _{\Gamma }^{\prime
},\theta ,Z,Z^{\prime },C,D\right)  \label{wgD}
\end{equation}%
\begin{equation}
S_{\Gamma }^{\left( 2\right) }=\int \Gamma ^{\dag }\left( T,\hat{T},\omega
_{\Gamma },\omega _{\Gamma }^{\prime },\theta ,Z,Z^{\prime },C,D\right)
\nabla _{\hat{T}}\left( \frac{\sigma _{\hat{T}}^{2}}{2}\nabla _{\hat{T}}+O_{%
\hat{T}}^{\omega }\right) \Gamma \left( T,\hat{T},\omega _{\Gamma },\omega
_{\Gamma }^{\prime },\theta ,Z,Z^{\prime },C,D\right)  \label{wgT}
\end{equation}%
\begin{equation}
S_{\Gamma }^{\left( 3\right) }=\int \Gamma ^{\dag }\left( T,\hat{T},\omega
_{\Gamma },\omega _{\Gamma }^{\prime },\theta ,Z,Z^{\prime },C,D\right)
\nabla _{C}\left( \frac{\sigma _{C}^{2}}{2}\nabla _{C}+O_{C}^{\omega
}\right) \Gamma \left( T,\hat{T},\omega _{\Gamma },\omega _{\Gamma }^{\prime
},\theta ,Z,Z^{\prime },C,D\right)  \label{wgQ}
\end{equation}%
\begin{equation}
S_{\Gamma }^{\left( 4\right) }=\int \Gamma ^{\dag }\left( T,\hat{T},\omega
_{\Gamma },\omega _{\Gamma }^{\prime },\theta ,Z,Z^{\prime },C,D\right)
\nabla _{D}\left( \frac{\sigma _{D}^{2}}{2}\nabla _{D}+O_{D}^{\omega
}\right) \Gamma \left( T,\hat{T},\omega _{\Gamma },\omega _{\Gamma }^{\prime
},\theta ,Z,Z^{\prime },C,D\right)  \label{wgC}
\end{equation}%
with:%
\begin{eqnarray}
O_{C}^{\omega } &=&\left( \frac{C}{\tau _{C}\bar{\omega}}-\frac{\alpha
_{C}\left( 1-C\right) \int \omega ^{\prime }\left\vert \Psi \left( \theta -%
\frac{\left\vert Z-Z^{\prime }\right\vert }{c},Z^{\prime },\omega ^{\prime
}\right) \right\vert ^{2}d\omega ^{\prime }}{\bar{\omega}}\right) \delta
\left( \left( \omega _{\Gamma },\omega _{\Gamma }^{\prime }\right) -\left( 
\bar{\omega},\bar{\omega}^{\prime }\right) \right)  \label{Gm} \\
O_{D}^{\omega } &=&\frac{D}{\tau _{D}\bar{\omega}}-\frac{\alpha _{D}\left(
1-D\right) \int \omega \left\vert \Psi \left( \theta ,Z,\omega \right)
\right\vert ^{2}d\omega }{\bar{\omega}}\delta \left( \left( \omega _{\Gamma
},\omega _{\Gamma }^{\prime }\right) -\left( \bar{\omega},\bar{\omega}%
^{\prime }\right) \right)  \notag \\
O_{\hat{T}}^{\omega } &=&-\frac{\rho }{\bar{\omega}}\left( \left( h\left(
Z,Z^{\prime }\right) -\hat{T}\right) C\int \left\vert \Psi \left( \theta
,Z,\omega \right) \right\vert ^{2}h_{C}\left( \omega \right) d\omega \right.
\notag \\
&&\left. -D\hat{T}\int \left\vert \Psi \left( \theta -\frac{\left\vert
Z-Z^{\prime }\right\vert }{c},Z^{\prime },\omega ^{\prime }\right)
\right\vert ^{2}h_{D}\left( \omega ^{\prime }\right) d\omega ^{\prime
}\right) \delta \left( \left( \omega _{\Gamma },\omega _{\Gamma }^{\prime
}\right) -\left( \bar{\omega},\bar{\omega}^{\prime }\right) \right)  \notag
\\
O_{T}^{\omega } &=&-\left( -\frac{1}{\tau \bar{\omega}}T+\frac{\lambda }{%
\bar{\omega}}\hat{T}\right) \delta \left( \left( \omega _{\Gamma },\omega
_{\Gamma }^{\prime }\right) -\left( \bar{\omega},\bar{\omega}^{\prime
}\right) \right)  \notag
\end{eqnarray}%
Here:%
\begin{eqnarray*}
\bar{\omega} &=&\frac{\int \omega \left\vert \Psi \left( \theta ,Z,\omega
\right) \right\vert ^{2}d\omega }{\int \left\vert \Psi \left( \theta
,Z,\omega \right) \right\vert ^{2}d\omega } \\
\bar{\omega}^{\prime } &=&\frac{\int \omega ^{\prime }\left\vert \Psi \left(
\theta -\frac{\left\vert Z-Z^{\prime }\right\vert }{c},Z^{\prime },\omega
^{\prime }\right) \right\vert ^{2}d\omega ^{\prime }}{\int \left\vert \Psi
\left( \theta -\frac{\left\vert Z-Z^{\prime }\right\vert }{c},Z^{\prime
},\omega ^{\prime }\right) \right\vert ^{2}d\omega ^{\prime }}
\end{eqnarray*}

\subsection{Full action for the system}

The full action for the system is obtained by gathering the different terms:%
\begin{equation}
-\frac{1}{2}\int \Psi ^{\dagger }\left( \theta ,Z,\omega \right) \nabla
\left( \frac{\sigma _{\theta }^{2}}{2}\nabla -\omega ^{-1}\right) \Psi
\left( \theta ,Z,\omega \right) +\frac{1}{2\eta ^{2}}\left( S_{\Gamma
}^{\left( 1\right) }+S_{\Gamma }^{\left( 2\right) }+S_{\Gamma }^{\left(
3\right) }+S_{\Gamma }^{\left( 4\right) }\right)  \label{FCN}
\end{equation}%
with $S_{\Gamma }^{\left( 1\right) }$, $S_{\Gamma }^{\left( 2\right) }$, $%
S_{\Gamma }^{\left( 3\right) }$, $S_{\Gamma }^{\left( 4\right) }$ given by (%
\ref{wgD}), (\ref{wgT}), (\ref{wgQ}), (\ref{wgC}).

\subsubsection{Remark: interpretation of the various field}

The action functional depends on two fields: $\Psi \left( \theta ,Z,\omega
\right) $ and $\Gamma \left( T,\hat{T},\omega _{\Gamma },\omega _{\Gamma
}^{\prime },\theta ,Z,Z^{\prime },C,D\right) $. These two abstract
quantities will enable us to derive the dynamic state of the entire system
and subsequently study transitions between different states. However, the
squared modulus of the two functions can be interpreted in terms of
statistical distribution, depending on the chosen description. If we
consider a system of simple cells spread along the thread, the function $%
\left\vert \Psi \left( \theta ,Z,\omega \right) \right\vert ^{2}$ measures
at time $\theta $, the density of active cells at point $Z$ with activity $%
\omega $. In the perspective of complex cells with multiple axons and
dendrites, we can consider that one cell stands at $Z$, and $\left\vert \Psi
\left( \theta ,Z,\omega \right) \right\vert ^{2}$ measures for that cell the
density of axons with activity $\omega $. A similar interpretation works for 
$\Gamma \left( T,\hat{T},\omega _{\Gamma },\omega _{\Gamma }^{\prime
},\theta ,Z,Z^{\prime },C,D\right) $. In the perspective of system of simple
cells "accumulated" in the neighborhood of $Z$, $\left\vert \Gamma \left( T,%
\hat{T},\omega _{\Gamma },\omega _{\Gamma }^{\prime },\theta ,Z,Z^{\prime
},C,D\right) \right\vert ^{2}$ measures the density of connections of value $%
T$ \ (and auxiliary variabls $\hat{T}$, $C,D$) between the set of cells
located at points $Z$ and $Z^{\prime }$ with activity $\omega _{\Gamma }$
and $\omega _{\Gamma }^{\prime }$. In the context of complex cells, it
describes the density of connections with strength $T$ between sets of axons
nd dendrites of cells with activity $\omega _{\Gamma },\omega _{\Gamma
}^{\prime }$.

\subsection{Projection on dependent activity states and effective action:}

We have shown in (\cite{GL}) that some simplifications arise in the action
functional. Using the fact that $\eta ^{2}<<1$, and noting that in this
case, field configurations $\Psi \left( \theta ,Z,\omega \right) $ such that:%
\begin{equation*}
\omega ^{-1}-G\left( J\left( \theta ,Z\right) +\int \frac{\kappa }{N}\frac{%
\omega _{1}}{\omega }\left\vert \Psi \left( \theta -\frac{\left\vert
Z-Z_{1}\right\vert }{c},Z_{1},\omega _{1}\right) \right\vert ^{2}T\left(
Z,\theta ,Z_{1}\right) dZ_{1}d\omega _{1}\right) \neq 0
\end{equation*}%
have negligible statistical weight, we can simplify (\ref{lfS}) and restrict
the fields to those of the form: 
\begin{equation}
\Psi \left( \theta ,Z\right) \delta \left( \omega ^{-1}-\omega ^{-1}\left(
J,\theta ,Z,\left\vert \Psi \right\vert ^{2}\right) \right)  \label{prt}
\end{equation}%
where $\omega ^{-1}\left( J,\theta ,Z,\Psi \right) $ satisfies:%
\begin{eqnarray*}
\omega ^{-1}\left( J,\theta ,Z,\left\vert \Psi \right\vert ^{2}\right)
&=&G\left( J\left( \theta ,Z\right) +\int \frac{\kappa }{N}\frac{\omega
_{1}T\left( Z,\theta ,Z_{1},\theta -\frac{\left\vert Z-Z_{1}\right\vert }{c}%
\right) }{\omega \left( J,\theta ,Z,\left\vert \Psi \right\vert ^{2}\right) }%
\left\vert \Psi \left( \theta -\frac{\left\vert Z-Z_{1}\right\vert }{c}%
,Z_{1},\omega _{1}\right) \right\vert ^{2}dZ_{1}d\omega _{1}\right) \\
&=&G\left( J\left( \theta ,Z\right) +\int \frac{\kappa }{N}\frac{\omega
_{1}T\left( Z,\theta ,Z_{1},\theta -\frac{\left\vert Z-Z_{1}\right\vert }{c}%
\right) }{\omega \left( J,\theta ,Z,\left\vert \Psi \right\vert ^{2}\right) }%
\left\vert \Psi \left( \theta -\frac{\left\vert Z-Z_{1}\right\vert }{c}%
,Z_{1}\right) \right\vert ^{2}\right. \\
&&\times \left. \delta \left( \omega _{1}^{-1}-\omega ^{-1}\left( J,\theta -%
\frac{\left\vert Z-Z_{1}\right\vert }{c},Z_{1},\left\vert \Psi \right\vert
^{2}\right) \right) dZ_{1}d\omega _{1}\right)
\end{eqnarray*}%
The last equation simplifies to yield:%
\begin{eqnarray}
&&\omega ^{-1}\left( J,\theta ,Z,\left\vert \Psi \right\vert ^{2}\right)
\label{qf} \\
&=&G\left( J\left( \theta ,Z\right) +\int \frac{\kappa }{N}\frac{\omega
\left( J,\theta -\frac{\left\vert Z-Z_{1}\right\vert }{c},Z_{1},\Psi \right)
T\left( Z,\theta ,Z_{1},\theta -\frac{\left\vert Z-Z_{1}\right\vert }{c}%
\right) }{\omega \left( J,\theta ,Z,\left\vert \Psi \right\vert ^{2}\right) }%
\left\vert \Psi \left( \theta -\frac{\left\vert Z-Z_{1}\right\vert }{c}%
,Z_{1}\right) \right\vert ^{2}dZ_{1}\right)  \notag
\end{eqnarray}%
The configurations $\Psi \left( \theta ,Z,\omega \right) $ that minimize the
potential (\ref{ptntl}) can now be considered: the field $\Psi \left( \theta
,Z,\omega \right) $ is projected on the subspace (\ref{prt}) of functions of
two variables, and we can therefore replace in (\ref{ptntl}):%
\begin{equation}
\omega \rightarrow \omega \left( J,\theta ,Z,\left\vert \Psi \right\vert
^{2}\right)  \label{DF}
\end{equation}%
\begin{equation}
\omega ^{\prime }\rightarrow \omega \left( J,\theta -\frac{\left\vert
Z-Z^{\prime }\right\vert }{c},Z^{\prime },\left\vert \Psi \right\vert
^{2}\right)  \label{DH}
\end{equation}%
The "classical" effective action becomes (see appendix 1):%
\begin{equation}
-\frac{1}{2}\Psi ^{\dagger }\left( \theta ,Z\right) \left( \nabla _{\theta
}\left( \frac{\sigma ^{2}}{2}\nabla _{\theta }-\omega ^{-1}\left( J,\theta
,Z,\left\vert \Psi \right\vert ^{2}\right) \right) \right) \Psi \left(
\theta ,Z\right)  \label{nmR}
\end{equation}%
with $\omega ^{-1}\left( J,\theta ,Z,\left\vert \Psi \right\vert ^{2}\right) 
$ given by equation (\ref{qf}). As in (\cite{GL}) we add to this action a
stabilization potential $V\left( \Psi \right) $ ensuring an average activity
of the system. The precise form of this potential is irrelevant here, but we
assume that it has a minimum $\Psi _{0}\left( \theta ,Z\right) $.

The projection on dependent activity also applies to connectivity action
terms. We can thus replace $\Gamma \left( T,\hat{T},\omega _{\Gamma },\omega
_{\Gamma }^{\prime },\theta ,Z,Z^{\prime },C,D\right) $ by $\Gamma \left( T,%
\hat{T},\theta ,Z,Z^{\prime },C,D\right) $ and the action becomes:%
\begin{eqnarray}
S_{full} &=&-\frac{1}{2}\Psi ^{\dagger }\left( \theta ,Z,\omega \right)
\nabla \left( \frac{\sigma _{\theta }^{2}}{2}\nabla -\omega ^{-1}\left(
J,\theta ,Z,\left\vert \Psi \right\vert ^{2}\right) \right) \Psi \left(
\theta ,Z\right) +V\left( \Psi \right)  \label{flt} \\
&&+\frac{1}{2\eta ^{2}}\left( S_{\Gamma }^{\left( 0\right) }+S_{\Gamma
}^{\left( 1\right) }+S_{\Gamma }^{\left( 2\right) }+S_{\Gamma }^{\left(
3\right) }+S_{\Gamma }^{\left( 4\right) }\right) +U\left( \left\{ \left\vert
\Gamma \left( \theta ,Z,Z^{\prime },C,D\right) \right\vert ^{2}\right\}
\right)  \notag
\end{eqnarray}%
with $S_{\Gamma }^{\left( 1\right) }$, $S_{\Gamma }^{\left( 2\right) }$, $%
S_{\Gamma }^{\left( 3\right) }$, $S_{\Gamma }^{\left( 4\right) }$ now given
by: 
\begin{equation}
S_{\Gamma }^{\left( 1\right) }=\int \Gamma ^{\dag }\left( T,\hat{T},\theta
,Z,Z^{\prime },C,D\right) \nabla _{T}\left( \frac{\sigma _{T}^{2}}{2}\nabla
_{T}+O_{T}\right) \Gamma \left( T,\hat{T},\theta ,Z,Z^{\prime },C,D\right)
\label{wGD}
\end{equation}%
\begin{equation}
S_{\Gamma }^{\left( 2\right) }=\int \Gamma ^{\dag }\left( T,\hat{T},\theta
,Z,Z^{\prime },C,D\right) \nabla _{\hat{T}}\left( \frac{\sigma _{\hat{T}}^{2}%
}{2}\nabla _{\hat{T}}+O_{\hat{T}}\right) \Gamma \left( T,\hat{T},\theta
,Z,Z^{\prime },C,D\right)  \label{wGT}
\end{equation}%
\begin{equation}
S_{\Gamma }^{\left( 3\right) }=\Gamma ^{\dag }\left( T,\hat{T},\theta
,Z,Z^{\prime },C,D\right) \nabla _{C}\left( \frac{\sigma _{C}^{2}}{2}\nabla
_{C}+O_{C}\right) \Gamma \left( T,\hat{T},\theta ,Z,Z^{\prime },C,D\right)
\label{wGQ}
\end{equation}%
\begin{equation}
S_{\Gamma }^{\left( 4\right) }=\Gamma ^{\dag }\left( T,\hat{T},\theta
,Z,Z^{\prime },C,D\right) \nabla _{D}\left( \frac{\sigma _{D}^{2}}{2}\nabla
_{D}+O_{D}\right) \Gamma \left( T,\hat{T},\theta ,Z,Z^{\prime },C,D\right)
\label{wGC}
\end{equation}%
where:%
\begin{eqnarray}
O_{C} &=&\frac{C}{\tau _{C}\omega \left( J,\theta ,Z,\left\vert \Psi
\right\vert ^{2}\right) }-\frac{\alpha _{C}\left( 1-C\right) \omega \left(
J,\theta -\frac{\left\vert Z-Z^{\prime }\right\vert }{c},Z^{\prime
},\left\vert \Psi \right\vert ^{2}\right) \left\vert \Psi \left( \theta -%
\frac{\left\vert Z-Z^{\prime }\right\vert }{c},Z^{\prime }\right)
\right\vert ^{2}}{\omega \left( J,\theta ,Z,\left\vert \Psi \right\vert
^{2}\right) }  \label{DP} \\
O_{D} &=&\frac{D}{\tau _{D}\omega \left( J,\theta ,Z,\left\vert \Psi
\right\vert ^{2}\right) }-\alpha _{D}\left( 1-D\right) \left\vert \Psi
\left( \theta ,Z\right) \right\vert ^{2}  \notag \\
O_{\hat{T}} &=&-\frac{\rho }{\omega \left( J,\theta ,Z,\left\vert \Psi
\right\vert ^{2}\right) }\left( \left( h\left( Z,Z^{\prime }\right) -\hat{T}%
\right) C\left\vert \Psi \left( \theta ,Z\right) \right\vert ^{2}h_{C}\left(
\omega \left( J,\theta ,Z,\left\vert \Psi \right\vert ^{2}\right) \right)
\right.  \notag \\
&&\left. -D\hat{T}\left\vert \Psi \left( \theta -\frac{\left\vert
Z-Z^{\prime }\right\vert }{c},Z^{\prime }\right) \right\vert ^{2}h_{D}\left(
\omega \left( J,\theta -\frac{\left\vert Z-Z^{\prime }\right\vert }{c}%
,Z^{\prime },\left\vert \Psi \right\vert ^{2}\right) \right) \right)  \notag
\\
O_{T} &=&-\left( -\frac{1}{\tau \omega \left( J,\theta ,Z,\left\vert \Psi
\right\vert ^{2}\right) }T+\frac{\lambda }{\omega \left( J,\theta
,Z,\left\vert \Psi \right\vert ^{2}\right) }\hat{T}\right)  \notag
\end{eqnarray}%
In these equations, the averages $\bar{\omega}$ and $\bar{\omega}^{\prime }$
have been replaced by\ $\omega \left( J,\theta ,Z,\left\vert \Psi
\right\vert ^{2}\right) $ and $\omega \left( J,\theta -\frac{\left\vert
Z-Z^{\prime }\right\vert }{c},Z^{\prime },\left\vert \Psi \right\vert
^{2}\right) $ as a consequence of the projection.

In (\ref{flt}), we added a potential:%
\begin{equation}
U\left( \left\{ \left\vert \Gamma \left( \theta ,Z,Z^{\prime },C,D\right)
\right\vert ^{2}\right\} \right) =U\left( \int T\left\vert \Gamma \left( T,%
\hat{T},\theta ,Z,Z^{\prime },C,D\right) \right\vert ^{2}dTd\hat{T}\right)
\label{pcn}
\end{equation}%
that models the constraint about the number of active connections in the
system.

\part*{Part II. Structural aspects of the system. Background fields, averages%
}

The following sections concentrate on solving the saddle-point equations for
(\ref{flt}). Considering that the time scale for cell activity is shorter
than that for connectivities, we solve in first approximation the
saddle-point equation for the neuron field action, given the connectivity
field. Subsequently, we calculate the background field for connectivities.
This ultimately leads to the equilibrium equations for average connectivity
variables and cell activities.

\section{Background states equations for neuron field and activities
depending on connectivty functions}

Our goal is to find the possible background states of action (\ref{flt}). In
principle we should minimize $S_{full}$ both over the neurons field $\left(
\Psi ,\Psi ^{\dag }\right) $ and the connectivity field $\left( \Gamma
,\Gamma ^{\dag }\right) $. However, the time scale of the neuron field is
lower than that of the connectivity field. To obtain an effective action for 
$\left( \Gamma ,\Gamma ^{\dag }\right) $, we intend to integrate over the
degrees of freedom for the field $\Psi $ in the partition function defined
by $S_{full}$ in (\ref{flt}). In first approximation, this corresponds to
set $\left( \Psi ,\Psi ^{\dag }\right) $ to its backround obtained through
the minimization of the effective action, written $S_{\Psi }\left( \Psi
,\Psi ^{\dag }\right) $. The series expansion of the effective action has
been computed in (\cite{GL}). At the lowest order in perturbation, this
corresponds to modify the action by a translation in $\left\vert \Psi
\right\vert ^{2}$: 
\begin{equation*}
S_{\Psi }\left( \Psi ,\Psi ^{\dag }\right) \mathcal{=}-\frac{1}{2}\Psi
^{\dagger }\left( \theta ,Z,\omega \right) \nabla \left( \frac{\sigma
_{\theta }^{2}}{2}\nabla -\omega ^{-1}\left( J,\theta ,Z,\mathcal{G}%
_{0}+\left\vert \Psi \right\vert ^{2}\right) \right) \Psi \left( \theta
,Z\right) +V\left( \Psi \right)
\end{equation*}%
where $\mathcal{G}_{0}$ is computed in (\cite{GL}), it is given by:%
\begin{equation*}
\mathcal{G}_{0}=\mathcal{G}_{0}\left( Z,Z\right)
\end{equation*}%
where $\mathcal{G}_{0}\left( Z,Z\right) $ is the static Green function for
the field $\Psi $. At the first order corrections, the activities defined
classically by formula (\ref{qf}) are now defined by the equation:

\begin{eqnarray}
&&\omega ^{-1}\left( J,\theta ,Z,\left\vert \Psi \right\vert ^{2}\right)
\label{nqf} \\
&=&G\left( J\left( \theta ,Z\right) +\int \frac{\kappa }{N}\frac{\omega
^{\prime }T\left( Z,\theta ,Z_{1},\theta -\frac{\left\vert
Z-Z_{1}\right\vert }{c}\right) }{\omega }\left( \mathcal{G}_{0}+\left\vert
\Psi \left( \theta -\frac{\left\vert Z-Z_{1}\right\vert }{c},Z_{1}\right)
\right\vert ^{2}\right) dZ_{1}\right)  \notag
\end{eqnarray}%
This equation depends on both on the external currents $J\left( \theta
,Z\right) $ and the fields $\Psi $ and $\Gamma $ through the expression (\ref%
{tpr}) of $T\left( Z,\theta ,Z_{1},\theta -\frac{\left\vert
Z-Z_{1}\right\vert }{c}\right) $. At the scale of activities, $T\left(
Z,\theta ,Z_{1},\theta -\frac{\left\vert Z-Z_{1}\right\vert }{c}\right) $
can be considered quasi static, allowing us to find quasi-static equilibria
for $\omega ^{-1}\left( J,\theta ,Z,\left\vert \Psi \right\vert ^{2}\right) $
and $\left\vert \Psi \left( \theta -\frac{\left\vert Z-Z_{1}\right\vert }{c}%
,Z_{1}\right) \right\vert ^{2}$ as functions of $T\left( Z,\theta
,Z_{1}\right) $. This result will be reintroduced in the dynamics for the
background field $\Gamma $.

The minimization equation for the background field becomes:%
\begin{equation}
-\nabla \left( \frac{\sigma _{\theta }^{2}}{2}\nabla -\omega ^{-1}\left(
J,\theta ,Z,\mathcal{G}_{0}+\left\vert \Psi \right\vert ^{2}\right) \right) +%
\frac{\delta V\left( \Psi \right) }{\delta \Psi \left( \theta ,Z\right) }=0
\label{sdp}
\end{equation}%
and the solutions of (\ref{sdp}) are functions of the connectivities.

\section{Background state equations for $\Gamma $ under simplifying
assumptions}

The minimization equation of (\ref{flt}) for the connectivity functions
background states have the form:%
\begin{equation}
0=\frac{\delta }{\Gamma ^{\dagger }\left( T,\hat{T},\theta ,Z,Z^{\prime
},C,D\right) }\left( S_{\Gamma }^{\left( 1\right) }+S_{\Gamma }^{\left(
2\right) }+S_{\Gamma }^{\left( 3\right) }+S_{\Gamma }^{\left( 4\right)
}+U\left( \left\{ \left\vert \Gamma \left( \theta ,Z,Z^{\prime },C,D\right)
\right\vert ^{2}\right\} \right) +S_{\Psi }\left( \Psi ,\Psi ^{\dag }\right)
\right)  \label{SDQ}
\end{equation}%
and:%
\begin{equation}
0=\frac{\delta }{\Gamma \left( T,\hat{T},\theta ,Z,Z^{\prime },C,D\right) }%
\left( S_{\Gamma }^{\left( 1\right) }+S_{\Gamma }^{\left( 2\right)
}+S_{\Gamma }^{\left( 3\right) }+S_{\Gamma }^{\left( 4\right) }+U\left(
\left\{ \left\vert \Gamma \left( \theta ,Z,Z^{\prime },C,D\right)
\right\vert ^{2}\right\} \right) +S_{\Psi }\left( \Psi ,\Psi ^{\dag }\right)
\right)  \label{SDR}
\end{equation}%
where $\left\vert \Psi \left( \theta ,Z\right) \right\vert ^{2}$ is defined
by (\ref{sdp}).

To solve these equations, we introduce some simplifying assumptions. A more
formal treatment of these equations is given in appendix 2.

\subsection{Use of background neuron field}

The first simplification uses equation (\ref{sdp}) to set the background
field $\Psi \left( \theta ,Z\right) $ to its average. As a consequence, we
will replace in the sequel $\left\vert \Psi \left( \theta ,Z\right)
\right\vert ^{2}$ by its average $\left\langle \left\vert \Psi \left( \theta
,Z\right) \right\vert ^{2}\right\rangle $. To simplify the notations, we
will write:%
\begin{equation*}
\left\langle \left\vert \Psi \left( \theta ,Z\right) \right\vert
^{2}\right\rangle \rightarrow \left\vert \Psi \left( \theta ,Z\right)
\right\vert ^{2}
\end{equation*}%
The activities are thus estimated for these averages. Moreover, using (\ref%
{tpr}), (\ref{nqf}) and (\ref{sdp}), the averages depend on:%
\begin{equation}
\left\{ T\left( Z,\theta ,Z_{1},\theta -\frac{\left\vert Z-Z_{1}\right\vert 
}{c}\right) \right\} _{\left( Z,\theta ,Z_{1}\right) }  \label{DNC}
\end{equation}%
where:%
\begin{equation}
T\left( Z,\theta ,Z_{1},\theta -\frac{\left\vert Z-Z_{1}\right\vert }{c}%
\right) =\int T\left\vert \Gamma \left( T,\hat{T},\omega ,\omega _{1},\theta
,Z,Z_{1},C,D\right) \right\vert ^{2}dTd\hat{T}dCdD\equiv T\left( Z,\theta
,Z_{1}\right)
\end{equation}

\subsection{Neglecting the derivatives of $S_{\Psi }\left( \Psi ,\Psi ^{\dag
}\right) $}

Given that the neuron fields $\left( \Psi ,\Psi ^{\dag }\right) $ are set to
their background field values, the derivatives od $S_{\Psi }\left( \Psi
,\Psi ^{\dag }\right) $ simplify. Actually in this case:%
\begin{equation*}
\frac{\delta S_{\Psi }\left( \Psi ,\Psi ^{\dag }\right) }{\delta \Psi }=%
\frac{\delta S_{\Psi }\left( \Psi ,\Psi ^{\dag }\right) }{\delta \Psi ^{\dag
}}=0
\end{equation*}%
and the derivatives with respect to connectivity fields reduce to partial
derivatives:%
\begin{eqnarray*}
\frac{\delta S_{\Psi }\left( \Psi ,\Psi ^{\dag }\right) }{\delta \Gamma
^{\dagger }\left( T,\hat{T},\theta ,Z,Z^{\prime },C,D\right) } &=&\frac{%
\partial S_{\Psi }\left( \Psi ,\Psi ^{\dag }\right) }{\partial \Gamma
^{\dagger }\left( T,\hat{T},\theta ,Z,Z^{\prime },C,D\right) } \\
\frac{\delta }{\delta \Gamma \left( T,\hat{T},\theta ,Z,Z^{\prime
},C,D\right) } &=&\frac{\partial S_{\Psi }\left( \Psi ,\Psi ^{\dag }\right) 
}{\partial \Gamma \left( T,\hat{T},\theta ,Z,Z^{\prime },C,D\right) }
\end{eqnarray*}%
These partial derivatives involve the derivatives of $G$ in (\ref{qf}) with
respect to the connectivity field. However, due to the disparity in time
scales between activities and connectivities, a modification of $\Gamma $
and $\Gamma ^{\dagger }$ initially modifies the density of active
axons/dentrite, subsequently influencing the level of activity. As a result,
the partial derivatives can thus be neglected in first approximation%
\footnote{%
Including these derivatives in the saddle point equations for connectivities
would modify these equation by a quasi linear contribution $\frac{1}{2}\frac{%
\kappa }{N}\Psi ^{\dagger }\left( \theta ,Z,\omega \right) \nabla G^{\prime
}\left( G^{-1}\left( \omega ^{-1}\left( J,\theta ,Z,\left\vert \Psi
\right\vert ^{2}\right) \right) \right) \Psi \left( \theta ,Z\right) T\Gamma 
$. This contribution shifts slightly the average connectivities. In a quasi
static approximation, it can be neglected.}.

\subsection{Neglecting backreaction contributions}

equation (\ref{SDQ}) (there is a similar treatment for (\ref{SDR})) becomes:%
\begin{eqnarray}
0 &=&\left( \nabla _{C}\left( \frac{\sigma _{C}^{2}}{2}\nabla
_{C}+O_{C}\right) +\nabla _{D}\left( \frac{\sigma _{C}^{2}}{2}\nabla
_{D}+O_{D}\right) +\nabla _{T}\left( \frac{\sigma _{C}^{2}}{2}\nabla
_{T}+O_{T}\right) +\nabla _{\hat{T}}\left( \frac{\sigma _{C}^{2}}{2}\nabla _{%
\hat{T}}+O_{\hat{T}}\right) \right.  \label{GSP} \\
&&+\left. K\left( \theta ,Z,Z^{\prime },\left\Vert \Psi \right\Vert
^{2},\left\Vert \Gamma \right\Vert ^{2}\right) T+\frac{\delta U\left(
\left\{ \left\vert \Gamma \left( \theta ,Z,Z^{\prime },C,D\right)
\right\vert ^{2}\right\} \right) }{\delta \Gamma ^{\dagger }\left( T,\hat{T}%
,\theta ,Z,Z^{\prime },C,D\right) }\right) \Gamma \left( T,\hat{T},\theta
,Z,Z^{\prime },C,D\right)  \notag
\end{eqnarray}%
and $K$ is given by: 
\begin{eqnarray}
&&K\left( \theta ,Z,Z^{\prime },\left\Vert \Psi \right\Vert ^{2},\left\Vert
\Gamma \right\Vert ^{2}\right) =\int \Gamma ^{\dagger }\left( T_{1},\hat{T}%
_{1},\theta _{1},Z_{1},Z_{1}^{\prime },C_{1},D_{1}\right)  \label{KQT} \\
&&\frac{\delta W\left( T_{1},\hat{T}_{1},\theta _{1},Z_{1},Z_{1}^{\prime
},C_{1},D_{1}\right) }{\delta T\left\vert \Gamma \left( T,\hat{T},\theta
,Z,Z^{\prime },C,D\right) \right\vert ^{2}}\Gamma \left( T_{1},\hat{T}%
_{1},\theta _{1},Z_{1},Z_{1}^{\prime },C_{1},D_{1}\right) d\left( T_{1},\hat{%
T}_{1},\theta _{1},Z_{1},Z_{1}^{\prime },C_{1},D_{1}\right)  \notag
\end{eqnarray}%
with:%
\begin{equation*}
W\left( T,\hat{T},\theta ,Z,Z^{\prime },C,D\right) =\nabla _{C}O_{C}+\nabla
_{D}O_{D}+\nabla _{T}O_{T}+\nabla _{\hat{T}}O_{\hat{T}}
\end{equation*}%
The last term in (\ref{GSP}) arises from the dependency of averages in $%
\Gamma \left( T,\hat{T},\theta ,Z,Z^{\prime },C,D\right) $ as given in (\ref%
{DNC}). It represents the backreaction of the system as a whole when a
variation in $\Gamma \left( T,\hat{T},\theta ,Z,Z^{\prime },C,D\right) $
occurs. This term can be considered as a correction and will be disregarded
in the sequel. Appendix 2 includes this contributions and computes its
impact on the background state.

Note that neglecting $W\left( T,\hat{T},\theta ,Z,Z^{\prime },C,D\right) $
amounts to consider in (\ref{GMN}) that the action:%
\begin{equation*}
S_{\Gamma }^{\left( 1\right) }+S_{\Gamma }^{\left( 2\right) }+S_{\Gamma
}^{\left( 3\right) }+S_{\Gamma }^{\left( 4\right) }
\end{equation*}%
is quadratic in $\Gamma \left( T,\hat{T},\theta ,Z,Z^{\prime },C,D\right) $.

\subsection{Separating variables in the connectivty field}

The third simplification arises from the fact that in (\ref{GMN}) the
dependency of the field in the variables $C$ and $D$ is independent from the
dependency in $T$ and $\hat{T}$. We thus start by the minimization of $%
S_{\Gamma }^{\left( 3\right) }+S_{\Gamma }^{\left( 4\right) }$. We assume
the existence of non-trivial minima $\Gamma \left( T,\hat{T},\theta
,Z,Z^{\prime },C,D\right) $, and we will determine, in the end, a condition
for the existence of such states.

Furthermore, the solutions for $\Gamma $ will depend on $\left\vert \Psi
\left( \theta ,Z\right) \right\vert ^{2}$ and $\omega \left( J,\theta
,Z,\left\vert \Psi \right\vert ^{2}\right) $. These fields, in turn, depend
as functionals on the entire collection $\left\{ \Gamma \left( T,\hat{T}%
,\theta ,Z,Z^{\prime },C,D\right) \right\} _{\left( Z,Z^{\prime }\right) }$,
or, in first approximation, on the norm $\left\Vert \Gamma \right\Vert ^{2}$%
. This implies that $\Gamma \left( T,\hat{T},\theta ,Z,Z^{\prime
},C,D\right) $ wil satisfy some compatibility conditions that will define
the equilibrium of the system. This equilibrium will be computed in the next
section.

\section{Solutions for the background state equations}

\subsection{Principle}

As a consequence of our simplifying assumptions, we can factor the solutions
of the minimization equations as:%
\begin{equation*}
\Gamma \left( T,\hat{T},\theta ,Z,Z^{\prime },C,D\right) =\Gamma _{1}\left(
Z,Z^{\prime },C\right) \Gamma _{2}\left( Z,Z^{\prime },D\right) \Gamma
\left( T,\hat{T},\theta ,Z,Z^{\prime }\right)
\end{equation*}%
and minimize first:%
\begin{equation*}
S_{\Gamma }^{\left( 3\right) }+S_{\Gamma }^{\left( 4\right) }
\end{equation*}%
to find $\Gamma _{1}\left( Z,Z^{\prime },C\right) $ and $\Gamma _{2}\left(
Z,Z^{\prime },D\right) $ along with the average values of $C$ and $D$ in
these states.When these functions are determined, we substitute their
expression in:%
\begin{equation*}
S_{\Gamma }^{\left( 1\right) }+S_{\Gamma }^{\left( 2\right) }+S_{\Gamma
}^{\left( 3\right) }+S_{\Gamma }^{\left( 4\right) }
\end{equation*}%
and minimize this action with respect to $\Gamma \left( T,\hat{T},\theta
,Z,Z^{\prime }\right) $. The equations for these functions allow to find the
consistency equations for the average values of $\left( T,\hat{T}\right) $
in the state $\Gamma \left( T,\hat{T},\theta ,Z,Z^{\prime }\right) $. Given
the threshold to creating connections, several possible solutions arise.
These values are then used to derive the final form of the possible states $%
\Gamma \left( T,\hat{T},\theta ,Z,Z^{\prime }\right) $.

\subsection{Background state for $C$ and $D$}

We first transform the terms involving the gradients $\nabla _{C}$ and $%
\nabla _{D}$ in $S_{\Gamma }^{\left( 3\right) }+S_{\Gamma }^{\left( 4\right)
}$\ by changing the variables. Starting with:%
\begin{eqnarray}
&&S_{\Gamma }^{\left( 3\right) }+S_{\Gamma }^{\left( 4\right) }=\Gamma
^{\dag }\left( T,\hat{T},\theta ,Z,Z^{\prime },C,D\right)  \notag \\
&&\times \left( \nabla _{C}\left( \frac{\sigma _{C}^{2}}{2}\nabla
_{C}+O_{C}\right) +\nabla _{D}\left( \frac{\sigma _{D}^{2}}{2}\nabla
_{D}+O_{D}\right) \right) \Gamma \left( T,\hat{T},\theta ,Z,Z^{\prime
},C,D\right)
\end{eqnarray}%
where $O_{C}$ and \ $O_{D}$ are defined in (\ref{DP}).

We considerthechange of variables:%
\begin{eqnarray*}
&&\Gamma \left( T,\hat{T},\theta ,Z,Z^{\prime },C,D\right) \\
&\rightarrow &\Gamma \left( T,\hat{T},\theta ,Z,Z^{\prime },C,D\right) \exp
\left( \int \frac{1}{\sigma _{D}^{2}}O_{D}dD\right) \exp \left( \frac{1}{%
\sigma _{C}^{2}}\int O_{C}dC\right)
\end{eqnarray*}%
and:%
\begin{eqnarray*}
&&\Gamma ^{\dag }\left( T,\hat{T},\theta ,Z,Z^{\prime },C,D\right) \\
&\rightarrow &\Gamma ^{\dag }\left( T,\hat{T},\theta ,Z,Z^{\prime
},C,D\right) \exp \left( -\int \frac{1}{\sigma _{D}^{2}}O_{D}dD\right) \exp
\left( -\frac{1}{\sigma _{C}^{2}}\int O_{C}dC\right)
\end{eqnarray*}%
As a consequence, the terms involving the gradients $\nabla _{C}$ and $%
\nabla _{D}$ in (\ref{flt}) rewrite:%
\begin{eqnarray*}
&&\Gamma ^{\dag }\left( T,\hat{T},\theta ,Z,Z^{\prime },C,D\right) \left( 
\frac{\sigma _{C}^{2}}{2}\nabla _{C}^{2}-\frac{1}{2\sigma _{C}^{2}}%
O_{C}^{2}\right. \\
&&+\frac{1}{2}\left( \frac{1}{\tau _{C}\omega }+\alpha _{C}\frac{\omega
^{\prime }\left\vert \Psi \left( \theta -\frac{\left\vert Z-Z^{\prime
}\right\vert }{c},Z^{\prime },\omega ^{\prime }\right) \right\vert ^{2}}{%
\omega }\right) \Gamma \left( T,\hat{T},\theta ,Z,Z^{\prime },C,D\right)
\end{eqnarray*}%
and:%
\begin{eqnarray*}
&&\Gamma ^{\dag }\left( T,\hat{T},\theta ,Z,Z^{\prime },C,D\right) \\
&&\times \left( \frac{\sigma _{D}^{2}}{2}\nabla _{D}^{2}-\frac{1}{2\sigma
_{D}^{2}}O_{D}^{2}+\frac{1}{2}\left( \frac{1}{\tau _{D}\omega }+\alpha
_{D}\left\vert \Psi \left( \theta ,Z\right) \right\vert ^{2}\right) \right)
\Gamma \left( T,\hat{T},\theta ,Z,Z^{\prime },C,D\right)
\end{eqnarray*}%
where we use the notation:%
\begin{eqnarray*}
\omega &\equiv &\omega \left( J,\theta ,Z,\left\vert \Psi \right\vert
^{2}\right) \\
\omega ^{\prime } &\equiv &\omega \left( J,\theta -\frac{\left\vert
Z-Z^{\prime }\right\vert }{c},Z^{\prime },\left\vert \Psi \right\vert
^{2}\right)
\end{eqnarray*}%
and $\omega $ and $\omega ^{\prime }$ are defined by equation (\ref{qf}):%
\begin{eqnarray*}
&&\omega ^{-1}\left( J,\theta ,Z,\left\vert \Psi \right\vert ^{2}\right) \\
&=&G\left( J\left( \theta ,Z\right) +\int \frac{\kappa }{N}\frac{\omega
\left( J,\theta -\frac{\left\vert Z-Z_{1}\right\vert }{c},Z_{1},\Psi \right)
T\left\vert \Gamma \left( T,\hat{T},\theta ,Z,Z_{1}\right) \right\vert ^{2}}{%
\omega \left( J,\theta ,Z,\left\vert \Psi \right\vert ^{2}\right) }%
\left\vert \Psi \left( \theta -\frac{\left\vert Z-Z_{1}\right\vert }{c}%
,Z_{1}\right) \right\vert ^{2}dZ_{1}\right)
\end{eqnarray*}%
The field $\Gamma \left( T,\hat{T},\theta ,Z,Z^{\prime },C,D\right) $ can be
written as a product:%
\begin{equation*}
\Gamma \left( T,\hat{T},\theta ,Z,Z^{\prime },C,D\right) =\Gamma _{1}\left(
Z,Z^{\prime },C\right) \Gamma _{2}\left( Z,Z^{\prime },D\right) \Gamma
\left( T,\hat{T},\theta ,Z,Z^{\prime }\right)
\end{equation*}%
and since we are looking for non trivial background states, i.e. states with
positive norm, we can constrain $\Gamma _{1}\left( Z,Z^{\prime },C\right) $
and $\Gamma _{2}\left( Z,Z^{\prime },D\right) $ to have a norm equal to $1$,
so that $\Gamma _{1}\left( Z,Z^{\prime },C\right) $ and $\Gamma _{2}\left(
Z,Z^{\prime },D\right) $ satisfy:%
\begin{equation}
\left( \frac{\sigma _{C}^{2}}{2}\nabla _{C}^{2}-\frac{1}{2\sigma _{C}^{2}}%
O_{C}^{2}-\frac{1}{2}a_{C}\left( Z\right) +\lambda _{1}\left( Z\right)
\right) \Gamma _{1}\left( Z,Z^{\prime },C\right) =0  \label{gmn}
\end{equation}%
\begin{equation}
\left( \frac{\sigma _{D}^{2}}{2}\nabla _{D}^{2}-\frac{1}{2\sigma _{D}^{2}}%
O_{D}^{2}-\frac{1}{2}a_{D}\left( Z\right) +\lambda _{2}\left( Z\right)
\right) \Gamma _{2}\left( Z,Z^{\prime },D\right) =0  \label{gmt}
\end{equation}%
with:%
\begin{eqnarray}
a_{C}\left( Z\right) &=&\frac{1}{\tau _{C}\omega }+\alpha _{C}\frac{\omega
^{\prime }\left\vert \Psi \left( \theta -\frac{\left\vert Z-Z^{\prime
}\right\vert }{c},Z^{\prime },\omega ^{\prime }\right) \right\vert ^{2}}{%
\omega }  \label{CD} \\
a_{D}\left( Z\right) &=&\frac{1}{\tau _{D}\omega }+\alpha _{D}\left\vert
\Psi \left( \theta ,Z\right) \right\vert ^{2}  \notag
\end{eqnarray}

These equations can be rewritten by defining the averages $\left\langle
C\left( \theta \right) \right\rangle $ and $\left\langle D\left( \theta
\right) \right\rangle $: 
\begin{eqnarray}
C &\rightarrow &\left\langle C\left( \theta \right) \right\rangle =\frac{%
\alpha _{C}\frac{\omega ^{\prime }\left\vert \Psi \left( \theta -\frac{%
\left\vert Z-Z^{\prime }\right\vert }{c},Z^{\prime }\right) \right\vert ^{2}%
}{\omega }}{\frac{1}{\tau _{C}\omega }+\alpha _{C}\frac{\omega ^{\prime
}\left\vert \Psi \left( \theta -\frac{\left\vert Z-Z^{\prime }\right\vert }{c%
},Z^{\prime }\right) \right\vert ^{2}}{\omega }}=\frac{\alpha _{C}\omega
^{\prime }\left\vert \Psi \left( \theta -\frac{\left\vert Z-Z^{\prime
}\right\vert }{c},Z^{\prime }\right) \right\vert ^{2}}{\frac{1}{\tau _{C}}%
+\alpha _{C}\omega ^{\prime }\left\vert \Psi \left( \theta -\frac{\left\vert
Z-Z^{\prime }\right\vert }{c},Z^{\prime }\right) \right\vert ^{2}}\equiv
C\left( \theta \right)  \label{vrG} \\
D &\rightarrow &\left\langle D\left( \theta \right) \right\rangle =\frac{%
\alpha _{D}\omega \left\vert \Psi \left( \theta ,Z\right) \right\vert ^{2}}{%
\frac{1}{\tau _{D}}+\alpha _{D}\omega \left\vert \Psi \left( \theta
,Z\right) \right\vert ^{2}}\equiv D\left( \theta \right)  \label{vRG}
\end{eqnarray}%
so that:%
\begin{equation}
\left( \frac{\sigma _{C}^{2}}{2}\nabla _{C}^{2}-\frac{1}{2\sigma _{C}^{2}}%
\left( a_{C}\left( Z\right) \left( C-C\left( \theta \right) \right) \right)
^{2}-\frac{1}{2}a_{C}\left( Z\right) +\lambda _{1}\left( Z\right) \right)
\Gamma _{1}\left( Z,Z^{\prime },C\right) =0  \label{GMN}
\end{equation}%
and:%
\begin{equation}
\left( \frac{\sigma _{D}^{2}}{2}\nabla _{D}^{2}-\frac{1}{2\sigma _{D}^{2}}%
\left( a_{D}\left( Z\right) \left( D-D\left( \theta \right) \right) \right)
^{2}-\frac{1}{2}a_{D}\left( Z\right) +\lambda _{2}\left( Z,Z^{\prime
}\right) \right) \Gamma _{2}\left( Z,Z^{\prime },D\right) =0  \label{GMT}
\end{equation}%
Note that we can consider $C\left( \theta \right) $ and $D\left( \theta
\right) $ as slowly varying, given that $\frac{1}{\tau _{C}}<<1$, $\frac{1}{%
\tau _{D}}<<1$. Moreover, the value $\left\vert \Psi \left( \theta ,Z\right)
\right\vert ^{2}$ may also be considered as slowly varying in time, as this
background field represents the average activity of the cells at point $Z$.
Consequently, $\Gamma _{1}\left( Z,Z^{\prime },C\right) $ and $\Gamma
_{2}\left( Z,Z^{\prime },D\right) $ are slowly varying as required for the
connectivity background field.

The solutions of equations (\ref{GMN}) and (\ref{GMT})\ are parabolic
cylinder function at each pair of points $\left( Z,Z^{\prime }\right) $.
Imposing a unit norm to these functions for $C$ and $D$ varying over $%
\mathbb{R}
$ yields the fundamental state for $\Gamma _{1}\left( Z,Z^{\prime },C\right) 
$ and $\Gamma _{2}\left( Z,Z^{\prime },D\right) $ which implies the
condition: 
\begin{eqnarray}
a_{C}\left( Z\right) -\lambda _{1}\left( Z,Z^{\prime }\right) &=&-\frac{1}{2}%
\left( \frac{1}{\tau _{C}\omega }+\alpha _{C}\frac{\omega ^{\prime
}\left\vert \Psi \left( \theta -\frac{\left\vert Z-Z^{\prime }\right\vert }{c%
},Z^{\prime }\right) \right\vert ^{2}}{\omega }\right) =-\frac{1}{2}%
a_{C}\left( Z\right)  \label{HRc} \\
a_{D}\left( Z\right) -\lambda _{2}\left( Z,Z^{\prime }\right) &=&-\frac{1}{2}%
\left( \frac{1}{\tau _{D}\omega }+\alpha _{D}\left\vert \Psi \left( \theta
,Z\right) \right\vert ^{2}\right) =-\frac{1}{2}a_{D}\left( Z\right)
\label{Hrc}
\end{eqnarray}%
The possibility to adjust the value of $\lambda _{1}\left( Z,Z^{\prime
}\right) $ translates that the determination of $C$ and $D$ represents local
activity, with the constraint on global activity being borne by the variable 
$T$ (see below).

Practically, constraint (\ref{HRc}) corresponds to the fundamental state of
a harmonic oscillator whose lowest eigenvalue is $\frac{1}{2}$ times the
fundamental frequency. If we relax the simplifying hypothesis that $C$ and $%
D $ vary over $%
\mathbb{R}
$, the eigenvalue constraint would not hold anymore. Actually, we should
also impose $\Gamma _{1}\left( Z,Z^{\prime },C\right) =\Gamma _{2}\left(
Z,Z^{\prime },C\right) =0$ for $C<0$ and for $D<0$. However, the appearance
of $C\left( \theta \right) $ and $D\left( \theta \right) $ imply that for $%
\left\vert \Psi \left( \theta ,Z\right) \right\vert ^{2}>1$, these
constraints are satisfied in first approximation. Moreover, assuming
relatively low variances in the variables, and given that the variables $%
C-\left\langle C\right\rangle $ and $D-\left\langle D\right\rangle $ are
centered justifies our assumption.

As a consequence, introducing normalization factors $\mathcal{N}_{i}$, $%
i=1,2 $, we find:%
\begin{eqnarray}
\Gamma _{1}\left( Z,Z^{\prime },C\right) &=&\mathcal{N}_{1}\exp \left( -%
\frac{a_{C}\left( Z\right) }{8\sigma _{C}^{2}}\left( C-C\left( \theta
\right) \right) ^{2}\right)  \label{GMTH} \\
\Gamma _{2}\left( Z,Z^{\prime },D\right) &=&\mathcal{N}_{2}\exp \left( -%
\frac{a_{D}\left( Z\right) }{8\sigma _{D}^{2}}\left( D-D\left( \theta
\right) \right) ^{2}\right)  \notag
\end{eqnarray}%
and $\Gamma \left( T,\hat{T},\theta ,Z,Z^{\prime },C,D\right) $ factors as: 
\begin{eqnarray*}
&&\Gamma \left( T,\hat{T},\theta ,Z,Z^{\prime },C,D\right) \\
&\simeq &\Gamma \left( T,\hat{T},\theta ,Z,Z^{\prime }\right) \exp \left( -%
\frac{1}{8\sigma _{C}^{2}}a_{C}\left( Z\right) \left( \left( C-C\left(
\theta \right) \right) \right) ^{2}\right) \exp \left( -\frac{a_{D}\left(
Z\right) }{8\sigma _{D}^{2}}\left( D-D\left( \theta \right) \right)
^{2}\right)
\end{eqnarray*}%
These solutions yield the following action for the field:%
\begin{eqnarray*}
&&\Gamma _{1}^{\dagger }\left( Z,Z^{\prime },C\right) \Gamma _{2}^{\dagger
}\left( Z,Z^{\prime },D\right) \left( \frac{1}{2}+\frac{1}{2}a_{C}\left(
Z\right) +\frac{1}{2}+\frac{1}{2}a_{D}\left( Z\right) \right) \Gamma
_{1}\left( Z,Z^{\prime },C\right) \Gamma _{2}\left( Z,Z^{\prime },D\right) \\
&=&\left( 1+\frac{1}{2}\left( a_{C}\left( Z\right) +a_{D}\left( Z\right)
\right) \right)
\end{eqnarray*}%
Iintroducing these expressions in the full action for $\Gamma \left( T,\hat{T%
},\theta ,Z,Z^{\prime },C,D\right) $ yields the contribution:%
\begin{equation}
S_{\Gamma }^{\left( 3\right) }+S_{\Gamma }^{\left( 4\right) }=\Gamma
^{\dagger }\left( T,\hat{T},\theta ,Z,Z^{\prime }\right) \left( a_{C}\left(
Z\right) +a_{D}\left( Z\right) \right) \Gamma \left( T,\hat{T},\theta
,Z,Z^{\prime }\right)  \label{Rsn}
\end{equation}%
Given our approximations, this is a constant that will not affect the
minimisation of the action for $\Gamma \left( T,\hat{T},\theta ,Z,Z^{\prime
}\right) $. However it will impact the condition of existence of a state
with $\left\Vert \Gamma \left( T,\hat{T},\theta ,Z,Z^{\prime },C,D\right)
\right\Vert ^{2}>0$.

\subsection{Minimization equation for $T,\hat{T}$}

\subsubsection{Action for $T,\hat{T}$}

Given the previous projection on the background states for $C$ and $D$, and
introducing the additional term 
\begin{equation*}
K\left( \theta ,Z,Z^{\prime },\left\Vert \Psi \right\Vert ^{2},\left\Vert
\Gamma \right\Vert ^{2}\right) T\Gamma \left( T,\hat{T},\theta ,Z,Z^{\prime
},C,D\right)
\end{equation*}%
up to contribution (\ref{Rsn}), the effective action for $\Gamma \left( T,%
\hat{T},\theta ,Z,Z^{\prime }\right) $ reduces to two terms:%
\begin{eqnarray}
&&S_{\Gamma }^{\left( 1\right) }+S_{\Gamma }^{\left( 2\right) }  \label{Tcf}
\\
&=&\Gamma ^{\dag }\left( T,\hat{T},\theta ,Z,Z^{\prime }\right) \left(
\nabla _{T}\left( \frac{\sigma _{T}^{2}}{2}\nabla _{T}+O_{T}\right) +K\left(
\theta ,Z,Z^{\prime },\left\Vert \Psi \right\Vert ^{2},\left\Vert \Gamma
\right\Vert ^{2}\right) T\right) \Gamma \left( T,\hat{T},\theta ,Z,Z^{\prime
}\right)  \notag \\
&&+\Gamma ^{\dag }\left( T,\hat{T},\theta ,Z,Z^{\prime }\right) \nabla _{%
\hat{T}}\left( \frac{\sigma _{\hat{T}}^{2}}{2}\nabla _{\hat{T}}+\bar{O}_{%
\hat{T}}\right) \Gamma \left( T,\hat{T},\theta ,Z,Z^{\prime }\right)  \notag
\end{eqnarray}%
with:%
\begin{eqnarray*}
O_{T} &=&-\left( -\frac{1}{\tau \omega }T+\frac{\lambda }{\omega }\hat{T}%
\right) \\
\bar{O}_{\hat{T}} &=&-\frac{\rho }{\omega \left( J,\theta ,Z,\left\vert \Psi
\right\vert ^{2}\right) }\left( \left( h\left( Z,Z^{\prime }\right) -\hat{T}%
\right) C\left( \theta \right) \left\vert \Psi \left( \theta ,Z\right)
\right\vert ^{2}h_{C}\left( \omega \left( \theta ,Z,\left\vert \Psi
\right\vert ^{2}\right) \right) -\eta H\left( \delta -T\right) \right. \\
&&\left. -D\left( \theta \right) \hat{T}\left\vert \Psi \left( \theta -\frac{%
\left\vert Z-Z^{\prime }\right\vert }{c},Z^{\prime }\right) \right\vert
^{2}h_{D}\left( \omega \left( \theta -\frac{\left\vert Z-Z^{\prime
}\right\vert }{c},Z^{\prime },\left\vert \Psi \right\vert ^{2}\right)
\right) \right) \\
&=&\bar{O}_{\hat{T}}+\frac{\rho \eta H\left( \delta -T\right) }{\omega
\left( J,\theta ,Z,\left\vert \Psi \right\vert ^{2}\right) }
\end{eqnarray*}%
and $C\left( \theta \right) $ and $D\left( \theta \right) $ defined in (\ref%
{vrG}) and (\ref{vRG}). We will consider them as relatively constant while
finding the background state. This amounts to projecting onto states of
fields $\Gamma \left( T,\hat{T},\theta ,Z,Z^{\prime }\right) $ and $\Gamma
^{\dagger }\left( T,\hat{T},\theta ,Z,Z^{\prime }\right) $ that are slowly
varying. As for $C$ and $D$, we will neglect the condition $T<0$ holds.
However, as long that $\left\langle T\right\rangle $ is significant enough,
this case is irrelevant.

\subsubsection{Change of variable}

We perform the following change of variable on the fields:%
\begin{eqnarray}
&&\Gamma \left( T,\hat{T},\theta ,Z,Z^{\prime }\right) \\
&\rightarrow &\exp \left( \int \frac{\bar{O}_{\hat{T}}}{\sigma _{\hat{T}%
}^{2}\omega \left( \theta ,Z,\left\vert \Psi \right\vert ^{2}\right) }d\hat{T%
}\right) \exp \left( \int \frac{O_{T}}{\sigma _{T}^{2}}dT\right) \Gamma
\left( T,\hat{T},\theta ,Z,Z^{\prime }\right)  \notag
\end{eqnarray}%
\begin{eqnarray*}
&&\Gamma ^{\dag }\left( T,\hat{T},\theta ,Z,Z^{\prime }\right) \\
&\rightarrow &\exp \left( -\int \frac{\bar{O}_{\hat{T}}}{\sigma _{\hat{T}%
}^{2}\omega \left( \theta ,Z,\left\vert \Psi \right\vert ^{2}\right) }d\hat{T%
}\right) \exp \left( -\int \frac{O_{T}}{\sigma _{T}^{2}}dT\right) \Gamma
^{\dag }\left( T,\hat{T},\theta ,Z,Z^{\prime }\right)
\end{eqnarray*}%
and $S_{\Gamma }^{\left( 1\right) }$ and $S_{\Gamma }^{\left( 2\right) }$ (%
\ref{Tcf}) write:%
\begin{eqnarray}
&&S_{\Gamma }^{\left( 1\right) }=\Gamma ^{\dag }\left( T,\hat{T},\theta
,Z,Z^{\prime }\right) \left( \frac{\sigma _{T}^{2}}{2}\nabla _{T}^{2}-\frac{%
O_{T}^{2}}{2\sigma _{T}^{2}}+K\left( \theta ,Z,Z^{\prime },\left\Vert \Psi
\right\Vert ^{2},\left\Vert \Gamma \right\Vert ^{2}\right) T\right.
\label{FFF} \\
&&\left. -\frac{\sigma _{\hat{T}}^{2}}{\sigma _{T}^{2}}\left( \nabla _{T}%
\frac{\lambda }{\omega }\left( \hat{T}-\lambda \tau \hat{T}\right) \right) -%
\frac{1}{2\tau \omega \left( Z\right) }\right) \Gamma \left( T,\hat{T}%
,\theta ,Z,Z^{\prime }\right)  \notag
\end{eqnarray}%
\begin{eqnarray}
S_{\Gamma }^{\left( 2\right) } &=&\Gamma ^{\dag }\left( T,\hat{T},\theta
,Z,Z^{\prime }\right) \left( \frac{\sigma _{\hat{T}}^{2}}{2}\nabla _{\hat{T}%
}^{2}-\frac{1}{2\sigma _{\hat{T}}^{2}}\bar{O}_{\hat{T}}^{2}\right.
\label{FC)} \\
&&-\left. \frac{\rho \left( C\left( \theta \right) \left\vert \Psi \left(
\theta ,Z\right) \right\vert ^{2}h_{C}-\eta H\left( \delta -T\right)
+D\left( \theta \right) \left\vert \Psi \left( \theta -\frac{\left\vert
Z-Z^{\prime }\right\vert }{c},Z^{\prime }\right) \right\vert
^{2}h_{D}\right) }{2\omega \left( \theta ,Z,\left\vert \Psi \right\vert
^{2}\right) }\right) \Gamma \left( T,\hat{T},\theta ,Z,Z^{\prime }\right) 
\notag
\end{eqnarray}%
with:%
\begin{eqnarray*}
h_{C}\left( \omega \left( \theta ,Z,\left\vert \Psi \right\vert ^{2}\right)
\right) &\equiv &h_{C} \\
h_{D}\left( \omega \left( \theta -\frac{\left\vert Z-Z^{\prime }\right\vert 
}{c},Z^{\prime },\left\vert \Psi \right\vert ^{2}\right) \right) &\equiv
&h_{D}
\end{eqnarray*}%
We aim at factoring:%
\begin{eqnarray}
\Gamma \left( T,\hat{T},\theta ,Z,Z^{\prime }\right) &=&\Gamma \left(
T,\theta ,Z,Z^{\prime }\right) \Gamma \left( \hat{T},\theta ,Z,Z^{\prime
}\right)  \label{FC!} \\
\Gamma ^{\dag }\left( T,\hat{T},\theta ,Z,Z^{\prime }\right) &=&\Gamma
^{\dag }\left( T,\theta ,Z,Z^{\prime }\right) \Gamma ^{\dag }\left( \hat{T}%
,\theta ,Z,Z^{\prime }\right)  \notag
\end{eqnarray}%
but the presence of the term:%
\begin{equation*}
\Gamma ^{\dag }\left( T,\hat{T},\theta ,Z,Z^{\prime }\right) \frac{\sigma _{%
\hat{T}}^{2}}{\sigma _{T}^{2}}\nabla _{T}\frac{\lambda }{\omega }\left( \hat{%
T}-\lambda \tau \hat{T}\right) \Gamma \left( T,\hat{T},\theta ,Z,Z^{\prime
}\right)
\end{equation*}%
in (\ref{FFF}) prevents this factorization since this contribution mixes
both variables $T,\hat{T}$. However, we may assume the variance of $T$ is
larger than the variance of $\hat{T}$. Actually, the value of connectivity
may depend on exogenous factors, whereas $\hat{T}$ is a variable counting
the output and input spikes. Thus, we can consider that $\frac{\sigma _{\hat{%
T}}^{2}}{\sigma _{T}^{2}}<<1$ and that%
\begin{equation*}
\left\vert \Gamma ^{\dag }\left( T,\hat{T},\theta ,Z,Z^{\prime }\right) 
\frac{\sigma _{\hat{T}}^{2}}{\sigma _{T}^{2}}\nabla _{T}\frac{\lambda }{%
\omega }\left( \hat{T}-\lambda \tau \hat{T}\right) \left\vert \Psi \left(
\theta ,Z\right) \right\vert ^{2}\Gamma \left( T,\hat{T},\theta ,Z,Z^{\prime
}\right) \right\vert <<1
\end{equation*}%
As a consequence the factorization (\ref{FC!}) holds in first approximation,
so that we can minimize (\ref{FFF}) in two parts, as we did for $C$ and $D$.

In Appendix 2, we show how to remove the hypothesis concerning $\hat{T}%
-\lambda \tau \hat{T}$. It amounts to solve directly the background state of
(\ref{Tcf}) both for $\hat{T}$ and $T$ , but the approximation we made above
is sufficient to understand the properties of $\Gamma _{0}\left( T,\theta
,Z,Z^{\prime }\right) $.

\subsection{Average values for $T$ and $\hat{T}$}

\subsubsection{General case}

Given that $C\left( \theta \right) $ and $D\left( \theta \right) $ given by (%
\ref{vrG}) and (\ref{vRG}).can be considered as slowly varying, we can look
for average values for $T$ and $\hat{T}$. They are obtained by setting the
quadratic potentials in the effective actions (\ref{FFF}) and (\ref{FC)}) to 
$0$. They are thus defined by:%
\begin{equation*}
-\frac{1}{\tau \omega \left( \theta ,Z\right) }\left\langle T\left(
Z,Z^{\prime }\right) \right\rangle +\frac{\lambda }{\omega \left( \theta
,Z\right) }\left\langle \hat{T}\left( Z,Z^{\prime }\right) \right\rangle =0
\end{equation*}%
and:%
\begin{eqnarray*}
0 &=&\left( h\left( Z,Z^{\prime }\right) -\left\langle \hat{T}\left(
Z,Z^{\prime }\right) \right\rangle \right) C\left( \theta \right) \left\vert
\Psi \left( \theta ,Z\right) \right\vert ^{2}h_{C}\left( \omega \left(
Z\right) \right) -\eta H\left( \delta -\left\langle T\left( Z,Z^{\prime
}\right) \right\rangle \right) \\
&&-\left\langle \hat{T}\left( Z,Z^{\prime }\right) \right\rangle D\left(
\theta \right) \left\vert \Psi \left( \theta -\frac{\left\vert Z-Z^{\prime
}\right\vert }{c},Z^{\prime }\right) \right\vert ^{2}h_{D}\left( \omega
\left( \theta -\frac{\left\vert Z-Z^{\prime }\right\vert }{c},Z^{\prime
}\right) \right)
\end{eqnarray*}

given by: 
\begin{equation*}
\left\langle T\left( Z,Z^{\prime }\right) \right\rangle =\lambda \tau
\left\langle \hat{T}\left( Z,Z^{\prime }\right) \right\rangle =\frac{\lambda
\tau \left( h\left( Z,Z^{\prime }\right) C_{Z,Z^{\prime }}\left( \theta
\right) h_{C}\left( \omega \left( \theta ,Z\right) \right) \left\vert \Psi
\left( \theta ,Z\right) \right\vert ^{2}-\eta H\left( \delta -\left\langle
T\left( Z,Z^{\prime }\right) \right\rangle \right) \right) }{h_{C}\left(
\omega \left( \theta ,Z\right) \right) \left\vert \bar{\Psi}\left( \theta
,Z,Z^{\prime }\right) \right\vert ^{2}}
\end{equation*}%
where:%
\begin{equation}
\left\vert \bar{\Psi}\left( \theta ,Z,Z^{\prime }\right) \right\vert ^{2}=%
\frac{C_{Z,Z^{\prime }}\left( \theta \right) \left\vert \Psi \left( \theta
,Z\right) \right\vert ^{2}h_{C}\left( \omega \left( \theta ,Z\right) \right)
+D_{Z,Z^{\prime }}\left( \theta \right) \left\vert \Psi \left( \theta -\frac{%
\left\vert Z-Z^{\prime }\right\vert }{c},Z^{\prime }\right) \right\vert
^{2}h_{D}\left( \omega \left( \theta -\frac{\left\vert Z-Z^{\prime
}\right\vert }{c},Z^{\prime }\right) \right) }{h_{C}\left( \omega \left(
\theta ,Z\right) \right) }  \label{PB}
\end{equation}%
is a weighted sum of the values of field $\left\vert \Psi \left( \theta
,Z\right) \right\vert ^{2}$ and $\left\vert \Psi \left( \theta -\frac{%
\left\vert Z-Z^{\prime }\right\vert }{c},Z^{\prime }\right) \right\vert ^{2}$%
.

The threshold $\delta $ implies three possibilities.

First, if:%
\begin{equation*}
\frac{\lambda \tau \left( h\left( Z,Z^{\prime }\right) C_{Z,Z^{\prime
}}\left( \theta \right) h_{C}\left( \omega \left( \theta ,Z\right) \right)
\left\vert \Psi \left( \theta ,Z\right) \right\vert ^{2}\right) }{%
h_{C}\left( \omega \left( \theta ,Z\right) \right) \left\vert \bar{\Psi}%
\left( \theta ,Z,Z^{\prime }\right) \right\vert ^{2}}<\delta
\end{equation*}%
then the average connectivity is given by:%
\begin{equation*}
\left\langle T\left( Z,Z^{\prime }\right) \right\rangle =\sup \left( \frac{%
\lambda \tau \left( h\left( Z,Z^{\prime }\right) C_{Z,Z^{\prime }}\left(
\theta \right) h_{C}\left( \omega \left( \theta ,Z\right) \right) \left\vert
\Psi \left( \theta ,Z\right) \right\vert ^{2}-\eta \right) }{h_{C}\left(
\omega \left( \theta ,Z\right) \right) \left\vert \bar{\Psi}\left( \theta
,Z,Z^{\prime }\right) \right\vert ^{2}},0\right)
\end{equation*}%
Second, if on the contrary:%
\begin{equation*}
\frac{\lambda \tau \left( \left( h\left( Z,Z^{\prime }\right) C_{Z,Z^{\prime
}}\left( \theta \right) h_{C}\left( \omega \left( \theta ,Z\right) \right)
\left\vert \Psi \left( \theta ,Z\right) \right\vert ^{2}\right) -\eta
\right) }{h_{C}\left( \omega \left( \theta ,Z\right) \right) \left\vert \bar{%
\Psi}\left( \theta ,Z,Z^{\prime }\right) \right\vert ^{2}}>\delta
\end{equation*}%
then the average connectivity is:%
\begin{equation*}
\left\langle T\left( Z,Z^{\prime }\right) \right\rangle =\frac{\lambda \tau
\left( h\left( Z,Z^{\prime }\right) C_{Z,Z^{\prime }}\left( \theta \right)
h_{C}\left( \omega \left( \theta ,Z\right) \right) \left\vert \Psi \left(
\theta ,Z\right) \right\vert ^{2}-\eta \right) }{h_{C}\left( \omega \left(
\theta ,Z\right) \right) \left\vert \bar{\Psi}\left( \theta ,Z,Z^{\prime
}\right) \right\vert ^{2}}
\end{equation*}%
Third, if:%
\begin{eqnarray*}
\frac{\lambda \tau \left( h\left( Z,Z^{\prime }\right) C_{Z,Z^{\prime
}}\left( \theta \right) h_{C}\left( \omega \left( \theta ,Z\right) \right)
\left\vert \Psi \left( \theta ,Z\right) \right\vert ^{2}\right) }{%
h_{C}\left( \omega \left( \theta ,Z\right) \right) \left\vert \bar{\Psi}%
\left( \theta ,Z,Z^{\prime }\right) \right\vert ^{2}} &>&\delta \\
\frac{\lambda \tau \left( h\left( Z,Z^{\prime }\right) C_{Z,Z^{\prime
}}\left( \theta \right) h_{C}\left( \omega \left( \theta ,Z\right) \right)
\left\vert \Psi \left( \theta ,Z\right) \right\vert ^{2}-\eta \right) }{%
h_{C}\left( \omega \left( \theta ,Z\right) \right) \left\vert \bar{\Psi}%
\left( \theta ,Z,Z^{\prime }\right) \right\vert ^{2}} &<&\delta
\end{eqnarray*}%
both solutions:%
\begin{eqnarray*}
\left\langle T\right\rangle &=&\sup \left( \frac{\lambda \tau \left( h\left(
Z,Z^{\prime }\right) C_{Z,Z^{\prime }}\left( \theta \right) h_{C}\left(
\omega \left( \theta ,Z\right) \right) \left\vert \Psi \left( \theta
,Z\right) \right\vert ^{2}-\eta \right) }{h_{C}\left( \omega \left( \theta
,Z\right) \right) \left\vert \bar{\Psi}\left( \theta ,Z,Z^{\prime }\right)
\right\vert ^{2}},0\right) \\
\left\langle T\right\rangle &=&\frac{\lambda \tau h\left( Z,Z^{\prime
}\right) C_{Z,Z^{\prime }}\left( \theta \right) h_{C}\left( \omega \left(
\theta ,Z\right) \right) \left\vert \Psi \left( \theta ,Z\right) \right\vert
^{2}}{h_{C}\left( \omega \left( \theta ,Z\right) \right) \left\vert \bar{\Psi%
}\left( \theta ,Z,Z^{\prime }\right) \right\vert ^{2}}
\end{eqnarray*}%
are possible.

\subsubsection{Average values for $T$ and $\hat{T}$ for sharp threshold}

The previous averages simplify for $\delta <<1$. If: 
\begin{equation*}
\lambda \tau \left( h\left( Z,Z^{\prime }\right) C_{Z,Z^{\prime }}\left(
\theta \right) h_{C}\left( \omega \left( \theta ,Z\right) \right) \left\vert
\Psi \left( \theta ,Z\right) \right\vert ^{2}-\eta \right) >0
\end{equation*}%
then:%
\begin{equation}
\left\langle T\left( Z,Z^{\prime }\right) \right\rangle =\frac{\lambda \tau
h\left( Z,Z^{\prime }\right) C_{Z,Z^{\prime }}\left( \theta \right)
\left\vert \Psi \left( \theta ,Z\right) \right\vert ^{2}}{\left\vert \bar{%
\Psi}\left( \theta ,Z,Z^{\prime }\right) \right\vert ^{2}}  \label{vrg}
\end{equation}%
and if:%
\begin{equation*}
\lambda \tau \left( h\left( Z,Z^{\prime }\right) C_{Z,Z^{\prime }}\left(
\theta \right) h_{C}\left( \omega \left( \theta ,Z\right) \right) \left\vert
\Psi \left( \theta ,Z\right) \right\vert ^{2}-\eta \right) <0
\end{equation*}%
\begin{equation*}
\left\langle T\left( Z,Z^{\prime }\right) \right\rangle =0
\end{equation*}%
otherwise.

\subsection{Background state for $T$ and $\hat{T}$}

Once the average values for $\left\langle T\left( Z,Z^{\prime }\right)
\right\rangle $ are obtained, we can close the resolution for the background 
$\Gamma \left( T,\hat{T},\theta ,Z,Z^{\prime }\right) $.We will consider a
sharp threshold only and consider two cases separately.

\subsubsection{First case, cleared threshold $T\left( Z,Z^{\prime }\right)
>0 $}

For points such that:%
\begin{equation*}
h\left( Z,Z^{\prime }\right) \left\langle C_{Z,Z^{\prime }}\left( \theta
\right) h_{C}\left( \omega \left( \theta ,Z\right) \right) \left\vert \Psi
\left( \theta ,Z\right) \right\vert ^{2}\right\rangle -\eta >0
\end{equation*}%
then the averages are:%
\begin{equation}
\left\langle T\left( Z,Z^{\prime }\right) \right\rangle =\lambda \tau
\left\langle \hat{T}\left( Z,Z^{\prime }\right) \right\rangle =\frac{\lambda
\tau h\left( Z,Z^{\prime }\right) \left\langle C_{Z,Z^{\prime }}\left(
\theta \right) \left\vert \Psi \left( \theta ,Z\right) \right\vert
^{2}\right\rangle }{\left\vert \bar{\Psi}\left( \theta ,Z,Z^{\prime }\right)
\right\vert ^{2}}  \label{THt}
\end{equation}%
Whereas, for points such that:%
\begin{equation*}
h\left( Z,Z^{\prime }\right) \left\langle C_{Z,Z^{\prime }}\left( \theta
\right) h_{C}\left( \omega \left( \theta ,Z\right) \right) \left\vert \Psi
\left( \theta ,Z\right) \right\vert ^{2}\right\rangle -\eta <0
\end{equation*}%
then:%
\begin{equation*}
\left\langle T\left( Z,Z^{\prime }\right) \right\rangle =0
\end{equation*}%
\begin{equation}
\left\langle \hat{T}\left( Z,Z^{\prime }\right) \right\rangle =\frac{h\left(
Z,Z^{\prime }\right) \left\langle C_{Z,Z^{\prime }}\left( \theta \right)
h_{C}\left( \omega \left( \theta ,Z\right) \right) \left\vert \Psi \left(
\theta ,Z\right) \right\vert ^{2}\right\rangle -\eta }{\left\langle
h_{C}\left( \omega \left( \theta ,Z\right) \right) \left\vert \bar{\Psi}%
\left( \theta ,Z,Z^{\prime }\right) \right\vert ^{2}\right\rangle }<0
\label{Thn}
\end{equation}

To compute the background state, we proceed in two steps as for $C$ and $D$.

We consider the two cases separately. If $T\left( Z,Z^{\prime }\right) $ is
given by (\ref{THt}), rxpression (\ref{FC)}) is:%
\begin{eqnarray}
S_{\Gamma }^{\left( 2\right) } &=&\Gamma ^{\dag }\left( T,\hat{T},\theta
,Z,Z^{\prime }\right) \left( \frac{\sigma _{\hat{T}}^{2}}{2}\nabla _{\hat{T}%
}^{2}-\frac{1}{2\sigma _{\hat{T}}^{2}}\left( \frac{\rho \left( h_{C}\left(
\omega \left( \theta ,Z\right) \right) \left\vert \bar{\Psi}\left( \theta
,Z,Z^{\prime }\right) \right\vert ^{2}\left( \hat{T}-\left\langle \hat{T}%
\right\rangle \right) \right) }{\omega \left( \theta ,Z,\left\vert \Psi
\right\vert ^{2}\right) }\right) ^{2}\right.  \label{RDT} \\
&&-\left. \frac{\rho h_{C}\left( \omega \left( \theta ,Z\right) \right)
\left\vert \bar{\Psi}\left( \theta ,Z,Z^{\prime }\right) \right\vert ^{2}}{%
2\omega \left( \theta ,Z,\left\vert \Psi \right\vert ^{2}\right) }\right)
\Gamma \left( T,\hat{T},\theta ,Z,Z^{\prime }\right)  \notag
\end{eqnarray}%
with $\left\langle \hat{T}\right\rangle $ given by (\ref{THt}).

As explained in the previous paragraph, given the form (\ref{FFF}) of $%
S_{\Gamma }^{\left( 1\right) }$ and (\ref{RDT}) of $S_{\Gamma }^{\left(
2\right) }$, $\Gamma \left( T,\hat{T},\theta ,Z,Z^{\prime }\right) $ factors
in first approximation as:%
\begin{equation}
\Gamma \left( T,\hat{T},\theta ,Z,Z^{\prime }\right) =\Gamma _{0}\left(
T,\theta ,Z,Z^{\prime }\right) \Gamma _{0}\left( \hat{T},\theta ,Z,Z^{\prime
}\right)  \label{FCt}
\end{equation}

Introducing a constraint normalizing $\left\Vert \Gamma _{0}\left( \hat{T}%
,\theta ,Z,Z^{\prime }\right) \right\Vert ^{2}$ to $1$ allows to proceed as
for the background state for $C$ and $D$ and minimizing (\ref{RDT}) under
the constraint allows to project $\Gamma \left( T,\hat{T},\theta
,Z,Z^{\prime }\right) $ on the background state:

\begin{eqnarray}
&&\Gamma _{0}\left( T,\hat{T},\theta ,Z,Z^{\prime }\right)  \label{GRTH} \\
&=&\Gamma _{0}\left( T,\theta ,Z,Z^{\prime }\right) \Gamma _{0}\left( \hat{T}%
,\theta ,Z,Z^{\prime }\right)  \notag \\
&=&\Gamma _{0}\left( T,\hat{T},\theta ,Z\right) \exp \left( -\frac{\rho
h_{C}\left( \omega \left( \theta ,Z\right) \right) \left\vert \bar{\Psi}%
\left( \theta ,Z,Z^{\prime }\right) \right\vert ^{2}}{4\sigma _{\hat{T}%
}^{2}\omega \left( \theta ,Z,\left\vert \Psi \right\vert ^{2}\right) }\left(
\left( \hat{T}-\left\langle \hat{T}\right\rangle \right) \right) ^{2}\right)
\notag
\end{eqnarray}%
and in this state, (\ref{RDT}) becomes:%
\begin{equation}
S_{\Gamma }^{\left( 2\right) }=\Gamma _{0}^{\dag }\left( T,\theta
,Z,Z^{\prime }\right) \left( \frac{\rho h_{C}\left( \omega \left( \theta
,Z\right) \right) \left\vert \bar{\Psi}\left( \theta ,Z,Z^{\prime }\right)
\right\vert ^{2}}{\omega \left( \theta ,Z,\left\vert \Psi \right\vert
^{2}\right) }\right) \Gamma _{0}\left( T,\theta ,Z,Z^{\prime }\right)
\label{rst}
\end{equation}%
Ultimately, using (\ref{FFF}) and (\ref{Tht}) we can rewrite $S_{\Gamma
}^{\left( 1\right) }$:%
\begin{equation*}
S_{\Gamma }^{\left( 1\right) }=\Gamma _{0}^{\dag }\left( T,\theta
,Z,Z^{\prime }\right) \left( \frac{\sigma _{T}^{2}}{2}\nabla _{T}^{2}-\frac{1%
}{2\sigma _{T}^{2}}\left( \left( \frac{1}{\tau \omega }\left( T-\left\langle
T\right\rangle \right) \right) \right) ^{2}-\frac{1}{2\tau \omega \left(
Z\right) }\right) \Gamma _{0}\left( T,\theta ,Z,Z^{\prime }\right)
\end{equation*}%
To obtain the complete action we have to add the contributions (\ref{Rsn})
and (\ref{rst}). We find ultimately the action for $\Gamma _{0}\left(
T,\theta ,Z,Z^{\prime }\right) $: 
\begin{eqnarray}
&&\Gamma _{0}^{\dag }\left( T,\theta ,Z,Z^{\prime }\right) \left( \frac{%
\sigma _{T}^{2}}{2}\nabla _{T}^{2}-\frac{1}{2\sigma _{T}^{2}}\left( \left( 
\frac{1}{\tau \omega }\left( T-\left\langle T\right\rangle \right) \right)
\left\vert \Psi \left( \theta ,Z\right) \right\vert ^{2}\right) ^{2}-\frac{1%
}{2\tau \omega \left( Z\right) }\right) \Gamma _{0}\left( T,\theta
,Z,Z^{\prime }\right)  \label{Ftn} \\
&&-\Gamma _{0}^{\dag }\left( T,\theta ,Z,Z^{\prime }\right) \left(
a_{C}\left( Z\right) +a_{D}\left( Z\right) +\frac{\rho h_{C}\left( \omega
\left( \theta ,Z\right) \right) \left\vert \bar{\Psi}\left( \theta
,Z,Z^{\prime }\right) \right\vert ^{2}}{\omega \left( \theta ,Z,\left\vert
\Psi \right\vert ^{2}\right) }\right) \Gamma _{0}\left( T,\theta
,Z,Z^{\prime }\right)  \notag
\end{eqnarray}%
This action results from the successive projections of partial background
states and becomes a function of $T$ only.

At this point we aim at minimizing (\ref{Ftn}), as we did for the other
parts of the background state. However, a difference appears. The potential:%
\begin{equation}
U\left( \left\{ \left\vert \Gamma \left( T,\hat{T},\theta ,Z,Z^{\prime
},C,D\right) \right\vert ^{2}\right\} \right)  \label{PTl}
\end{equation}%
introduced in the full action imposes an overall constraint for the whole
set of connections in the system. We assume for the sake of simplicity that $%
U$ depends only on the connectivities at points $\left( Z,Z^{\prime }\right) 
$, i.e. on:%
\begin{equation*}
\left\{ \int \left\vert \Gamma \left( T,\hat{T},\theta ,Z,Z^{\prime
},C,D\right) \right\vert ^{2}d\left( T,\hat{T},\theta ,C,D\right) \right\}
=\left\{ \int \left\vert \Gamma \left( T,\hat{T},\theta ,Z,Z^{\prime
},C,D\right) \right\vert ^{2}d\left( T,\hat{T},\theta ,C,D\right) \right\}
\end{equation*}%
Given our assumption about the norms of the fields arising in the
decomposition of $\Gamma \left( T,\hat{T},\theta ,Z,Z^{\prime },C,D\right) $%
, we have:%
\begin{equation*}
\int \left\vert \Gamma \left( T,\hat{T},\theta ,Z,Z^{\prime },C,D\right)
\right\vert ^{2}d\left( T,\hat{T},C,D\right) =\int \left\Vert \Gamma
_{0}\left( T,\theta ,Z,Z^{\prime }\right) \right\Vert ^{2}dT=\left\Vert
\Gamma _{0}\left( Z,Z^{\prime }\right) \right\Vert ^{2}
\end{equation*}%
and the potential becomes:%
\begin{equation*}
U\left( \left\{ \left\vert \Gamma \left( T,\hat{T},\theta ,Z,Z^{\prime
},C,D\right) \right\vert ^{2}\right\} \right) =U\left( \left\{ \left\Vert
\Gamma _{0}\left( Z,Z^{\prime }\right) \right\Vert ^{2}\right\} \right)
\end{equation*}%
Including the potential in (\ref{Ftn}) leads thus to the saddle point
equation:%
\begin{eqnarray}
&&\left( \frac{\sigma _{T}^{2}}{2}\nabla _{T}^{2}-\frac{1}{2\sigma _{T}^{2}}%
\left( \left( \frac{1}{\tau \omega }\left( T-\left\langle T\right\rangle
\right) \right) \left\vert \Psi \left( \theta ,Z\right) \right\vert
^{2}\right) ^{2}-\frac{1}{\tau \omega \left( Z\right) }-2a_{C}\left(
Z\right) -2a_{D}\left( Z\right) \right.  \label{SDq} \\
&&-\left. \frac{\rho h_{C}\left( \omega \left( \theta ,Z\right) \right)
\left\vert \bar{\Psi}\left( \theta ,Z,Z^{\prime }\right) \right\vert ^{2}}{%
\omega \left( \theta ,Z,\left\vert \Psi \right\vert ^{2}\right) }-\frac{%
\delta U\left( \left\{ \left\Vert \Gamma _{0}\left( \theta ,Z,Z^{\prime
}\right) \right\Vert ^{2}\right\} \right) }{\delta \left\Vert \Gamma
_{0}\left( \theta ,Z,Z^{\prime }\right) \right\Vert ^{2}}\right) \Gamma
_{0}\left( T,\theta ,Z,Z^{\prime }\right)  \notag
\end{eqnarray}%
This saddle point equation has a minimum at $\left( Z,Z^{\prime }\right) $
for:%
\begin{eqnarray}
0 &=&\frac{1}{\tau \omega \left( Z\right) }+\frac{1}{2}a_{C}\left( Z\right) +%
\frac{1}{2}a_{D}\left( Z\right)  \label{SLt} \\
&&+\frac{\rho h_{C}\left( \omega \left( \theta ,Z\right) \right) \left\vert 
\bar{\Psi}\left( \theta ,Z,Z^{\prime }\right) \right\vert ^{2}}{2\omega
\left( \theta ,Z,\left\vert \Psi \right\vert ^{2}\right) }+\frac{\delta
U\left( \left\{ \left\Vert \Gamma _{0}\left( \theta ,Z,Z^{\prime }\right)
\right\Vert ^{2}\right\} \right) }{\delta \left\Vert \Gamma _{0}\left(
\theta ,Z,Z^{\prime }\right) \right\Vert ^{2}}  \notag
\end{eqnarray}%
and this set of equations yields the norm $\left\Vert \Gamma _{0}\left(
\theta ,Z,Z^{\prime }\right) \right\Vert ^{2}$ at each point.

The background $\Gamma _{0}\left( T,\theta ,Z,Z^{\prime }\right) $ is given
by:%
\begin{equation}
\Gamma _{0}\left( T,\theta ,Z,Z^{\prime }\right) =\left\Vert \Gamma
_{0}\left( \theta ,Z,Z^{\prime }\right) \right\Vert \exp \left( -\frac{1}{%
4\sigma _{T}^{2}\tau \omega }\left( \left( T-\left\langle T\right\rangle
\right) \right) ^{2}\right)  \label{gMR}
\end{equation}%
Using (\ref{SLt}), the sum of (\ref{Ftn}) and (\ref{PTl}) at point ($%
Z,Z^{\prime }$) for this state writes:%
\begin{equation}
S\left( \left\Vert \Gamma _{0}\left( Z,Z^{\prime }\right) \right\Vert
^{2}\right) =U\left( \left\{ \left\Vert \Gamma _{0}\left( Z,Z^{\prime
}\right) \right\Vert ^{2}\right\} \right) -\frac{\delta U\left( \left\{
\left\Vert \Gamma _{0}\left( \theta ,Z,Z^{\prime }\right) \right\Vert
^{2}\right\} \right) }{\delta \left\Vert \Gamma _{0}\left( \theta
,Z,Z^{\prime }\right) \right\Vert ^{2}}\left\Vert \Gamma _{0}\left(
Z,Z^{\prime }\right) \right\Vert ^{2}  \label{CTf}
\end{equation}%
This expression yields the condition for the existence of a non trivial
minimum for the action at $\left( Z,Z^{\prime }\right) $.

If $S\left( \left\Vert \Gamma _{0}\left( Z,Z^{\prime }\right) \right\Vert
^{2}\right) <0$, the state:%
\begin{equation*}
\Gamma \left( T,\hat{T},\theta ,Z,Z^{\prime },C,D\right) =\Gamma _{1}\left(
Z,Z^{\prime },C\right) \Gamma _{2}\left( Z,Z^{\prime },D\right) \Gamma
_{0}\left( T,\theta ,Z,Z^{\prime }\right) \Gamma _{0}\left( \hat{T},\theta
,Z,Z^{\prime }\right)
\end{equation*}%
with the various contributions defined by (\ref{GMTH}), (\ref{GRTH}) and (%
\ref{gMR}), is a non trivial minimum. Otherwise the minimum of the action
for the connectivity field is:%
\begin{equation*}
\Gamma \left( T,\hat{T},\theta ,Z,Z^{\prime },C,D\right) =0
\end{equation*}%
As a consequence, given the parameters arising in (\ref{SLt}), that is, the
average activity at some points, the density of active neurons or axons $%
\left\vert \Psi \left( \theta ,Z\right) \right\vert ^{2}$, and depending on
the potential, the non-trivial state (\ref{gMR}) $\Gamma \left( T,\hat{T}%
,\theta ,Z,Z^{\prime },C,D\right) $ may be a minimum of (\ref{CTf}). Since
the parameters describing the activity vary across space defined by the
points $\left( Z,Z^{\prime }\right) $, we can expect some islands of
connectivity. Submanifolds satisfying $S\left( \left\Vert \Gamma _{0}\left(
Z,Z^{\prime }\right) \right\Vert ^{2}\right) <0$ present connectivity
patterns, whereas those satisfying $S\left( \left\Vert \Gamma _{0}\left(
Z,Z^{\prime }\right) \right\Vert ^{2}\right) >0$ present low levels of
connectivity.

\subsubsection{Second case: $T\left( Z,Z^{\prime }\right) =0$}

In the second case, i.e. $T\left( Z,Z^{\prime }\right) =0$ we find the $\hat{%
T}$ dependency of the background state as in the first case, and we obtain:%
\begin{eqnarray*}
&&\Gamma \left( T,\hat{T},\theta ,Z,Z^{\prime }\right) \\
&=&\Gamma _{0}\left( T,\hat{T},\theta ,Z\right) \\
&&\times \exp \left( -\left( \frac{\rho h_{C}\left( \omega \left( \theta
,Z\right) \right) \left\vert \bar{\Psi}\left( \theta ,Z,Z^{\prime }\right)
\right\vert ^{2}\left( \hat{T}-\left\langle \hat{T}\right\rangle \right) }{%
4\sigma _{\hat{T}}^{2}\omega \left( \theta ,Z,\left\vert \Psi \right\vert
^{2}\right) }\right) ^{2}\right)
\end{eqnarray*}%
but now, $\left\langle \hat{T}\right\rangle $ is given by formula (\ref{Thn}%
).

The difference with the first case arises when looking for the dependency in 
$\left\langle T\right\rangle $. When $\left\langle T\left( Z,Z^{\prime
}\right) \right\rangle =0$, the threshold makes the background state is
peaked at $T=0$. Only a significative change in activities and field $%
\left\vert \Psi \right\vert ^{2}$, i.e. a change in background state, allows
to depart from $T=0$. We thus have:%
\begin{equation*}
\Gamma _{0}\left( T,\hat{T},\theta ,Z\right) =\delta \left( T\right)
\end{equation*}

\subsubsection{Extension global constraint}

In addition to the potential $U\left( \left\{ \left\Vert \Gamma _{0}\left(
Z,Z^{\prime }\right) \right\Vert ^{2}\right\} \right) $, a global constraint
on the overall level of connectivity could be introduced, either with a
potential $U\left( \left\{ \left\Vert \Gamma _{0}\right\Vert ^{2}\right\}
\right) $ with $\left\Vert \Gamma _{0}\right\Vert ^{2}=\int \left\Vert
\Gamma _{0}\left( Z,Z^{\prime }\right) \right\Vert ^{2}d\left( Z,Z^{\prime
}\right) $, or with a constraint of the form:%
\begin{equation*}
\left\Vert \Gamma _{0}\right\Vert ^{2}=\overline{\left\Vert \Gamma
\right\Vert }^{2}
\end{equation*}%
This constraint may be introduced through a Lagrange multiplier $\alpha _{0}$
in (\ref{SDq}). The resolution is identical as in the previous paragraph and
amounts to add a term $\alpha _{0}$ in (\ref{SLt}). This implies that the
solution of this equation becomes a function of $\alpha _{0}$ $\left\Vert
\Gamma _{0}\left( \theta ,Z,Z^{\prime },\alpha _{0}\right) \right\Vert ^{2}$%
, the value of $\alpha _{0}$ being obtained by implementing:%
\begin{equation*}
\int \left\Vert \Gamma _{0}\left( \theta ,Z,Z^{\prime },\alpha _{0}\right)
\right\Vert ^{2}=\overline{\left\Vert \Gamma \right\Vert }^{2}
\end{equation*}%
Again depending on the sign of the action:%
\begin{equation}
S\left( \left\Vert \Gamma _{0}\left( Z,Z^{\prime }\right) \right\Vert
^{2}\right) =U\left( \left\{ \left\Vert \Gamma _{0}\left( Z,Z^{\prime
}\right) \right\Vert ^{2}\right\} \right) +\left( \alpha _{0}-\frac{\delta
U\left( \left\{ \left\Vert \Gamma _{0}\left( \theta ,Z,Z^{\prime }\right)
\right\Vert ^{2}\right\} \right) }{\delta \left\Vert \Gamma _{0}\left(
\theta ,Z,Z^{\prime }\right) \right\Vert ^{2}}\right) \left\Vert \Gamma
_{0}\left( Z,Z^{\prime }\right) \right\Vert ^{2}
\end{equation}%
i.e. for $S\left( \left\Vert \Gamma _{0}\left( Z,Z^{\prime }\right)
\right\Vert ^{2}\right) <0$, a nontrivial state for $\Gamma _{0}\left( T,%
\hat{T},\theta ,Z\right) $ will exist.

\section{Full background state for connectivity and static averages}

In the previous section, we determined the form of the background
connectivity state between two points, $Z$ and $Z^{\prime }$. The full
background state is, therefore, the tensor product of such states for every
pair $\left( Z,Z^{\prime }\right) $. However, to complete the description,
the average values of connectivity in these states need to be derived. The
entire background state is depicted by a set of equations for these averages.

In this section, by performing the integrations arising from the various
changes of variables, we can consolidate the previous results by providing
the background state of the system and the equations for the averages. Then,
as an example, we will solve these equations under some simplifying
assumptions.

\subsection{Fields}

To write the background state of the system in a compact form, we define:%
\begin{eqnarray*}
\Gamma \left( T,\hat{T},\theta ,C,D\right) &=&\left\{ \Gamma \left( T,\hat{T}%
,\theta ,Z,Z^{\prime },C,D\right) \right\} _{\left( Z,Z^{\prime }\right) } \\
\Gamma ^{\dagger }\left( T,\hat{T},\theta ,C,D\right) &=&\left\{ \Gamma
^{\dagger }\left( T,\hat{T},\theta ,Z,Z^{\prime },C,D\right) \right\}
_{\left( Z,Z^{\prime }\right) } \\
\left\vert \Gamma \left( T,\hat{T},\theta ,C,D\right) \right\vert ^{2}
&=&\left\{ \left\vert \Gamma \left( T,\hat{T},\theta ,Z,Z^{\prime
},C,D\right) \right\vert ^{2}\right\} _{\left( Z,Z^{\prime }\right) }
\end{eqnarray*}%
These fields decompose as:%
\begin{eqnarray*}
\Gamma \left( T,\hat{T},\theta ,C,D\right) &=&\left( \Gamma _{a}\left( T,%
\hat{T},\theta ,C,D\right) ,\Gamma _{u}\left( T,\hat{T},\theta ,C,D\right)
\right) \\
\Gamma ^{\dagger }\left( T,\hat{T},\theta ,C,D\right) &=&\left( \Gamma
_{a}^{\dagger }\left( T,\hat{T},\theta ,C,D\right) ,\Gamma _{u}^{\dagger
}\left( T,\hat{T},\theta ,C,D\right) \right) \\
\left\vert \Gamma \right\vert ^{2}\left( T,\hat{T},\theta ,C,D\right)
&=&\left( \left\vert \Gamma \right\vert _{a}^{2}\left( T,\hat{T},\theta
,C,D\right) ,\left\vert \Gamma \right\vert _{u}^{2}\left( T,\hat{T},\theta
,C,D\right) \right)
\end{eqnarray*}%
where%
\begin{eqnarray*}
\Gamma _{a/u}\left( T,\hat{T},\theta ,C,D\right) &=&\left\{ \Gamma \left( T,%
\hat{T},\theta ,Z,Z^{\prime },C,D\right) \right\} _{\left( Z,Z^{\prime
}\right) ,\left\langle T\left( Z,Z^{\prime }\right) \right\rangle \neq
0/\left\langle T\left( Z,Z^{\prime }\right) \right\rangle =0} \\
\Gamma _{a/u}^{\dagger }\left( T,\hat{T},\theta ,C,D\right) &=&\left\{
\Gamma ^{\dagger }\left( T,\hat{T},\theta ,Z,Z^{\prime },C,D\right) \right\}
_{\left( Z,Z^{\prime }\right) ,\left\langle T\left( Z,Z^{\prime }\right)
\right\rangle \neq 0/\left\langle T\left( Z,Z^{\prime }\right) \right\rangle
=0} \\
\left\vert \Gamma _{a/u}\left( T,\hat{T},\theta ,C,D\right) \right\vert ^{2}
&=&\left\{ \left\vert \Gamma \left( T,\hat{T},\theta ,Z,Z^{\prime
},C,D\right) \right\vert ^{2}\right\} _{\left( Z,Z^{\prime }\right)
,\left\langle T\left( Z,Z^{\prime }\right) \right\rangle \neq 0/\left\langle
T\left( Z,Z^{\prime }\right) \right\rangle =0}
\end{eqnarray*}%
The subscripts refer to active or unactive doublet $\left( Z,Z^{\prime
}\right) $. The expression for $\Gamma _{a},\Gamma _{u}$ and their
conjugates are: 
\begin{eqnarray}
&&\Gamma _{a}\left( T,\hat{T},\theta ,C,D\right)  \label{gmcc} \\
&\simeq &\left\{ \mathcal{N}\exp \left( -\frac{1}{2\sigma _{C}^{2}}\left( 
\frac{1}{\tau _{C}\omega }+\alpha _{C}\frac{\omega ^{\prime }\left\vert \Psi
\left( \theta -\frac{\left\vert Z-Z^{\prime }\right\vert }{c},Z^{\prime
}\right) \right\vert ^{2}}{\omega }\right) \left( C-C\left( \theta \right)
\right) ^{2}\right) \right.  \notag \\
&&\times \exp \left( -\frac{\left( \frac{1}{\tau _{D}\omega }+\alpha
_{D}\left\vert \Psi \left( \theta ,Z\right) \right\vert ^{2}\right) }{%
2\sigma _{D}^{2}}\left( D-D\left( \theta \right) \right) ^{2}\right)  \notag
\\
&&\times \exp \left( -\frac{\rho h_{C}\left( \omega \left( \theta ,Z\right)
\right) \left\vert \bar{\Psi}\left( \theta ,Z,Z^{\prime }\right) \right\vert
^{2}}{2\sigma _{\hat{T}}^{2}\omega \left( \theta ,Z,\left\vert \Psi
\right\vert ^{2}\right) }\left( \left( \hat{T}-\left\langle \hat{T}%
\right\rangle \right) \right) ^{2}\right)  \notag \\
&&\times \left. \left\Vert \Gamma _{0}\left( \theta ,Z,Z^{\prime }\right)
\right\Vert \exp \left( -\frac{\left\vert \Psi \left( \theta ,Z\right)
\right\vert ^{2}}{2\sigma _{T}^{2}\tau \omega }\left( \left( T-\left\langle
T\right\rangle \right) \right) ^{2}\right) \right\} _{\left( Z,Z^{\prime
}\right) ,\left\langle T\left( Z,Z^{\prime }\right) \right\rangle \neq 0} 
\notag
\end{eqnarray}%
\begin{eqnarray}
&&\Gamma _{u}\left( T,\hat{T},\theta ,C,D\right)  \label{GMuu} \\
&\simeq &\left\{ \mathcal{N}\exp \left( -\frac{1}{2}a_{C}\left( Z\right)
\left( C-C\left( \theta \right) \right) ^{2}\right) \right. \exp \left( -%
\frac{a_{D}\left( Z\right) }{2}\left( D-\left\langle D\right\rangle \right)
^{2}\right)  \notag \\
&&\left. \times \exp \left( -\frac{\rho h_{C}\left( \omega \left( \theta
,Z\right) \right) \left\vert \bar{\Psi}\left( \theta ,Z,Z^{\prime }\right)
\right\vert ^{2}}{2\omega \left( \theta ,Z,\left\vert \Psi \right\vert
^{2}\right) }\left( \left( \hat{T}-\left\langle \hat{T}\right\rangle \right)
\right) ^{2}\right) \delta \left( T\right) \right\} _{\left( Z,Z^{\prime
}\right) ,\left\langle T\left( Z,Z^{\prime }\right) \right\rangle =0}  \notag
\end{eqnarray}%
with (see (\ref{CD})):%
\begin{eqnarray}
a_{C}\left( Z\right) &=&\frac{1}{\tau _{C}\omega }+\alpha _{C}\frac{\omega
^{\prime }\left\vert \Psi \left( \theta -\frac{\left\vert Z-Z^{\prime
}\right\vert }{c},Z^{\prime },\omega ^{\prime }\right) \right\vert ^{2}}{%
\omega } \\
a_{D}\left( Z\right) &=&\frac{1}{\tau _{D}\omega }+\alpha _{D}\left\vert
\Psi \left( \theta ,Z\right) \right\vert ^{2}  \notag
\end{eqnarray}%
and where $\mathcal{N}$ is a normalization factor ensuring that the
constraint over the number of connections is satisfied.%
\begin{equation*}
\Gamma _{a}^{\dagger }\left( T,\hat{T},\theta ,C,D\right) \simeq \left\{
1\right\} _{\left( Z,Z^{\prime }\right) ,\left\langle T\left( Z,Z^{\prime
}\right) \right\rangle \neq 0}
\end{equation*}%
\begin{equation*}
\Gamma _{u}^{\dagger }\left( T,\hat{T},\theta ,C,D\right) \simeq \left\{
\delta \left( T\right) \right\} _{\left( Z,Z^{\prime }\right) ,\left\langle
T\left( Z,Z^{\prime }\right) \right\rangle =0}
\end{equation*}%
The modulus of these fields define the density of the various variables in
the state defined by $\Gamma _{a}$ or $\Gamma _{u}$. 
\begin{eqnarray}
&&\left\vert \Gamma \right\vert _{a}^{2}\left( T,\hat{T},\theta ,C,D\right)
\label{gmvv} \\
&\simeq &\left\{ \mathcal{N}\exp \left( -\frac{1}{2}a_{C}\left( Z\right)
\left( C-C\left( \theta \right) \right) ^{2}\right) \right. \exp \left( -%
\frac{a_{D}\left( Z\right) }{2}\left( D-\left\langle D\right\rangle \right)
^{2}\right)  \notag \\
&&\times \exp \left( -\frac{\rho h_{C}\left( \omega \left( \theta ,Z\right)
\right) \left\vert \bar{\Psi}\left( \theta ,Z,Z^{\prime }\right) \right\vert
^{2}}{2\omega \left( \theta ,Z,\left\vert \Psi \right\vert ^{2}\right) }%
\left( \hat{T}-\left\langle \hat{T}\right\rangle \right) ^{2}\right)  \notag
\\
&&\times \left. \left\Vert \Gamma _{0}\left( \theta ,Z,Z^{\prime }\right)
\right\Vert \exp \left( -\frac{\left\vert \Psi \left( \theta ,Z\right)
\right\vert ^{2}}{2\tau \omega }\left( T-\left\langle T\right\rangle \right)
^{2}\right) \right\} _{\left( Z,Z^{\prime }\right) ,\left\langle T\left(
Z,Z^{\prime }\right) \right\rangle \neq 0}  \notag
\end{eqnarray}%
\begin{eqnarray}
&&\left\vert \Gamma \right\vert _{u}^{2}\left( T,\hat{T},\theta ,C,D\right)
\label{GC} \\
&\simeq &\left\{ \mathcal{N}\exp \left( -\frac{1}{2}a_{C}\left( Z\right)
\left( C-C\left( \theta \right) \right) ^{2}\right) \right. \exp \left( -%
\frac{a_{D}\left( Z\right) }{2}\left( D-\left\langle D\right\rangle \right)
^{2}\right)  \notag \\
&&\left. \times \exp \left( -\frac{\rho h_{C}\left( \omega \left( \theta
,Z\right) \right) \left\vert \bar{\Psi}\left( \theta ,Z,Z^{\prime }\right)
\right\vert ^{2}}{2\omega \left( \theta ,Z,\left\vert \Psi \right\vert
^{2}\right) }\left( \hat{T}-\left\langle \hat{T}\right\rangle \right)
^{2}\right) \times \delta \left( T\right) \right\} _{\left( Z,Z^{\prime
}\right) ,\left\langle T\left( Z,Z^{\prime }\right) \right\rangle \neq 0} 
\notag
\end{eqnarray}

\subsection{Average values for connctivities}

The average values in this background states are given by:

\begin{eqnarray*}
C_{Z,Z^{\prime }} &=&\frac{\alpha _{C}\omega ^{\prime }\left\vert \Psi
\left( \theta -\frac{\left\vert Z-Z^{\prime }\right\vert }{c},Z^{\prime
}\right) \right\vert ^{2}}{\frac{1}{\tau _{C}}+\alpha _{C}\omega ^{\prime
}\left\vert \Psi \left( \theta -\frac{\left\vert Z-Z^{\prime }\right\vert }{c%
},Z^{\prime }\right) \right\vert ^{2}} \\
D_{Z,Z^{\prime }} &=&\frac{\alpha _{D}\omega \left\vert \Psi \left( \theta
,Z\right) \right\vert ^{2}}{\frac{1}{\tau _{D}}+\alpha _{D}\omega \left\vert
\Psi \left( \theta ,Z\right) \right\vert ^{2}}
\end{eqnarray*}%
\begin{eqnarray}
\left\langle T\left( Z,Z^{\prime }\right) \right\rangle &=&\lambda \tau
\left\langle \hat{T}\left( Z,Z^{\prime }\right) \right\rangle  \label{Cnv} \\
&=&\frac{\lambda \tau h\left( Z,Z^{\prime }\right) \left\langle
C_{Z,Z^{\prime }}\left( \theta \right) \left\vert \Psi \left( \theta
,Z\right) \right\vert ^{2}\right\rangle }{\left\vert \bar{\Psi}\left( \theta
,Z,Z^{\prime }\right) \right\vert ^{2}}  \notag
\end{eqnarray}%
for $\left( Z,Z^{\prime }\right) $ an $"a"$ (active) doublet, and:%
\begin{equation*}
\left\langle T\left( Z,Z^{\prime }\right) \right\rangle =0
\end{equation*}%
\begin{equation}
\left\langle \hat{T}\left( Z,Z^{\prime }\right) \right\rangle =\frac{h\left(
Z,Z^{\prime }\right) \left\langle C_{Z,Z^{\prime }}\left( \theta \right)
h_{C}\left( \omega \left( \theta ,Z\right) \right) \left\vert \Psi \left(
\theta ,Z\right) \right\vert ^{2}\right\rangle -\eta }{\left\langle
h_{C}\left( \omega \left( \theta ,Z\right) \right) \left\vert \bar{\Psi}%
\left( \theta ,Z,Z^{\prime }\right) \right\vert ^{2}\right\rangle }<0
\end{equation}

for an $"u"$ (unactive) doublet.

Choosing an exponential form for the function $h\left( Z,Z^{\prime }\right) $%
: 
\begin{equation*}
h\left( Z,Z^{\prime }\right) =\exp \left( -\frac{\left\vert Z-Z^{\prime
}\right\vert }{\nu c}\right)
\end{equation*}%
the average simplifies as:%
\begin{equation}
\left\langle T\left( Z,Z^{\prime }\right) \right\rangle =\frac{\lambda \tau
\exp \left( -\frac{\left\vert Z-Z^{\prime }\right\vert }{\nu c}\right) }{1+%
\frac{\alpha _{D}\omega h_{D}}{\alpha _{C}\omega ^{\prime }h_{C}}\frac{\frac{%
1}{\tau _{C}}+\alpha _{C}\omega ^{\prime }\left\vert \Psi \left( \theta -%
\frac{\left\vert Z-Z^{\prime }\right\vert }{c},Z^{\prime }\right)
\right\vert ^{2}}{\frac{1}{\tau _{D}}+\alpha _{D}\omega \left\vert \Psi
\left( \theta ,Z\right) \right\vert ^{2}}}  \label{mnt}
\end{equation}%
in an $a$ doublet.

Note that the background state obtained above is quasi-static. As explained
in the previous paragraph, the variations in $\theta $\ of the background
field are slow. Some modifications in the parameters may induce a switch at
some point from an '$a$' to a '$u$' doublet or from a '$u$' to an '$a$'
doublet. To consider a static background, the quantities involved in the
definition of (\ref{gmcc}), (\ref{GMuu}) and (\ref{gmvv}) can be averaged
over time.

\subsubsection{Interpretation of the background states}

Formulas (\ref{gmvv}) and (\ref{GC}) for the densities of connectivities may
be interpreted as follows: Regardless of the system's interpretation,
whether as a description of groups of simple cells or a single complex cell
at each point, the stable backgrounds are not defined with a given value of
connectivity. On the contrary, the background states are described by a
normal distribution around some average value. That is, the cells or groups
of axons/dendrites are connected with connectivities that are spread around
this average.

\section{System's background states averages}

The full system background state average is given by several equations
defining the averages connectivits, the neural background stt and the
neurons' activities. Equations (\ref{Cnv}) and (\ref{mnt}) 
\begin{equation*}
\left\langle T\left( Z,Z^{\prime }\right) \right\rangle =\frac{\lambda \tau
\exp \left( -\frac{\left\vert Z-Z^{\prime }\right\vert }{\nu c}\right) }{1+%
\frac{\alpha _{D}\omega h_{D}}{\alpha _{C}\omega ^{\prime }h_{C}}\frac{\frac{%
1}{\tau _{C}}+\alpha _{C}\omega ^{\prime }\left\vert \Psi \left( \theta -%
\frac{\left\vert Z-Z^{\prime }\right\vert }{c},Z^{\prime }\right)
\right\vert ^{2}}{\frac{1}{\tau _{D}}+\alpha _{D}\omega \left\vert \Psi
\left( \theta ,Z\right) \right\vert ^{2}}}=\lambda \tau \left\langle \hat{T}%
\left( Z,Z^{\prime }\right) \right\rangle
\end{equation*}%
are considered together with (\ref{qf}) determining the activit at the
lowest order in perturbation:%
\begin{eqnarray}
&&\omega ^{-1}\left( J,\theta ,Z,\left\vert \Psi \right\vert ^{2}\right)
\label{fq} \\
&=&G\left( J\left( \theta ,Z\right) +\int \frac{\kappa }{N}\frac{\omega
\left( J,\theta -\frac{\left\vert Z-Z_{1}\right\vert }{c},Z_{1},\Psi \right)
T\left( Z,\theta ,Z_{1},\theta -\frac{\left\vert Z-Z_{1}\right\vert }{c}%
\right) }{\omega \left( J,\theta ,Z,\left\vert \Psi \right\vert ^{2}\right) }%
\left\vert \Psi \left( \theta -\frac{\left\vert Z-Z_{1}\right\vert }{c}%
,Z_{1}\right) \right\vert ^{2}dZ_{1}\right)  \notag
\end{eqnarray}%
and the minimizing equation for the field $\left\vert \Psi \left( \theta
,Z\right) \right\vert ^{2}$%
\begin{eqnarray*}
0 &=&\frac{\delta }{\delta \left\vert \Psi \left( \theta ,Z\right)
\right\vert ^{2}}\left[ \int \Psi ^{\dagger }\left( \theta ,Z\right) \left(
-\nabla _{\theta }\left( \frac{\sigma _{\theta }^{2}}{2}\nabla _{\theta }-%
\frac{1}{\hat{G}\left( \left( T\left( Z,Z_{1}\right) \right)
_{Z_{1}},\left\vert \Psi \left( Z\right) \right\vert ^{2}\right) }\right)
\right) \Psi \left( \theta ,Z\right) \right] \\
&&+\frac{\delta }{\delta \left\vert \Psi \left( \theta ,Z\right) \right\vert
^{2}}\left[ \int U\left( \left\vert \Psi \left( \theta ,Z\right) \right\vert
^{2}\right) \right]
\end{eqnarray*}

where we will assume that the potential has the specific form:%
\begin{equation}
U\left( \left\vert \Psi \left( \theta ,Z\right) \right\vert ^{2}\right)
=V\left( \left\vert \Psi \left( \theta ,Z\right) \right\vert ^{2}-\int
T\left( Z^{\prime },Z_{1}\right) \left\vert \Psi _{0}\left( Z\right)
\right\vert ^{2}dZ_{1}\right)  \label{ptl}
\end{equation}%
This potential encompasses the fact that the field is constrained by a
potential limiting the activity around some average\footnote{%
This characteristic in line with the litterature about homeostatic activit
quoted in the literature review.} $\left\vert \Psi _{0}\left( Z\right)
\right\vert ^{2}$. It also implies that the activity at some point depends
on the average activity in the neighborhood of the point. We will look for
static solutions of these equations under some simplifying assumptions in
the next section. However, given the form of the various equations, we can
draw some general patterns for the solutions. Equation (\ref{Cnv}) shows
that groups of mutually connected states arise in localized region, i.e. for 
$\left\vert Z-Z^{\prime }\right\vert <1$, and bewteen cells such that:%
\begin{equation*}
\omega \simeq \omega ^{\prime }
\end{equation*}%
and:%
\begin{equation*}
\left\vert \Psi \left( \theta ,Z\right) \right\vert ^{2}\simeq \left\vert
\Psi \left( \theta -\frac{\left\vert Z-Z^{\prime }\right\vert }{c},Z^{\prime
}\right) \right\vert ^{2}
\end{equation*}%
Given (\ref{ptl}) and (\ref{fq}), this is also realized in localized
regions. As a consequence, assuming an implicit global potential limiting
the overall activity, we may expect groups of connected points, with
relatively close activities. We inspect this possibility more precisely for
the static case in the next section.

\section{Static background state for the system}

We look for a static background state for the whole system. It corresponds
to averaging over time in the background fields (\ref{gmcc}) (\ref{GMuu})
and (\ref{gmvv}). The equilibrium we look for is obtained as consistency
conditions for (\ref{gmcc}). Actually, the background $\Gamma $ depends on $%
\omega \left( \theta ,Z\right) $ and $\left\vert \Psi \left( \theta
,Z\right) \right\vert ^{2}$ that depend themselves functionally on $\Gamma $%
, through (\ref{nqf}) (\ref{sdp}).

\subsection{General equations}

An approximate static solution of (\ref{nqf}) can be found for the constant
background and a constant current, i.e. $J=\bar{J}$. We also set:%
\begin{equation*}
T\left( Z,Z_{1}\right) =\bar{T}\left( Z,\theta ,Z_{1},\theta -\frac{%
\left\vert Z-Z_{1}\right\vert }{c}\right)
\end{equation*}%
From now on, the quantity $T\left( Z,Z_{1}\right) $ refers to the average of
the connectivity function at points $\left( Z,Z_{1}\right) $, in the
background state defined above, i.e. $T\left( Z,Z_{1}\right) $ refers to $%
\left\langle T\left( Z,Z_{1}\right) \right\rangle $ defined as: 
\begin{equation*}
\left\langle T\left( Z,Z_{1}\right) \right\rangle =\int T\left\vert \Gamma
\left( T,\hat{T},\theta ,Z,Z_{1}\right) \right\vert ^{2}dT
\end{equation*}%
For points such that $T\left( Z,Z^{\prime }\right) \neq 0$, it is defined by
the set of equations (\ref{qf}) or (\ref{nqf}), (\ref{vrg}) (\ref{mnt}) and
the minimizatn equatn (\ref{sdp}) if:%
\begin{equation*}
h\left( Z,Z^{\prime }\right) C_{Z,Z^{\prime }}\left( \theta \right)
h_{C}\left( \omega \left( \theta ,Z\right) \right) \left\vert \Psi \left(
\theta ,Z\right) \right\vert ^{2}>0
\end{equation*}%
We choose, as in (\cite{IFR}), $h_{C}\left( \omega \right) =\omega $ and $%
h_{C}\left( \omega ^{\prime }\right) =\omega ^{\prime }$. The static
equilibrium for $\omega \left( Z\right) $, $T\left( Z,Z_{1}\right) $ and $%
\left\vert \Psi \left( Z\right) \right\vert ^{2}$ is found in three steps.

\subsubsection{First step: finding $\protect\omega \left( Z\right) $}

We first solve for $\omega \left( Z\right) $ as a function of other
variables using (\ref{mnt}) that allows to replace $\omega ^{\prime
}\left\vert \Psi \left( Z^{\prime }\right) \right\vert ^{2}$:

\begin{equation*}
\omega ^{\prime }\left\vert \Psi \left( Z^{\prime }\right) \right\vert ^{2}=%
\frac{\lambda \tau \exp \left( -\frac{\left\vert Z-Z^{\prime }\right\vert }{%
\nu c}\right) -T\left( Z,Z^{\prime }\right) }{T\left( Z,Z^{\prime }\right) }%
\left( \frac{1}{\alpha _{D}\tau _{D}}+\omega \left\vert \Psi \left( Z\right)
\right\vert ^{2}\right) -\frac{1}{\alpha _{C}\tau _{C}}
\end{equation*}%
We show in appendix 3 that this allows to rewrite (\ref{nqf}) as an equation
for $\omega ^{-1}\left( Z\right) $:%
\begin{eqnarray}
&&\omega \left( Z\right)  \label{frn} \\
&\simeq &G\left( \int \frac{\kappa }{N}\left( \left( \lambda \tau \exp
\left( -\frac{\left\vert Z-Z_{1}\right\vert }{\nu c}\right) -T\left(
Z,Z_{1}\right) \right) \left( \left( \frac{1}{\alpha _{D}\tau _{D}}-\frac{%
T\left( Z,Z_{1}\right) }{\alpha _{C}\tau _{C}}\right) \omega
^{-1}+\left\vert \Psi \left( Z\right) \right\vert ^{2}\right) \right)
dZ_{1}\right)  \notag
\end{eqnarray}%
with the solution defined by a function:%
\begin{equation*}
\omega \left( Z\right) =\hat{G}\left( T\left( Z\right) ,\left\vert \Psi
\left( Z\right) \right\vert ^{2}\right)
\end{equation*}%
where:%
\begin{equation*}
\frac{1}{V}T\left( Z\right) =\frac{1}{V}\int T\left( Z,Z_{1}\right) dZ_{1}
\end{equation*}

\subsubsection{Second step: finding $T\left( Z\right) $ and $T\left(
Z,Z^{\prime }\right) $}

In a second step, once $\omega \left( Z\right) $ found, we use (\ref{vrg}) (%
\ref{mnt}) to obtain $T\left( Z\right) $ and $T\left( Z,Z^{\prime }\right) $
as a function of $\hat{G}\left( T\left( Z^{\prime }\right) ,\left\vert \Psi
\left( Z^{\prime }\right) \right\vert ^{2}\right) \left\vert \Psi \left(
Z,\omega \right) \right\vert ^{2}$. We show in appendix 3 that:%
\begin{equation}
T\left( Z\right) =\frac{\lambda \tau \nu c}{1+\frac{\frac{1}{\tau _{C}\alpha
_{C}}+\Omega }{\frac{1}{\tau _{D}\alpha _{D}}+\hat{G}\left( T\left( Z\right)
,\left\vert \Psi \left( Z\right) \right\vert ^{2}\right) \left\vert \Psi
\left( Z\right) \right\vert ^{2}}}=\frac{\lambda \tau \nu c\left( \frac{1}{%
\tau _{D}\alpha _{D}}+\hat{G}\left( T\left( Z\right) ,\left\vert \Psi \left(
Z\right) \right\vert ^{2}\right) \left\vert \Psi \left( Z\right) \right\vert
^{2}\right) }{\frac{1}{\tau _{D}\alpha _{D}}+\frac{1}{\tau _{C}\alpha _{C}}%
+\Omega +\hat{G}\left( T\left( Z\right) ,\left\vert \Psi \left( Z\right)
\right\vert ^{2}\right) \left\vert \Psi \left( Z\right) \right\vert ^{2}}
\label{Trr}
\end{equation}%
where:%
\begin{equation}
\Omega =\frac{1}{V}\int \hat{G}\left( T\left( Z^{\prime }\right) ,\left\vert
\Psi \left( Z^{\prime }\right) \right\vert ^{2}\right) \left\vert \Psi
\left( Z^{\prime }\right) \right\vert ^{2}dZ^{\prime }  \label{MGg}
\end{equation}%
and this leads to the following formula for $T\left( Z,Z^{\prime }\right) $: 
\begin{equation}
T\left( Z,Z^{\prime }\right) =\frac{\lambda \tau \exp \left( -\frac{%
\left\vert Z-Z^{\prime }\right\vert }{\nu c}\right) }{1+\frac{\frac{1}{\tau
_{C}\alpha _{C}}+\omega ^{\prime }\left\vert \Psi \left( Z^{\prime },\omega
^{\prime }\right) \right\vert ^{2}}{\frac{1}{\tau _{D}\alpha _{D}}+\omega
\left\vert \Psi \left( Z,\omega \right) \right\vert ^{2}}}=\frac{\lambda
\tau \exp \left( -\frac{\left\vert Z-Z^{\prime }\right\vert }{\nu c}\right) 
}{1+\frac{\frac{1}{\tau _{C}\alpha _{C}}+\hat{G}\left( T\left( Z^{\prime
}\right) ,\left\vert \Psi \left( Z^{\prime }\right) \right\vert ^{2}\right)
\left\vert \Psi \left( Z^{\prime },\omega ^{\prime }\right) \right\vert ^{2}%
}{\frac{1}{\tau _{D}\alpha _{D}}+\hat{G}\left( T\left( Z\right) ,\left\vert
\Psi \left( Z\right) \right\vert ^{2}\right) \left\vert \Psi \left( Z,\omega
\right) \right\vert ^{2}}}  \label{Ttpp}
\end{equation}

\subsubsection{Third step: finding $\Psi \left( \protect\theta ,Z\right) $}

In a third step, the system is closed by minimizing the action for the field 
$\Psi \left( \theta ,Z\right) $: 
\begin{eqnarray*}
&&\int \Psi ^{\dagger }\left( \theta ,Z\right) \left( -\nabla _{\theta
}\left( \frac{\sigma _{\theta }^{2}}{2}\nabla _{\theta }-\frac{1}{\hat{G}%
\left( \left( T\left( Z,Z_{1}\right) \right) _{Z_{1}},\left\vert \Psi \left(
Z\right) \right\vert ^{2}\right) }\right) \right) \Psi \left( \theta
,Z\right) \\
&&+\int V\left( \left\vert \Psi \left( \theta ,Z\right) \right\vert
^{2}-\int T\left( Z^{\prime },Z_{1}\right) \left\vert \Psi _{0}\left(
Z\right) \right\vert ^{2}dZ_{1}\right)
\end{eqnarray*}%
Where we have assumed that the field is constrained by a potential limiting
the activit around some average $\left\vert \Psi _{0}\left( Z\right)
\right\vert ^{2}$. We choose:%
\begin{equation*}
V=\frac{1}{2}\left( \left\vert \Psi \left( Z\right) \right\vert ^{2}-\int
T\left( Z^{\prime },Z_{1}\right) \left\vert \Psi _{0}\left( Z\right)
\right\vert ^{2}dZ_{1}\right) ^{2}
\end{equation*}%
We show in Appendix 3 that an equilibrium with static $\left\vert \Psi
\left( Z\right) \right\vert ^{2}$ exist and minimizes:%
\begin{equation*}
\int \left( \frac{1}{\hat{G}\left( \left( T\left( Z,Z_{1}\right) \right)
_{Z_{1}},\left\vert \Psi \left( Z\right) \right\vert ^{2}\right) }\right)
^{2}\left\vert \Psi \left( Z\right) \right\vert ^{2}+\int V\left( \left\vert
\Psi \left( \theta ,Z\right) \right\vert ^{2}-\int T\left( Z^{\prime
},Z_{1}\right) \left\vert \Psi _{0}\left( Z\right) \right\vert
^{2}dZ_{1}\right)
\end{equation*}%
with saddle point equation:%
\begin{equation*}
0\simeq \left( \left( \frac{1}{\hat{G}\left( \left( T\left( Z,Z_{1}\right)
\right) _{Z_{1}},\left\vert \Psi \left( Z\right) \right\vert ^{2}\right) }%
\right) ^{2}+\left( \left\vert \Psi \left( Z\right) \right\vert ^{2}-\int
T\left( Z^{\prime },Z_{1}\right) \left\vert \Psi _{0}\left( Z\right)
\right\vert ^{2}dZ_{1}\right) \right) \Psi \left( \theta ,Z\right)
\end{equation*}%
with solutions:%
\begin{equation*}
\Psi \left( \theta ,Z\right) =0
\end{equation*}%
or approximately:%
\begin{eqnarray}
\left( \frac{1}{\hat{G}\left( \left( T\left( Z,Z_{1}\right) \right)
_{Z_{1}},\left\vert \Psi \left( Z\right) \right\vert ^{2}\right) }\right)
^{2}+\left\vert \Psi \left( Z\right) \right\vert ^{2} &\simeq &T\left(
Z\right) \frac{\int \left\vert \Psi _{0}\left( Z^{\prime }\right)
\right\vert ^{2}k\left( Z,Z^{\prime }\right) dZ^{\prime }}{V}  \label{Pll} \\
&\equiv &T\left( Z\right) \left\langle \left\vert \Psi _{0}\left( Z^{\prime
}\right) \right\vert ^{2}\right\rangle _{Z}  \notag
\end{eqnarray}%
Remark that $\left\langle \left\vert \Psi _{0}\left( Z^{\prime }\right)
\right\vert ^{2}\right\rangle _{Z}$ is a function of $Z$ given by an average
over the points $Z^{\prime }$ surrounding $Z$. It is defind in appendix 3
and represents the average activity in the neighbourhood of $Z$.

The system is now reduced to two variables $T\left( Z\right) $ and $%
\left\vert \Psi \left( Z\right) \right\vert ^{2}$ together with equations (%
\ref{Trr}) and (\ref{Pll}). The average connectivity being then retrieved by
(\ref{Ttpp}).

\subsection{Solving for $\left\vert \Psi \left( Z\right) \right\vert ^{2}$
and $T\left( Z\right) $}

Appendix 3 solves the system for $T\left( Z\right) $ and $\left\vert \Psi
\left( Z\right) \right\vert ^{2}$. We show that:

\begin{equation}
\left\vert \Psi \left( Z\right) \right\vert ^{2}=\frac{2T\left( Z\right)
\left\langle \left\vert \Psi _{0}\left( Z^{\prime }\right) \right\vert
^{2}\right\rangle _{Z}}{\left( 1+\sqrt{1+4\left( \frac{\lambda \tau \nu
c-T\left( Z\right) }{\left( \frac{1}{\tau _{D}\alpha _{D}}+\frac{1}{\tau
_{C}\alpha _{C}}+\Omega \right) T\left( Z\right) -\frac{1}{\tau _{D}\alpha
_{D}}\lambda \tau \nu c}\right) ^{2}T\left( Z\right) \left\langle \left\vert
\Psi _{0}\left( Z^{\prime }\right) \right\vert ^{2}\right\rangle _{Z}}%
\right) }  \label{Ps}
\end{equation}%
Ultimately, inserting this result in (\ref{Tng}) yields the following
equation for $T\left( Z\right) $. 
\begin{eqnarray}
\frac{\hat{\Omega}T\left( Z\right) -\frac{1}{\tau _{D}\alpha _{D}}\lambda
\tau \nu c}{\lambda \tau \nu c-T\left( Z\right) } &=&\hat{G}\left( T\left(
Z\right) ,\frac{2T\left( Z\right) \left\langle \left\vert \Psi _{0}\left(
Z^{\prime }\right) \right\vert ^{2}\right\rangle _{Z}}{\left( 1+\sqrt{%
1+4\left( \frac{\lambda \tau \nu c-T\left( Z\right) }{\hat{\Omega}T\left(
Z\right) -\frac{1}{\tau _{D}\alpha _{D}}\lambda \tau \nu c}\right)
^{2}T\left( Z\right) \left\langle \left\vert \Psi _{0}\left( Z^{\prime
}\right) \right\vert ^{2}\right\rangle _{Z}}\right) }\right)  \notag \\
&&\times \frac{2T\left( Z\right) \left\langle \left\vert \Psi _{0}\left(
Z^{\prime }\right) \right\vert ^{2}\right\rangle _{Z}}{\left( 1+\sqrt{%
1+4\left( \frac{\lambda \tau \nu c-T\left( Z\right) }{\hat{\Omega}T\left(
Z\right) -\frac{1}{\tau _{D}\alpha _{D}}\lambda \tau \nu c}\right)
^{2}T\left( Z\right) \left\langle \left\vert \Psi _{0}\left( Z^{\prime
}\right) \right\vert ^{2}\right\rangle _{Z}}\right) }  \label{Tqn}
\end{eqnarray}%
with:%
\begin{eqnarray*}
\Omega &=&\frac{1}{V}\int \hat{G}\left( T\left( Z\right) ,\left\vert \Psi
\left( Z\right) \right\vert ^{2}\right) \left\vert \Psi \left( Z\right)
\right\vert ^{2} \\
\hat{\Omega} &=&\left( \frac{1}{\tau _{D}\alpha _{D}}+\frac{1}{\tau
_{C}\alpha _{C}}+\Omega \right)
\end{eqnarray*}

Equation (\ref{Tqn}) has in general several solutions (see below)
corresponding to several regime of activity, depending on the point. Once $%
T\left( Z\right) $ is found, one can obtain $\left\vert \Psi \left( Z\right)
\right\vert ^{2}$ using (\ref{Ps}). To obtain more precise formula for these
solutions, we will detail a particular case below. The system is ultimately
determined by finding $\Omega $. Details are given in appendix 3.

\subsection{Particular case}

For $G$ an increasing function of the form $G\left( x\right) \simeq b_{0}x$
for $x<1$, we can solve the system. In our preliminary set up, this
corresponds, for $b_{0}=1$, to consider $\gamma \simeq 0$, that is, the
activity of a cell depends only on the external currents. The resolution
starts by finding $\Omega $, then ultimately $T\left( Z\right) $ and $%
T\left( Z,Z^{\prime }\right) $.

Setting $b=b_{0}\frac{\kappa V}{N}$, $\bar{T}=\frac{\lambda \tau \nu c}{2}$
and $\left\langle \left\vert \Psi _{0}\right\vert ^{2}\right\rangle
=\left\langle \left\langle \left\vert \Psi _{0}\left( Z^{\prime }\right)
\right\vert ^{2}\right\rangle _{Z}\right\rangle $, the average of $%
\left\langle \left\vert \Psi _{0}\left( Z^{\prime }\right) \right\vert
^{2}\right\rangle _{Z}$, we find in appendix 3 the equation for $\Omega $:%
\begin{equation}
1=\frac{b\bar{T}\Omega ^{3}}{\left( b\bar{T}^{2}\left\langle \left\vert \Psi
_{0}\right\vert ^{2}\right\rangle \Omega -1\right) ^{2}}
\end{equation}%
Several cases arise:

For $d=\left( b\bar{T}\right) ^{2}\left( \bar{T}\left\langle \left\vert \Psi
_{0}\right\vert ^{2}\right\rangle \right) ^{3}<\frac{27}{4}$ there is one
solution:%
\begin{equation}
\Omega \simeq \left( b\bar{T}\right) ^{-\frac{1}{3}}<<1  \label{AA}
\end{equation}

For $d=\left( b\bar{T}\right) ^{2}\left\langle \bar{T}\left\vert \Psi
_{0}\right\vert ^{2}\right\rangle ^{3}>\frac{27}{4}$ there are three
solutions. The first one is:%
\begin{equation}
\Omega \simeq b\bar{T}\left( \bar{T}\left\langle \left\vert \Psi
_{0}\right\vert ^{2}\right\rangle \right) ^{2}  \label{BB}
\end{equation}%
The two other solutions are centered around $\frac{1}{b\bar{T}%
^{2}\left\langle \left\vert \Psi _{0}\right\vert ^{2}\right\rangle }$. They
are given by:%
\begin{equation}
\Omega =\frac{1\pm \sqrt{\frac{1}{\left( b\bar{T}\right) ^{2}\left( \bar{T}%
\left\langle \left\vert \Psi _{0}\right\vert ^{2}\right\rangle \right) ^{3}}}%
}{b\bar{T}^{2}\left\langle \left\vert \Psi _{0}\right\vert ^{2}\right\rangle 
}  \label{CC}
\end{equation}

Solution (\ref{AA}) corresponds to relatively low activity, i.e. $%
\left\langle \left\vert \Psi _{0}\right\vert ^{2}\right\rangle <<1$, so that
we only consider solutions (\ref{BB}) and (\ref{CC}) in the sequel.
Moreover, since the solutions of (\ref{CC}) are centered around$\frac{1}{%
b\left\langle \left\vert \Psi _{0}\right\vert ^{2}\right\rangle }$, we can
consider only the two solutions:%
\begin{equation*}
\Omega _{\pm }=\left( \Omega _{+},\Omega _{-}\right)
\end{equation*}%
with:%
\begin{equation}
\Omega _{+}=b\bar{T}\left( \bar{T}\left\langle \left\vert \Psi
_{0}\right\vert ^{2}\right\rangle \right) ^{2},\Omega _{-}=\frac{1}{b\bar{T}%
^{2}\left\langle \left\vert \Psi _{0}\right\vert ^{2}\right\rangle }
\label{MH}
\end{equation}%
To these solutions for $\Omega $, we associate the $Z$ dependent parameters,
which are at the lwst rdr\footnote{%
First order corrections are given in appendix 3.}:%
\begin{equation}
Y_{+}\left( Z\right) \simeq b\bar{T}\left( \bar{T}\left\langle \left\vert
\Psi _{0}\left( Z^{\prime }\right) \right\vert ^{2}\right\rangle
_{Z}^{2}\right) ^{2}  \label{Sltdn1}
\end{equation}

\begin{equation}
Y_{-}\left( Z\right) =\frac{1}{b\bar{T}^{2}\left\langle \left\vert \Psi
_{0}\left( Z^{\prime }\right) \right\vert ^{2}\right\rangle _{Z}}
\label{Sltdd1}
\end{equation}%
where $Y_{\pm }$ gathers the possibilities $Y_{+}$ and $Y$. Appendix 3 shows
that $T\left( Z\right) $ is a function of these parameters:

\begin{equation*}
T\left( Z_{\pm }\right) =\frac{\lambda \tau \nu cY_{\pm }\left( Z\right)
+\lambda \tau \nu c\frac{1}{\tau _{D}\alpha _{D}}}{Y_{\pm }\left( Z\right) +%
\frac{1}{\tau _{D}\alpha _{D}}+\frac{1}{\alpha _{C}\tau _{C}}+\Omega _{\pm }}
\end{equation*}%
and takes the values:%
\begin{eqnarray*}
T_{+}\left( Z\right) &\simeq &\bar{T}-\frac{3\bar{T}\left( \left\langle
\left\vert \Psi _{0}\right\vert ^{2}\right\rangle ^{2}-\left\langle
\left\vert \Psi _{0}\left( Z^{\prime }\right) \right\vert ^{2}\right\rangle
_{Z}^{2}\right) }{2\left\langle \left\vert \Psi _{0}\left( Z^{\prime
}\right) \right\vert ^{2}\right\rangle _{Z}} \\
T_{-}\left( Z\right) &\simeq &\bar{T}+\frac{5}{2}\sqrt{\frac{1}{\bar{T}%
^{2}b^{2}\left\langle \bar{T}\left\vert \Psi _{0}\left( Z^{\prime }\right)
\right\vert ^{2}\right\rangle _{Z}^{5}}}\bar{T}\left( \left\langle
\left\vert \Psi _{0}\right\vert ^{2}\right\rangle ^{2}-\left\langle
\left\vert \Psi _{0}\left( Z^{\prime }\right) \right\vert ^{2}\right\rangle
_{Z}^{2}\right)
\end{eqnarray*}

Ultimately, we show that there are four possibilities for the connectivity
functions that are written:%
\begin{equation*}
T\left( Z_{\pm },Z_{\pm }^{\prime }\right) =\frac{\lambda \tau \exp \left( -%
\frac{\left\vert Z-Z^{\prime }\right\vert }{\nu c}\right) \left( \frac{1}{%
\tau _{D}\alpha _{D}}+Y_{\pm }\left( Z\right) \right) }{\frac{1}{\tau
_{D}\alpha _{D}}+Y_{\pm }\left( Z\right) +\frac{1}{\alpha _{C}\tau _{C}}%
+Y_{\pm }\left( Z^{\prime }\right) }
\end{equation*}%
Given our assumptions on $d$, in most cases\footnote{%
In general $\bar{T}=\frac{\lambda \tau \nu c}{2}>>1$}:%
\begin{equation*}
Y_{+}\left( Z\right) >>Y_{-}\left( Z\right)
\end{equation*}%
Moreover:%
\begin{equation*}
\frac{1}{\tau _{D}\alpha _{D}}<<1\text{, }\frac{1}{\alpha _{C}\tau _{C}}<<1
\end{equation*}%
so that due to the threshold in connectivity, we can write:%
\begin{eqnarray}
T\left( Z_{-},Z_{+}^{\prime }\right) &=&\frac{\lambda \tau \exp \left( -%
\frac{\left\vert Z-Z^{\prime }\right\vert }{\nu c}\right) \left( \frac{1}{%
\tau _{D}\alpha _{D}}+\frac{1}{b\bar{T}^{2}\left\langle \left\vert \Psi
_{0}\left( Z^{\prime }\right) \right\vert ^{2}\right\rangle _{Z}}\right) }{%
\frac{1}{\tau _{D}\alpha _{D}}+\frac{1}{\alpha _{C}\tau _{C}}+\frac{1}{b\bar{%
T}^{2}\left\langle \left\vert \Psi _{0}\left( Z^{\prime }\right) \right\vert
^{2}\right\rangle _{Z}}+b\bar{T}\left( \bar{T}\left\langle \left\vert \Psi
_{0}\left( Z^{\prime }\right) \right\vert ^{2}\right\rangle _{Z^{\prime
}}^{2}\right) ^{2}}\simeq 0  \label{cnvv} \\
T\left( Z_{+},Z_{+}^{\prime }\right) &=&\frac{\lambda \tau \exp \left( -%
\frac{\left\vert Z-Z^{\prime }\right\vert }{\nu c}\right) \left( \frac{1}{%
\tau _{D}\alpha _{D}}+b\bar{T}\left( \bar{T}\left\langle \left\vert \Psi
_{0}\left( Z^{\prime }\right) \right\vert ^{2}\right\rangle _{Z}^{2}\right)
^{2}\right) }{\frac{1}{\tau _{D}\alpha _{D}}+\frac{1}{\alpha _{C}\tau _{C}}+b%
\bar{T}\left( \bar{T}\left\langle \left\vert \Psi _{0}\left( Z^{\prime
}\right) \right\vert ^{2}\right\rangle _{Z}^{2}\right) ^{2}+b\bar{T}\left( 
\bar{T}\left\langle \left\vert \Psi _{0}\left( Z^{\prime }\right)
\right\vert ^{2}\right\rangle _{Z^{\prime }}^{2}\right) ^{2}}\simeq \frac{%
\lambda \tau \exp \left( -\frac{\left\vert Z-Z^{\prime }\right\vert }{\nu c}%
\right) }{2}  \notag \\
T\left( Z_{+},Z_{-}^{\prime }\right) &=&\frac{\lambda \tau \exp \left( -%
\frac{\left\vert Z-Z^{\prime }\right\vert }{\nu c}\right) \left( \frac{1}{%
\tau _{D}\alpha _{D}}+b\bar{T}\left( \bar{T}\left\langle \left\vert \Psi
_{0}\left( Z^{\prime }\right) \right\vert ^{2}\right\rangle _{Z}^{2}\right)
^{2}\right) }{\frac{1}{\tau _{D}\alpha _{D}}+\frac{1}{\alpha _{C}\tau _{C}}+b%
\bar{T}\left( \bar{T}\left\langle \left\vert \Psi _{0}\left( Z^{\prime
}\right) \right\vert ^{2}\right\rangle _{Z}^{2}\right) ^{2}+\frac{1}{b\bar{T}%
^{2}\left\langle \left\vert \Psi _{0}\left( Z^{\prime }\right) \right\vert
^{2}\right\rangle _{Z^{\prime }}}}\simeq \lambda \tau \exp \left( -\frac{%
\left\vert Z-Z^{\prime }\right\vert }{\nu c}\right)  \notag \\
T\left( Z_{-},Z_{-}^{\prime }\right) &\simeq &\frac{\lambda \tau \exp \left(
-\frac{\left\vert Z-Z^{\prime }\right\vert }{\nu c}\right) +\frac{1}{b\bar{T}%
^{2}\left\langle \left\vert \Psi _{0}\left( Z^{\prime }\right) \right\vert
^{2}\right\rangle _{Z}}}{1+\frac{\tau _{D}\alpha _{D}}{\alpha _{C}\tau _{C}}+%
\frac{1}{b\bar{T}^{2}\left\langle \left\vert \Psi _{0}\left( Z^{\prime
}\right) \right\vert ^{2}\right\rangle _{Z}}+\frac{1}{b\bar{T}%
^{2}\left\langle \left\vert \Psi _{0}\left( Z^{\prime }\right) \right\vert
^{2}\right\rangle _{Z^{\prime }}}}\simeq \frac{\lambda \tau \exp \left( -%
\frac{\left\vert Z-Z^{\prime }\right\vert }{\nu c}\right) }{2}  \notag
\end{eqnarray}%
Given equation (\ref{MGg}) and (\ref{MH}), points $Z_{+},Z_{+}^{\prime }$
correspond to cells with high activity while $Z_{-},Z_{-}^{\prime }$
describe points with low activity. This is confirmed in appendix 3, where we
provide the following formula for activities:%
\begin{eqnarray*}
\omega _{+}\left( Z\right) &\simeq &b_{0}\frac{\kappa }{N}\bar{T}\left( 1-%
\frac{3\left( \left\langle \left\vert \Psi _{0}\left( Z^{\prime }\right)
\right\vert ^{2}\right\rangle _{Z}^{2}-\left\langle \left\vert \Psi
_{0}\right\vert ^{2}\right\rangle ^{2}\right) }{2\left\langle \left\vert
\Psi _{0}\left( Z^{\prime }\right) \right\vert ^{2}\right\rangle _{Z}^{2}}%
\right) \left\vert \Psi _{+}\left( Z\right) \right\vert ^{2} \\
\omega _{-}\left( Z\right) &\simeq &b_{0}\frac{\kappa }{N}\bar{T}\left(
\left( 1+\frac{5}{2}\sqrt{\frac{1}{\bar{T}^{2}b^{2}\left\langle \bar{T}%
\left\vert \Psi _{0}\left( Z^{\prime }\right) \right\vert ^{2}\right\rangle
_{Z}^{5}}}\left( \left\langle \left\vert \Psi _{0}\left( Z^{\prime }\right)
\right\vert ^{2}\right\rangle _{Z}^{2}-\left\langle \left\vert \Psi
_{0}\right\vert ^{2}\right\rangle ^{2}\right) \right) \left\vert \Psi
_{-}\left( Z\right) \right\vert ^{2}\right)
\end{eqnarray*}%
and given our assumptions, we show that:%
\begin{eqnarray*}
\left\vert \Psi _{+}\left( Z\right) \right\vert ^{2} &\simeq &2\bar{T}%
\left\langle \left\vert \Psi _{0}\left( Z^{\prime }\right) \right\vert
^{2}\right\rangle _{Z}^{2} \\
\left\vert \Psi _{-}\left( Z\right) \right\vert ^{2} &<&<1
\end{eqnarray*}%
so that:%
\begin{equation*}
\omega _{-}\left( Z\right) <<\omega _{+}\left( Z\right)
\end{equation*}

As a consequence, equation (\ref{cnvv}) shows that cells with high activity
bind together, as do points with low activity, provided that the distance
between them is moderate. However if low activity cells connect to high
activity neighbors, the high activity cells do not bind to low activity
neighbors. If we assume an overall bounded activity (which could be included
as a global potential in the action, or equivalently, as a constraint), this
may favour islands of connected cells with high activity, relatively
independent from the rest of the thread. Actually, the activity is defined
by regions. Either in some zone, activity and the level of connectivity are
high, or activity is low, with the possibility of bonded groups. Given that
connectivity decreases with distance, groups of high activity are of bounded
extension, and given that global activity is limited, a discrete set of such
groups exists.

\section{Generalization: Background state for $n$ interacting fields}

The previous description may be generalized to describe $n$ interacting
types of cells, with arbitrary interactions. Each type of cells is
caracterized by its activity $i=1,...,n$, and interacts either positively or
negatively with each other. Each type is defined by a field $\Psi _{i}$ and
activities $\omega _{i}\left( \theta ,Z\right) $. The general version of (%
\ref{flt}), that includes (\ref{pcn}), becomes:%
\begin{eqnarray}
S_{full} &=&-\frac{1}{2}\sum_{i}\Psi _{i}^{\dagger }\left( \theta ,Z\right)
\nabla _{\theta }\left( \frac{\sigma _{\theta }^{2}}{2}\nabla _{\theta
}-\omega _{i}^{-1}\left( J,\theta ,Z,\left[ \Psi _{j}\right]
_{j=1,...,n}\right) \right) \Psi _{i}\left( \theta ,Z\right)  \label{tcn} \\
&&+\frac{1}{2\eta ^{2}}\left( S_{\Gamma }^{\left( 0\right) }+S_{\Gamma
}^{\left( 1\right) }+S_{\Gamma }^{\left( 2\right) }+S_{\Gamma }^{\left(
3\right) }+S_{\Gamma }^{\left( 4\right) }\right) +U\left( \left\{ \left\vert
\Gamma _{ij}\left( \theta ,Z,Z^{\prime },C,D\right) \right\vert ^{2}\right\}
\right)
\end{eqnarray}%
and equations for activities are defined by:

\begin{eqnarray}
\omega _{i}\left( \theta ,Z\right) &=&F_{i}\left( J\left( \theta \right) +%
\frac{\kappa }{N}\int \frac{\omega _{j}\left( \theta -\frac{\left\vert
Z-Z_{1}\right\vert }{c},Z_{1}\right) }{\omega _{i}\left( \theta ,Z\right) }%
G^{ij}\right.  \label{ftm} \\
&&\times \left. T_{ij}\left( Z,Z_{1}\right) \left( \mathcal{\bar{G}}%
_{0j}\left( 0,Z_{1}\right) +\left\vert \Psi _{j}\left( \theta -\frac{%
\left\vert Z-Z_{1}\right\vert }{c},Z_{1}\right) \right\vert ^{2}\right)
dZ_{1}\right)  \notag
\end{eqnarray}

The $n\times n$ matrix $G$ has coefficients in the interval $\left[ -1,1%
\right] $. In the sequel, the sum over index $j$ is implicit. For instance,
if $n=2$, the matrix $g$:%
\begin{equation*}
G=\left( 
\begin{array}{cc}
1 & -g \\ 
-g & 1%
\end{array}%
\right)
\end{equation*}%
represents inhibitory interactions between the two populations of cells.

As in the one field case, we define:

\begin{equation*}
T_{ij}\left( Z,Z_{1}\right) =\int T_{ij}\left\vert \Gamma _{ij}\left( T_{ij},%
\hat{T}_{ij},\theta ,Z,Z^{\prime },C_{ij},D_{ij}\right) \right\vert ^{2}
\end{equation*}%
and $S_{\Gamma }^{\left( 1\right) }$, $S_{\Gamma }^{\left( 2\right) }$, $%
S_{\Gamma }^{\left( 3\right) }$, $S_{\Gamma }^{\left( 4\right) }$ are given
by: 
\begin{eqnarray}
S_{\Gamma }^{\left( 1\right) } &=&\int \Gamma _{ij}^{\dag }\left( T_{ij},%
\hat{T}_{ij},\theta ,Z,Z^{\prime },C_{ij},D_{ij}\right) \nabla
_{T_{ij}}\left( \frac{\sigma _{T}^{2}}{2}\nabla _{T_{ij}}-\left( -\frac{1}{%
\tau \omega _{i}}T_{ij}+\frac{\lambda }{\omega _{i}}\hat{T}_{ij}\right)
\left\vert \Psi _{i}\left( \theta ,Z\right) \right\vert ^{2}\right) \\
&&\times \Gamma _{ij}\left( T_{ij},\hat{T}_{ij},\theta ,Z,Z^{\prime
},C_{ij},D_{ij}\right)  \notag
\end{eqnarray}%
\begin{eqnarray}
S_{\Gamma }^{\left( 2\right) } &=&\int \Gamma _{ij}^{\dag }\left( T_{ij},%
\hat{T}_{ij},\theta ,Z,Z^{\prime },C_{ij},D_{ij}\right) \\
&&\times \nabla _{\hat{T}_{ij}}\left( \frac{\sigma _{\hat{T}_{ij}}^{2}}{2}%
\nabla _{\hat{T}_{ij}}-\frac{\rho }{\omega _{i}}\left( \left( h_{ij}\left(
Z,Z^{\prime }\right) -\hat{T}_{ij}\right) C_{ij}\left\vert \Psi _{i}\left(
\theta ,Z\right) \right\vert ^{2}h_{C}\left( \omega _{i}\right) \right.
\right.  \notag \\
&&\left. \left. -D_{ij}\hat{T}_{ij}\left\vert \Psi _{j}\left( \theta -\frac{%
\left\vert Z-Z^{\prime }\right\vert }{c},Z^{\prime }\right) \right\vert
^{2}h_{D}\left( \omega _{j}^{\prime }\right) \right) \right) \Gamma
_{ij}\left( T_{ij},\hat{T}_{ij},\theta ,Z,Z^{\prime },C_{ij},D_{ij}\right) 
\notag
\end{eqnarray}%
\begin{eqnarray}
S_{\Gamma }^{\left( 3\right) } &=&\Gamma _{ij}^{\dag }\left( T,\hat{T}%
,\theta ,Z,Z^{\prime },C_{ij},D_{ij}\right) \\
&&\times \nabla _{C_{ij}}\left( \frac{\sigma _{C_{ij}}^{2}}{2}\nabla
_{C_{ij}}+\left( \frac{C_{ij}}{\tau _{C}\omega _{i}}-\alpha _{C}\left(
1-C_{ij}\right) \frac{\omega _{j}^{\prime }\left\vert \Psi _{j}\left( \theta
-\frac{\left\vert Z-Z^{\prime }\right\vert }{c},Z^{\prime }\right)
\right\vert ^{2}}{\omega _{i}}\right) \left\vert \Psi _{i}\left( \theta
,Z\right) \right\vert ^{2}\right)  \notag \\
&&\times \Gamma _{ij}\left( T,\hat{T},\theta ,Z,Z^{\prime
},C_{ij},D_{ij}\right)  \notag
\end{eqnarray}%
\begin{eqnarray}
S_{\Gamma }^{\left( 4\right) } &=&\Gamma _{ij}^{\dag }\left( T,\hat{T}%
,\theta ,Z,Z^{\prime },C_{ij},D_{ij}\right) \nabla _{D_{ij}}\left( \frac{%
\sigma _{D}^{2}}{2}\nabla _{D_{ij}}+\left( \frac{D_{ij}}{\tau _{D}\omega _{i}%
}-\alpha _{D_{ij}}\left( 1-D_{ij}\right) \left\vert \Psi _{i}\left( \theta
,Z\right) \right\vert ^{2}\right) \right) \\
&&\times \Gamma _{ij}\left( T,\hat{T},\theta ,Z,Z^{\prime
},C_{ij},D_{ij}\right)  \notag
\end{eqnarray}%
where;%
\begin{eqnarray*}
\omega _{i} &=&\omega _{i}\left( J,\theta ,Z,\left\vert \Psi \right\vert
^{2}\right) \\
\omega _{j}^{\prime } &=&\omega _{j}\left( J,\theta -\frac{\left\vert
Z-Z^{\prime }\right\vert }{c},Z^{\prime },\left\vert \Psi \right\vert
^{2}\right)
\end{eqnarray*}%
In (\ref{flt}), we added a potential:%
\begin{equation*}
U\left( \left\{ \left\vert \Gamma \left( \theta ,Z,Z^{\prime },C,D\right)
\right\vert ^{2}\right\} \right) =U\left( \int T\left\vert \Gamma \left( T,%
\hat{T},\theta ,Z,Z^{\prime },C,D\right) \right\vert ^{2}dTd\hat{T}\right)
\end{equation*}%
that models the constraint about the number of active connections in the
system. The resolution for the bckground fld follows several stps.

\subsection{Equilibrium \ activities}

The equilibrium activities (\ref{ftm}) can be rewritten by replacing $%
T_{ij}\left( Z,Z_{1}\right) $ with its average values depending on the
connectivity field: 
\begin{equation*}
\omega _{i}\left( Z\right) =G\left( \sum_{j}\int \frac{\kappa }{N}\frac{%
G^{ij}T_{ij}\left\vert \Gamma _{ij}\left( T,\hat{T},Z,Z_{1}\right)
\right\vert ^{2}\omega _{j}\left( J,Z_{1}\right) }{\omega _{i}\left(
Z\right) }\left( \mathcal{G}_{j0}+\left\vert \Psi _{j}\left( Z_{1}\right)
\right\vert ^{2}\right) dZ_{1}\right)
\end{equation*}%
Considering the interactions between the different types of fields as
relatively weak, we can perform an expansion of this equation around the
non-interacting firing rates:%
\begin{eqnarray*}
\omega _{i}\left( Z\right) &\simeq &G\left( \int \frac{\kappa }{N}\frac{%
G^{ii}T_{ii}\left\vert \Gamma _{ii}\left( T,\hat{T},Z,Z_{1}\right)
\right\vert ^{2}\omega _{i}\left( Z_{1}\right) }{\omega _{i}\left( Z\right) }%
\left( \mathcal{G}_{i0}+\left\vert \Psi _{i}\left( Z_{1}\right) \right\vert
^{2}\right) dZ_{1}\right) \\
&&+\sum_{j}G^{\prime }\left( \int \frac{\kappa }{N}\frac{G^{ii}T_{ii}\left%
\vert \Gamma _{ii}\left( T,\hat{T},Z,Z_{1}\right) \right\vert ^{2}\omega
_{i}\left( Z_{1}\right) }{\omega _{i}\left( Z\right) }\left( \mathcal{G}%
_{i0}+\left\vert \Psi _{i}\left( Z_{1}\right) \right\vert ^{2}\right)
dZ_{1}\right) \\
&&\times \sum_{j\neq i}\int \frac{\kappa }{N}\frac{G^{ij}T_{ij}\left\vert
\Gamma _{ij}\left( T,\hat{T},Z,Z_{1}\right) \right\vert ^{2}\omega
_{j}\left( J,Z_{1}\right) }{\omega _{i}\left( Z\right) }\left( \mathcal{G}%
_{j0}+\left\vert \Psi _{j}\left( Z_{1}\right) \right\vert ^{2}\right) dZ_{1}
\end{eqnarray*}%
Given that:%
\begin{equation*}
\omega _{0i}\left( Z\right) =G\left( \int \frac{\kappa }{N}\frac{%
G^{ii}T_{ii}\left\vert \Gamma _{ii}\left( T,\hat{T},Z,Z_{1}\right)
\right\vert ^{2}\omega _{i}\left( Z_{1}\right) }{\omega _{i}\left( Z\right) }%
\left( \mathcal{G}_{i0}+\left\vert \Psi _{i}\left( Z_{1}\right) \right\vert
^{2}\right) dZ_{1}\right)
\end{equation*}%
and that:%
\begin{equation*}
G^{\prime }\left( \int \frac{\kappa }{N}\frac{G^{ii}T_{ii}\left\vert \Gamma
_{ii}\left( T,\hat{T},Z,Z_{1}\right) \right\vert ^{2}\omega _{i}\left(
Z_{1}\right) }{\omega _{i}\left( Z\right) }\left( \mathcal{G}%
_{i0}+\left\vert \Psi _{i}\left( Z_{1}\right) \right\vert ^{2}\right)
dZ_{1}\right) =G^{\prime }\left( G^{-1}\left( \omega _{0i}\left( Z\right)
\right) \right)
\end{equation*}%
we find the solution for $\omega _{i}\left( Z\right) $ at the first order: 
\begin{eqnarray*}
&&\omega _{i}\left( Z\right) =\sum_{j}\left( 1-G^{\prime }\left(
G^{-1}\left( \omega _{0i}\left( Z\right) \right) \right) \right. \\
&&\times \left. \left( \int \frac{\kappa }{N}\frac{G^{ij}T_{ij}\left\vert
\Gamma _{ij}\left( T,\hat{T},Z,Z_{1}\right) \right\vert ^{2}\omega
_{j}\left( J,Z_{1}\right) }{\omega _{0i}\left( Z\right) }\left( \mathcal{G}%
_{j0}+\left\vert \Psi _{j}\left( Z_{1}\right) \right\vert ^{2}\right)
dZ_{1}\right) _{j\neq i}\right) _{ij}^{-1}\omega _{0j}\left( Z\right)
\end{eqnarray*}%
for inhibitory interactions $\omega _{i}\left( Z\right) <\omega _{0i}\left(
Z\right) $.

\subsection{Background fields for connectivity functions}

The formula are similar to the case of one type of cells. We find:%
\begin{eqnarray}
&&\Gamma _{aij}\left( T_{ij},\hat{T}_{ij},\theta ,C_{ij},D_{i}\right)
\label{gmc} \\
&\simeq &\left\{ \mathcal{N}\exp \left( -\frac{1}{2\sigma _{C}^{2}}\left( 
\frac{1}{\tau _{C}\omega _{i}}+\alpha _{C}\frac{\omega _{j}^{\prime
}\left\vert \Psi _{j}\left( \theta -\frac{\left\vert Z-Z^{\prime
}\right\vert }{c},Z^{\prime }\right) \right\vert ^{2}}{\omega _{i}}\right)
\left\vert \Psi _{i}\left( \theta ,Z\right) \right\vert ^{2}\left(
C_{ij}-C_{ij}\left( \theta \right) \right) ^{2}\right) \right.  \notag \\
&&\times \exp \left( -\frac{\left( \frac{1}{\tau _{D}\omega _{i}}+\alpha
_{D}\left\vert \Psi _{i}\left( \theta ,Z\right) \right\vert ^{2}\right) }{%
2\sigma _{D}^{2}}\left( D_{i}-D_{i}\left( \theta \right) \right) ^{2}\right)
\exp \left( -\frac{\rho \left\vert \bar{\Psi}_{ij}\left( \theta ,Z,Z^{\prime
}\right) \right\vert ^{2}}{2\sigma _{\hat{T}}^{2}\omega _{i}}\left( \hat{T}%
_{ij}-\left\langle \hat{T}_{ij}\right\rangle \right) ^{2}\right)  \notag \\
&&\times \left. \left\Vert \Gamma _{0ij}\left( \theta ,Z,Z^{\prime }\right)
\right\Vert \exp \left( -\frac{\left\vert \Psi _{i}\left( \theta ,Z\right)
\right\vert ^{2}}{2\sigma _{T}^{2}\tau \omega _{i}}\left(
T_{ij}-\left\langle T_{ij}\right\rangle \right) ^{2}\right) \right\}
_{\left( Z,Z^{\prime }\right) ,\left\langle T_{ij}\left( Z,Z^{\prime
}\right) \right\rangle \neq 0}  \notag
\end{eqnarray}%
\begin{eqnarray}
&&\Gamma _{u}\left( T_{ij},\hat{T}_{ij},\theta ,C_{ij},D_{i}\right)
\label{GMu} \\
&\simeq &\left\{ \mathcal{N}\exp \left( -\frac{1}{2\sigma _{C}^{2}}\left( 
\frac{1}{\tau _{C}\omega _{i}}+\alpha _{C}\frac{\omega _{j}^{\prime
}\left\vert \Psi _{j}\left( \theta -\frac{\left\vert Z-Z^{\prime
}\right\vert }{c},Z^{\prime }\right) \right\vert ^{2}}{\omega _{i}}\right)
\left\vert \Psi _{i}\left( \theta ,Z\right) \right\vert ^{2}\left(
C_{ij}-C_{ij}\left( \theta \right) \right) ^{2}\right) \right.  \notag \\
&&\times \exp \left( -\frac{\left( \frac{1}{\tau _{D}\omega _{i}}+\alpha
_{D}\left\vert \Psi _{i}\left( \theta ,Z\right) \right\vert ^{2}\right) }{%
2\sigma _{D}^{2}}\left( D_{i}-\left\langle D_{i}\right\rangle \right)
^{2}\right)  \notag \\
&&\left. \exp \left( -\frac{\rho \left\vert \bar{\Psi}_{ij}\left( \theta
,Z,Z^{\prime }\right) \right\vert ^{2}}{2\sigma _{\hat{T}}^{2}}\left( \hat{T}%
_{ij}-\left\langle \hat{T}_{ij}\right\rangle \right) ^{2}\right) \delta
\left( T_{ij}\right) \right\} _{\left( Z,Z^{\prime }\right) ,\left\langle
T_{ij}\left( Z,Z^{\prime }\right) \right\rangle =0}  \notag
\end{eqnarray}%
with:%
\begin{equation*}
\left\vert \bar{\Psi}_{ij}\left( \theta ,Z,Z^{\prime }\right) \right\vert
^{2}=\frac{\left( C_{ij}\left( \theta \right) \left\vert \Psi _{i}\left(
\theta ,Z\right) \right\vert ^{2}h_{C}\left( \omega _{i}\left( \theta
,Z\right) \right) +D_{i}\left( \theta \right) \left\vert \Psi _{j}\left(
\theta -\frac{\left\vert Z-Z^{\prime }\right\vert }{c},Z^{\prime }\right)
\right\vert ^{2}h_{D}\left( \omega _{j}\left( \theta -\frac{\left\vert
Z-Z^{\prime }\right\vert }{c},Z^{\prime }\right) \right) \right) }{2\omega
_{i}\left( \theta ,Z\right) }
\end{equation*}%
and where $\mathcal{N}$ is a normalization factor ensuring that the
constraint over the number of connections is satisfied. For the conjugate
fields, we find:%
\begin{equation*}
\Gamma _{a}^{\dagger }\left( T,\hat{T},\theta ,C,D\right) \simeq \left\{
1\right\} _{\left( Z,Z^{\prime }\right) ,\left\langle T\left( Z,Z^{\prime
}\right) \right\rangle \neq 0}
\end{equation*}%
\begin{equation*}
\Gamma _{u}^{\dagger }\left( T,\hat{T},\theta ,C,D\right) \simeq \left\{
\delta \left( T\right) \right\} _{\left( Z,Z^{\prime }\right) ,\left\langle
T\left( Z,Z^{\prime }\right) \right\rangle =0}
\end{equation*}

\subsection{averages for the background field}

To complete the resolution of the minimization equation, we also need the
averages of the dynamic variables in the state defined by the background
field. The generalization from the one-field case is straightforward:%
\begin{eqnarray}
C_{ij} &\rightarrow &\left\langle C\left( \theta \right) \right\rangle =%
\frac{\alpha _{C}\frac{\omega _{j}^{\prime }\left\vert \Psi _{j}\left(
\theta -\frac{\left\vert Z-Z^{\prime }\right\vert }{c},Z^{\prime }\right)
\right\vert ^{2}}{\omega _{i}}}{\frac{1}{\tau _{C}\omega _{i}}+\alpha _{C}%
\frac{\omega _{j}^{\prime }\left\vert \Psi _{j}\left( \theta -\frac{%
\left\vert Z-Z^{\prime }\right\vert }{c},Z^{\prime }\right) \right\vert ^{2}%
}{\omega _{i}}} \\
&=&\frac{\alpha _{C}\omega _{j}^{\prime }\left\vert \Psi _{j}\left( \theta -%
\frac{\left\vert Z-Z^{\prime }\right\vert }{c},Z^{\prime }\right)
\right\vert ^{2}}{\frac{1}{\tau _{C}}+\alpha _{C}\omega _{j}^{\prime
}\left\vert \Psi _{j}\left( \theta -\frac{\left\vert Z-Z^{\prime
}\right\vert }{c},Z^{\prime }\right) \right\vert ^{2}}\equiv C_{j}\left(
\theta ,Z,Z^{\prime }\right)  \notag \\
D_{i} &\rightarrow &\left\langle D\left( \theta \right) \right\rangle =\frac{%
\alpha _{D}\omega _{i}\left\vert \Psi \left( \theta ,Z\right) \right\vert
^{2}}{\frac{1}{\tau _{D}}+\alpha _{D}\omega _{j}^{\prime }\left\vert \Psi
\left( \theta ,Z\right) \right\vert ^{2}}\equiv D\left( \theta \right)
\end{eqnarray}%
\begin{eqnarray}
\left\langle T_{ij}\left( Z,Z^{\prime }\right) \right\rangle &=&\lambda \tau
\left\langle \hat{T}_{ij}\left( Z,Z^{\prime }\right) \right\rangle
\label{Tht} \\
&=&\frac{\lambda \tau h\left( Z,Z^{\prime }\right) \left\langle C_{j}\left(
\theta ,Z,Z^{\prime }\right) h_{C}\left( \omega _{i}\left( \theta ,Z\right)
\right) \left\vert \Psi _{i}\left( \theta ,Z\right) \right\vert
^{2}\right\rangle }{C_{j}\left( \theta ,Z,Z^{\prime }\right) \left\vert \Psi
_{i}\left( \theta ,Z\right) \right\vert ^{2}h_{C}\left( \omega _{i}\left(
\theta ,Z\right) \right) +D_{i}\left( \theta ,Z,Z^{\prime }\right)
\left\vert \Psi _{i}\left( \theta -\frac{\left\vert Z-Z^{\prime }\right\vert 
}{c},Z^{\prime }\right) \right\vert ^{2}h_{D}\left( \omega _{j}\left( \theta
-\frac{\left\vert Z-Z^{\prime }\right\vert }{c},Z^{\prime }\right) \right) }
\notag
\end{eqnarray}

We look for a static background state for the whole system. In the static
case, we assume that the static background field $\Psi _{0j}\left( Z\right) $
is the minimum of $V\left( \Psi _{j}\right) $.

\subsection{General equations and methdod of resolution}

Derivation of equations for connectivity functions is similar to one field
case. As before, we first solve for the activities, then for the
connectivities and the neurons fields.

\subsubsection{Expression of $\protect\omega \left( Z\right) $}

We use (\ref{mnt}) to replace $\omega _{j}^{\prime }\left\vert \Psi
_{j}\left( Z^{\prime }\right) \right\vert ^{2}$:

\begin{equation*}
\omega _{j}\left( Z^{\prime }\right) \left\vert \Psi _{j}\left( Z^{\prime
}\right) \right\vert ^{2}=\frac{\lambda \tau \exp \left( -\frac{\left\vert
Z-Z^{\prime }\right\vert }{\nu c}\right) -T_{ij}\left( Z,Z^{\prime }\right) 
}{T_{ij}\left( Z,Z^{\prime }\right) }\left( \frac{1}{\alpha _{D}\tau _{D}}%
+\omega _{i}\left\vert \Psi _{i}\left( Z\right) \right\vert ^{2}\right) -%
\frac{1}{\alpha _{C}\tau _{C}}
\end{equation*}%
which yilds:%
\begin{eqnarray*}
\omega _{i}\left( Z\right) &=&G\left( \sum_{j}\int \frac{\kappa }{N}\frac{%
G^{ij}T_{ij}\left\vert \Gamma _{ij}\left( T,\hat{T},Z,Z_{1}\right)
\right\vert ^{2}\omega _{j}\left( J,Z_{1}\right) }{\omega _{i}\left(
Z\right) }\left( \mathcal{G}_{j0}+\left\vert \Psi _{j}\left( Z_{1}\right)
\right\vert ^{2}\right) dZ_{1}\right) \\
&\rightarrow &G\left( \sum_{j}G^{ij}\int \frac{\kappa }{N}\left( \frac{%
\lambda \tau \exp \left( -\frac{\left\vert Z-Z_{1}\right\vert }{\nu c}%
\right) -T_{ij}\left( Z,Z_{1}\right) }{T_{ij}\left( Z,Z_{1}\right) }\left( 
\frac{1}{\alpha _{D}\tau _{D}}+\omega _{i}\left\vert \Psi _{i}\left(
Z\right) \right\vert ^{2}\right) -\frac{1}{\alpha _{C}\tau _{C}}\right)
\right. \\
&&\times \left. \frac{T_{ij}\left\vert \Gamma _{ij}\left( T,\hat{T}%
,Z,Z_{1}\right) \right\vert ^{2}}{\omega _{i}\left( Z\right) }dZ_{1}\right)
\\
&\simeq &G\left( \sum_{j}G^{ij}\int \frac{\kappa }{N}\left( \left( \lambda
\tau \exp \left( -\frac{\left\vert Z-Z_{1}\right\vert }{\nu c}\right)
-T_{ij}\left( Z,Z_{1}\right) \right) \left( \left( \frac{1}{\alpha _{D}\tau
_{D}}-\frac{T_{ij}\left( Z,Z_{1}\right) }{\alpha _{C}\tau _{C}}\right)
\omega _{i}^{-1}+\left\vert \Psi _{i}\left( Z\right) \right\vert ^{2}\right)
\right) dZ_{1}\right)
\end{eqnarray*}%
We can replace $T\left( Z,Z_{1}\right) $ in the integral by its average:%
\begin{equation*}
\frac{1}{V}T_{ij}\left( Z\right) =\frac{1}{V}\int T_{ij}\left(
Z,Z_{1}\right) dZ_{1}
\end{equation*}%
so that:%
\begin{equation*}
\omega _{i}\left( Z\right) =G\left( \sum_{j}G^{ij}\int \frac{\kappa }{N}%
\left( \left( \lambda \tau \nu c-T_{ij}\left( Z\right) \right) \left( \left( 
\frac{1}{\alpha _{D}\tau _{D}}-\frac{T_{ij}\left( Z\right) }{V\alpha
_{C}\tau _{C}}\right) \omega _{i}^{-1}+\left\vert \Psi _{i}\left( Z\right)
\right\vert ^{2}\right) \right) \right)
\end{equation*}%
The solution for the firing rates is defined by a function:%
\begin{equation*}
\omega _{i}\left( Z\right) =\hat{G}\left( \left( G^{ij}\right) ,\left(
T_{ij}\left( Z\right) \right) ,\left\vert \Psi _{i}\left( Z\right)
\right\vert ^{2}\right) =\hat{G}_{i}\left( \left( T_{ij}\left( Z\right)
\right) ,\left\vert \Psi _{i}\left( Z\right) \right\vert ^{2}\right)
\end{equation*}%
Here, $\left( T_{ij}\left( Z\right) \right) $ denotes the set of all the $%
T_{ij}\left( Z\right) $ with $j$ running over the whole space.

To find $\omega _{i}\left( Z\right) $, we thus have to determine $%
T_{ij}\left( Z\right) $ and $\left\vert \Psi _{i}\left( Z\right) \right\vert
^{2}$.

\subsubsection{Equations for $T_{ij}\left( Z\right) $, $T_{ij}\left(
Z,Z^{\prime }\right) $ and $\left\vert \Psi _{i}\left( Z\right) \right\vert
^{2}$}

As in the one field case, the average connectivities at $Z$ are given by:%
\begin{equation}
T_{ij}\left( Z\right) =\frac{\lambda \tau \nu c}{1+\frac{\frac{1}{\tau
_{C}\alpha _{C}}+\Omega _{j}}{\frac{1}{\tau _{D}\alpha _{D}}+\hat{G}\left(
\left( T_{ij}\left( Z\right) \right) ,\left\vert \Psi _{i}\left( Z\right)
\right\vert ^{2}\right) \left\vert \Psi _{i}\left( Z\right) \right\vert ^{2}}%
}=\frac{\lambda \tau \nu c\left( \frac{1}{\tau _{D}\alpha _{D}}+\hat{G}%
_{i}\left( \left( T_{ij}\left( Z\right) \right) ,\left\vert \Psi _{i}\left(
Z\right) \right\vert ^{2}\right) \left\vert \Psi _{i}\left( Z\right)
\right\vert ^{2}\right) }{\frac{1}{\tau _{D}\alpha _{D}}+\frac{1}{\tau
_{C}\alpha _{C}}+\Omega _{j}+\hat{G}_{j}\left( \left( T_{ij}\left( Z\right)
\right) ,\left\vert \Psi _{i}\left( Z\right) \right\vert ^{2}\right)
\left\vert \Psi _{i}\left( Z\right) \right\vert ^{2}}  \label{TR}
\end{equation}%
where:%
\begin{equation}
\Omega _{j}=\frac{1}{V}\int \hat{G}_{j}\left( \left( T_{jk}\left( Z^{\prime
}\right) \right) ,\left\vert \Psi _{j}\left( Z^{\prime }\right) \right\vert
^{2}\right) \left\vert \Psi _{j}\left( Z^{\prime }\right) \right\vert
^{2}dZ^{\prime }  \label{GJ}
\end{equation}%
and this leads to the following formula for $T_{ij}\left( Z,Z^{\prime
}\right) $: 
\begin{equation}
T_{ij}\left( Z,Z^{\prime }\right) =\frac{\lambda \tau \exp \left( -\frac{%
\left\vert Z-Z^{\prime }\right\vert }{\nu c}\right) }{1+\frac{\frac{1}{\tau
_{C}\alpha _{C}}+\omega _{j}^{\prime }\left\vert \Psi _{j}\left( Z^{\prime
}\right) \right\vert ^{2}}{\frac{1}{\tau _{D}\alpha _{D}}+\omega
_{i}\left\vert \Psi _{i}\left( Z\right) \right\vert ^{2}}}=\frac{\lambda
\tau \exp \left( -\frac{\left\vert Z-Z^{\prime }\right\vert }{\nu c}\right) 
}{1+\frac{\frac{1}{\tau _{C}\alpha _{C}}+\hat{G}_{j}\left( \left(
T_{jk}\left( Z\right) \right) ,\left\vert \Psi _{j}\left( Z^{\prime }\right)
\right\vert ^{2}\right) \left\vert \Psi _{j}\left( Z^{\prime }\right)
\right\vert ^{2}}{\frac{1}{\tau _{D}\alpha _{D}}+\hat{G}_{i}\left( \left(
T_{ij}\left( Z\right) \right) ,\left\vert \Psi _{i}\left( Z\right)
\right\vert ^{2}\right) \left\vert \Psi _{i}\left( Z\right) \right\vert ^{2}}%
}  \label{FT}
\end{equation}%
On the other hand, the minimization equation for the field $\Psi _{i}$ is: 
\begin{equation*}
0\simeq \left( \left( \frac{1}{\hat{G}_{i}\left( \left( T_{ij}\left(
Z,Z_{1}\right) \right) _{Z_{1}},\left\vert \Psi _{i}\left( Z\right)
\right\vert ^{2}\right) }\right) ^{2}+\left( \left\vert \Psi _{i}\left(
Z\right) \right\vert ^{2}-\int T_{ij}\left( Z^{\prime },Z_{1}\right)
\left\vert \Psi _{j0}\left( Z\right) \right\vert ^{2}dZ_{1}\right) \right)
\Psi _{i}\left( \theta ,Z\right)
\end{equation*}%
with solutions:%
\begin{equation*}
\Psi _{i}\left( \theta ,Z\right) =0
\end{equation*}%
or $\left\vert \Psi _{i}\left( Z\right) \right\vert ^{2}$ satisfying:%
\begin{equation*}
\left( \frac{1}{\hat{G}_{i}\left( \left( T_{ij}\left( Z,Z_{1}\right) \right)
_{Z_{1}},\left\vert \Psi _{i}\left( Z\right) \right\vert ^{2}\right) }%
\right) ^{2}+\left\vert \Psi _{i}\left( Z\right) \right\vert ^{2}\simeq
\sum_{j}\int T_{ij}\left( Z,Z^{\prime }\right) \left\vert \Psi _{j0}\left(
Z^{\prime }\right) \right\vert ^{2}k_{ij}\left( Z,Z^{\prime }\right)
dZ^{\prime }
\end{equation*}%
This equation can be approximated by:%
\begin{eqnarray}
\left( \frac{1}{\hat{G}_{i}\left( \left( T_{ij}\left( Z,Z_{1}\right) \right)
_{Z_{1}},\left\vert \Psi _{i}\left( Z\right) \right\vert ^{2}\right) }%
\right) ^{2}+\left\vert \Psi _{i}\left( Z\right) \right\vert ^{2} &\simeq
&\sum_{j}T_{ij}\left( Z\right) \frac{\int \left\vert \Psi _{j0}\left(
Z^{\prime }\right) \right\vert ^{2}k_{ij}\left( Z,Z^{\prime }\right)
dZ^{\prime }}{V}  \label{PL} \\
&\equiv &\sum_{j}T_{ij}\left( Z\right) \left\langle \left\vert \Psi
_{j0}\left( Z^{\prime }\right) \right\vert ^{2}\right\rangle _{Z}  \notag
\end{eqnarray}

\subsubsection{Expression of $\left\vert \Psi _{i}\left( Z\right)
\right\vert ^{2}$ and $T_{ij}\left( Z\right) $ as functions of average values%
}

To solve (\ref{TR}) and (\ref{PL}) for $T_{ij}\left( Z\right) $ and $%
\left\vert \Psi _{i}\left( Z\right) \right\vert ^{2}$, we first use (\ref{TR}%
) to express $\hat{G}\left( \left( T_{ij}\left( Z\right) \right) ,\left\vert
\Psi _{i}\left( Z\right) \right\vert ^{2}\right) \left\vert \Psi _{i}\left(
Z\right) \right\vert ^{2}$ as a function of $T_{ij}\left( Z\right) $: 
\begin{equation}
\hat{G}_{i}\left( \left( T_{ij}\left( Z\right) \right) ,\left\vert \Psi
_{i}\left( Z\right) \right\vert ^{2}\right) \left\vert \Psi _{i}\left(
Z\right) \right\vert ^{2}=\frac{\left( \frac{1}{\tau _{D}\alpha _{D}}+\frac{1%
}{\tau _{C}\alpha _{C}}+\Omega _{j}\right) T_{ij}\left( Z\right) -\frac{1}{%
\tau _{D}\alpha _{D}}\lambda \tau \nu c}{\lambda \tau \nu c-T_{ij}\left(
Z\right) }  \label{TV}
\end{equation}%
Inserting this result in (\ref{Pl}) leads to thefollowng equation for $%
\left\vert \Psi \left( Z\right) \right\vert ^{2}$:

\begin{equation}
\left( \frac{\lambda \tau \nu c-T_{ij}\left( Z\right) }{\left( \frac{1}{\tau
_{D}\alpha _{D}}+\frac{1}{\tau _{C}\alpha _{C}}+\Omega _{j}\right)
T_{ij}\left( Z\right) -\frac{1}{\tau _{D}\alpha _{D}}\lambda \tau \nu c}%
\left\vert \Psi _{i}\left( Z\right) \right\vert ^{2}\right) ^{2}+\left\vert
\Psi _{i}\left( Z\right) \right\vert ^{2}\simeq \sum_{j}T_{ij}\left(
Z\right) \left\langle \left\vert \Psi _{j0}\left( Z^{\prime }\right)
\right\vert ^{2}\right\rangle _{Z}
\end{equation}%
with solution:%
\begin{equation}
\left\vert \Psi _{i}\left( Z\right) \right\vert ^{2}=\frac{%
2\sum_{j}T_{ij}\left( Z\right) \left\langle \left\vert \Psi _{j0}\left(
Z^{\prime }\right) \right\vert ^{2}\right\rangle _{Z}}{\left( 1+\sqrt{%
1+4\left( \frac{\lambda \tau \nu c-T_{ij}\left( Z\right) }{\left( \frac{1}{%
\tau _{D}\alpha _{D}}+\frac{1}{\tau _{C}\alpha _{C}}+\Omega _{j}\right)
T_{ij}\left( Z\right) -\frac{1}{\tau _{D}\alpha _{D}}\lambda \tau \nu c}%
\right) ^{2}\sum_{j}T_{ij}\left( Z\right) \left\langle \left\vert \Psi
_{j0}\left( Z^{\prime }\right) \right\vert ^{2}\right\rangle _{Z}}\right) }
\label{PS}
\end{equation}%
Ultimately, inserting this result in (\ref{TV}) writes $\hat{G}_{i}$ as a
function of $\left( T_{ij}\left( Z\right) \right) $, $\left( \hat{\Omega}%
_{j}\right) $:%
\begin{equation}
\hat{G}_{i}\left( \left( T_{ij}\left( Z\right) \right) ,\left\vert \Psi
_{i}\left( Z\right) \right\vert ^{2}\right) \left\vert \Psi _{i}\left(
Z\right) \right\vert ^{2}=\hat{G}_{i}\left( \left( T_{ij}\left( Z\right)
\right) ,\left\vert \Psi _{i}\left( Z,\left( \hat{\Omega}_{j}\right)
,T_{ij}\left( Z\right) \right) \right\vert ^{2}\right)  \label{TN}
\end{equation}%
with:%
\begin{equation*}
\hat{\Omega}_{j}=\left( \frac{1}{\tau _{D}\alpha _{D}}+\frac{1}{\tau
_{C}\alpha _{C}}+\Omega _{j}\right)
\end{equation*}%
Averaging this relation will yield a consistency condtn for the $\hat{\Omega}%
_{j}$.

\subsubsection{Identification of $\Omega _{j}$ and $\bar{T}_{ij}\left(
Z\right) $}

The resolution is finalized by using (\ref{TN}) to identify the constant $%
\Omega _{j}$. Averagng (\ref{TN}) and using (\ref{GJ}) yields: 
\begin{eqnarray}
\Omega _{i} &=&\hat{G}_{i}\left( \left( \bar{T}_{ij}\right) ,\frac{2\sum_{j}%
\bar{T}_{ij}\left( Z\right) \left\langle \left\vert \Psi _{j0}\left(
Z^{\prime }\right) \right\vert ^{2}\right\rangle _{Z}}{\left( 1+\sqrt{1+%
\frac{4}{\Omega _{i}^{2}}\sum_{j}\bar{T}_{ij}\left( Z\right) \left\langle
\left\vert \Psi _{j0}\left( Z^{\prime }\right) \right\vert ^{2}\right\rangle
_{Z}}\right) }\right)  \label{PV} \\
&&\times \frac{2\sum_{j}\bar{T}_{ij}\left( Z\right) \left\langle \left\vert
\Psi _{j0}\left( Z^{\prime }\right) \right\vert ^{2}\right\rangle _{Z}}{%
\left( 1+\sqrt{1+\frac{4}{\Omega _{i}^{2}}\sum_{j}\bar{T}_{ij}\left(
Z\right) \left\langle \left\vert \Psi _{j0}\left( Z^{\prime }\right)
\right\vert ^{2}\right\rangle _{Z}}\right) }  \notag
\end{eqnarray}%
The $\bar{T}_{ij}\left( Z\right) $ can be replaced using (\ref{TV}) and (\ref%
{GJ}), and computing averages:

\begin{eqnarray}
\Omega _{j} &\simeq &\frac{1}{V}\int \hat{G}\left( \left( T_{jk}\left(
Z\right) \right) ,\left\vert \Psi _{j}\left( Z\right) \right\vert
^{2}\right) \left\vert \Psi _{j}\left( Z\right) \right\vert ^{2}  \label{GM}
\\
&=&\frac{1}{V}\int \frac{\left( \frac{1}{\tau _{D}\alpha _{D}}+\frac{1}{\tau
_{C}\alpha _{C}}+\Omega _{k}\right) T_{jk}\left( Z\right) -\frac{1}{\tau
_{D}\alpha _{D}}\lambda \tau \nu c}{\lambda \tau \nu c-T_{jk}\left( Z\right) 
}  \notag \\
&\simeq &\frac{\left( \frac{1}{\tau _{D}\alpha _{D}}+\frac{1}{\tau
_{C}\alpha _{C}}+\Omega _{k}\right) \bar{T}_{jk}-\frac{1}{\tau _{D}\alpha
_{D}}\lambda \tau \nu c}{\lambda \tau \nu c-\bar{T}_{jk}}  \notag
\end{eqnarray}%
where: 
\begin{equation}
\bar{T}_{ij}=\frac{1}{V}\int T_{ij}\left( Z\right) dZ  \label{CR}
\end{equation}%
is the average cnnctvt of the system. This can be solved for $\bar{T}_{ij}$:%
\begin{equation}
\bar{T}_{jk}=\frac{\lambda \tau \nu c\Omega _{j}+\frac{1}{\tau _{D}\alpha
_{D}}\lambda \tau \nu c}{\left( \frac{1}{\tau _{D}\alpha _{D}}+\frac{1}{\tau
_{C}\alpha _{C}}+\Omega _{j}+\Omega _{k}\right) }\simeq \frac{\lambda \tau
\nu c\Omega _{j}}{\Omega _{j}+\Omega _{k}}=\frac{2\bar{T}\Omega _{j}}{\Omega
_{j}+\Omega _{k}}  \label{VN}
\end{equation}%
Inserted in (\ref{PV}), this yields an equation for the $\Omega _{j}$. Once
found (\ref{VN}) yields $\bar{T}_{ij}$.

We can assume that the several types have approximatively the constant
ration of activt, so that $\frac{\Omega _{j}}{\Omega _{j}+\Omega _{k}}\simeq 
\frac{1}{2}g_{jk}$ so that:%
\begin{equation*}
\bar{T}_{jk}\simeq g_{jk}\bar{T}
\end{equation*}%
so that equation (\ref{PV}) bcms:%
\begin{equation}
\Omega _{i}=\hat{G}_{i}\left( \bar{T},\frac{2\bar{T}\sum_{j}g_{ij}\left%
\langle \left\vert \Psi _{j0}\left( Z^{\prime }\right) \right\vert
^{2}\right\rangle _{Z}}{\left( 1+\sqrt{1+\frac{4\bar{T}}{\Omega _{i}^{2}}%
\sum_{j}g_{ij}\left\langle \left\vert \Psi _{j0}\left( Z^{\prime }\right)
\right\vert ^{2}\right\rangle _{Z}}\right) }\right) \frac{2\bar{T}%
\sum_{j}g_{ij}\left\langle \left\vert \Psi _{j0}\left( Z^{\prime }\right)
\right\vert ^{2}\right\rangle _{Z}}{\left( 1+\sqrt{1+\frac{4\bar{T}}{\Omega
_{i}^{2}}\sum_{j}g_{ij}\left\langle \left\vert \Psi _{j0}\left( Z^{\prime
}\right) \right\vert ^{2}\right\rangle _{Z}}\right) }
\end{equation}%
and solvng ths equation ylds $\Omega _{i}$.

\subsubsection{Deriving $T_{ij}\left( Z\right) $}

Once the $\Omega _{j}$ is determined,, we can use (\ref{PS}) to substitute $%
\left\vert \Psi _{i}\left( Z\right) \right\vert ^{2}$ in (\ref{TR}). This
results in an equation for $T_{ij}\left( Z\right) $. Substituting the
various expressions in (\ref{FT}), we obtain the connectivity $T_{ij}\left(
Z,Z^{\prime }\right) $.

\subsection{Particular case}

We can solve the system for $G$ an increasing function of the form $%
G_{i}\left( x_{j}\right) =G\left( \sum_{j}G_{ij}x_{j}\right) \simeq
b_{0}G_{ij}x$ for $x_{j}<1$. We start with the derivation of $\Omega $:%
\begin{eqnarray*}
\omega _{i}\left( Z\right) &=&G\left( \sum_{j}G^{ij}\int \frac{\kappa }{N}%
\left( \left( \lambda \tau \nu c-T_{ij}\left( Z\right) \right) \left( \left( 
\frac{1}{\alpha _{D}\tau _{D}}-\frac{T_{ij}\left( Z\right) }{V\alpha
_{C}\tau _{C}}\right) \omega _{i}^{-1}+\left\vert \Psi _{i}\left( Z\right)
\right\vert ^{2}\right) \right) \right) \\
&\simeq &G\left( \sum_{j}VG^{ij}\frac{\kappa }{N}\left( \left( \lambda \tau
\nu c-\bar{T}_{ij}\right) \left( \left( \frac{1}{\alpha _{D}\tau _{D}}-\frac{%
\bar{T}_{ij}}{V\alpha _{C}\tau _{C}}\right) \omega _{i}^{-1}+\left\vert \Psi
_{i}\left( Z\right) \right\vert ^{2}\right) \right) \right)
\end{eqnarray*}%
In first approximation, we thus find for the activity:%
\begin{eqnarray*}
\omega _{i}\left( Z\right) &\simeq &G\left( \sum_{j}VG^{ij}\frac{\kappa }{N}%
\left( \left( \lambda \tau \nu c-\bar{T}_{ij}\right) \left( \left\vert \Psi
_{i}\left( Z\right) \right\vert ^{2}\right) \right) \right) \\
&\simeq &b_{0}\left( \sum_{j}VG^{ij}\frac{\kappa }{N}\left( \left( \lambda
\tau \nu c-\bar{T}_{ij}\right) \left( \left\vert \Psi _{i}\left( Z\right)
\right\vert ^{2}\right) \right) \right)
\end{eqnarray*}%
Given that the average $\bar{T}_{ij}$ is given by (\ref{GM}): 
\begin{equation*}
\bar{T}_{ij}=\frac{\lambda \tau \nu c\Omega _{i}+\frac{1}{\tau _{D}\alpha
_{D}}\lambda \tau \nu c}{\frac{1}{\tau _{D}\alpha _{D}}+\frac{1}{\tau
_{C}\alpha _{C}}+\Omega _{j}+\Omega _{i}}\simeq \frac{\lambda \tau \nu
c\Omega _{i}}{\Omega _{j}+\Omega _{i}}\simeq g_{ij}\bar{T}
\end{equation*}%
we obtain:%
\begin{equation*}
\omega _{i}\left( Z\right) \simeq bG^{i}\bar{T}\left\vert \Psi _{i}\left(
Z\right) \right\vert ^{2}
\end{equation*}%
with $G^{i}=\sum_{j}G^{ij}\left( 1-g_{ij}\right) $ and $b=b_{0}\frac{\kappa 
}{N}V$

\subsubsection{Derivation of the $\Omega _{i}$}

In this particular case, equation (\ref{mgb}) writes:

\begin{equation*}
\Omega _{i}=G^{i}b\bar{T}\left( \frac{2\sum_{j}\bar{T}_{ij}\left\langle
\left\vert \Psi _{j0}\left( Z^{\prime }\right) \right\vert ^{2}\right\rangle
_{Z}}{\left( 1+\sqrt{1+\frac{4}{\Omega _{i}^{2}}\sum_{j}\bar{T}%
_{ij}\left\langle \left\vert \Psi _{j0}\left( Z^{\prime }\right) \right\vert
^{2}\right\rangle _{Z}}\right) }\right) ^{2}
\end{equation*}%
that can be rewritten as:%
\begin{equation*}
1\simeq \frac{bG^{i}\bar{T}\Omega _{i}^{3}}{\left( bG^{i}\bar{T}\left(
\sum_{j}\bar{T}_{ij}\left\langle \left\vert \Psi _{j0}\left( Z^{\prime
}\right) \right\vert ^{2}\right\rangle _{Z}\right) \Omega _{i}-1\right) ^{2}}
\end{equation*}%
Defining:%
\begin{equation*}
\bar{T}\left\langle \left\vert \bar{\Psi}_{i0}\left( Z^{\prime }\right)
\right\vert ^{2}\right\rangle _{Z}=\sum_{j}\bar{T}_{ij}\left\langle
\left\vert \Psi _{j0}\left( Z^{\prime }\right) \right\vert ^{2}\right\rangle
_{Z}
\end{equation*}%
and its average:%
\begin{equation*}
\bar{T}\left\langle \left\vert \bar{\Psi}_{i0}\right\vert ^{2}\right\rangle
=\left\langle \sum_{j}\bar{T}_{ij}\left\langle \left\vert \Psi _{j0}\left(
Z^{\prime }\right) \right\vert ^{2}\right\rangle _{Z}\right\rangle
\end{equation*}%
we find:

For $d=\left( bG^{i}\bar{T}\right) ^{2}\left( \bar{T}\left\langle \left\vert 
\bar{\Psi}_{i0}\right\vert ^{2}\right\rangle \right) ^{3}<\frac{27}{4}$
there is one solution:%
\begin{equation}
\Omega _{i}\simeq \left( bG^{i}\bar{T}\right) ^{-\frac{1}{3}}<<1  \label{SLN}
\end{equation}

For $d=\left( bG^{i}\bar{T}\right) ^{2}\left( \bar{T}\left\langle \left\vert 
\bar{\Psi}_{i0}\right\vert ^{2}\right\rangle \right) ^{3}>\frac{27}{4}$
there are three solutions. The first one is:%
\begin{equation}
\Omega _{i}\simeq \left( bG^{i}\bar{T}\right) \left( \bar{T}\left\langle
\left\vert \bar{\Psi}_{i0}\right\vert ^{2}\right\rangle _{Z}\right) ^{2}
\label{SLD}
\end{equation}%
The two other solutions are centered around $\frac{1}{bG^{i}\bar{T}%
\left\langle \bar{T}\left\vert \bar{\Psi}_{i0}\right\vert ^{2}\right\rangle }
$. We set: 
\begin{equation*}
\Omega _{i}=\frac{1\pm \delta }{bG^{i}\bar{T}\left\langle \bar{T}\left\vert 
\bar{\Psi}_{i0}\right\vert ^{2}\right\rangle }
\end{equation*}
and (\ref{RCDTQT}) becomes: 
\begin{equation*}
\left( bG^{i}\bar{T}\right) ^{2}\left( \bar{T}\left\langle \left\vert \bar{%
\Psi}_{i0}\right\vert ^{2}\right\rangle \right) ^{3}\simeq \frac{1}{\delta
^{2}}
\end{equation*}%
so that:%
\begin{equation}
\Omega _{i}=\frac{1\pm \sqrt{\frac{1}{\left( bG^{i}\bar{T}\right) ^{2}\left( 
\bar{T}\left\langle \left\vert \bar{\Psi}_{i0}\right\vert ^{2}\right\rangle
\right) ^{3}}}}{bG^{i}\bar{T}\left\langle \bar{T}\left\vert \bar{\Psi}%
_{i0}\right\vert ^{2}\right\rangle }  \label{SLDF}
\end{equation}

Solution (\ref{SLN}) corresponds to relatively low activity, i.e. $%
\left\langle \left\vert \bar{\Psi}_{i0}\right\vert ^{2}\right\rangle <<1$,
so we only consider solutions (\ref{SLD}) and (\ref{SLDF}) in the sequel.
For $\lambda \tau \nu cb^{2}>>1$ the solutions of (\ref{SLDF}) are both
approximatively given by:%
\begin{equation*}
\Omega _{i}=\frac{1}{bG^{i}\bar{T}\left\langle \bar{T}\left\vert \bar{\Psi}%
_{i0}\right\vert ^{2}\right\rangle }
\end{equation*}%
Writing $\Omega _{i-}$ this solution and $\Omega _{i+}$ the solution (\ref%
{SLD}), we gather them and write $\Omega _{i\pm }$. To these solutions for $%
\Omega $, we associate the $Z$ dependent parameters:%
\begin{equation}
Y_{i+}\left( Z\right) \simeq \frac{\left( bG^{i}\bar{T}\right) \left( \bar{T}%
\left\langle \left\vert \bar{\Psi}_{i0}\left( Z^{\prime }\right) \right\vert
^{2}\right\rangle _{Z}\right) ^{2}}{2}  \label{Sltdn}
\end{equation}

\begin{equation}
Y_{i-}\left( Z\right) =\frac{1}{bG^{i}\bar{T}\left\langle \bar{T}\left\vert 
\bar{\Psi}_{i0}\left( Z^{\prime }\right) \right\vert ^{2}\right\rangle _{Z}}
\label{Sltdd}
\end{equation}%
where $Y_{i\pm }$ gathers the possibilities $Y_{i+}$ and $Y_{i-}$. Similarly
to the derivation in Appendix 3 of $T\left( Z\right) $, $T_{ij}\left( Z_{\pm
}\right) $ is a function of these parameters and can take the values:

\begin{equation*}
T_{ij}\left( Z_{\pm }\right) =\frac{\lambda \tau \nu cY_{i\pm }\left(
Z\right) +\frac{1}{\tau _{D}\alpha _{D}}\lambda \tau \nu c}{\frac{1}{\tau
_{D}\alpha _{D}}+\frac{1}{\tau _{C}\alpha _{C}}+\Omega _{j\pm }+Y_{i\pm
}\left( Z\right) }
\end{equation*}%
Ultimately, we show that there are four possibilities for the connectivity
functions that are written:%
\begin{equation*}
T_{ij}\left( Z_{\pm },Z_{\pm }^{\prime }\right) =\frac{\lambda \tau \exp
\left( -\frac{\left\vert Z-Z^{\prime }\right\vert }{\nu c}\right) \left( 
\frac{1}{\tau _{D}\alpha _{D}}+Y_{i\pm }\left( Z\right) \right) }{\frac{1}{%
\tau _{D}\alpha _{D}}+Y_{i\pm }\left( Z\right) +\frac{1}{\alpha _{C}\tau _{C}%
}+Y_{j\pm }\left( Z^{\prime }\right) }
\end{equation*}%
Given our assumptions on $d$, in most cases:%
\begin{equation*}
Y_{+}\left( Z\right) >>Y_{-}\left( Z\right)
\end{equation*}%
Moreover:%
\begin{equation*}
\frac{1}{\tau _{D}\alpha _{D}}<<1\text{, }\frac{1}{\alpha _{C}\tau _{C}}<<1
\end{equation*}%
so that due to the threshold in connectivity, we have:%
\begin{eqnarray}
T_{ij}\left( Z_{-},Z_{+}^{\prime }\right) &=&\frac{\lambda \tau \exp \left( -%
\frac{\left\vert Z-Z^{\prime }\right\vert }{\nu c}\right) \left( \frac{1}{%
\tau _{D}\alpha _{D}}+\frac{1}{bG^{i}\bar{T}\left\langle \bar{T}\left\vert 
\bar{\Psi}_{i0}\left( Z^{\prime }\right) \right\vert ^{2}\right\rangle _{Z}}%
\right) }{\frac{1}{\tau _{D}\alpha _{D}}+\frac{1}{\alpha _{C}\tau _{C}}+%
\frac{1}{bG^{i}\bar{T}\left\langle \bar{T}\left\vert \bar{\Psi}_{i0}\left(
Z^{\prime }\right) \right\vert ^{2}\right\rangle _{Z}}+\frac{\left( bG^{j}%
\bar{T}\right) \left( \bar{T}\left\langle \left\vert \bar{\Psi}_{j0}\left(
Z^{\prime }\right) \right\vert ^{2}\right\rangle _{Z^{\prime }}\right) ^{2}}{%
2}}\simeq 0  \label{cnv} \\
T_{ij}\left( Z_{+},Z_{+}^{\prime }\right) &=&\frac{\lambda \tau \exp \left( -%
\frac{\left\vert Z-Z^{\prime }\right\vert }{\nu c}\right) \left( \frac{1}{%
\tau _{D}\alpha _{D}}+\frac{\left( bG^{i}\bar{T}\right) \left( \bar{T}%
\left\langle \left\vert \bar{\Psi}_{i0}\left( Z^{\prime }\right) \right\vert
^{2}\right\rangle _{Z}\right) ^{2}}{2}\right) }{\frac{1}{\tau _{D}\alpha _{D}%
}+\frac{1}{\alpha _{C}\tau _{C}}+\frac{\left( bG^{i}\bar{T}\right) \left( 
\bar{T}\left\langle \left\vert \bar{\Psi}_{i0}\left( Z^{\prime }\right)
\right\vert ^{2}\right\rangle _{Z}\right) ^{2}}{2}+\frac{\left( bG^{j}\bar{T}%
\right) \left( \bar{T}\left\langle \left\vert \bar{\Psi}_{j0}\left(
Z^{\prime }\right) \right\vert ^{2}\right\rangle _{Z^{\prime }}\right) ^{2}}{%
2}}\simeq \frac{G^{i}\lambda \tau \exp \left( -\frac{\left\vert Z-Z^{\prime
}\right\vert }{\nu c}\right) }{\left( G^{i}+G^{j}\right) }  \notag \\
T_{ij}\left( Z_{+},Z_{-}^{\prime }\right) &=&\frac{\lambda \tau \exp \left( -%
\frac{\left\vert Z-Z^{\prime }\right\vert }{\nu c}\right) \left( \frac{1}{%
\tau _{D}\alpha _{D}}+\frac{\left( bG^{i}\bar{T}\right) \left( \bar{T}%
\left\langle \left\vert \bar{\Psi}_{i0}\left( Z^{\prime }\right) \right\vert
^{2}\right\rangle _{Z}\right) ^{2}}{2}\right) }{\frac{1}{\tau _{D}\alpha _{D}%
}+\frac{1}{\alpha _{C}\tau _{C}}+\frac{\left( bG^{i}\bar{T}\right) \left( 
\bar{T}\left\langle \left\vert \bar{\Psi}_{i0}\left( Z^{\prime }\right)
\right\vert ^{2}\right\rangle _{Z}\right) ^{2}}{2}+\frac{1}{bG^{j}\bar{T}%
\left\langle \bar{T}\left\vert \bar{\Psi}_{j0}\left( Z^{\prime }\right)
\right\vert ^{2}\right\rangle _{Z}}}\simeq \lambda \tau \exp \left( -\frac{%
\left\vert Z-Z^{\prime }\right\vert }{\nu c}\right)  \notag \\
T\left( Z_{-},Z_{-}^{\prime }\right) &\simeq &\frac{\lambda \tau \exp \left(
-\frac{\left\vert Z-Z^{\prime }\right\vert }{\nu c}\right) +\frac{1}{bG^{i}%
\bar{T}\left\langle \bar{T}\left\vert \bar{\Psi}_{i0}\left( Z^{\prime
}\right) \right\vert ^{2}\right\rangle _{Z}}}{1+\frac{\tau _{D}\alpha _{D}}{%
\alpha _{C}\tau _{C}}+\frac{1}{bG^{i}\bar{T}\left\langle \bar{T}\left\vert 
\bar{\Psi}_{i0}\left( Z^{\prime }\right) \right\vert ^{2}\right\rangle _{Z}}+%
\frac{1}{bG^{j}\bar{T}\left\langle \bar{T}\left\vert \bar{\Psi}_{j0}\left(
Z^{\prime }\right) \right\vert ^{2}\right\rangle _{Z}}}\simeq \frac{%
G^{j}\lambda \tau \exp \left( -\frac{\left\vert Z-Z^{\prime }\right\vert }{%
\nu c}\right) }{\left( G^{i}+G^{j}\right) }  \notag
\end{eqnarray}

\part*{Part III Dynamic aspects of the system: modifications of background
fields due to external sources.}

Having derived the possible forms for background, i.e. equilibrium states,
we turn to the study of the dynamical aspects of the system. External
sources may induce fluctuations around static values. We first present the
results of (\cite{GL}) concerning flutuations in cell background field and
activities: we derive the dynamic corrections for the neurons background
field $\Psi \left( \theta ,Z\right) $ and obtain a formula for these
corrections $\delta \Psi \left( \theta ,Z\right) $ as a function of the
corrections in the activities $\omega \left( J\left( \theta ,Z\right)
,\theta ,Z\right) $. We then derive a wave equation for these frequencies,
and show that stable oscillations may occur. We also inspect the
interferences of waves of activities induced by several oscillating sources
and their effect on the connectivities. Some patterns of bound cells arise
from these interferences so that, the propagation of periodic perturbations
may change the static background field for connectivity functions in the
long run.

We also consider the association of signals through synchronized stimuli
and, ultimately, the recovery of a combined state by the reactivation of
part of the full state.

The approach in this section is local and relies only partly on the field
formalsm. We show in the next papers how to derive more generally the
presented result via a field-theoretic based description.

\section{Source induced fluctuations of neurons background field}

We first study dynamic fluctuations in cells background field $\Psi \left(
\theta ,Z\right) $ and activities $\omega \left( J\left( \theta ,Z\right)
,\theta ,Z,\mathcal{G}_{0}+\left\vert \Psi \right\vert ^{2}\right) $ around
some static background state. These fluctuations may be induced by some
time-dependent current $J\left( \theta ,Z\right) $. The connectivity
function are assumed to remain constant, since the time scale of their
modification is slower. We follow the derivations of the fluctuating
background state $\Psi \left( \theta ,Z\right) $ in (\cite{GL}).

\subsection{Minimization equation for dynamic fields}

In (\cite{GL}) we show that the background field $\Psi \left( \theta
,Z\right) $ can be decomposed in a static part and a fluctuation part:%
\begin{eqnarray*}
\Psi \left( \theta ,Z\right) &\simeq &\Psi _{0}\left( Z\right) +\delta \Psi
\left( \theta ,Z\right) \\
\Psi ^{\dag }\left( \theta ,Z\right) &\simeq &\Psi _{0}^{\dag }\left(
Z\right)
\end{eqnarray*}%
where:%
\begin{equation*}
\left\vert \delta \Psi \left( \theta ^{\left( j\right) },Z_{j}\right)
\right\vert <\left\vert \Psi _{0}\left( Z_{j}\right) \right\vert
\end{equation*}

The static part $\Psi _{0}\left( Z\right) $ is the minimum of $V\left( \Psi
\right) $. To simplify, we could consider $\left\vert \Psi \left( Z\right)
\right\vert ^{2}$ as exogenous and minimizing a stabilizing potential with
minimum $\Psi _{0}\left( \theta ,Z\right) =X_{0}$.

Expanding the potential around $\Psi _{0}\left( \theta ,Z\right) $ and
setting $V=1$, yields at the second order the effective action:%
\begin{eqnarray*}
S_{\Psi }\left( \Psi ,\Psi ^{\dag }\right) &=&-\frac{1}{2}\int \delta \Psi
^{\dagger }\left( \theta ,Z\right) \left( \nabla _{\theta }\left( \frac{%
\sigma _{\theta }^{2}}{2}\nabla _{\theta }-\omega ^{-1}\left( J\left( \theta
\right) ,\theta ,Z,\mathcal{G}_{0}+\left\vert \Psi \right\vert ^{2}\right)
\right) \right) X_{0} \\
&&-\frac{1}{2}\int \delta \Psi ^{\dagger }\left( \theta ,Z\right) \left(
\nabla _{\theta }\left( \frac{\sigma _{\theta }^{2}}{2}\nabla _{\theta
}-\omega ^{-1}\left( J\left( \theta \right) ,\theta ,Z,\mathcal{G}%
_{0}+\left\vert \Psi \right\vert ^{2}\right) \right) \right) \delta \Psi
\left( \theta ,Z\right) \\
&&+\frac{1}{2}\int \delta \Psi ^{\dagger }\left( \theta ,Z\right) U^{\prime
\prime }\left( X_{0}\right) \delta \Psi \left( \theta ,Z\right)
\end{eqnarray*}%
with $\left\vert \Psi \right\vert ^{2}=X_{0}+\sqrt{X_{0}}\left( \delta
\left( \Psi ^{\dagger }+\delta \Psi \right) \right) $. This leads to the
first order condition for $\delta \Psi \left( \theta _{1},Z_{1}\right) $:%
\begin{eqnarray*}
0 &=&\frac{1}{2}\delta \Psi ^{\dagger }\left( \theta ,Z\right) \left(
-\nabla _{\theta }\left( \frac{\sigma _{\theta }^{2}}{2}\nabla _{\theta
}-\omega ^{-1}\left( J\left( \theta \right) ,\theta ,Z,\mathcal{G}%
_{0}+X_{0}\right) \right) +U^{\prime \prime }\left( X_{0}\right) \right) \\
&&-\frac{1}{2}\int \delta \Psi ^{\dagger }\left( \theta _{1},Z_{1}\right) 
\sqrt{X_{0}}\left( \nabla _{\theta }\frac{\delta \omega ^{-1}\left( J\left(
\theta _{1}\right) ,\theta _{1},Z_{1},\mathcal{G}_{0}+X_{0}\right) }{\delta
\left\vert \Psi \left( \theta ,Z\right) \right\vert ^{2}}\right)
X_{0}d\theta _{1}dZ_{1}
\end{eqnarray*}%
with solution $\delta \Psi ^{\dagger }\left( \theta ,Z\right) =0$. This
implies that the first order condition for $\delta \Psi ^{\dag }\left(
\theta ,Z\right) $ becomes: 
\begin{eqnarray}
&&0=-\frac{1}{2}\left( \nabla _{\theta }\left( \frac{\sigma _{\theta }^{2}}{2%
}\nabla _{\theta }-\omega ^{-1}\left( J\left( \theta \right) ,\theta ,Z,%
\mathcal{G}_{0}+\left\vert \Psi \right\vert ^{2}\right) \right) \right) X_{0}
\label{sDP} \\
&&-\frac{1}{2}\left( \nabla _{\theta }\left( \frac{\sigma _{\theta }^{2}}{2}%
\nabla _{\theta }-\omega ^{-1}\left( J\left( \theta \right) ,\theta ,Z,%
\mathcal{G}_{0}+\left\vert \Psi \right\vert ^{2}\right) \right) \right)
\delta \Psi \left( \theta ,Z\right)  \notag \\
&&+\frac{1}{2}U^{\prime \prime }\left( X_{0}\right) \delta \Psi \left(
\theta ,Z\right)  \notag
\end{eqnarray}%
In first approximation, for $U^{\prime \prime }\left( X_{0}\right) >>1$ and $%
\sigma _{\theta }^{2}<<1$, this yields\footnote{%
Note that for a slowly background field $\Psi _{0}\left( \theta ,Z\right) $,
equation (\ref{psG}) remains valid and becomes:%
\begin{equation}
\delta \Psi \left( \theta ,Z\right) \simeq -\frac{\nabla _{\theta }\omega
^{-1}\left( J\left( \theta \right) ,\theta ,Z,\mathcal{G}_{0}+\left\vert
\Psi \right\vert ^{2}\right) }{U^{\prime \prime }\left( X_{0}\right) }\Psi
_{0}\left( \theta ,Z\right)  \label{pSG}
\end{equation}%
}: 
\begin{eqnarray}
\delta \Psi \left( \theta ,Z\right) &\simeq &-\frac{\nabla _{\theta }\omega
^{-1}\left( J\left( \theta \right) ,\theta ,Z,\mathcal{G}_{0}+\left\vert
\Psi \right\vert ^{2}\right) }{U^{\prime \prime }\left( X_{0}\right) +\nabla
_{\theta }\omega ^{-1}\left( J\left( \theta \right) ,\theta ,Z,\mathcal{G}%
_{0}+\left\vert \Psi \right\vert ^{2}\right) }X_{0}  \label{psG} \\
&\simeq &-\frac{\nabla _{\theta }\omega ^{-1}\left( J\left( \theta \right)
,\theta ,Z,\mathcal{G}_{0}+\left\vert \Psi \right\vert ^{2}\right) }{%
U^{\prime \prime }\left( X_{0}\right) }X_{0}  \notag
\end{eqnarray}%
Equation (\ref{sDP}) also rewrites:%
\begin{equation}
\left( -\left( \nabla _{\theta }\left( \frac{\sigma _{\theta }^{2}}{2}\nabla
_{\theta }-\omega ^{-1}\left( J\left( \theta \right) ,\theta ,Z,\mathcal{G}%
_{0}+\left\vert \Psi \right\vert ^{2}\right) \right) \right) +U^{\prime
\prime }\left( X_{0}\right) \right) \left( \delta \Psi \left( \theta
,Z\right) +X_{0}\right) =U^{\prime \prime }\left( X_{0}\right) X_{0}
\label{sDT}
\end{equation}%
Equation (\ref{sDT}) can be used to write $\delta \Psi \left( \theta
,Z\right) $ as a function of $\omega ^{-1}\left( J\left( \theta \right)
,\theta ,Z,\mathcal{G}_{0}+\left\vert \Psi \right\vert ^{2}\right) $:%
\begin{eqnarray}
\delta \Psi \left( \theta ,Z\right) &=&\left( \frac{\left( \nabla _{\theta
}\left( \frac{\sigma _{\theta }^{2}}{2}\nabla _{\theta }-\omega ^{-1}\left(
J\left( \theta \right) ,\theta ,Z,\mathcal{G}_{0}+\left\vert \Psi
\right\vert ^{2}\right) \right) \right) }{U^{\prime \prime }\left(
X_{0}\right) -\left( \nabla _{\theta }\left( \frac{\sigma _{\theta }^{2}}{2}%
\nabla _{\theta }-\omega ^{-1}\left( J\left( \theta \right) ,\theta ,Z,%
\mathcal{G}_{0}+\left\vert \Psi \right\vert ^{2}\right) \right) \right) }%
\right) X_{0}  \label{psv} \\
&=&-\frac{\nabla _{\theta }\left( \omega ^{-1}\left( J\left( \theta \right)
,\theta ,Z,\mathcal{G}_{0}+\left\vert \Psi \right\vert ^{2}\right) \right) }{%
U^{\prime \prime }\left( X_{0}\right) -\left( \nabla _{\theta }\left( \frac{%
\sigma _{\theta }^{2}}{2}\nabla _{\theta }-\omega ^{-1}\left( J\left( \theta
\right) ,\theta ,Z,\mathcal{G}_{0}+\left\vert \Psi \right\vert ^{2}\right)
\right) \right) }X_{0}  \notag \\
&\simeq &\frac{\nabla _{\theta }\left( \omega \left( J\left( \theta \right)
,\theta ,Z,\mathcal{G}_{0}+\left\vert \Psi \right\vert ^{2}\right) \right) }{%
\left( \omega \left( J\left( \theta \right) ,\theta ,Z,\mathcal{G}%
_{0}+\left\vert \Psi \right\vert ^{2}\right) \right) ^{2}U^{\prime \prime
}\left( X_{0}\right) }X_{0}  \notag
\end{eqnarray}%
leading to system of equation for activities and field:%
\begin{eqnarray*}
&&\omega ^{-1}\left( J,\theta ,Z,\left\vert \Psi \right\vert ^{2}\right) \\
&=&G\left( J\left( \theta ,Z\right) +\int \frac{\kappa }{N}\frac{\omega
\left( J,\theta -\frac{\left\vert Z-Z_{1}\right\vert }{c},Z_{1},\Psi \right)
T\left( Z,\theta ,Z_{1},\theta -\frac{\left\vert Z-Z_{1}\right\vert }{c}%
\right) }{\omega \left( J,\theta ,Z,\left\vert \Psi \right\vert ^{2}\right) }%
\left( \mathcal{G}_{0}+\left\vert \Psi \left( \theta -\frac{\left\vert
Z-Z_{1}\right\vert }{c},Z_{1}\right) \right\vert ^{2}\right) dZ_{1}\right)
\end{eqnarray*}%
and:%
\begin{equation}
\delta \Psi \left( \theta ,Z\right) =\frac{\nabla _{\theta }\left( \omega
\left( J\left( \theta \right) ,\theta ,Z,\mathcal{G}_{0}+\left\vert \Psi
\right\vert ^{2}\right) \right) }{\left( \omega \left( J\left( \theta
\right) ,\theta ,Z,\mathcal{G}_{0}+\left\vert \Psi \right\vert ^{2}\right)
\right) ^{2}U^{\prime \prime }\left( X_{0}\right) }X_{0}  \label{FP}
\end{equation}

Relation (\ref{psv}) is sufficient to derive the next section's activities
equations, but can however be used to find $\delta \Psi \left( \theta
,Z\right) $, at our order of approximation (see appendix 4). In the local
approximation and for slowly varying currents, we show that the fluctuation $%
\delta \Psi \left( \theta ,Z\right) $ (\ref{FP}) is equal to:%
\begin{eqnarray}
\delta \Psi \left( \theta ,Z\right) &=&\left( G^{-1}\left( -\frac{U^{\prime
\prime }\left( X_{0}\right) }{X_{0}}\exp \left( H^{-1}\left( \frac{\theta }{%
\Gamma \left( \mathcal{G}_{0}\left( Z_{1}\right) +\sqrt{X_{0}}\right) }%
+d\right) \right) \right) -J\left( \theta ,Z\right) \right)  \label{sdt} \\
&&\times \exp \left( H^{-1}\left( \frac{\theta }{\Gamma \left( \mathcal{G}%
_{0}\left( Z_{1}\right) +\sqrt{X_{0}}\right) }+d\right) \right)  \notag
\end{eqnarray}%
with: 
\begin{equation*}
H\left( Y\right) =\int \frac{dY}{G^{-1}\left( -\frac{U^{\prime \prime
}\left( X_{0}\right) }{X_{0}}\exp Y\right) -J\left( \theta ,Z\right) }
\end{equation*}%
and:%
\begin{equation*}
\Gamma =\int \frac{\kappa }{N}\frac{\left\vert Z-Z_{1}\right\vert }{c}%
T\left( Z,\theta ,Z_{1},\theta -\frac{\left\vert Z-Z_{1}\right\vert }{c}%
\right) dZ_{1}
\end{equation*}%
The constant $d$ is chosen so that $\lim_{\theta \rightarrow \infty }\delta
\Psi \left( \theta ,Z\right) =0$.

The field $\Psi \left( \theta ^{\left( j\right) },Z_{j}\right) $ is the -
phase-dependent - background field. It is null in the trivial phase, so that
the effective action is the "classical" one. In a non-trivial phase, $\Psi
\left( \theta ^{\left( j\right) },Z_{j}\right) $ is not null and may be
time-dependent. It describes the accumulation of currents or signals that
shapes the long-term dynamics of activities. Incidentally, we note that a
non-trivial minimum that depends on the system parameters should allow for
phase transition in the system of activities.

\section{Dynamic wave equation for activities}

This section studies the dynamic solutions of (\ref{qf}). We use relation (%
\ref{psv}) to replace the non-static part of the field $\Psi $ as a function
of the activities and then deduce a wave equation for the activities.

\subsection{ Differential equation for activities in the local approximation}

A local approximation of (\ref{nqf}) around some position-independent static
equilibrium can be derived for non static activities. Assuming a static
background field $\Psi _{0}$, we derived above the relation between $\delta
\Psi \left( \theta ,Z\right) $ and $\omega \left( J,Z,\left\vert \Psi
\right\vert ^{2}\right) $ (see (\ref{FP})):%
\begin{equation}
\delta \Psi \left( \theta ,Z\right) \simeq \frac{\nabla _{\theta }\omega
\left( J,Z,\left\vert \Psi \right\vert ^{2}\right) }{V^{\prime \prime
}\left( \Psi _{0}\left( Z\right) \right) \omega _{0}^{2}\left(
J,Z,\left\vert \Psi \right\vert ^{2}\right) }\Psi _{0}  \label{psf}
\end{equation}%
where $\omega \left( J,Z,\left\vert \Psi \right\vert ^{2}\right) $ is the
time-dependent firing rate, or activity.

We can find a local approximation of (\ref{qf}) if we expand $\omega \left(
J\left( \theta \right) ,\theta ,Z,\mathcal{G}_{0}+\left\vert \Psi
\right\vert ^{2}\right) $ to the second-order in $Z-Z_{1}$, and consider the
other terms in the right-hand side of (\ref{qf}) as corrections. The
equation for $\omega \left( J\left( \theta \right) ,\theta ,Z,\mathcal{G}%
_{0}+\left\vert \Psi \right\vert ^{2}\right) $ is:%
\begin{eqnarray}
F^{-1}\left( \omega \left( J\left( \theta \right) ,\theta \right) \right)
&=&J\left( \theta ,Z\right) +\int \frac{\kappa }{N}\frac{\omega \left(
J,\theta -\frac{\left\vert Z-Z_{1}\right\vert }{c},Z_{1},\Psi \right)
T\left( Z,\theta ,Z_{1},\theta -\frac{\left\vert Z-Z_{1}\right\vert }{c}%
\right) }{\omega \left( J,\theta ,Z,\left\vert \Psi \right\vert ^{2}\right) }
\label{vnq} \\
&&\times \left( \left\vert \Psi _{0}+\delta \Psi \left( \theta -\frac{%
\left\vert Z-Z_{1}\right\vert }{c},Z_{1}\right) \right\vert ^{2}\right)
dZ_{1}  \notag
\end{eqnarray}

where $F=\frac{1}{G}$ and $F^{-1}$ the reciprocal function of $F$. We then
expand $\omega \left( \theta -\frac{\left\vert Z-Z_{1}\right\vert }{c}%
,Z_{1}\right) $ around $\omega \left( \theta ,Z\right) $ to the second-order
in $Z-Z_{1}$\ and compute the integrals, which yields for the right-hand
side of (\ref{vnq}):%
\begin{eqnarray}
&&J\left( \theta \right) +\int \frac{\kappa }{N}\frac{\omega \left( \theta -%
\frac{\left\vert Z-Z_{1}\right\vert }{c},Z_{1}\right) }{\omega \left( \theta
,Z\right) }T\left( Z,\theta ,Z_{1},\theta -\frac{\left\vert
Z-Z_{1}\right\vert }{c}\right)  \label{RS} \\
&&\times \left\vert \Psi _{0}\left( Z_{1}\right) +\delta \Psi \left( \theta -%
\frac{\left\vert Z-Z_{1}\right\vert }{c},Z_{1}\right) \right\vert ^{2}dZ_{1}
\notag \\
&\simeq &J\left( \theta \right) +\frac{TW\left( 1\right) }{\bar{\Lambda}}+%
\frac{\hat{f}_{1}\nabla _{\theta }\omega \left( \theta ,Z\right) }{\omega
\left( \theta ,Z\right) }+\frac{\hat{f}_{3}\nabla _{\theta }^{2}\omega
\left( \theta ,Z\right) }{\omega \left( \theta ,Z\right) }+c^{2}\frac{\hat{f}%
_{3}\nabla _{Z}^{2}\omega \left( \theta ,Z\right) }{\omega \left( \theta
,Z\right) }+T\Psi _{0}\delta \Psi \left( \theta ,Z\right)  \notag
\end{eqnarray}%
where we defined:%
\begin{eqnarray}
\hat{f}_{1} &=&-\frac{\Gamma _{1}}{c}\text{, }\hat{f}_{3}=\frac{\Gamma _{2}}{%
c^{2}}  \label{ctf} \\
\Gamma _{1} &=&\frac{\kappa }{NX_{r}}\int \left\vert Z-Z_{1}\right\vert
T\left( Z,Z_{1}\right) \left\vert \Psi _{0}\left( Z_{1}\right) \right\vert
^{2}dZ_{1}  \notag \\
\Gamma _{2} &=&\frac{\kappa }{2NX_{r}}\int \left( Z-Z_{1}\right) ^{2}T\left(
Z,Z_{1}\right) \left\vert \Psi _{0}\left( Z_{1}\right) \right\vert ^{2}dZ_{1}
\notag
\end{eqnarray}%
and:%
\begin{equation}
T\Psi _{0}\delta \Psi \left( \theta ,Z\right) =\int \frac{\kappa T\left(
Z,Z_{1}\right) }{N}\Psi _{0}\left( Z_{1}\right) \delta \Psi \left( \theta -%
\frac{\left\vert Z-Z_{1}\right\vert }{c},Z_{1}\right) dZ_{1}  \label{TP}
\end{equation}%
Using (\ref{psf}) we can rewrite (\ref{TP}) as:%
\begin{eqnarray}
T\delta \Psi \left( \theta ,Z\right) &\simeq &\delta \Psi \left( \theta
,Z\right) -\Gamma _{1}\nabla _{\theta }\delta \Psi \left( \theta ,Z\right)
\label{TQ} \\
&\simeq &N_{1}\nabla _{\theta }\omega _{0}\left( J,Z,\left\vert \Psi
_{0}\right\vert ^{2}\right) -N_{2}\nabla _{\theta }\omega _{0}\left(
J,Z,\left\vert \Psi _{0}\right\vert ^{2}\right)  \notag
\end{eqnarray}%
with:%
\begin{eqnarray*}
N_{1} &=&\frac{\Psi _{0}\left( Z\right) }{U^{\prime \prime }\left(
X_{0}\right) \omega ^{2}\left( J,Z,\left\vert \Psi _{0}\right\vert
^{2}\right) } \\
N_{2} &=&\frac{\Gamma _{1}\Psi _{0}\left( Z\right) }{U^{\prime \prime
}\left( X_{0}\right) \omega ^{2}\left( J,Z,\left\vert \Psi _{0}\right\vert
^{2}\right) }
\end{eqnarray*}%
Then, replacing (\ref{TQ}) in (\ref{RS}), equation (\ref{vnq}) becomes:%
\begin{eqnarray}
&&F^{-1}\left( \omega \left( J\left( \theta \right) ,\theta \right) \right)
-F^{-1}\left( \omega _{0}\right)  \label{fsr} \\
&=&J\left( \theta ,Z\right) +\left( \frac{\hat{f}_{1}}{\omega \left( \theta
,Z\right) }+N_{1}\right) \nabla _{\theta }\omega \left( \theta ,Z\right)
+\left( \frac{\hat{f}_{3}}{\omega \left( \theta ,Z\right) }-N_{2}\right)
\nabla _{\theta }^{2}\omega \left( \theta ,Z\right) +c^{2}\hat{f}_{3}\frac{%
\nabla _{Z}^{2}\omega \left( \theta ,Z\right) }{\omega \left( \theta
,Z\right) }  \notag
\end{eqnarray}

To obtain an equation for the fluctuations of activities around the
background state values, we assume that $F^{-1}$ is slowly varying, so that:%
\begin{equation*}
F^{-1}\left( \omega \left( J\left( \theta \right) ,\theta \right) \right)
-F^{-1}\left( \omega _{0}\right) \simeq \Gamma _{0}\left( \omega \left(
J\left( \theta \right) ,\theta \right) -\omega _{0}\right)
\end{equation*}%
with\footnote{%
Given our assumption that $F$ is an increasing function, $f>0$.
\par
{}}:%
\begin{equation*}
f=\left( F^{-1}\right) ^{\prime }\left( \frac{\kappa }{N}\int T\left(
Z,Z_{1}\right) W\left( 1\right) dZ_{1}\mathcal{\bar{G}}_{0}\left(
0,Z_{1}\right) \right)
\end{equation*}%
and define:%
\begin{equation*}
\Omega \left( \theta ,Z\right) =\omega \left( \theta ,Z\right) -\omega _{0}
\end{equation*}%
As a result, the expansion of (\ref{fsr}) for a non-static current is then:%
\begin{equation}
f\Omega \left( \theta ,Z\right) =J\left( \theta ,Z\right) +\left( \frac{\hat{%
f}_{1}}{\omega \left( \theta ,Z\right) }+N_{1}\right) \nabla _{\theta
}\Omega \left( \theta ,Z\right) +\left( \frac{\hat{f}_{3}}{\omega \left(
\theta ,Z\right) }-N_{2}\right) \nabla _{\theta }^{2}\Omega \left( \theta
,Z\right) +\frac{c^{2}\hat{f}_{3}}{\omega \left( \theta ,Z\right) }\nabla
_{Z}^{2}\Omega \left( \theta ,Z\right)  \label{pw}
\end{equation}

A careful study of this equation is performed in (\cite{GL}). We show that
this equation has non sinusoidal stable traveling wave solutions and that in
first approximation it can be replaced by a usual wave equation:%
\begin{equation}
f\Omega \left( \theta ,Z\right) -\left( \frac{\hat{f}_{3}}{\omega _{0}}%
-N_{2}\right) \nabla _{\theta }^{2}\Omega \left( \theta ,Z\right) -\frac{%
c^{2}\hat{f}_{3}}{\omega _{0}}\nabla _{Z}^{2}\Omega \left( \theta ,Z\right)
=J\left( \theta ,Z\right)  \label{wvn}
\end{equation}%
where $\omega _{0}$ is the average of the static activity.

\subsection{Perturbative corrections to the local frequency equations}

The perturbative expansion of the path integral for the field action (\ref%
{flt}) local modifies the activities equation. We computed this effective
action, written $\Gamma \left( \Psi ^{\dagger },\Psi \right) $, in (\cite{GL}%
). In the local approximation it is given by:%
\begin{equation}
\Gamma \left( \Psi ^{\dagger },\Psi \right) \simeq \int \Psi ^{\dagger
}\left( \theta ,Z\right) \left( -\nabla _{\theta }\left( \frac{\sigma
_{\theta }^{2}}{2}\nabla _{\theta }-\omega ^{-1}\left( J\left( \theta
\right) ,\theta ,Z,\mathcal{G}_{0}+\left\vert \Psi \right\vert ^{2}\right)
\right) \delta \left( \theta _{f}-\theta _{i}\right) +\Omega \left( \theta
,Z\right) \right) \Psi \left( \theta ,Z\right)  \label{gmf}
\end{equation}%
where $\Omega \left( \theta ,Z\right) $ is a corrective term depending on
the successive\ derivatives of the field. The term $\mathcal{G}_{0}$ is a
function of $Z$ and represents a two points free Green function (see (\cite%
{GL})).

The previous equation (\ref{gmf}) defines an effective activity that can be
identified as:%
\begin{equation}
\omega _{e}^{-1}\left( J\left( \theta \right) ,\theta ,Z,\mathcal{G}%
_{0}+\left\vert \Psi \right\vert ^{2}\right) =\omega ^{-1}\left( J\left(
\theta \right) ,\theta ,Z,\mathcal{G}_{0}+\left\vert \Psi \right\vert
^{2}\right) +\int^{\theta }\Omega \left( \theta ,Z\right)  \label{fcv}
\end{equation}%
where $\omega \left( J\left( \theta \right) ,\theta ,Z,\mathcal{\bar{G}}%
_{0}+\left\vert \Psi \right\vert ^{2}\right) $ is the solution of:%
\begin{eqnarray*}
\omega ^{-1}\left( J,\theta ,Z,\left\vert \Psi \right\vert ^{2}\right)
&=&G\left( J\left( \theta ,Z\right) +\int \frac{\kappa }{N}\frac{\omega
\left( J,\theta -\frac{\left\vert Z-Z_{1}\right\vert }{c},Z_{1},\Psi \right)
T\left( Z,\theta ,Z_{1},\theta -\frac{\left\vert Z-Z_{1}\right\vert }{c}%
\right) }{\omega \left( J,\theta ,Z,\left\vert \Psi \right\vert ^{2}\right) }%
\right. \\
&&\times \left. \left( \mathcal{\bar{G}}_{0}\left( 0,Z_{1}\right)
+\left\vert \Psi \left( \theta -\frac{\left\vert Z-Z_{1}\right\vert }{c}%
,Z_{1}\right) \right\vert ^{2}\right) dZ_{1}\right)
\end{eqnarray*}%
Wich is the classical activity equation, up to the inclusion of the Green
function $\mathcal{\bar{G}}_{0}\left( 0,Z_{1}\right) $.

The second term $\int^{\theta }\Omega \left( \theta ,Z\right) $ in (\ref{fcv}%
) represents corrections due to the interactions. Using (\ref{gmf}), we can
find its expression as a series expansion in terms of activities and field.
The computations of these corrections to the classical equation are
presented in (\cite{GL}) and confirm the possibility of traveling wave
solutions. To sum up, the perturbative corrections account for interactions
between the classical solutions and the whole thread and these interactions
stabilize the traveling waves.

\section{Sources induced activities, interferences.}

\subsection{Local approximation}

In the perspective of this work, we are looking at the solutions of (\ref%
{wvn}) induced by some ponctual sources. Assume several signals arising at
some points $\left( Z_{1},\theta _{1}\right) ,...,\left( Z_{N},\theta
_{N}\right) $.

The solution to (\ref{wvn}) are then:%
\begin{equation}
\Omega \left( Z,\theta \right) =\sum_{i=1}^{N}\emph{G}\left( \left( Z,\theta
\right) ,\left( Z_{i},\theta _{i}\right) \right) J\left( Z_{i},\theta
_{i}\right)  \label{lcs}
\end{equation}%
Where $\emph{G}\left( \left( Z,\theta \right) ,\left( Z_{i},\theta
_{i}\right) \right) $ is the Green function of the operator involvd in (\ref%
{wvn}):%
\begin{equation*}
f-\left( \frac{\hat{f}_{3}}{\omega _{0}}-N_{2}\right) \nabla _{\theta }^{2}-%
\frac{c^{2}\hat{f}_{3}}{\omega _{0}}\nabla _{Z}^{2}
\end{equation*}%
Solutions of (\ref{lcs})\ present interference phenomena. When the number of
sources is large, we may expect that solutions of (\ref{lcs}) locate mainly
at some maxima depending both on the connectivity field $\left\vert \Gamma
\left( T,\hat{T},\theta ,Z,Z^{\prime }\right) \right\vert ^{2}$ and neuron
field. In the sequel, we will write:

\begin{equation*}
Z_{M}^{\left( \varepsilon \right) }\left( \left\vert \Gamma \right\vert
^{2},\left\vert \Psi _{0}\right\vert ^{2}\right)
\end{equation*}%
the location of these maxima, with $\varepsilon =1,...$ indexing these
maxima. We will also assume that at these maxima, the activities are all
equal to some value:%
\begin{equation*}
\omega \simeq \omega ^{\prime }\simeq \omega _{M}
\end{equation*}%
so that:%
\begin{eqnarray*}
h_{C}\left( \omega \right) &\simeq &h_{C}\left( \omega _{M}\right) \\
h_{D}\left( \omega \right) &\simeq &h_{D}\left( \omega _{M}\right)
\end{eqnarray*}%
The precise derivation of the interference phenomenom will be presented in a
field theoretic context in part II. It is sufficient for the rest of this
article to build on the previous qualititative argument.

\subsection{Non local propagation and interferences}

Note that more generally, we can go farther than the local equation (\ref%
{wvn}) by considering the non local equation (\ref{vnq}) in presence of
sources, this equation has solutions:%
\begin{equation}
\Omega \left( \theta ,Z\right) \simeq \sum_{i=1}^{N}\emph{G}_{T}\left(
\left( Z,\theta \right) ,\left( Z_{i},\theta _{i}\right) \right) J\left(
Z_{i},\theta _{i}\right)  \label{nlss}
\end{equation}%
where $\emph{G}_{T}\left( \left( Z,\theta \right) ,\left( Z_{i},\theta
_{i}\right) \right) $ is defined by:%
\begin{equation*}
\emph{G}_{T}\left( \left( Z,\theta \right) ,\left( Z_{i},\theta _{i}\right)
\right) =\left( \frac{1}{1-G_{T}}\right) \left( \left( Z,\theta \right)
,\left( Z_{i},\theta _{i}\right) \right)
\end{equation*}%
and $G_{T}$ is an operator whose kernel $G_{T,\Psi _{0}}\left( Z,\theta
,Z_{1},\theta -\frac{\left\vert Z-Z_{1}\right\vert }{c}\right) $ depending
on $T$ and $\Psi _{0}$ will be studied in the second article of this series.
Remark that formula (\ref{nlss}) is the non-local version of (\ref{lcs}).
Both solutions present interference phenomena with maxima located at specifc
points $Z_{M}^{\left( \varepsilon \right) }\left( \left\vert \Gamma
\right\vert ^{2},\left\vert \Psi _{0}\right\vert ^{2}\right) $.

\section{Effective action and background state for given sources
perturbations}

We have seen that for a given external source state, interferences occur in
cells activity, leading to localized activities at specific points:%
\begin{equation*}
Z_{M}^{\left( \varepsilon \right) }\left( \left\vert \Gamma \right\vert
^{2},\left\vert \Psi _{0}\right\vert ^{2}\right)
\end{equation*}%
with $\varepsilon =1,...$ indexing these points. To derive the connctivity
background field associated to these source induced interferences, we have
to proceed as we did in section 8 and start with the level of activity at
each points of the system. We will assume that the effect of activity is
large at the points of positive interferences and neglect at these points
the level of static equilibrium level $\omega _{0}\left( Z\right) $ in the
static background field. This corresponds to study states with source
induced additional connectivity and activation. The background field
obtained will thus described this modification "above" the static background
field, similar to an activated state above some vacuum.

We will also assume, for the sake of simplicity, that that at these maxima $%
Z_{M}^{\left( \varepsilon \right) }\left( \left\vert \Gamma \right\vert
^{2},\left\vert \Psi _{0}\right\vert ^{2}\right) $, the activities are all
equal to some value:%
\begin{equation}
\omega \simeq \omega ^{\prime }\simeq \omega _{M}  \label{BC}
\end{equation}%
so that:%
\begin{eqnarray*}
h_{C}\left( \omega \right) &\simeq &h_{C}\left( \omega _{M}\right) \\
h_{D}\left( \omega \right) &\simeq &h_{D}\left( \omega _{M}\right)
\end{eqnarray*}%
Assuming that functions $h_{C}\left( \omega \right) $ and $h_{D}\left(
\omega \right) $ are proportional to some positive power of $\omega $
implies that outside the set of points 
\begin{equation*}
U_{M}=\left\{ Z_{M}^{\left( \varepsilon \right) }\left( \left\vert \Gamma
\right\vert ^{2},\left\vert \Psi _{0}\right\vert ^{2}\right) \right\}
\end{equation*}%
the functions $h_{C}\left( \omega \right) $ and $h_{D}\left( \omega \right) $
can be considered as nul. We will write $Z$ the generic points of the
complementary set of $U_{M}$, written $CU_{M}$.

We compute average connectivity between points of $U_{M}$, between points of 
$CU_{M}$, and between points of $U_{M}$ and $CU_{M}$.

\subsection{Connectivity between points of $U_{M}$}

Using (\ref{BC}), we can compute the average connectivities for points of $%
U_{M}$, the points with constructive interferences. We use that the
background state at points $\left( Z_{M}^{\left( \varepsilon _{1}\right)
},Z_{M}^{\left( \varepsilon _{2}\right) }\right) \subset U_{M}$ is:%
\begin{eqnarray*}
&&\Gamma \left( T,\hat{T},\theta ,Z_{M}^{\left( \varepsilon _{1}\right)
},Z_{M}^{\left( \varepsilon _{2}\right) }\right) \\
&=&\exp \left( -\left( \left( -\frac{1}{\tau \omega _{M}}T+\frac{\lambda }{%
\omega _{M}}\left\langle \hat{T}\right\rangle \right) \left\vert \Psi \left(
\theta ,Z_{M}^{\left( \varepsilon _{1}\right) }\right) \right\vert
^{2}\right) ^{2}\right) \\
&&\times \exp \left( -\left( \frac{\rho }{\omega _{M}}H\left( Z_{M}^{\left(
\varepsilon _{1}\right) },Z_{M}^{\left( \varepsilon _{2}\right) }\right)
\left\vert \Psi \left( \theta ,Z_{M}^{\left( \varepsilon _{1}\right)
}\right) \right\vert ^{2}\left\vert \Psi \left( \theta -\frac{\left\vert
Z_{M}^{\left( \varepsilon _{1}\right) }-Z_{M}^{\left( \varepsilon
_{2}\right) }\right\vert }{c},Z_{M}^{\left( \varepsilon _{2}\right) }\right)
\right\vert ^{2}\right) ^{2}\right)
\end{eqnarray*}%
where:%
\begin{equation*}
H\left( Z_{M}^{\left( \varepsilon _{1}\right) },Z_{M}^{\left( \varepsilon
_{2}\right) }\right) =\left( \left( h\left( Z_{M}^{\left( \varepsilon
_{1}\right) },Z_{M}^{\left( \varepsilon _{2}\right) }\right) -\hat{T}\right)
C\left( \theta \right) h_{C}-D\left( \theta \right) \hat{T}h_{D}\right)
\end{equation*}%
and the average values of $C$, $D$, $\hat{T}$ and $T$in this background
states are:

\begin{eqnarray*}
C_{Z_{M}^{\left( \varepsilon _{1}\right) },Z_{M}^{\left( \varepsilon
_{2}\right) }} &=&\frac{\alpha _{C}\omega _{M}\left\vert \Psi \left( \theta -%
\frac{\left\vert Z_{M}^{\left( \varepsilon _{1}\right) }-Z_{M}^{\left(
\varepsilon _{2}\right) }\right\vert }{c},Z_{M}^{\left( \varepsilon
_{2}\right) }\right) \right\vert ^{2}}{\frac{1}{\tau _{C}}+\alpha _{C}\omega
_{M}\left\vert \Psi \left( \theta -\frac{\left\vert Z_{M}^{\left(
\varepsilon _{1}\right) }-Z_{M}^{\left( \varepsilon _{2}\right) }\right\vert 
}{c},Z_{M}^{\left( \varepsilon _{2}\right) }\right) \right\vert ^{2}} \\
D_{Z_{M}^{\left( \varepsilon _{1}\right) },Z_{M}^{\left( \varepsilon
_{2}\right) }} &=&\frac{\alpha _{D}\omega _{M}}{\frac{1}{\tau _{D}}+\alpha
_{D}\omega _{M}\left\vert \Psi \left( \theta ,Z_{M}^{\left( \varepsilon
_{1}\right) }\right) \right\vert ^{2}}
\end{eqnarray*}

\begin{equation*}
T\left( Z_{M}^{\left( \varepsilon _{1}\right) },Z_{M}^{\left( \varepsilon
_{2}\right) }\right) =\lambda \tau \hat{T}\left( Z_{M}^{\left( \varepsilon
_{1}\right) },Z_{M}^{\left( \varepsilon _{2}\right) }\right) =\lambda \tau 
\frac{h\left( Z_{M}^{\left( \varepsilon _{1}\right) },Z_{M}^{\left(
\varepsilon _{2}\right) }\right) C_{Z_{M}^{\left( \varepsilon _{1}\right)
},Z_{M}^{\left( \varepsilon _{2}\right) }}\left( \theta \right) h_{C}}{%
C_{Z_{M}^{\left( \varepsilon _{1}\right) },Z_{M}^{\left( \varepsilon
_{2}\right) }}\left( \theta \right) h_{C}+D_{Z_{M}^{\left( \varepsilon
_{1}\right) },Z_{M}^{\left( \varepsilon _{2}\right) }}\left( \theta \right)
h_{D}}
\end{equation*}%
Then assuming an exponential decreasing dependency in distance for the
connectivities: 
\begin{equation*}
h\left( Z_{M}^{\left( \varepsilon _{1}\right) },Z_{M}^{\left( \varepsilon
_{2}\right) }\right) \simeq \exp \left( -\frac{\left\vert Z_{M}^{\left(
\varepsilon _{1}\right) }-Z_{M}^{\left( \varepsilon _{2}\right) }\right\vert 
}{\nu c}\right)
\end{equation*}%
we obtain:%
\begin{eqnarray}
&&T\left( Z_{M}^{\left( \varepsilon _{1}\right) },Z_{M}^{\left( \varepsilon
_{2}\right) }\right)  \label{dfM} \\
&=&\frac{\lambda \tau \exp \left( -\frac{\left\vert Z_{M}^{\left(
\varepsilon _{1}\right) }-Z_{M}^{\left( \varepsilon _{2}\right) }\right\vert 
}{\nu c}\right) \left\vert \Psi \left( \theta -\frac{\left\vert
Z_{M}^{\left( \varepsilon _{1}\right) }-Z_{M}^{\left( \varepsilon
_{2}\right) }\right\vert }{c},Z_{M}^{\left( \varepsilon _{2}\right) }\right)
\right\vert ^{2}h_{C}}{\left\vert \Psi \left( \theta -\frac{\left\vert
Z_{M}^{\left( \varepsilon _{1}\right) }-Z_{M}^{\left( \varepsilon
_{2}\right) }\right\vert }{c},Z_{M}^{\left( \varepsilon _{2}\right) }\right)
\right\vert ^{2}h_{C}+\left( \frac{1}{\alpha _{C}\tau _{C}}+\omega
_{M}\left\vert \Psi \left( \theta -\frac{\left\vert Z_{M}^{\left(
\varepsilon _{1}\right) }-Z_{M}^{\left( \varepsilon _{2}\right) }\right\vert 
}{c},Z_{M}^{\left( \varepsilon _{2}\right) }\right) \right\vert ^{2}\right) 
\frac{\alpha _{D}h_{D}}{\frac{1}{\tau _{D}}+\alpha _{D}\omega _{M}\left\vert
\Psi \left( \theta ,Z_{M}^{\left( \varepsilon _{1}\right) }\right)
\right\vert ^{2}}}  \notag
\end{eqnarray}%
In a long run static perspective, this formula reduces to:%
\begin{equation*}
T\left( Z_{M}^{\left( \varepsilon _{1}\right) },Z_{M}^{\left( \varepsilon
_{2}\right) }\right) =\frac{\lambda \tau \exp \left( -\frac{\left\vert
Z_{M}^{\left( \varepsilon _{1}\right) }-Z_{M}^{\left( \varepsilon
_{2}\right) }\right\vert }{\nu c}\right) \left\vert \Psi _{0}\left(
Z_{M}^{\left( \varepsilon _{2}\right) }\right) \right\vert ^{2}h_{C}}{%
\left\vert \Psi _{0}\left( Z_{M}^{\left( \varepsilon _{2}\right) }\right)
\right\vert ^{2}h_{C}+\left( \frac{1}{\alpha _{C}\tau _{C}}+\omega
_{M}\left\vert \Psi _{0}\left( Z_{M}^{\left( \varepsilon _{2}\right)
}\right) \right\vert ^{2}\right) \frac{\alpha _{D}h_{D}}{\frac{1}{\tau _{D}}%
+\alpha _{D}\omega _{M}\left\vert \Psi _{0}\left( Z_{M}^{\left( \varepsilon
_{1}\right) }\right) \right\vert ^{2}}}
\end{equation*}%
The decription of the set $U_{M}$ is achieved by adding the long term
determination of activities $\omega _{M}$:%
\begin{eqnarray*}
\omega _{M}^{-1}\left( Z_{M}^{\left( \varepsilon _{1}\right) },\left\vert
\Psi \right\vert ^{2}\right) &\simeq &G\left( \frac{\kappa }{N}%
\sum_{Z_{M}^{\left( \varepsilon _{2}\right) }}T\left( Z_{M}^{\left(
\varepsilon _{1}\right) },Z_{M}^{\left( \varepsilon _{2}\right) }\right)
\left\vert \Psi _{0}\left( Z_{M}^{\left( \varepsilon _{2}\right) }\right)
\right\vert ^{2}\right) \\
&\simeq &G\left( C\frac{\left\vert \Psi _{0}\left( Z_{M}\right) \right\vert
^{4}h_{C}}{\left\vert \Psi _{0}\left( Z_{M}\right) \right\vert
^{2}h_{C}+\left( \frac{1}{\alpha _{C}\tau _{C}}+\omega _{M}\left\vert \Psi
_{0}\left( Z_{M}\right) \right\vert ^{2}\right) \frac{\alpha _{D}h_{D}}{%
\frac{1}{\tau _{D}}+\alpha _{D}\omega _{M}\left\vert \Psi _{0}\left(
Z_{M}\right) \right\vert ^{2}}}\right)
\end{eqnarray*}%
whr:%
\begin{equation*}
C=\frac{\kappa \lambda \tau }{N\left( \sharp \left\{ Z_{M}^{\left(
\varepsilon _{1}\right) }\right\} \right) }\sum_{Z_{M}^{\left( \varepsilon
_{1}\right) },Z_{M}^{\left( \varepsilon _{2}\right) }}\exp \left( -\frac{%
\left\vert Z_{M}^{\left( \varepsilon _{1}\right) }-Z_{M}^{\left( \varepsilon
_{2}\right) }\right\vert }{\nu c}\right)
\end{equation*}%
nd $\left\vert \Psi _{0}\left( Z_{M}\right) \right\vert ^{2}$ is the average
of $\left\vert \Psi _{0}\left( Z_{M}^{\left( \varepsilon _{2}\right)
}\right) \right\vert ^{2}$ over $\left\{ Z_{M}^{\left( \varepsilon
_{2}\right) }\right\} $. The value of $\left\vert \Psi _{0}\left(
Z_{M}\right) \right\vert ^{2}$ can be approximated in the following way.

In the field-theoretic approach to interferences, we will see that the
signals modify the potential for $\left\vert \Psi _{0}\left( Z\right)
\right\vert ^{2}$ but that in first approximation, this modification can be
neglected. Thus, the value of $\left\vert \Psi _{0}\left( Z\right)
\right\vert ^{2}$ after interferences may be computed by the background
field before interferences. This is formula (\ref{Ps}):%
\begin{equation*}
\left\vert \Psi \left( Z\right) \right\vert ^{2}=\frac{2T\left( Z\right)
\left\langle \left\vert \Psi _{0}\left( Z^{\prime }\right) \right\vert
^{2}\right\rangle _{Z}}{\left( 1+\sqrt{1+4\left( \frac{\lambda \tau \nu
c-T\left( Z\right) }{\left( \frac{1}{\tau _{D}\alpha _{D}}+\frac{1}{\tau
_{C}\alpha _{C}}+\Omega \right) T\left( Z\right) -\frac{1}{\tau _{D}\alpha
_{D}}\lambda \tau \nu c}\right) ^{2}T\left( Z\right) \left\langle \left\vert
\Psi _{0}\left( Z^{\prime }\right) \right\vert ^{2}\right\rangle _{Z}}%
\right) }
\end{equation*}%
where all quantities are computed in the initial background state. The
system emrging from the interferences thus depends on the whole initial
structure.

\subsection{Connectivity between points of $CU_{M}$}

The connectivity function for two points in $CU_{M}$ is obtained by setting $%
\omega <<1$ and $\omega ^{\prime }<<1$:

\begin{equation*}
T\left( Z,Z^{\prime }\right) \simeq \frac{\alpha _{C}\omega \lambda \tau
\exp \left( -\frac{\left\vert Z-Z^{\prime }\right\vert }{\nu c}\right)
\left\vert \Psi \left( \theta -\frac{\left\vert Z-Z^{\prime }\right\vert }{c}%
,Z^{\prime }\right) \right\vert ^{2}h_{C}}{\alpha _{C}\omega \left\vert \Psi
\left( \theta -\frac{\left\vert Z-Z^{\prime }\right\vert }{c},Z^{\prime
}\right) \right\vert ^{2}h_{C}+\left( \frac{\omega ^{\prime }}{\tau _{C}}%
\right) \frac{\alpha _{D}h_{D}}{\frac{\left\vert \Psi \left( \theta
,Z\right) \right\vert ^{2}}{\tau _{D}}}}
\end{equation*}%
and these values are identitical to those computed for the static background
state in the previous section (see (\ref{cnv})), up to some global
modifications of the system by the interfering signals. These modifications
are encompassed in the values of the constants $\Omega $, $\bar{\Omega}$...
in (\ref{cnv}). These modifications are negligible in general.

\subsection{Connectivity between points of $U_{M}$ and points of $CU_{M}$}

Two cases arise. We have ton consider both:%
\begin{equation*}
T\left( Z_{M}^{\left( \varepsilon \right) },Z^{\prime }\right)
\end{equation*}%
describing the connectivity of points of $CU_{M}$ towards point of $U_{M}$,
measuring the strength of signals send from $CU_{M}$ to $U_{M}$, and:%
\begin{equation*}
T\left( Z,Z_{M}^{\left( \varepsilon \right) }\right)
\end{equation*}%
computing the connectivity of points of $U_{M}$ towards point of $CU_{M}$.

The connectivity function $T\left( Z_{M}^{\left( \varepsilon \right)
},Z^{\prime }\right) $ is obtained by setting $\omega =\omega _{M}$ and $%
\omega ^{\prime }<<1$ or $\omega <<1$ and $\omega ^{\prime }=\omega _{M}$.
We find%
\begin{equation}
T\left( Z_{M}^{\left( \varepsilon \right) },Z^{\prime }\right) \simeq \frac{%
\lambda \tau \exp \left( -\frac{\left\vert Z-Z^{\prime }\right\vert }{\nu c}%
\right) \left\vert \Psi \left( \theta -\frac{\left\vert Z-Z^{\prime
}\right\vert }{c},Z^{\prime }\right) \right\vert ^{2}h_{C}}{\left\vert \Psi
\left( \theta -\frac{\left\vert Z-Z^{\prime }\right\vert }{c},Z^{\prime
}\right) \right\vert ^{2}h_{C}+\left\vert \Psi \left( \theta -\frac{%
\left\vert Z-Z^{\prime }\right\vert }{c},Z^{\prime }\right) \right\vert ^{2}%
\frac{\alpha _{D}\omega _{M}h_{D}}{\frac{\left\vert \Psi \left( \theta
,Z\right) \right\vert ^{2}}{\tau _{D}}+\alpha _{D}\omega _{M}}}  \label{cnn}
\end{equation}%
The connectivity function $T\left( Z_{M}^{\left( \varepsilon \right)
},Z^{\prime }\right) $ is derived by setting $\omega <<1$ and $\omega
^{\prime }=\omega _{M}$:%
\begin{equation}
T\left( Z,Z_{M}^{\left( \varepsilon \right) }\right) \simeq \frac{\alpha
_{C}\omega \lambda \tau \exp \left( -\frac{\left\vert Z-Z^{\prime
}\right\vert }{\nu c}\right) \left\vert \Psi \left( \theta -\frac{\left\vert
Z-Z^{\prime }\right\vert }{c},Z^{\prime }\right) \right\vert ^{2}h_{C}}{%
\alpha _{C}\omega \left\vert \Psi \left( \theta -\frac{\left\vert
Z-Z^{\prime }\right\vert }{c},Z^{\prime }\right) \right\vert ^{2}h_{C}+\frac{%
\omega _{M}}{\tau _{C}}\frac{\alpha _{D}h_{D}\tau _{D}}{\left\vert \Psi
\left( \theta ,Z\right) \right\vert ^{2}}}<<1  \label{nT}
\end{equation}%
As a consequence, points of the set $U$ do not connect with elements of $CU$%
. On the contrary, elements of $CU$ send signals and connect to elements of $%
U$ but their firing rate being low, they do not influence the whole set that
remains unaffected.

\subsection{Equations defining the set of connected cells}

The results of the previous paragraph where derived considering a given set $%
U_{M}$. However, there is a priori no guarantee for the unicity of this set.
Actually, the set of constructive interferences points $Z_{M}^{\left(
\varepsilon \right) }$ are by definition, dependent on both fields $\Gamma
\left( T,\hat{T},\theta ,Z,Z^{\prime }\right) $ and $\Psi \left( \theta
,Z\right) $ through $T\left( Z,Z^{\prime }\right) $, so that the points $%
Z_{M}^{\left( \varepsilon \right) }$ are in fact themselves endogeneous:

\begin{eqnarray*}
Z_{M}^{\left( \varepsilon _{1}\right) } &\equiv &Z_{M}^{\left( \varepsilon
_{1}\right) }\left( T\left( Z_{M}^{\left( \varepsilon _{1}\right)
},Z_{M}^{\left( \varepsilon _{2}\right) }\right) ,\left\vert \Psi \left(
\theta ,Z_{M}^{\left( \varepsilon _{1}\right) }\right) \right\vert
^{2},\left\vert \Psi \left( \theta -\frac{\left\vert Z_{M}^{\left(
\varepsilon _{1}\right) }-Z_{M}^{\left( \varepsilon _{2}\right) }\right\vert 
}{c},Z_{M}^{\left( \varepsilon _{2}\right) }\right) \right\vert ^{2}\right)
\\
Z_{M}^{\left( \varepsilon _{2}\right) } &\equiv &Z_{M}^{\left( \varepsilon
_{2}\right) }\left( T\left( Z_{M}^{\left( \varepsilon _{1}\right)
},Z_{M}^{\left( \varepsilon _{2}\right) }\right) ,\left\vert \Psi \left(
\theta ,Z_{M}^{\left( \varepsilon _{1}\right) }\right) \right\vert
^{2},\left\vert \Psi \left( \theta -\frac{\left\vert Z_{M}^{\left(
\varepsilon _{1}\right) }-Z_{M}^{\left( \varepsilon _{2}\right) }\right\vert 
}{c},Z_{M}^{\left( \varepsilon _{2}\right) }\right) \right\vert ^{2}\right)
\end{eqnarray*}%
As a consequence, equation (\ref{dfM}) is a self consistent functionnal non
linear equation for $\left\vert \Psi \left( \theta ,Z_{M}^{\left(
\varepsilon _{1}\right) }\right) \right\vert ^{2}$, $\left\vert \Psi \left(
\theta -\frac{\left\vert Z_{M}^{\left( \varepsilon _{1}\right)
}-Z_{M}^{\left( \varepsilon _{2}\right) }\right\vert }{c},Z_{M}^{\left(
\varepsilon _{2}\right) }\right) \right\vert ^{2}$ and $T\left(
Z_{M}^{\left( \varepsilon _{1}\right) },Z_{M}^{\left( \varepsilon
_{2}\right) }\right) $. This implies multiple possible states of
constructive interferences. The full environment impacts the interference
pattern, and is itself modified by these interferences, leading possibly to
multiple equilibria. These solutions depend on $\Psi $. If thereisa
threshold for connectivity to be effective, given the form of $h\left(
Z_{M}^{\left( \varepsilon _{1}\right) },Z_{M}^{\left( \varepsilon
_{2}\right) }\right) $, connections arise along lines or branched lines.
This may induce some effective form of engrams, asset of branched lines,
described by activities and connectivity at nodes.

\section{Medium term state reactivation}

In this section, we focus on the possible reactivation of a state defined by
the points $Z_{M}^{\left( \varepsilon \right) }$. When the sources are off,
we have $J\left( Z\right) =0$ for $\theta >\theta _{t}$, so that in the
medium run:%
\begin{eqnarray*}
\omega \left( Z_{M}^{\left( \varepsilon _{1}\right) }\right) &\simeq &\omega
_{J=0}<<\omega _{M} \\
T\left( Z_{M}^{\left( \varepsilon _{1}\right) },Z_{M}^{\left( \varepsilon
_{2}\right) }\right) &\simeq &T\exp \left( -\lambda \left( \theta -\theta
_{t}\right) \right)
\end{eqnarray*}%
The activation at some point $Z_{M}^{\left( \varepsilon _{2}\right) }$, $%
\omega \left( Z_{M}^{\left( \varepsilon _{1}\right) }\right) =\omega _{M}$
leads to the dynamics between points $\left\{ Z_{M}^{\left( \varepsilon
_{1}\right) }\right\} $ with $T\left( Z_{M}^{\left( \varepsilon _{1}\right)
},Z_{M}^{\left( \varepsilon _{2}\right) }\right) =T$. Set $U$ do not connect
with set $CU$. Writing the equation for $\omega ^{-1}\left( J,\theta
,Z,\left\vert \Psi \right\vert ^{2}\right) $: 
\begin{eqnarray*}
&&\omega ^{-1}\left( J,\theta ,Z,\left\vert \Psi \right\vert ^{2}\right) \\
&=&G\left( J\left( \theta ,Z\right) +\int \frac{\kappa }{N}\frac{\omega
\left( J,\theta -\frac{\left\vert Z-Z_{1}\right\vert }{c},Z_{1},\Psi \right)
T\left\vert \Gamma \left( T,\hat{T},\theta ,Z,Z_{1}\right) \right\vert ^{2}}{%
\omega \left( J,\theta ,Z,\left\vert \Psi \right\vert ^{2}\right) }%
\left\vert \Psi \left( \theta -\frac{\left\vert Z-Z_{1}\right\vert }{c}%
,Z_{1}\right) \right\vert ^{2}dZ_{1}\right)
\end{eqnarray*}%
and using the condition about the average connectivit: 
\begin{equation*}
\int T\left\vert \Gamma \left( T,\hat{T},\theta ,Z,Z_{1}\right) \right\vert
^{2}dT=T\left( Z_{M}^{\left( \varepsilon _{1}\right) },Z_{M}^{\left(
\varepsilon _{2}\right) }\right)
\end{equation*}%
that connects only the points $Z_{M}^{\left( \varepsilon \right) }$ in first
approximation yields: 
\begin{equation*}
\omega _{M}^{-1}\left( Z_{M}^{\left( \varepsilon _{1}\right) },\left\vert
\Psi \right\vert ^{2},\theta \right) \simeq G\left( \frac{\kappa }{N}%
\sum_{Z_{M}^{\left( \varepsilon _{2}\right) }}\frac{\omega _{M}\left(
Z_{M}^{\left( \varepsilon _{2}\right) },\left\vert \Psi \right\vert
^{2},\theta -\frac{\left\vert Z_{M}^{\left( \varepsilon _{1}\right)
}-Z_{M}^{\left( \varepsilon _{2}\right) }\right\vert }{c}\right) }{\omega
_{M}\left( Z_{M}^{\left( \varepsilon _{1}\right) },\left\vert \Psi
\right\vert ^{2},\theta \right) }T\left( Z_{M}^{\left( \varepsilon
_{1}\right) },Z_{M}^{\left( \varepsilon _{2}\right) }\right) \left\vert \Psi
_{0}\left( Z_{M}^{\left( \varepsilon _{2}\right) }\right) \right\vert
^{2}\right)
\end{equation*}%
with the static lmt:%
\begin{eqnarray*}
\omega _{M}^{-1}\left( Z_{M}^{\left( \varepsilon _{1}\right) },\left\vert
\Psi \right\vert ^{2}\right) &\simeq &G\left( \frac{\kappa }{N}%
\sum_{Z_{M}^{\left( \varepsilon _{2}\right) }}T\left( Z_{M}^{\left(
\varepsilon _{1}\right) },Z_{M}^{\left( \varepsilon _{2}\right) }\right)
\left\vert \Psi _{0}\left( Z_{M}^{\left( \varepsilon _{2}\right) }\right)
\right\vert ^{2}\right) \\
&\simeq &G\left( C\frac{\left\vert \Psi _{0}\left( Z_{M}\right) \right\vert
^{4}h_{C}}{\left\vert \Psi _{0}\left( Z_{M}\right) \right\vert
^{2}h_{C}+\left( \frac{1}{\alpha _{C}\tau _{C}}+\omega _{M}\left\vert \Psi
_{0}\left( Z_{M}\right) \right\vert ^{2}\right) \frac{\alpha _{D}h_{D}}{%
\frac{1}{\tau _{D}}+\alpha _{D}\omega _{M}\left\vert \Psi _{0}\left(
Z_{M}\right) \right\vert ^{2}}}\right)
\end{eqnarray*}%
and th st is reactivated as a whole, with lower level activity compared to
the initial level.

\section{Background state for associated signals}

We can can consider that several disconnected state become associated by an
external source. When two states $U_{M}$ and $U_{M}^{\prime }$\ activated
simultaneously their average connectivities are:%
\begin{equation*}
T\left( Z_{M}^{\left( \varepsilon _{1}\right) },Z_{M}^{\left( \varepsilon
_{2}\right) }\right) =\frac{\lambda \tau \exp \left( -\frac{\left\vert
Z_{M}^{\left( \varepsilon _{1}\right) }-Z_{M}^{\left( \varepsilon
_{2}\right) }\right\vert }{\nu c}\right) \left\vert \Psi _{0}\left(
Z_{M}^{\left( \varepsilon _{2}\right) }\right) \right\vert ^{2}h_{C}}{%
\left\vert \Psi _{0}\left( Z_{M}^{\left( \varepsilon _{2}\right) }\right)
\right\vert ^{2}h_{C}+\left( \frac{1}{\alpha _{C}\tau _{C}}+\omega
_{M}\left\vert \Psi _{0}\left( Z_{M}^{\left( \varepsilon _{2}\right)
}\right) \right\vert ^{2}\right) \frac{\alpha _{D}h_{D}}{\frac{1}{\tau _{D}}%
+\alpha _{D}\omega _{M}\left\vert \Psi _{0}\left( Z_{M}^{\left( \varepsilon
_{1}\right) }\right) \right\vert ^{2}}}
\end{equation*}%
\begin{equation*}
T\left( Z_{M}^{\prime \left( \varepsilon _{1}\right) },Z_{M}^{\prime \left(
\varepsilon _{2}\right) }\right) =\frac{\lambda \tau \exp \left( -\frac{%
\left\vert Z_{M}^{\prime \left( \varepsilon _{1}\right) }-Z_{M}^{\prime
\left( \varepsilon _{2}\right) }\right\vert }{\nu c}\right) \left\vert \Psi
_{0}\left( Z_{M}^{\prime \left( \varepsilon _{2}\right) }\right) \right\vert
^{2}h_{C}}{\left\vert \Psi _{0}\left( Z_{M}^{\prime \left( \varepsilon
_{2}\right) }\right) \right\vert ^{2}h_{C}+\left( \frac{1}{\alpha _{C}\tau
_{C}}+\omega _{M}^{\prime }\left\vert \Psi _{0}\left( Z_{M}^{\prime \left(
\varepsilon _{2}\right) }\right) \right\vert ^{2}\right) \frac{\alpha
_{D}h_{D}}{\frac{1}{\tau _{D}}+\alpha _{D}\omega _{M}^{\prime }\left\vert
\Psi _{0}\left( Z_{M}^{\prime \left( \varepsilon _{1}\right) }\right)
\right\vert ^{2}}}
\end{equation*}%
But crssd connectivities have also to be considered:

\begin{equation*}
T\left( Z_{M}^{\left( \varepsilon _{1}\right) },Z_{M}^{\prime \left(
\varepsilon _{2}\right) }\right) =\frac{\lambda \tau \exp \left( -\frac{%
\left\vert Z_{M}^{\left( \varepsilon _{1}\right) }-Z_{M}^{\prime \left(
\varepsilon _{2}\right) }\right\vert }{\nu c}\right) \left\vert \Psi
_{0}\left( Z_{M}^{\prime \left( \varepsilon _{2}\right) }\right) \right\vert
^{2}h_{C}}{\left\vert \Psi _{0}\left( Z_{M}^{\prime \left( \varepsilon
_{2}\right) }\right) \right\vert ^{2}h_{C}+\left( \frac{1}{\alpha _{C}\tau
_{C}}+\omega _{M}^{\prime }\left\vert \Psi _{0}\left( Z_{M}^{\prime \left(
\varepsilon _{2}\right) }\right) \right\vert ^{2}\right) \frac{\alpha
_{D}h_{D}}{\frac{1}{\tau _{D}}+\alpha _{D}\omega _{M}\left\vert \Psi
_{0}\left( Z_{M}^{\left( \varepsilon _{1}\right) }\right) \right\vert ^{2}}}
\end{equation*}%
\begin{equation*}
T\left( Z_{M}^{\prime \left( \varepsilon _{2}\right) },Z_{M}^{\left(
\varepsilon _{1}\right) }\right) \simeq \frac{\lambda \tau \exp \left( -%
\frac{\left\vert Z_{M}^{\prime \left( \varepsilon _{1}\right)
}-Z_{M}^{\left( \varepsilon _{2}\right) }\right\vert }{\nu c}\right)
\left\vert \Psi _{0}\left( Z_{M}^{\left( \varepsilon _{2}\right) }\right)
\right\vert ^{2}h_{C}}{\left\vert \Psi _{0}\left( Z_{M}^{\left( \varepsilon
_{2}\right) }\right) \right\vert ^{2}h_{C}+\left( \frac{1}{\alpha _{C}\tau
_{C}}+\omega _{M}\left\vert \Psi _{0}\left( Z_{M}^{\left( \varepsilon
_{2}\right) }\right) \right\vert ^{2}\right) \frac{\alpha _{D}h_{D}}{\frac{1%
}{\tau _{D}}+\alpha _{D}\omega _{M}^{\prime }\left\vert \Psi _{0}\left(
Z_{M}^{\prime \left( \varepsilon _{1}\right) }\right) \right\vert ^{2}}}
\end{equation*}%
along with the associated activities:%
\begin{equation*}
\omega _{M}^{-1}\left( \hat{Z}_{M}^{\left( \varepsilon _{1}\right)
},\left\vert \Psi \right\vert ^{2},\theta \right) \simeq G\left( \frac{%
\kappa }{N}\sum_{\hat{Z}_{M}^{\left( \varepsilon _{2}\right) }}\frac{\omega
_{M}\left( \hat{Z}_{M}^{\left( \varepsilon _{2}\right) },\left\vert \Psi
\right\vert ^{2},\theta -\frac{\left\vert \hat{Z}_{M}^{\left( \varepsilon
_{1}\right) }-\hat{Z}_{M}^{\left( \varepsilon _{2}\right) }\right\vert }{c}%
\right) }{\omega _{M}\left( \hat{Z}_{M}^{\left( \varepsilon _{1}\right)
},\left\vert \Psi \right\vert ^{2},\theta \right) }T\left( \hat{Z}%
_{M}^{\left( \varepsilon _{1}\right) },\hat{Z}_{M}^{\left( \varepsilon
_{2}\right) }\right) \left\vert \Psi _{0}\left( \hat{Z}_{M}^{\left(
\varepsilon _{2}\right) }\right) \right\vert ^{2}\right)
\end{equation*}%
with:%
\begin{equation*}
\hat{Z}_{M}^{\left( \varepsilon \right) }\in \left\{ Z_{M}^{\left(
\varepsilon _{1}\right) },Z_{M}^{\prime \left( \varepsilon _{2}\right)
}\right\}
\end{equation*}%
nd sttc lmt:%
\begin{equation*}
\omega _{M}^{-1}\left( \hat{Z}_{M}^{\left( \varepsilon _{1}\right) }\right)
\simeq G\left( \frac{\kappa }{N}\sum_{\hat{Z}_{M}^{\left( \varepsilon
_{2}\right) }}\frac{\omega _{M}\left( \hat{Z}_{M}^{\left( \varepsilon
_{2}\right) },\left\vert \Psi \right\vert ^{2}\right) }{\omega _{M}\left( 
\hat{Z}_{M}^{\left( \varepsilon _{1}\right) },\left\vert \Psi \right\vert
^{2}\right) }T\left( \hat{Z}_{M}^{\left( \varepsilon _{1}\right) },\hat{Z}%
_{M}^{\left( \varepsilon _{2}\right) }\right) \left\vert \Psi _{0}\left( 
\hat{Z}_{M}^{\left( \varepsilon _{2}\right) }\right) \right\vert ^{2}\right)
\end{equation*}%
\begin{eqnarray*}
\omega _{M} &=&\omega _{M}\left( Z_{M}^{\left( \varepsilon \right) }\right)
\\
\omega _{M}^{\prime } &=&\omega _{M}\left( Z_{M}^{\prime \left( \varepsilon
\right) }\right)
\end{eqnarray*}

\section{Reactivation of associated signals}

As before, activation at one point for constant connectivities yields,
dynamic system:%
\begin{equation*}
\omega _{M}^{-1}\left( \hat{Z}_{M}^{\left( \varepsilon _{1}\right)
},\left\vert \Psi \right\vert ^{2},\theta \right) \simeq G\left( \frac{%
\kappa }{N}\sum_{\hat{Z}_{M}^{\left( \varepsilon _{2}\right) }}\frac{\omega
_{M}\left( \hat{Z}_{M}^{\left( \varepsilon _{2}\right) },\left\vert \Psi
\right\vert ^{2},\theta -\frac{\left\vert \hat{Z}_{M}^{\left( \varepsilon
_{1}\right) }-\hat{Z}_{M}^{\left( \varepsilon _{2}\right) }\right\vert }{c}%
\right) }{\omega _{M}\left( \hat{Z}_{M}^{\left( \varepsilon _{1}\right)
},\left\vert \Psi \right\vert ^{2},\theta \right) }T\left( \hat{Z}%
_{M}^{\left( \varepsilon _{1}\right) },\hat{Z}_{M}^{\left( \varepsilon
_{2}\right) }\right) \left\vert \Psi _{0}\left( \hat{Z}_{M}^{\left(
\varepsilon _{2}\right) }\right) \right\vert ^{2}\right)
\end{equation*}%
cnvrgng twrd qlbrm:%
\begin{equation*}
\omega _{M}^{-1}\left( \hat{Z}_{M}^{\left( \varepsilon _{1}\right)
},\left\vert \Psi \right\vert ^{2},\theta \right) \simeq G\left( \frac{%
\kappa }{N}\sum_{\hat{Z}_{M}^{\left( \varepsilon _{2}\right) }}\frac{\omega
_{M}\left( \hat{Z}_{M}^{\left( \varepsilon _{2}\right) },\left\vert \Psi
\right\vert ^{2},\theta -\frac{\left\vert \hat{Z}_{M}^{\left( \varepsilon
_{1}\right) }-\hat{Z}_{M}^{\left( \varepsilon _{2}\right) }\right\vert }{c}%
\right) }{\omega _{M}\left( \hat{Z}_{M}^{\left( \varepsilon _{1}\right)
},\left\vert \Psi \right\vert ^{2},\theta \right) }T\left( \hat{Z}%
_{M}^{\left( \varepsilon _{1}\right) },\hat{Z}_{M}^{\left( \varepsilon
_{2}\right) }\right) \left\vert \Psi _{0}\left( \hat{Z}_{M}^{\left(
\varepsilon _{2}\right) }\right) \right\vert ^{2}\right)
\end{equation*}

\section{Background state for sequence of signals}

The previous section allows to consider now the combined effects of two
different sources modifying the system at different moments. We will
consider two possibilities, distant and subsequent activations. We present
these two possibilities qualitatively and provide some details while
developping the field formalism dynamic for connectivities.

\subsection{Distant activation}

We assume that the system is in a static background state, as computed in
the previous section, and described as $\left\{ \omega _{0},T_{0}\right\} $.
The effect of two distant signals in time can be described in the sequence:%
\begin{eqnarray*}
\left\{ \omega _{0},T_{0}\right\} &\rightarrow &\left\{ T\left(
Z_{M}^{\left( \varepsilon _{1}\right) },Z_{M}^{\left( \varepsilon
_{2}\right) }\right) ,\omega _{M}\right\} \rightarrow \left\{ T\left(
Z_{M}^{\left( \varepsilon _{1}\right) },Z_{M}^{\left( \varepsilon
_{2}\right) }\right) ,\omega _{0}\right\} \\
&\rightarrow &\left\{ T\left( Z_{M}^{\left( \varepsilon _{1}\right)
},Z_{M}^{\left( \varepsilon _{2}\right) }\right) ,\omega _{0}\right\}
+\left\{ T\left( Z_{M}^{\prime \left( \varepsilon _{1}\right)
},Z_{M}^{\prime \left( \varepsilon _{2}\right) }\right) ,\omega _{M}^{\prime
}\right\} \\
&\rightarrow &\left\{ T\left( Z_{M}^{\left( \varepsilon _{1}\right)
},Z_{M}^{\left( \varepsilon _{2}\right) }\right) ,\omega _{0}\right\}
+\left\{ T\left( Z_{M}^{\prime \left( \varepsilon _{1}\right)
},Z_{M}^{\prime \left( \varepsilon _{2}\right) }\right) ,\omega _{0}\right\}
\end{eqnarray*}%
where $\left\{ T\left( Z_{M}^{\left( \varepsilon _{1}\right) },Z_{M}^{\left(
\varepsilon _{2}\right) }\right) ,\omega _{M}\right\} $ describes a modified
state where the connectivities are modified at points $\left( Z_{M}^{\left(
\varepsilon _{1}\right) },Z_{M}^{\left( \varepsilon _{2}\right) }\right) $
with values $T\left( Z_{M}^{\left( \varepsilon _{1}\right) },Z_{M}^{\left(
\varepsilon _{2}\right) }\right) $ and $\omega _{M}$ are the modified
activities. The sequence describe the modification by a first signal. This
implies modified connectivities and modified activities. As the signals
ends, and time increase, the activities, whose time scale can be considered
smaller than the connectivities time scale, come back to their equilibrium,
while the modifications $T\left( Z_{M}^{\left( \varepsilon _{1}\right)
},Z_{M}^{\left( \varepsilon _{2}\right) }\right) $ are persistent. In a
second step an other signal modifies the system at some points, which
induces a new set of modification in connectivities. As time increases the
frequencies come back to some equilibrium values, but connection remain
active. The main feature of the modification is that the two sets:%
\begin{equation*}
\left\{ T\left( Z_{M}^{\left( \varepsilon _{1}\right) },Z_{M}^{\left(
\varepsilon _{2}\right) }\right) ,\omega _{0}\right\} +\left\{ T\left(
Z_{M}^{\prime \left( \varepsilon _{1}\right) },Z_{M}^{\prime \left(
\varepsilon _{2}\right) }\right) ,\omega _{0}\right\}
\end{equation*}%
are not connected a priori. In the activation, the connections between
activated points increase, but connections with other points decrease as
seen from (\ref{cnn}) and (\ref{nT}). The two set remain independent, and in
a reactivation of one set, the following sequence applies (we consider
here,the reactivation of $\omega _{M}^{\prime }$):%
\begin{equation*}
\left\{ T\left( Z_{M}^{\left( \varepsilon _{1}\right) },Z_{M}^{\left(
\varepsilon _{2}\right) }\right) ,\omega _{0}\right\} +\left\{ T\left(
Z_{M}^{\prime \left( \varepsilon _{1}\right) },Z_{M}^{\prime \left(
\varepsilon _{2}\right) }\right) ,\omega _{0}\right\} \rightarrow \left\{
T\left( Z_{M}^{\left( \varepsilon _{1}\right) },Z_{M}^{\left( \varepsilon
_{2}\right) }\right) ,\omega _{0}\right\} +\left\{ T\left( Z_{M}^{\prime
\left( \varepsilon _{1}\right) },Z_{M}^{\prime \left( \varepsilon
_{2}\right) }\right) ,\omega _{M}^{\prime }\right\}
\end{equation*}%
While $\omega _{M}^{\prime }$ is activated at some points of $\left(
Z_{M}^{\prime \left( \varepsilon _{1}\right) },Z_{M}^{\prime \left(
\varepsilon _{2}\right) }\right) $, the activated points will send signals
to the points of the same set to which they are connected. It will lead to
reactivate the activities at these points. However, the other state $\left\{
T\left( Z_{M}^{\left( \varepsilon _{1}\right) },Z_{M}^{\left( \varepsilon
_{2}\right) }\right) ,\omega _{0}\right\} $ being independent, as a
consequenc of (\ref{cnn}) and (\ref{nT}), will remain silent. More about the
mechanism of diffusion of activities will be given in the dynamic field
theoretic approach.

\subsection{Subsequent activation}

When the activation are subsequent, i.e. when the two perturbations are
closed in time, the following sequence applies:%
\begin{eqnarray*}
\left\{ \omega _{0},T_{0}\right\} &\rightarrow &\left\{ T\left(
Z_{M}^{\left( \varepsilon _{1}\right) },Z_{M}^{\left( \varepsilon
_{2}\right) }\right) ,\omega _{M}\right\} \\
&\rightarrow &\left\{ T\left( Z_{M}^{\left( \varepsilon _{1}\right)
},Z_{M}^{\left( \varepsilon _{2}\right) }\right) ,\omega _{M},T\left(
Z_{M}^{\prime \left( \varepsilon _{1}\right) },Z_{M}^{\prime \left(
\varepsilon _{2}\right) }\right) ,\omega _{M}^{\prime },T\left(
Z_{M}^{\left( \varepsilon _{1}\right) },Z_{M}^{\prime \left( \varepsilon
_{2}\right) }\right) ,T\left( Z_{M}^{\prime \left( \varepsilon _{2}\right)
},Z_{M}^{\left( \varepsilon _{1}\right) }\right) \right\} \\
&\rightarrow &\left\{ T\left( Z_{M}^{\left( \varepsilon _{1}\right)
},Z_{M}^{\left( \varepsilon _{2}\right) }\right) ,\omega _{0},T\left(
Z_{M}^{\prime \left( \varepsilon _{1}\right) },Z_{M}^{\prime \left(
\varepsilon _{2}\right) }\right) ,\omega _{0},T\left( Z_{M}^{\left(
\varepsilon _{1}\right) },Z_{M}^{\prime \left( \varepsilon _{2}\right)
}\right) ,T\left( Z_{M}^{\prime \left( \varepsilon _{2}\right)
},Z_{M}^{\left( \varepsilon _{1}\right) }\right) \right\}
\end{eqnarray*}

While activating the system with second signal, the track of the first one
is still active and thus both sets $\left( Z_{M}^{\left( \varepsilon
_{1}\right) },Z_{M}^{\left( \varepsilon _{2}\right) }\right) $ and $\left(
Z_{M}^{\prime \left( \varepsilon _{2}\right) },Z_{M}^{\left( \varepsilon
_{1}\right) }\right) $ are connected. The activated system is thus made of a
set of connected points $\left\{ \left( Z_{M}^{\left( \varepsilon
_{1}\right) },Z_{M}^{\left( \varepsilon _{2}\right) }\right) ,\left(
Z_{M}^{\prime \left( \varepsilon _{2}\right) },Z_{M}^{\left( \varepsilon
_{1}\right) }\right) \right\} $. This is translated by the notation:%
\begin{equation*}
\left\{ T\left( Z_{M}^{\left( \varepsilon _{1}\right) },Z_{M}^{\left(
\varepsilon _{2}\right) }\right) ,\omega _{M},T\left( Z_{M}^{\prime \left(
\varepsilon _{1}\right) },Z_{M}^{\prime \left( \varepsilon _{2}\right)
}\right) ,\omega _{M}^{\prime },T\left( Z_{M}^{\left( \varepsilon
_{1}\right) },Z_{M}^{\prime \left( \varepsilon _{2}\right) }\right) ,T\left(
Z_{M}^{\prime \left( \varepsilon _{2}\right) },Z_{M}^{\left( \varepsilon
_{1}\right) }\right) \right\}
\end{equation*}

When both signals fade away, we are left with:%
\begin{equation*}
\left\{ T\left( Z_{M}^{\left( \varepsilon _{1}\right) },Z_{M}^{\left(
\varepsilon _{2}\right) }\right) ,\omega _{0},T\left( Z_{M}^{\prime \left(
\varepsilon _{1}\right) },Z_{M}^{\prime \left( \varepsilon _{2}\right)
}\right) ,\omega _{0},T\left( Z_{M}^{\left( \varepsilon _{1}\right)
},Z_{M}^{\prime \left( \varepsilon _{2}\right) }\right) ,T\left(
Z_{M}^{\prime \left( \varepsilon _{2}\right) },Z_{M}^{\left( \varepsilon
_{1}\right) }\right) \right\}
\end{equation*}

When one set is reactivated (here choose $\omega _{M}^{\prime }$), we can
write:%
\begin{eqnarray*}
&&\left\{ T\left( Z_{M}^{\left( \varepsilon _{1}\right) },Z_{M}^{\left(
\varepsilon _{2}\right) }\right) ,\omega _{0},T\left( Z_{M}^{\prime \left(
\varepsilon _{1}\right) },Z_{M}^{\prime \left( \varepsilon _{2}\right)
}\right) ,\omega _{0}^{\prime },T\left( Z_{M}^{\left( \varepsilon
_{1}\right) },Z_{M}^{\prime \left( \varepsilon _{2}\right) }\right) ,T\left(
Z_{M}^{\prime \left( \varepsilon _{2}\right) },Z_{M}^{\left( \varepsilon
_{1}\right) }\right) \right\} \\
&\rightarrow &\left\{ T\left( Z_{M}^{\left( \varepsilon _{1}\right)
},Z_{M}^{\left( \varepsilon _{2}\right) }\right) ,\omega _{0},T\left(
Z_{M}^{\prime \left( \varepsilon _{1}\right) },Z_{M}^{\prime \left(
\varepsilon _{2}\right) }\right) ,\omega _{M}^{\prime },T\left(
Z_{M}^{\left( \varepsilon _{1}\right) },Z_{M}^{\prime \left( \varepsilon
_{2}\right) }\right) ,T\left( Z_{M}^{\prime \left( \varepsilon _{2}\right)
},Z_{M}^{\left( \varepsilon _{1}\right) }\right) \right\} \\
&\rightarrow &\left\{ T\left( Z_{M}^{\left( \varepsilon _{1}\right)
},Z_{M}^{\left( \varepsilon _{2}\right) }\right) ,\omega _{M},T\left(
Z_{M}^{\prime \left( \varepsilon _{1}\right) },Z_{M}^{\prime \left(
\varepsilon _{2}\right) }\right) ,\omega _{M}^{\prime },T\left(
Z_{M}^{\left( \varepsilon _{1}\right) },Z_{M}^{\prime \left( \varepsilon
_{2}\right) }\right) ,T\left( Z_{M}^{\prime \left( \varepsilon _{2}\right)
},Z_{M}^{\left( \varepsilon _{1}\right) }\right) \right\}
\end{eqnarray*}%
Actually, while $\omega _{M}^{\prime }$ is activated at some point, the
whole system defined by the points $\left( Z_{M}^{\prime \left( \varepsilon
_{2}\right) },Z_{M}^{\left( \varepsilon _{1}\right) }\right) $ and their
connections $T\left( Z_{M}^{\prime \left( \varepsilon _{2}\right)
},Z_{M}^{\left( \varepsilon _{1}\right) }\right) $ will be activated, as in
the previous paragraph. However, given that both sets $T\left( Z_{M}^{\left(
\varepsilon _{1}\right) },Z_{M}^{\left( \varepsilon _{2}\right) }\right) $
are connected, the diffucion of the signal will reactive the activity
between the points $\left( Z_{M}^{\left( \varepsilon _{1}\right)
},Z_{M}^{\left( \varepsilon _{2}\right) }\right) $. The two structures are
synchronically activated.

\section{Conclusion}

Our results reveal that the primary entities emerging from our model are
sets of interconnected cells. The activity levels of these cells are jointly
defined with their interconnections within the set. We observed that
dynamically, such sets interact with each other and can engage in
associations, deactivations, and reactivations. While our results were
obtained in a qualitative manner, the subsequent article will provide a more
technical derivation, emphasizing the role of the field theoretic framework.
Nevertheless, our study already offers insights into two key characteristics
of the formalism.

Firstly, the interactions among interconnected sets imply the need to
develop a dynamic effective formalism for connectivity functions.
Integrating the degrees of freedom for neuronal activity fields should yield
such a formalism, directly describing activations, associations, and other
group-related phenomena. This effective formalism is expounded upon in the
third article of this series.

Secondly, the existence of additional states that may be activated due to
external signals suggests the necessity for a formalism to describe
interconnected groups. The coexistence, interaction, creation, or
deactivation of several groups prompts the consideration of a field
formalism for these groups. This is the objective of the fourth article in
this series, which will develop such an expanded model.

\pagebreak

\section*{Appendix 1 Effective action for $\Psi \left( \protect\theta %
,Z\right) $}

\subsection*{1.1 Projection on activities states}

When we restrict the fields to those of the form: 
\begin{equation}
\Psi \left( \theta ,Z\right) \delta \left( \omega ^{-1}-\omega ^{-1}\left(
J,\theta ,Z,\left\vert \Psi \right\vert ^{2}\right) \right)
\end{equation}%
where $\omega ^{-1}\left( J,\theta ,Z,\Psi \right) $ satisfies:%
\begin{eqnarray}
&&\omega ^{-1}\left( J,\theta ,Z,\left\vert \Psi \right\vert ^{2}\right) \\
&=&G\left( J\left( \theta ,Z\right) +\int \frac{\kappa }{N}\frac{\omega
\left( J,\theta -\frac{\left\vert Z-Z_{1}\right\vert }{c},Z_{1},\Psi \right)
T\left( Z,\theta ,Z_{1},\theta -\frac{\left\vert Z-Z_{1}\right\vert }{c}%
\right) }{\omega \left( J,\theta ,Z,\left\vert \Psi \right\vert ^{2}\right) }%
\left\vert \Psi \left( \theta -\frac{\left\vert Z-Z_{1}\right\vert }{c}%
,Z_{1}\right) \right\vert ^{2}dZ_{1}\right)  \notag
\end{eqnarray}%
The classical effective action writes: 
\begin{subequations}
\begin{equation}
-\frac{1}{2}\int \Psi ^{\dagger }\left( \theta ,Z\right) \Psi \left( \theta
,Z\right) \delta \left( \omega ^{-1}-\omega ^{-1}\left( J,\theta
,Z,\left\vert \Psi \right\vert ^{2}\right) \right) \left( \left( \frac{%
\sigma ^{2}}{2}\nabla _{\theta }-\omega ^{-1}\right) \nabla _{\theta
}\right) \Psi \left( \theta ,Z\right) \delta \left( \omega ^{-1}-\omega
^{-1}\left( J,\theta ,Z,\left\vert \Psi \right\vert ^{2}\right) \right)
\label{brt}
\end{equation}

We can replace the first $\delta $ function by $1$ to normalize the
projection on the activity dependent states.. The action of $\nabla _{\theta
}$ on $\Psi \left( \theta ,Z\right) \delta \left( \omega ^{-1}-\omega
^{-1}\left( J,\theta ,Z,\left\vert \Psi \right\vert ^{2}\right) \right) $
yields: 
\end{subequations}
\begin{eqnarray}
&&\nabla _{\theta }\left( \Psi \left( \theta ,Z\right) \delta \left( \omega
^{-1}-\omega ^{-1}\left( J,\theta ,Z,\left\vert \Psi \right\vert ^{2}\right)
\right) \right)  \label{rlT} \\
&=&\left( \nabla _{\theta }\Psi \left( \theta ,Z\right) \right) \delta
\left( \omega ^{-1}-\omega ^{-1}\left( J,\theta ,Z,\left\vert \Psi
\right\vert ^{2}\right) \right)  \notag \\
&&-\left( \nabla _{\theta }\omega ^{-1}\left( J,\theta ,Z,\left\vert \Psi
\right\vert ^{2}\right) \right) \Psi \left( \theta ,Z\right) \delta ^{\prime
}\left( \omega ^{-1}-\omega ^{-1}\left( J,\theta ,Z,\left\vert \Psi
\right\vert ^{2}\right) \right)  \notag
\end{eqnarray}%
Inserting the result (\ref{rlT}) in (\ref{brt}) leads to:%
\begin{eqnarray*}
&&-\frac{1}{2}\int \Psi ^{\dagger }\left( \theta ,Z\right) \Psi \left(
\theta ,Z\right) \left( \left( \frac{\sigma ^{2}}{2}\nabla _{\theta }-\omega
^{-1}\right) \right) \left( \nabla _{\theta }\Psi \left( \theta ,Z\right)
\right) \delta \left( \omega ^{-1}-\omega ^{-1}\left( J,\theta ,Z,\left\vert
\Psi \right\vert ^{2}\right) \right) \\
&&+\frac{1}{2}\int \Psi ^{\dagger }\left( \theta ,Z\right) \Psi \left(
\theta ,Z\right) \left( \left( \frac{\sigma ^{2}}{2}\nabla _{\theta }-\omega
^{-1}\right) \right) \Psi \left( \theta ,Z\right) \delta ^{\prime }\left(
\omega ^{-1}-\omega ^{-1}\left( J,\theta ,Z,\left\vert \Psi \right\vert
^{2}\right) \right) \\
&=&-\frac{1}{2}\int \Psi ^{\dagger }\left( \theta ,Z\right) \Psi \left(
\theta ,Z\right) \left( \left( \frac{\sigma ^{2}}{2}\nabla _{\theta }-\omega
^{-1}\left( J,\theta ,Z,\left\vert \Psi \right\vert ^{2}\right) \right)
\right) \nabla _{\theta }\Psi \left( \theta ,Z\right) \\
&&-\frac{1}{2}\int \Psi ^{\dagger }\left( \theta ,Z\right) \Psi \left(
\theta ,Z\right) \left( \left( \frac{\sigma ^{2}}{2}\nabla _{\theta }-\nabla
_{\theta }\omega ^{-1}\left( J,\theta ,Z,\left\vert \Psi \right\vert
^{2}\right) \right) \right) \Psi \left( \theta ,Z\right)
\end{eqnarray*}%
and the sum of the two last terms is, as in the text:%
\begin{equation*}
-\frac{1}{2}\int \Psi ^{\dagger }\left( \theta ,Z\right) \left( \nabla
_{\theta }\left( \frac{\sigma ^{2}}{2}\nabla _{\theta }-\omega ^{-1}\left(
J,\theta ,Z,\left\vert \Psi \right\vert ^{2}\right) \right) \right) \Psi
\left( \theta ,Z\right)
\end{equation*}

\subsection*{Appendix 1.2 Effective action for $\Psi \left( \protect\theta %
,Z\right) $ at the lowest order}

To find the effective action for field $\Psi $ at lowest order, we start
with the two points Green function and prove (\ref{fctnbs}). To do so, we
will expand the action functional in series of the field $\Psi $. The two
points Green functions will be computed by using the "free" action's
propagator, obtained by replacing $\omega ^{-1}\left( J,\theta ,Z,\Psi
\right) $ with $\omega ^{-1}\left( J,\theta ,Z,0\right) $ in (\ref{lcn}).
The free action is:%
\begin{equation}
S_{0}=-\frac{1}{2}\Psi ^{\dagger }\left( \theta ,Z\right) \nabla _{\theta
}\left( \frac{\sigma ^{2}}{2}\nabla _{\theta }-\omega ^{-1}\left( J,\theta
,Z,0\right) \right) \Psi \left( \theta ,Z\right)  \label{zs}
\end{equation}%
and the series in field of (\ref{lcn}) will be considered, as usual, as a
perturbation expansion.

\subsubsection*{1.2.1 "Free" action propagator}

Now, we compute the propagator associated to (\ref{zs}). We decompose the
external current into a static and a time dependent parts $\bar{J}+J\left(
\theta \right) $ where $\bar{J}$ can be thought as the time average of the
current. We will consider that $\left\vert \bar{J}\left( Z\right)
\right\vert >\left\vert J\left( \theta ,Z\right) \right\vert $. At zeroth
order in current $J\left( \theta \right) $, the function $\omega ^{-1}\left(
J,\theta ,Z,0\right) $ satisfies:%
\begin{eqnarray}
\omega ^{-1}\left( J,\theta ,Z,0\right) &=&G\left( \bar{J}+J\left( \theta
\right) \right)  \label{mgv} \\
&\simeq &G\left( \bar{J}\left( Z\right) \right) =\frac{\arctan \left( \left( 
\frac{1}{X_{r}}-\frac{1}{X_{p}}\right) \sqrt{\bar{J}\left( Z\right) }\right) 
}{\sqrt{\bar{J}\left( Z\right) }}=\frac{1}{\bar{X}_{r}\left( Z\right) }%
\equiv \frac{1}{\bar{X}_{r}}  \notag
\end{eqnarray}%
where the dependence in $Z$ of $\bar{X}_{r}$ will be understood. As a
consequence $\omega \left( \theta ,Z\right) $ is thus approximatively equal
to $\bar{X}_{r}$. Under this approximation:%
\begin{equation*}
S_{0}=-\Psi ^{\dagger }\left( \theta ,Z\right) \nabla _{\theta }\left( \frac{%
\sigma ^{2}}{2}\nabla _{\theta }-\frac{1}{\bar{X}_{r}}\right) \Psi \left(
\theta ,Z\right)
\end{equation*}%
and the Green function of the operator $\nabla _{\theta }\left( \frac{\sigma
^{2}}{2}\nabla _{\theta }-\frac{1}{\bar{X}_{r}}\right) $ is computed as:%
\begin{equation}
\left\langle \Psi ^{\dagger }\left( \theta ,Z\right) \Psi \left( \theta
^{\prime },Z\right) \right\rangle \equiv \mathcal{G}_{0}\left( \left( \theta
,Z\right) ,\left( \theta ^{\prime },Z^{\prime }\right) \right) \equiv 
\mathcal{G}_{0}\left( \theta ,\theta ^{\prime },Z\right) =\delta \left(
Z-Z^{\prime }\right) \int \frac{\exp \left( ik\left( \theta -\theta ^{\prime
}\right) \right) }{\frac{\sigma ^{2}}{2}k^{2}+ik\frac{1}{\bar{X}_{r}}+\alpha 
}dk  \label{Grnrr}
\end{equation}%
The right hand side of (\ref{Grnrr}) can be computed as: 
\begin{eqnarray}
\int \frac{\exp \left( ik\left( \theta -\theta ^{\prime }\right) \right) }{%
\frac{\sigma ^{2}}{2}k^{2}+ik\frac{1}{\bar{X}_{r}}+\alpha }dk &=&\exp \left( 
\frac{\theta -\theta ^{\prime }}{\sigma ^{2}\bar{X}_{r}}\right) \int \frac{%
\exp \left( ik\left( \theta -\theta ^{\prime }\right) \right) }{\frac{\sigma
^{2}}{2}k^{2}+\frac{1}{2}\left( \frac{1}{\sigma \bar{X}_{r}}\right)
^{2}+\alpha }dk  \notag \\
&=&\frac{1}{\sqrt{\frac{\pi }{2}}}\frac{\exp \left( -\sqrt{\left( \frac{1}{%
\sigma ^{2}\bar{X}_{r}}\right) ^{2}+\frac{2\alpha }{\sigma ^{2}}}\left\vert
\theta -\theta ^{\prime }\right\vert \right) }{\sqrt{\left( \frac{1}{\sigma
^{2}\bar{X}_{r}}\right) ^{2}+\frac{2\alpha }{\sigma ^{2}}}}\exp \left( \frac{%
\theta -\theta ^{\prime }}{\sigma ^{2}\bar{X}_{r}}\right)  \label{rdrrz}
\end{eqnarray}%
and this is quickly suppressed for $\theta -\theta ^{\prime }<0$. This is
the direct consequence of non-hermiticity of operator. In the sequel, for $%
\sigma ^{2}\bar{X}_{r}<<1$, we can thus consider that:%
\begin{equation}
\mathcal{G}_{0}\left( \theta ,\theta ^{\prime },Z\right) =\delta \left(
Z-Z^{\prime }\right) \frac{1}{\sqrt{\frac{\pi }{2}}}\frac{\exp \left(
-\left( \sqrt{\left( \frac{1}{\sigma ^{2}\bar{X}_{r}}\right) ^{2}+\frac{%
2\alpha }{\sigma ^{2}}}-\frac{1}{\sigma ^{2}\bar{X}_{r}}\right) \left(
\theta -\theta ^{\prime }\right) \right) }{\sqrt{\left( \frac{1}{\sigma ^{2}%
\bar{X}_{r}}\right) ^{2}+\frac{2\alpha }{\sigma ^{2}}}}H\left( \theta
-\theta ^{\prime }\right)  \label{rdrzr}
\end{equation}%
where $H$ is the Heaviside function:%
\begin{eqnarray*}
H\left( \theta -\theta ^{\prime }\right) &=&0\text{ for }\theta -\theta
^{\prime }<0 \\
&=&1\text{ for }\theta -\theta ^{\prime }>0
\end{eqnarray*}%
Formula (\ref{rdrzr}) for the propagator is sufficient to compute the graphs
expansion in the next paragraphs. We can check that the corrections due to a
non-static current do not modify the result at a good level of
approximation. Considering the following form for $G\left( J\left( \theta
,Z\right) \right) $: 
\begin{equation*}
G\left( J\left( \theta ,Z\right) \right) =\frac{\arctan \left( \left( \frac{1%
}{X_{r}}-\frac{1}{X_{p}}\right) \sqrt{J\left( \theta ,Z\right) }\right) }{%
\sqrt{J\left( \theta ,Z\right) }}
\end{equation*}%
For relatively high frequency firing rates, i.e., small periods of time
between two spikes, we can write in first approximation:%
\begin{eqnarray*}
G\left( \bar{J}+J\left( \theta ,Z\right) \right) &\simeq &G\left( \bar{J}%
\right) +J\left( \theta ,Z\right) G^{\prime }\left( \bar{J}\right) \\
&=&\frac{1}{\bar{X}_{r}}+J\left( \theta ,Z\right) G^{\prime }\left( \bar{J}%
\right)
\end{eqnarray*}%
and replace (\ref{Grnrr}) by the Green function of:%
\begin{equation*}
\nabla _{\theta }\left( \frac{\sigma ^{2}}{2}\nabla _{\theta }-G\left(
J\left( \theta ,Z\right) \right) \right) \simeq \nabla _{\theta }\left( 
\frac{\sigma ^{2}}{2}\nabla _{\theta }-\frac{1}{\bar{X}_{r}}-J\left( \theta
,Z\right) G^{\prime }\left( \bar{J}\right) \right)
\end{equation*}%
As a consequence, the inverse activity $\mathcal{G}_{0}\left( \theta ,\theta
^{\prime },Z\right) $ defined in (\ref{rdrzr}) is replaced by:%
\begin{eqnarray*}
\mathcal{G}_{0}\left( \left( \theta ,Z\right) ,\left( \theta ^{\prime
},Z^{\prime }\right) \right) &=&\delta \left( Z-Z^{\prime }\right) \frac{1}{%
\sqrt{\frac{\pi }{2}}}\frac{\exp \left( -\left( \sqrt{\left( \frac{1}{\sigma
^{2}\bar{X}_{r}}\right) ^{2}+\frac{2\alpha }{\sigma ^{2}}}-\frac{1}{\sigma
^{2}\bar{X}_{r}}\right) \left( \theta -\theta ^{\prime }\right) \right) }{%
\sqrt{\left( \frac{1}{\sigma ^{2}\bar{X}_{r}}\right) ^{2}+\frac{2\alpha }{%
\sigma ^{2}}}}H\left( \theta -\theta ^{\prime }\right) \\
&&\times \left( 1-\frac{1}{\sqrt{\frac{\pi }{2}}}\frac{G^{\prime }\left( 
\bar{J}\right) }{\sqrt{\left( \frac{1}{\sigma ^{2}\bar{X}_{r}}\right) ^{2}+%
\frac{2\alpha }{\sigma ^{2}}}}\int_{\theta }^{\theta ^{\prime }}J\left(
\theta ^{\prime \prime },Z\right) d\theta ^{\prime \prime }\right)
\end{eqnarray*}%
Since $J\left( \theta ,Z\right) $ is a deviation around the static part $%
\bar{J}$, the corrective term:%
\begin{equation*}
-\frac{1}{\sqrt{\frac{\pi }{2}}}\frac{G^{\prime }\left( \bar{J}\right) }{%
\sqrt{\left( \frac{1}{\sigma ^{2}\bar{X}_{r}}\right) ^{2}+\frac{2\alpha }{%
\sigma ^{2}}}}\int_{\theta }^{\theta ^{\prime }}J\left( \theta ^{\prime
\prime },Z\right) d\theta ^{\prime \prime }
\end{equation*}%
vanishes quickly as $\theta -\theta ^{\prime }$ increases, which justifies
approximation (\ref{rdrzr}).

\subsubsection*{1.2.2 perturbation expansion and the two points Green
function}

Formula (\ref{rdrzr}) allows to compute higher order contributions to the
Green function of action (\ref{lcn}) by using a graph expansion. Actually,
writing $\omega ^{-1}\left( \theta ,Z\right) $ for $\omega ^{-1}\left(
J,\theta ,Z,\Psi \right) $ when no ambiguity is possible, the higher order
contribution for the series expansion of $\omega ^{-1}\left( \theta
,Z\right) $\ in fields are obtained by solving recursively:%
\begin{equation}
\omega ^{-1}\left( J,\theta ,Z\right) =G\left( J\left( \theta ,Z\right)
+\int \frac{\kappa }{N}\frac{\omega \left( J,\theta -\frac{\left\vert
Z-Z_{1}\right\vert }{c},Z_{1}\right) }{\omega \left( J,\theta ,Z\right) }%
\left\vert \Psi \left( \theta -\frac{\left\vert Z-Z_{1}\right\vert }{c}%
,Z_{1}\right) \right\vert ^{2}T\left( Z,\theta ,Z_{1}\right) dZ_{1}d\omega
_{1}\right)  \label{gnf}
\end{equation}%
This will be done precisely in the next paragraph. For now, it is enough to
note that given (\ref{gnf}), the recursive expansion in $\omega ^{-1}\left(
J,\theta ,Z\right) $ of the potential term in (\ref{lcn}):

\begin{equation}
\frac{1}{2}\Psi ^{\dagger }\left( \theta ,Z\right) \nabla \left( G\left(
J\left( \theta ,Z\right) +\int \frac{\kappa }{N}\frac{\omega \left( J,\theta
-\frac{\left\vert Z-Z_{1}\right\vert }{c},Z_{1}\right) }{\omega \left(
J,\theta ,Z\right) }\left\vert \Psi \left( \theta -\frac{\left\vert
Z-Z_{1}\right\vert }{c},Z_{1}\right) \right\vert ^{2}T\left( Z,Z_{1}\right)
dZ_{1}\right) \right) \Psi \left( \theta ,Z\right)  \label{ptnlcn}
\end{equation}%
induces the presence of products in the series expansion of the two points
Green function:%
\begin{eqnarray}
&&\dprod\limits_{i=1}^{m}\int \Psi ^{\dagger }\left( \theta ^{\left(
i\right) },Z_{i}\right) \nabla _{\theta ^{\left( i\right)
}}\dprod\limits_{k=1}^{k_{i}}\left(
\dprod\limits_{l=1}^{l_{k}}\dprod\limits_{\alpha \left( l\right)
=1}^{n\left( \alpha \left( l\right) \right) }\int \left\vert \Psi \left(
\theta ^{\left( i\right) }-\frac{\left\vert Z_{i}-Z_{\alpha \left( l\right)
}^{\left( 1\right) }\right\vert +...+\left\vert Z_{\alpha \left( l\right)
}^{\left( l-1\right) }-Z_{\alpha \left( l\right) }^{\left( l\right)
}\right\vert }{c},Z_{\alpha \left( l\right) }^{\left( l\right) }\right)
\right\vert ^{2}\right)  \notag \\
&&\qquad \qquad \qquad \times dZ_{\alpha \left( l\right) }^{\left( 1\right)
}...dZ_{\alpha \left( l\right) }^{\left( l_{k}\right) }\Psi \left( \theta
^{\left( i\right) },Z_{i}\right) d\theta ^{\left( i\right) }dZ_{i}
\label{pdtc}
\end{eqnarray}%
with $n\left( \alpha \left( l\right) \right) \geqslant n\left( \alpha \left(
l^{\prime }\right) \right) $ for $l>l^{\prime }$ and $m\in 
\mathbb{N}
$. The function $\delta \left( Z-Z^{\prime }\right) $ in\ (\ref{Grnrr}) and
the use of Wick's theorem imply that all subgraphs drawn from this product
reduce to a product of\ free Green functions (\ref{rdrzr}) of the following
form (the gradient terms and the indices $\alpha \left( l\right) $ are not
included and do not impact the reasoning):%
\begin{eqnarray}
&&\int \dprod\limits_{i}\mathcal{G}_{0}\left( \theta ^{\left( i\right)
}-\sum_{l\leqslant n_{i}}\frac{\left\vert Z_{i}-Z_{i}^{\left( l\right)
}\right\vert }{c},\theta ^{\left( i+1\right) }-\sum_{k\leqslant n_{i+1}}%
\frac{\left\vert Z_{i+1}-Z_{i+1}^{\left( k\right) }\right\vert }{c}%
,Z_{i}^{\left( n_{i}\right) },Z_{i}^{\left( n_{i+1}\right) }\right)  \notag
\\
&&\times \delta \left( Z_{1}-Z_{i}^{\left( n_{i}\right) }\right) \delta
\left( Z_{1}-Z_{i+1}^{\left( n_{i+1}\right) }\right) dZ_{i}^{\left(
n_{i}\right) }dZ_{i+1}^{\left( n_{i+1}\right) }\dprod\limits_{i}d\theta
^{\left( i\right) }  \notag \\
&=&\int \dprod\limits_{i}\mathcal{G}_{0}\left( \theta ^{\left( i\right)
}-\sum_{l\leqslant n}\frac{\left\vert Z_{i}-Z_{1}^{\left( l\right)
}\right\vert }{c},\theta ^{\left( i+1\right) }-\sum_{k\leqslant m}\frac{%
\left\vert Z_{i+1}-Z_{1}^{\left( k\right) }\right\vert }{c},Z_{1}\right)
\dprod\limits_{i}d\theta ^{\left( i\right) }  \notag \\
&=&\int \dprod\limits_{i}\mathcal{G}_{0}\left( \theta ^{\left( i\right)
},\theta ^{\left( i+1\right) },Z_{1}\right) \dprod\limits_{i}d\theta
^{\left( i\right) }  \label{lp}
\end{eqnarray}%
by change of variable in the successive integrations. Moreover, the
cancelation of $\mathcal{G}_{0}\left( \theta ,\theta ^{\prime },Z\right) $
for $\theta <\theta ^{\prime }$ implies that this product is different from
zero only for $\theta ^{\left( i\right) }<\theta ^{\left( i+1\right) }$. As
a consequence, for all closed loops $\theta _{1}<...<\theta ^{\left(
i\right) }<\theta ^{\left( i+1\right) }<...\theta _{n}=\theta _{1}$, the
contribution (\ref{lp}) for loop graphs reduces to:%
\begin{equation*}
\dprod\limits_{i}\mathcal{G}_{0}\left( \theta _{1},\theta _{1},Z_{1}\right)
=\dprod\limits_{i}\mathcal{G}_{0}\left( 0,Z_{1}\right)
\end{equation*}%
with (see (\ref{rdrzr})):%
\begin{equation*}
\mathcal{G}_{0}\left( 0,Z\right) =\frac{1}{\sqrt{\frac{\pi }{2}\left( \frac{1%
}{\sigma ^{2}\bar{X}_{r}}\right) ^{2}+\frac{2\pi \alpha }{\sigma ^{2}}}}
\end{equation*}%
As a consequence, the contribution of (\ref{pdtc}) to the two points Green
function between an initial and final state:%
\begin{eqnarray}
&&\left\langle \Psi ^{\dagger }\left( \theta _{in},Z_{in}\right) \int
\dprod\limits_{i=1}^{m}\Psi ^{\dagger }\left( \theta ^{\left( i\right)
},Z_{i}\right) \right.  \notag \\
&&\times \nabla _{\theta ^{\left( i\right)
}}\dprod\limits_{k=1}^{k_{i}}\left( \left( \dprod\limits_{l=1}^{l_{k}}\int
\left\vert \Psi \left( \theta ^{\left( i\right) }-\frac{\left\vert
Z_{i}-Z^{\left( 1\right) }\right\vert +...+\left\vert Z^{\left( l-1\right)
}-Z^{\left( l\right) }\right\vert }{c},Z^{\left( l\right) }\right)
\right\vert ^{2}dZ^{\left( 1\right) }...dZ^{\left( l_{k}\right) }\right)
\right)  \notag \\
&&\times \left. \Psi \left( \theta ^{\left( i\right) },Z_{i}\right) d\theta
^{\left( i\right) }dZ_{i}\Psi \left( \theta _{fn},Z_{fn}\right) \right\rangle
\label{prtr}
\end{eqnarray}%
reduces to sums and integrals of the type:%
\begin{eqnarray}
&&\delta \left( Z_{in}-Z_{fn}\right) \sum_{p}\mathcal{G}_{0}\left( \theta
_{in},\theta _{1},Z_{in}\right) \mathcal{G}_{0}\left( \theta _{1},\theta
_{2},Z_{in}\right) ...\mathcal{G}_{0}\left( \theta _{p},\theta
_{fn},Z_{in}\right)  \label{grpn} \\
&&\times \left( \sum_{n}\sum_{\left\{ L_{1}^{\left( p\right)
},...,L_{n}^{\left( p\right) }\right\} }\dprod_{m=1}^{n}\left( \mathcal{G}%
_{0}\left( 0,0,Z_{m}\right) \right) ^{l\left( L_{m}^{\left( p\right)
}\right) }\right)  \notag
\end{eqnarray}%
where $\left\{ L_{1}^{\left( p\right) },...,L_{n}^{\left( p\right) }\right\} 
$ is the set of all $n$-uplet of possible closed loops that can be drawn
from the remaining variables in (\ref{prtr}) once $p$ variables have been
chosen.

The result (\ref{grpn}) is the same as if in (\ref{ptnlcn}) the potential
had been expanded to the second order in $\Psi $ and in all terms of higher
order, $\left\vert \Psi \left( \theta ,Z\right) \right\vert ^{2}$ had been
replaced by $\mathcal{G}_{0}\left( 0,Z\right) $.

Now, writing $\omega \left( J,\theta ,Z,\left\vert \Psi \right\vert
^{2}\right) $ for $\omega $ and $\omega \left( 0\right) =\omega \left(
J,\theta ,Z,0\right) $ (i.e. when we set $\Psi \equiv 0$), this means that
the $2$ points Green functions are computed using the free action:%
\begin{eqnarray}
&&-\frac{1}{2}\Psi ^{\dagger }\left( \theta ,Z\right) \nabla _{\theta
}\left( \frac{\sigma _{\theta }^{2}}{2}\nabla _{\theta }-\omega ^{-1}\left(
0\right) \right) \Psi \left( \theta ,Z\right)  \label{fctc} \\
&&+\frac{1}{2}\Psi ^{\dagger }\left( \theta ,Z\right) \sum_{n>0}\frac{\nabla
_{\theta }\left( \omega ^{-1}\right) ^{\left( \left[ n\right] \right)
}\left( 0\right) }{\left[ n\right] !}\left( \mathcal{G}_{0}\left( 0,Z\right)
\right) ^{n}\Psi \left( \theta ,Z\right)  \notag \\
&&+\sum_{n>0}\left( \nabla _{\theta }\frac{\left( \omega ^{-1}\right)
^{\left( \left[ n-1\right] \right) }\left( 0\right) \left\vert \Psi
\right\vert ^{2}}{\left[ n-1\right] !}\left( \mathcal{G}_{0}\left(
0,Z\right) \right) ^{n-1}\mathcal{G}_{0}\left( \theta ,\theta ^{\prime
},Z\right) \right) _{\theta ^{\prime }=\theta }  \notag \\
&=&-\frac{1}{2}\Psi ^{\dagger }\left( \theta ,Z\right) \nabla _{\theta
}\left( \frac{\sigma _{\theta }^{2}}{2}\nabla _{\theta }-\omega ^{-1}\left(
0\right) \right) \Psi \left( \theta ,Z\right) +\frac{1}{2}\Psi ^{\dagger
}\left( \theta ,Z\right) \sum_{n>0}\nabla _{\theta }\left( \left( \omega
^{-1}\right) \left( \mathcal{G}_{0}\left( 0,Z\right) \right) -\omega
^{-1}\left( 0\right) \right) \Psi \left( \theta ,Z\right)  \notag \\
&&+\Psi ^{\dagger }\left( \theta ,Z\right) \left( \nabla _{\theta ^{\prime
}}\left( \left( \omega ^{-1}\right) ^{\left( \left[ 1\right] \right) }\left( 
\mathcal{G}_{0}\left( 0,Z\right) \right) \Psi \left( \theta ^{\prime
},Z\right) \mathcal{G}_{0}\left( \theta ,\theta ^{\prime },Z\right) \right)
\right) _{\theta ^{\prime }=\theta }  \notag \\
&\equiv &-\frac{1}{2}\Psi ^{\dagger }\left( \theta ,Z\right) \left( \nabla
_{\theta }\frac{\sigma _{\theta }^{2}}{2}\nabla _{\theta }\right) \Psi
\left( \theta ,Z\right) +\frac{1}{2}\left\vert \Psi \right\vert ^{2}\left[ 
\frac{\delta \left[ \Psi ^{\dagger }\left( \theta ^{\prime },Z\right) \nabla
_{\theta }\omega ^{-1}\left( J,\theta ,Z,\left\vert \Psi \right\vert
^{2}\right) \Psi \left( \theta ,Z\right) \right] }{\delta \left\vert \Psi
\right\vert ^{2}}\right] _{\left\vert \Psi \left( \theta ,Z\right)
\right\vert ^{2}=\mathcal{G}_{0}\left( 0,Z\right) }  \notag
\end{eqnarray}%
where $\frac{\left( \omega ^{-1}\right) ^{\left( \left[ n\right] \right)
}\left( 0\right) }{\left[ n\right] !}$ is a short notation for:%
\begin{equation*}
\sum_{l_{i}}\int \dprod\limits_{i=1}^{n}dZ_{l_{i}}^{\left( 1\right)
}...dZ_{l_{i}}^{\left( l_{i}\right) }\left( \frac{\delta ^{n}\left[ \omega
^{-1}\left( J,\theta ,Z,\left\vert \Psi \right\vert ^{2}\right) \right] }{%
\dprod\limits_{i=1}^{n}\delta \left( \left\vert \Psi \left( \theta -\frac{%
\left\vert Z-Z_{l_{i}}^{\left( 1\right) }\right\vert +...+\left\vert
Z^{_{l_{i}}\left( l-1\right) }-Z_{l_{i}}^{\left( l_{i}\right) }\right\vert }{%
c},Z_{l_{i}}^{\left( l_{i}\right) }\right) \right\vert ^{2}\right) }\right)
_{\left\vert \Psi \right\vert =0}
\end{equation*}%
and $\frac{\left( \omega ^{-1}\right) ^{\left( \left[ n-1\right] \right)
}\left( 0\right) \left\vert \Psi \right\vert ^{2}}{\left[ n-1\right] !}$
stands for:%
\begin{eqnarray*}
&&\sum_{l_{i}}\int \dprod\limits_{i=1}^{n-1}dZ_{l_{i}}^{\left( 1\right)
}...dZ_{l_{i}}^{\left( l_{i}\right) }\left( \frac{\delta ^{n-1}\left[ \omega
^{-1}\left( J,\theta ,Z,\left\vert \Psi \right\vert ^{2}\right) \right] }{%
\dprod\limits_{i}\delta \left( \left\vert \Psi \left( \theta -\frac{%
\left\vert Z-Z_{l_{i}}^{\left( 1\right) }\right\vert +...+\left\vert
Z_{l_{i}}^{\left( l-1\right) }-Z_{l_{i}}^{\left( l_{i}\right) }\right\vert }{%
c},Z_{l_{i}}^{\left( l_{i}\right) }\right) \right\vert ^{2}\right)
^{k_{l_{i}}}}\right) _{\left\vert \Psi \right\vert =0} \\
&&\times \sum_{j=1}^{n-1}\left\vert \Psi \left( \theta -\frac{\left\vert
Z-Z_{l_{j}}^{\left( 1\right) }\right\vert +...+\left\vert Z_{l_{i}}^{\left(
l-1\right) }-Z_{l_{j}}^{\left( l_{j}\right) }\right\vert }{c}%
,Z_{l_{j}}^{\left( l_{j}\right) }\right) \right\vert ^{2}
\end{eqnarray*}%
Similar notation is valid for $\frac{\left( \omega ^{-1}\right) ^{\left( %
\left[ n\right] \right) }\left( \mathcal{G}_{0}\left( 0,0,Z\right) \right)
\left\vert \Psi \right\vert ^{2}}{\left[ n-1\right] !}$, the derivatives are
evaluated at $\left\vert \Psi \left( \theta ,Z\right) \right\vert ^{2}=%
\mathcal{G}_{0}\left( 0,0,Z\right) $.

We have also used $\left\vert \Psi \right\vert ^{2}\left[ \frac{\delta }{%
\delta \left\vert \Psi \right\vert ^{2}}\right] $ as a shorthand for:%
\begin{eqnarray}
&&\sum_{l}\int \left( \frac{dZ_{l}^{\left( 1\right) }...dZ_{l}^{\left(
l\right) }}{\left( k_{l}\right) !}\right) \left\vert \Psi \left( \theta -%
\frac{\left\vert Z-Z_{l}^{\left( 1\right) }\right\vert +...+\left\vert
Z_{l}^{\left( l-1\right) }-Z_{l}^{\left( l_{j}\right) }\right\vert }{c}%
,Z_{l}^{\left( l\right) }\right) \right\vert ^{2}  \label{drvf} \\
&&\times \frac{\delta }{\delta \left( \left\vert \Psi \left( \theta -\frac{%
\left\vert Z-Z_{l}^{\left( 1\right) }\right\vert +...+\left\vert
Z_{l}^{\left( l-1\right) }-Z_{l}^{\left( l\right) }\right\vert }{c}%
,Z_{l}^{\left( l\right) }\right) \right\vert ^{2}\right) }  \notag
\end{eqnarray}%
Ultimately, the computation of the Green function involves the series
expansion of the potential $V\left( \Psi \right) $. As shown earlier (see
equation(\ref{grpn})) the graphs generated by this expansion are equivalent
to those that would result if, in equation (\ref{ptnlcn}) the potential had
been expanded to the second order in $\Psi $ and if, in all terms of higher
order, $\left\vert \Psi \left( \theta ,Z\right) \right\vert ^{2}$ had been
replaced by $\mathcal{G}_{0}\left( 0,Z\right) $. As a consequence, the
second order Green functions are computed with the action:%
\begin{eqnarray*}
&&-\frac{1}{2}\Psi ^{\dagger }\left( \theta ,Z\right) \left( \nabla _{\theta
}\frac{\sigma _{\theta }^{2}}{2}\nabla _{\theta }\right) \Psi \left( \theta
,Z\right) \\
&&+\frac{1}{2}\left\vert \Psi \right\vert ^{2}\left[ \frac{\delta \left[
\Psi ^{\dagger }\left( \theta ^{\prime },Z\right) \nabla _{\theta }\left(
\omega ^{-1}\left( J,\theta ,Z,\left\vert \Psi \right\vert ^{2}\right) \Psi
\left( \theta ,Z\right) \right) \right] }{\delta \left\vert \Psi \right\vert
^{2}}\right] _{\substack{ \left\vert \Psi \left( \theta ,Z\right)
\right\vert ^{2}  \\ =\mathcal{G}_{0}\left( 0,Z\right) }}+\left\vert \Psi
\right\vert ^{2}\left[ \frac{\delta \left[ V\left( \Psi \right) \right] }{%
\delta \left\vert \Psi \right\vert ^{2}}\right] _{\substack{ \left\vert \Psi
\left( \theta ,Z\right) \right\vert ^{2}  \\ =\mathcal{G}_{0}\left(
0,Z\right) }}
\end{eqnarray*}%
Equivalently, this means that the $2$ points Green functions are the inverse
of the operator:%
\begin{equation*}
-\frac{1}{2}\nabla _{\theta }\frac{\sigma _{\theta }^{2}}{2}\nabla _{\theta
}+\frac{1}{2}\left[ \frac{\delta \left[ \Psi ^{\dagger }\left( \theta
^{\prime },Z\right) \nabla _{\theta }\left( \omega ^{-1}\left( J,\theta
,Z,\left\vert \Psi \right\vert ^{2}\right) \Psi \left( \theta ,Z\right)
\right) \right] }{\delta \left\vert \Psi \right\vert ^{2}}\right] 
_{\substack{ \left\vert \Psi \left( \theta ,Z\right) \right\vert ^{2}  \\ =%
\mathcal{G}_{0}\left( 0,Z\right) }}+\left[ \frac{\delta \left[ V\left( \Psi
\right) \right] }{\delta \left\vert \Psi \right\vert ^{2}}\right] 
_{\substack{ \left\vert \Psi \left( \theta ,Z\right) \right\vert ^{2}  \\ =%
\mathcal{G}_{0}\left( 0,Z\right) }}
\end{equation*}%
and, at the lowest order in $\left\vert \Psi \left( \theta ,Z\right)
\right\vert ^{2}$, this corresponds to the effective action of the text.

\section*{Appendix 2}

\subsection*{Corrections to background field}

\subsubsection*{Corrections to saddle point}

To compute the corrections to the background due to: 
\begin{equation*}
K=K\left( \theta ,Z,Z^{\prime },\left\Vert \Psi \right\Vert ^{2},\left\Vert
\Gamma \right\Vert ^{2}\right)
\end{equation*}%
we can ultimately rewrite (\ref{FFF}) by including the term $KT$ so that the
squared term writes:%
\begin{eqnarray*}
&&-\frac{\left( \left( \frac{\left\vert \Psi \left( \theta ,Z\right)
\right\vert ^{2}}{\tau \omega }\left( T-\lambda \tau \hat{T}\right) \right)
\right) ^{2}}{2\sigma _{T}^{2}}+KT \\
&=&-\frac{1}{2\sigma _{T}^{2}}\left( \frac{\left\vert \Psi \left( \theta
,Z\right) \right\vert ^{2}}{\tau \omega }\left( T-\lambda \tau \hat{T}%
-K\sigma _{T}^{2}\left( \frac{\tau \omega }{\left\vert \Psi \left( \theta
,Z\right) \right\vert ^{2}}\right) ^{2}\right) \right) ^{2}+\frac{1}{2}%
K^{2}\sigma _{T}^{2}\left( \frac{\tau \omega }{\left\vert \Psi \left( \theta
,Z\right) \right\vert ^{2}}\right) ^{2}+K\lambda \tau \hat{T}
\end{eqnarray*}%
the last term can be included in $S_{\Gamma }^{\left( 2\right) }$, and
defining:%
\begin{eqnarray*}
h_{C} &=&h_{C}\left( \omega \left( J,\theta ,Z,\left\vert \Psi \right\vert
^{2}\right) \right) \\
h_{D} &=&h_{D}\left( \omega \left( J,\theta -\frac{\left\vert Z-Z^{\prime
}\right\vert }{c},Z^{\prime },\left\vert \Psi \right\vert ^{2}\right) \right)
\end{eqnarray*}%
we find ultimately: 
\begin{eqnarray}
S_{\Gamma }^{\left( 1\right) } &=&\Gamma ^{\dag }\left( T,\hat{T},\theta
,Z,Z^{\prime }\right) \left( \frac{\sigma _{T}^{2}}{2}\nabla _{T}^{2}-\frac{1%
}{2\sigma _{T}^{2}}\left( \frac{\left\vert \Psi \left( \theta ,Z\right)
\right\vert ^{2}}{\tau \omega }\left( T-\lambda \tau \hat{T}-K\sigma
_{T}^{2}\left( \frac{\tau \omega }{\left\vert \Psi \left( \theta ,Z\right)
\right\vert ^{2}}\right) ^{2}\right) \right) ^{2}\right.  \label{SNC} \\
&&\left. +\frac{1}{2\tau \omega \left( Z\right) }+\frac{1}{2}K^{2}\sigma
_{T}^{2}\left( \frac{\tau \omega }{\left\vert \Psi \left( \theta ,Z\right)
\right\vert ^{2}}\right) ^{2}\right) \Gamma \left( T,\hat{T},\theta
,Z,Z^{\prime }\right)  \notag
\end{eqnarray}%
\begin{eqnarray}
S_{\Gamma }^{\left( 2\right) } &=&\Gamma ^{\dag }\left( T,\hat{T},\theta
,Z,Z^{\prime }\right) \left( \frac{\sigma _{\hat{T}}^{2}}{2}\nabla _{\hat{T}%
}^{2}-\frac{\left( \rho \left( C\left( \theta \right) \left\vert \Psi \left(
\theta ,Z\right) \right\vert ^{2}h_{C}+D\left( \theta \right) \left\vert
\Psi \left( \theta -\frac{\left\vert Z-Z^{\prime }\right\vert }{c},Z^{\prime
}\right) \right\vert ^{2}h_{D}\right) \left( \hat{T}-\left\langle \hat{T}%
\right\rangle _{0}\right) \right) ^{2}}{2\sigma _{\hat{T}}^{2}\omega
^{2}\left( \theta ,Z,\left\vert \Psi \right\vert ^{2}\right) }\right.
\label{STC} \\
&&+\left. \frac{\rho \left( C\left( \theta \right) \left\vert \Psi \left(
\theta ,Z\right) \right\vert ^{2}h_{C}-\eta H\left( \delta -T\right)
+D\left( \theta \right) \left\vert \Psi \left( \theta -\frac{\left\vert
Z-Z^{\prime }\right\vert }{c},Z^{\prime }\right) \right\vert
^{2}h_{D}\right) }{2\omega \left( \theta ,Z,\left\vert \Psi \right\vert
^{2}\right) }\right) \Gamma \left( T,\hat{T},\theta ,Z,Z^{\prime }\right) 
\notag \\
&&+\frac{1}{2}\sigma _{\hat{T}}^{2}\left( \frac{\omega \left( \theta
,Z,\left\vert \Psi \right\vert ^{2}\right) }{\rho \left( C\left( \theta
\right) \left\vert \Psi \left( \theta ,Z\right) \right\vert
^{2}h_{C}+D\left( \theta \right) \left\vert \Psi \left( \theta -\frac{%
\left\vert Z-Z^{\prime }\right\vert }{c},Z^{\prime }\right) \right\vert
^{2}h_{D}\right) }\right) ^{2}\left( K\lambda \tau \right) ^{2}+K\lambda
\tau \left\langle \hat{T}\right\rangle _{0}  \notag
\end{eqnarray}%
with:%
\begin{eqnarray*}
\left\langle \hat{T}\left( Z,Z\right) \right\rangle _{0} &=&\frac{\left(
h\left( Z,Z^{\prime }\right) C_{Z,Z^{\prime }}\left( \theta \right)
h_{C}\left\vert \Psi \left( \theta ,Z\right) \right\vert ^{2}-\eta H\left(
\delta -T\left( Z,Z^{\prime }\right) \right) \right) }{C_{Z,Z^{\prime
}}\left( \theta \right) \left\vert \Psi \left( \theta ,Z\right) \right\vert
^{2}h_{C}+D_{Z,Z^{\prime }}\left( \theta \right) \left\vert \Psi \left(
\theta -\frac{\left\vert Z-Z^{\prime }\right\vert }{c},Z^{\prime }\right)
\right\vert ^{2}h_{D}} \\
&&+\sigma _{\hat{T}}^{2}\left( \frac{\omega \left( \theta ,Z,\left\vert \Psi
\right\vert ^{2}\right) }{\rho \left( C\left( \theta \right) \left\vert \Psi
\left( \theta ,Z\right) \right\vert ^{2}h_{C}+D\left( \theta \right)
\left\vert \Psi \left( \theta -\frac{\left\vert Z-Z^{\prime }\right\vert }{c}%
,Z^{\prime }\right) \right\vert ^{2}h_{D}\right) }\right) ^{2}K\lambda \tau
\end{eqnarray*}%
The effective potentials are thus modified by shifts proportional \ to $%
\sigma _{T}^{2}$ and $\sigma _{\hat{T}}^{2}$ respectively. As a consequence
for $\frac{\sigma _{\hat{T}}^{2}}{\sigma _{T}^{2}}<<1$, $\sigma _{T}^{2}<<1$
these corrections can be treated purturbatively and neglected in first
approximation as quoted in the text.

\subsubsection*{Corrections to the averages}

The average equations are modified with terms proportional to $\left\langle
K\right\rangle $: 
\begin{eqnarray*}
\left\langle \hat{T}\left( Z,Z^{\prime }\right) \right\rangle
&=&\left\langle \hat{T}\right\rangle _{0}=\frac{\left( \left\langle h\left(
Z,Z^{\prime }\right) C_{Z,Z^{\prime }}\left( \theta \right) h_{C}\left\vert
\Psi \left( \theta ,Z\right) \right\vert ^{2}\right\rangle -\eta H\left(
\delta -\left\langle T\left( Z,Z^{\prime }\right) \right\rangle \right)
\right) }{\left\langle C_{Z,Z^{\prime }}\left( \theta \right) \left\vert
\Psi \left( \theta ,Z\right) \right\vert ^{2}h_{C}\right\rangle
+\left\langle D_{Z,Z^{\prime }}\left( \theta \right) \left\vert \Psi \left(
\theta -\frac{\left\vert Z-Z^{\prime }\right\vert }{c},Z^{\prime }\right)
\right\vert ^{2}h_{D}\right\rangle } \\
&&+\sigma _{\hat{T}}^{2}\left( \frac{\omega \left( \theta ,Z,\left\vert \Psi
\right\vert ^{2}\right) }{\rho \left( C\left( \theta \right) \left\vert \Psi
\left( \theta ,Z\right) \right\vert ^{2}h_{C}+D\left( \theta \right)
\left\vert \Psi \left( \theta -\frac{\left\vert Z-Z^{\prime }\right\vert }{c}%
,Z^{\prime }\right) \right\vert ^{2}h_{D}\right) }\right) ^{2}\left\langle
K\right\rangle \lambda \tau
\end{eqnarray*}%
and: 
\begin{equation*}
\left\langle T\left( Z,Z^{\prime }\right) \right\rangle =\lambda \tau
\left\langle \hat{T}\left( Z,Z^{\prime }\right) \right\rangle +\left\langle
K\right\rangle \sigma _{T}^{2}\left( \frac{\tau \omega }{\left\vert \Psi
\left( \theta ,Z\right) \right\vert ^{2}}\right) ^{2}
\end{equation*}

Given the definition of $K$, it is a function of the collection $\left\{
\left\langle T\left( Z,Z^{\prime }\right) \right\rangle \right\} _{\left(
Z,Z^{\prime }\right) }$. The corrections to $\left\langle T\left(
Z,Z^{\prime }\right) \right\rangle $ and $\left\langle \hat{T}\left(
Z,Z^{\prime }\right) \right\rangle $ are obtained purturbatively by
replacing $K\left( \left\{ \left\langle T\left( Z,Z^{\prime }\right)
\right\rangle \right\} _{\left( Z,Z^{\prime }\right) }\right) $ with $%
\left\langle T\left( Z,Z^{\prime }\right) \right\rangle $ computed for $K=0$%
. As a first approximation, it is possible to replace $\left\langle T\left(
Z,Z^{\prime }\right) \right\rangle $ by its space average, so that:%
\begin{equation*}
K\left( \left\{ \left\langle T\left( Z,Z^{\prime }\right) \right\rangle
\right\} _{\left( Z,Z^{\prime }\right) }\right) \simeq K\left( \left\langle
T\right\rangle \right)
\end{equation*}%
where:%
\begin{equation*}
\left\langle T\right\rangle \simeq \frac{\lambda \tau }{V}\int \frac{\left(
\left\langle h\left( Z,Z^{\prime }\right) C_{Z,Z^{\prime }}\left( \theta
\right) h_{C}\left\vert \Psi \left( \theta ,Z\right) \right\vert
^{2}\right\rangle -\eta H\left( \delta -\left\langle T\left( Z,Z^{\prime
}\right) \right\rangle \right) \right) }{\left\langle C_{Z,Z^{\prime
}}\left( \theta \right) \left\vert \Psi \left( \theta ,Z\right) \right\vert
^{2}h_{C}\right\rangle +\left\langle D_{Z,Z^{\prime }}\left( \theta \right)
\left\vert \Psi \left( \theta -\frac{\left\vert Z-Z^{\prime }\right\vert }{c}%
,Z^{\prime }\right) \right\vert ^{2}h_{D}\right\rangle }d\left( Z,Z^{\prime
}\right)
\end{equation*}%
and $V$ is the thread's volume.

The correction terms proportional to $K$ in (\ref{SNC}) and (\ref{STC})
modify the condition for minima (\ref{SLt}) that writes:

\begin{eqnarray}
2 &=&\frac{1}{2\tau \omega \left( Z\right) }+a_{C}\left( Z\right)
+a_{D}\left( Z\right) +a_{K}\left( Z\right) \\
&&+\frac{\rho \left( C\left( \theta \right) \left\vert \Psi \left( \theta
,Z\right) \right\vert ^{2}h_{C}+D\left( \theta \right) \left\vert \Psi
\left( \theta -\frac{\left\vert Z-Z^{\prime }\right\vert }{c},Z^{\prime
}\right) \right\vert ^{2}h_{D}\right) }{2\omega \left( \theta ,Z,\left\vert
\Psi \right\vert ^{2}\right) }-\frac{\delta U\left( \left\{ \left\Vert
\Gamma _{0}\left( \theta ,Z,Z^{\prime }\right) \right\Vert ^{2}\right\}
\right) }{\delta \left\Vert \Gamma _{0}\left( \theta ,Z,Z^{\prime }\right)
\right\Vert ^{2}}  \notag
\end{eqnarray}%
with:%
\begin{eqnarray*}
a_{K}\left( Z\right) &=&\frac{1}{2}K^{2}\sigma _{T}^{2}\left( \frac{\tau
\omega }{\left\vert \Psi \left( \theta ,Z\right) \right\vert ^{2}}\right)
^{2} \\
&&+\frac{1}{2}\sigma _{\hat{T}}^{2}\left( \frac{\omega \left( \theta
,Z,\left\vert \Psi \right\vert ^{2}\right) }{\rho \left( C\left( \theta
\right) \left\vert \Psi \left( \theta ,Z\right) \right\vert
^{2}h_{C}+D\left( \theta \right) \left\vert \Psi \left( \theta -\frac{%
\left\vert Z-Z^{\prime }\right\vert }{c},Z^{\prime }\right) \right\vert
^{2}h_{D}\right) }\right) ^{2}\left( K\lambda \tau \right) ^{2}+K\lambda
\tau \left\langle \hat{T}\right\rangle _{0}
\end{eqnarray*}%
and this set of equations shifts the norm $\left\Vert \Gamma _{0}\left(
\theta ,Z,Z^{\prime }\right) \right\Vert ^{2}$ at each point.

As a consequence, the condition (\ref{CTF}) for non trivial background state
at $\left( Z,Z^{\prime }\right) $:%
\begin{equation}
S\left( \left\Vert \Gamma _{0}\left( Z,Z^{\prime }\right) \right\Vert
^{2}\right) =U\left( \left\{ \left\Vert \Gamma _{0}\left( Z,Z^{\prime
}\right) \right\Vert ^{2}\right\} \right) -\frac{\delta U\left( \left\{
\left\Vert \Gamma _{0}\left( \theta ,Z,Z^{\prime }\right) \right\Vert
^{2}\right\} \right) }{\delta \left\Vert \Gamma _{0}\left( \theta
,Z,Z^{\prime }\right) \right\Vert ^{2}}\left\Vert \Gamma _{0}\left(
Z,Z^{\prime }\right) \right\Vert ^{2}<0
\end{equation}%
is modified by the shift of the norm. The correction due to the backreaction
modifies the set of points with non trivial background.

The expression for $K\left( \theta ,Z,Z^{\prime },\left\Vert \Psi
\right\Vert ^{2},\left\Vert \Gamma \right\Vert ^{2}\right) $ can be obtained
by using (\ref{KQT}):%
\begin{eqnarray}
&&K\left( \theta ,Z,Z^{\prime },\left\Vert \Psi \right\Vert ^{2},\left\Vert
\Gamma \right\Vert ^{2}\right)  \label{CFK} \\
&=&\int \Gamma ^{\dagger }\left( T_{1},\hat{T}_{1},\theta
_{1},Z_{1},Z_{1}^{\prime },C_{1},D_{1}\right) \frac{\delta W\left( T_{1},%
\hat{T}_{1},\theta _{1},Z_{1},Z_{1}^{\prime },C_{1},D_{1}\right) }{\delta
T\left\vert \Gamma \left( T,\hat{T},\theta ,Z,Z^{\prime },C,D\right)
\right\vert ^{2}}  \notag \\
&&\times \Gamma \left( T_{1},\hat{T}_{1},\theta _{1},Z_{1},Z_{1}^{\prime
},C_{1},D_{1}\right) d\left( T_{1},\hat{T}_{1},\theta
_{1},Z_{1},Z_{1}^{\prime },C_{1},D_{1}\right)  \notag
\end{eqnarray}%
where:

\begin{eqnarray*}
&&W\left( T,\hat{T},\theta ,Z,Z^{\prime },C,D\right) \\
&=&\nabla _{C}\left( \frac{C}{\tau _{C}\omega \left( J,\theta ,Z,\left\vert
\Psi \right\vert ^{2}\right) }-\frac{\alpha _{C}\left( 1-C\right) \omega
\left( J,\theta -\frac{\left\vert Z-Z^{\prime }\right\vert }{c},Z^{\prime
},\left\vert \Psi \right\vert ^{2}\right) \left\vert \Psi \left( \theta -%
\frac{\left\vert Z-Z^{\prime }\right\vert }{c},Z^{\prime },\omega ^{\prime
}\right) \right\vert ^{2}}{\omega \left( J,\theta ,Z,\left\vert \Psi
\right\vert ^{2}\right) }\right) \left\vert \Psi \left( \theta ,Z\right)
\right\vert ^{2} \\
&&+\nabla _{D}\left( \frac{D}{\tau _{D}\omega \left( J,\theta ,Z,\left\vert
\Psi \right\vert ^{2}\right) }-\alpha _{D}\left( 1-D\right) \left\vert \Psi
\left( \theta ,Z\right) \right\vert ^{2}\right) \\
&&-\nabla _{\hat{T}}\frac{\rho \left( \left( h\left( Z,Z^{\prime }\right) -%
\hat{T}\right) C\left\vert \Psi \left( \theta ,Z\right) \right\vert
^{2}h_{C}-D\hat{T}\left\vert \Psi \left( \theta -\frac{\left\vert
Z-Z^{\prime }\right\vert }{c},Z^{\prime }\right) \right\vert
^{2}h_{D}\right) }{\omega \left( J,\theta ,Z,\left\vert \Psi \right\vert
^{2}\right) } \\
&&-\nabla _{T}\left( -\frac{1}{\tau \omega }T+\frac{\lambda }{\omega }\hat{T}%
\right) \left\vert \Psi \left( \theta ,Z,\omega \right) \right\vert ^{2}
\end{eqnarray*}%
\bigskip Given that $\tau _{C}>>1$ and $\tau _{D}>>1$ and that $C$ and $D$
are close to $1$, the two first contributions are negligibles. Given that $%
\left\vert \Psi \left( \theta ,Z,\omega \right) \right\vert ^{2}$ depends on
the potential $V\left( \left\vert \Psi \left( \theta ,Z,\omega \right)
\right\vert ^{2}\right) $, we can assume that:%
\begin{equation*}
\left\vert \frac{\delta \left\vert \Psi \left( \theta ,Z,\omega \right)
\right\vert ^{2}}{\delta \left\vert \Gamma \left( T,\hat{T},\theta
,Z,Z^{\prime },C,D\right) \right\vert ^{2}}\right\vert <<\left\vert \frac{%
\delta \omega \left( J,\theta ,Z,\left\vert \Psi \right\vert ^{2}\right) }{%
\delta \left\vert \Gamma \left( T,\hat{T},\theta ,Z,Z^{\prime },C,D\right)
\right\vert ^{2}}\right\vert
\end{equation*}%
the last contribution can be neglected, and we can consider that $K\left(
\theta ,Z,Z^{\prime },\left\Vert \Psi \right\Vert ^{2},\left\Vert \Gamma
\right\Vert ^{2}\right) $ can be computed with:%
\begin{eqnarray}
&&W\left( T,\hat{T},\theta ,Z,Z^{\prime },C,D\right)  \label{RCW} \\
&=&-\nabla _{\hat{T}}\frac{\rho \left( \left( h\left( Z,Z^{\prime }\right) -%
\hat{T}\right) C\left\vert \Psi \left( \theta ,Z\right) \right\vert
^{2}h_{C}-D\hat{T}\left\vert \Psi \left( \theta -\frac{\left\vert
Z-Z^{\prime }\right\vert }{c},Z^{\prime }\right) \right\vert
^{2}h_{D}\right) }{\omega \left( J,\theta ,Z,\left\vert \Psi \right\vert
^{2}\right) }  \notag
\end{eqnarray}%
In the computation of (\ref{CFK}), the contribution proportional to:%
\begin{equation*}
\int \Gamma ^{\dagger }\left( T_{1},\hat{T}_{1},\theta
_{1},Z_{1},Z_{1}^{\prime },C_{1},D_{1}\right) \nabla _{\hat{T}}\Gamma \left(
T_{1},\hat{T}_{1},\theta _{1},Z_{1},Z_{1}^{\prime },C_{1},D_{1}\right)
d\left( T_{1},\hat{T}_{1},\theta _{1},Z_{1},Z_{1}^{\prime
},C_{1},D_{1}\right)
\end{equation*}%
is equal to $0$, since $\Gamma $ is gaussian. As a consequence, the gradient 
$\nabla _{\hat{T}}$ in (\ref{RCW}) acts on the function at its right in (\ref%
{RCW}).

We thus replace:%
\begin{eqnarray}
&&W\left( T,\hat{T},\theta ,Z,Z^{\prime },C,D\right) \rightarrow \\
&=&\frac{\rho \left( C\left\vert \Psi \left( \theta ,Z\right) \right\vert
^{2}h_{C}+D\left\vert \Psi \left( \theta -\frac{\left\vert Z-Z^{\prime
}\right\vert }{c},Z^{\prime }\right) \right\vert ^{2}h_{D}\right) }{\omega
\left( J,\theta ,Z,\left\vert \Psi \right\vert ^{2}\right) }  \notag \\
&\simeq &\rho C\left\vert \Psi \left( \theta ,Z\right) \right\vert ^{2}+%
\frac{\rho D\left\vert \Psi \left( \theta -\frac{\left\vert Z-Z^{\prime
}\right\vert }{c},Z^{\prime }\right) \right\vert ^{2}\omega \left( J,\theta -%
\frac{\left\vert Z-Z^{\prime }\right\vert }{c},Z^{\prime },\left\vert \Psi
\right\vert ^{2}\right) }{\omega \left( J,\theta ,Z,\left\vert \Psi
\right\vert ^{2}\right) }  \notag
\end{eqnarray}%
for $h_{C}\left( \omega \left( J,\theta ,Z,\left\vert \Psi \right\vert
^{2}\right) \right) \simeq \omega \left( J,\theta ,Z,\left\vert \Psi
\right\vert ^{2}\right) $ and $h_{D}\left( \omega \left( J,\theta -\frac{%
\left\vert Z-Z^{\prime }\right\vert }{c},Z^{\prime },\left\vert \Psi
\right\vert ^{2}\right) \right) \simeq \left( \omega \left( J,\theta -\frac{%
\left\vert Z-Z^{\prime }\right\vert }{c},Z^{\prime },\left\vert \Psi
\right\vert ^{2}\right) \right) $.

Given our approximations, we then deduce that (\ref{CFK}) reduces to:%
\begin{eqnarray}
&&K\left( \theta ,Z,Z^{\prime },\left\Vert \Psi \right\Vert ^{2},\left\Vert
\Gamma \right\Vert ^{2}\right) \\
&=&\int \rho D\left\vert \Psi \left( \theta -\frac{\left\vert
Z_{1}-Z_{1}^{\prime }\right\vert }{c},Z_{1}^{\prime }\right) \right\vert ^{2}%
\frac{\delta \frac{\omega \left( J,\theta -\frac{\left\vert
Z_{1}-Z_{1}^{\prime }\right\vert }{c},Z_{1}^{\prime },\left\vert \Psi
\right\vert ^{2}\right) }{\omega \left( J,\theta ,Z_{1},\left\vert \Psi
\right\vert ^{2}\right) }}{\delta T\left\vert \Gamma \left( T,\hat{T},\theta
,Z,Z^{\prime },C,D\right) \right\vert ^{2}}  \notag \\
&&\times \left\vert \Gamma \left( T_{1},\hat{T}_{1},\theta
_{1},Z_{1},Z_{1}^{\prime },C_{1},D_{1}\right) \right\vert ^{2}d\left( T_{1},%
\hat{T}_{1},\theta _{1},Z_{1},Z_{1}^{\prime },C_{1},D_{1}\right)  \notag
\end{eqnarray}%
where we approximate:%
\begin{equation*}
\delta \frac{\omega \left( J,\theta -\frac{\left\vert Z_{1}-Z_{1}^{\prime
}\right\vert }{c},Z_{1}^{\prime },\left\vert \Psi \right\vert ^{2}\right) }{%
\omega \left( J,\theta ,Z_{1},\left\vert \Psi \right\vert ^{2}\right) }%
\simeq \delta \frac{\left\vert Z_{1}-Z_{1}^{\prime }\right\vert ^{2}\nabla
_{Z_{1}}^{2}\omega \left( J,\theta ,Z_{1},\left\vert \Psi \right\vert
^{2}\right) }{\omega \left( J,\theta ,Z_{1},\left\vert \Psi \right\vert
^{2}\right) }
\end{equation*}%
The last expression being obtained in the limit of slowly varying
activities. In the same approximation:%
\begin{eqnarray*}
\frac{\delta \frac{\left\vert Z_{1}-Z_{1}^{\prime }\right\vert ^{2}\nabla
_{Z_{1}}^{2}\omega \left( J,\theta ,Z_{1},\left\vert \Psi \right\vert
^{2}\right) }{\omega \left( J,\theta ,Z_{1},\left\vert \Psi \right\vert
^{2}\right) }}{\delta T\left\vert \Gamma \left( T,\hat{T},\theta
,Z,Z^{\prime },C,D\right) \right\vert ^{2}} &=&\frac{\delta \left(
\left\vert Z_{1}-Z_{1}^{\prime }\right\vert ^{2}\left( \nabla
_{Z_{1}}^{2}\ln \left( \omega \left( J,\theta ,Z_{1},\left\vert \Psi
\right\vert ^{2}\right) \right) -\left( \nabla _{Z_{1}}\ln \left( \omega
\left( J,\theta ,Z_{1},\left\vert \Psi \right\vert ^{2}\right) \right)
\right) ^{2}\right) \right) }{\delta T\left\vert \Gamma \left( T,\hat{T}%
,\theta ,Z,Z^{\prime },C,D\right) \right\vert ^{2}} \\
&\simeq &\left\vert Z_{1}-Z_{1}^{\prime }\right\vert ^{2}\nabla _{Z_{1}}^{2}%
\frac{\delta \left( \ln \left( \omega \left( J,\theta ,Z_{1},\left\vert \Psi
\right\vert ^{2}\right) \right) \right) }{\delta T\left\vert \Gamma \left( T,%
\hat{T},\theta ,Z,Z^{\prime },C,D\right) \right\vert ^{2}}
\end{eqnarray*}

We show in appendix 5 that the derivatives:%
\begin{equation*}
\frac{\delta \omega \left( J,\theta ,Z_{1},\left\vert \Psi \right\vert
^{2}\right) }{\delta T\left\vert \Gamma \left( T,\hat{T},\theta ,Z,Z^{\prime
},C,D\right) \right\vert ^{2}}
\end{equation*}%
are proportional to some exponential:%
\begin{equation*}
\exp \left( -a\left\vert Z_{1}-Z\right\vert \right)
\end{equation*}%
so that, by averaging over the entire space, we find:%
\begin{eqnarray}
&&K\left( \theta ,Z,Z^{\prime },\left\Vert \Psi \right\Vert ^{2},\left\Vert
\Gamma \right\Vert ^{2}\right)  \label{KNM} \\
&\simeq &\frac{1}{a}\rho D\left\vert \Psi \left( \theta -\frac{\left\vert
Z-Z^{\prime }\right\vert }{c},Z^{\prime }\right) \right\vert ^{2}\left\vert
Z-Z^{\prime }\right\vert ^{2}\nabla _{Z}^{2}\frac{\delta \left( \ln \left(
\omega \left( J,\theta ,Z,\left\vert \Psi \right\vert ^{2}\right) \right)
\right) }{\delta T\left\vert \Gamma \left( T,\hat{T},\theta ,Z,Z^{\prime
},C,D\right) \right\vert ^{2}}\left\Vert \Gamma \right\Vert ^{2}  \notag
\end{eqnarray}%
Given (\ref{nqf}):%
\begin{equation}
\omega ^{-1}\left( J,\theta ,Z,\left\vert \Psi \right\vert ^{2}\right)
=G\left( J\left( \theta ,Z\right) +\int \frac{\kappa }{N}\frac{\omega \left(
J,\theta -\frac{\left\vert Z-Z_{1}\right\vert }{c},Z_{1},\Psi \right)
T\left( Z,\theta ,Z_{1},\theta -\frac{\left\vert Z-Z_{1}\right\vert }{c}%
\right) }{\omega \left( J,\theta ,Z,\left\vert \Psi \right\vert ^{2}\right) }%
\left( \mathcal{G}_{0}+\left\vert \Psi \left( \theta -\frac{\left\vert
Z-Z_{1}\right\vert }{c},Z_{1}\right) \right\vert ^{2}\right) dZ_{1}\right)
\label{NQ!}
\end{equation}%
and using:%
\begin{eqnarray*}
&&\frac{\delta \int \frac{\kappa }{N}\frac{\omega \left( J,\theta -\frac{%
\left\vert Z-Z_{1}\right\vert }{c},Z_{1},\Psi \right) T\left( Z,\theta
,Z_{1},\theta -\frac{\left\vert Z-Z_{1}\right\vert }{c}\right) }{\omega
\left( J,\theta ,Z,\left\vert \Psi \right\vert ^{2}\right) }}{\delta
T\left\vert \Gamma \left( T,\hat{T},\theta ,Z,Z^{\prime },C,D\right)
\right\vert ^{2}} \\
&\simeq &\frac{\kappa }{N}\frac{\omega \left( J,\theta -\frac{\left\vert
Z-Z^{\prime }\right\vert }{c},Z^{\prime },\Psi \right) }{\omega \left(
J,\theta ,Z,\left\vert \Psi \right\vert ^{2}\right) }+\frac{\kappa }{N}\int 
\frac{\delta \frac{\omega \left( J,\theta -\frac{\left\vert
Z-Z_{1}\right\vert }{c},Z_{1},\Psi \right) }{\omega \left( J,\theta
,Z,\left\vert \Psi \right\vert ^{2}\right) }}{\delta T\left\vert \Gamma
\left( T,\hat{T},\theta ,Z,Z^{\prime },C,D\right) \right\vert ^{2}}T\left(
Z,\theta ,Z_{1},\theta -\frac{\left\vert Z-Z_{1}\right\vert }{c}\right) \\
&\simeq &\frac{\kappa }{N}\frac{\omega \left( J,\theta -\frac{\left\vert
Z-Z^{\prime }\right\vert }{c},Z^{\prime },\Psi \right) }{\omega \left(
J,\theta ,Z,\left\vert \Psi \right\vert ^{2}\right) }+\frac{1}{a}\frac{%
\delta \frac{\kappa }{N}\omega \left( J,\theta -\frac{\left\vert Z-Z^{\prime
}\right\vert }{c},Z^{\prime },\Psi \right) }{\delta T\left\vert \Gamma
\left( T,\hat{T},\theta ,Z,Z^{\prime },C,D\right) \right\vert ^{2}}T\left(
Z,\theta ,Z^{\prime },\theta -\frac{\left\vert Z-Z^{\prime }\right\vert }{c}%
\right) \\
&\simeq &\frac{\kappa }{N}\frac{\omega \left( J,\theta -\frac{\left\vert
Z-Z^{\prime }\right\vert }{c},Z^{\prime },\Psi \right) }{\omega \left(
J,\theta ,Z,\left\vert \Psi \right\vert ^{2}\right) }
\end{eqnarray*}%
the differentiation of (\ref{NQ!}) is:%
\begin{eqnarray*}
&&\frac{\delta \omega ^{-1}\left( J,\theta ,Z,\left\vert \Psi \right\vert
^{2}\right) }{\delta T\left\vert \Gamma \left( T,\hat{T},\theta ,Z,Z^{\prime
},C,D\right) \right\vert ^{2}} \\
&=&\frac{\kappa }{N}\frac{\omega \left( J,\theta -\frac{\left\vert
Z-Z_{1}\right\vert }{c},Z_{1},\Psi \right) }{\omega \left( J,\theta
,Z,\left\vert \Psi \right\vert ^{2}\right) }\left( \mathcal{G}%
_{0}+\left\vert \Psi \left( \theta -\frac{\left\vert Z-Z^{\prime
}\right\vert }{c},Z^{\prime }\right) \right\vert ^{2}\right) \\
&&G^{\prime }\left( J\left( \theta ,Z\right) +\int \frac{\kappa }{N}\frac{%
\omega \left( J,\theta -\frac{\left\vert Z-Z_{1}\right\vert }{c},Z_{1},\Psi
\right) T\left( Z,\theta ,Z_{1},\theta -\frac{\left\vert Z-Z_{1}\right\vert 
}{c}\right) }{\omega \left( J,\theta ,Z,\left\vert \Psi \right\vert
^{2}\right) }\left( \mathcal{G}_{0}+\left\vert \Psi \left( \theta -\frac{%
\left\vert Z-Z_{1}\right\vert }{c},Z_{1}\right) \right\vert ^{2}\right)
dZ_{1}\right)
\end{eqnarray*}

so that:%
\begin{eqnarray*}
\frac{\delta \omega \left( J,\theta ,Z,\left\vert \Psi \right\vert
^{2}\right) }{\delta T\left\vert \Gamma \left( T,\hat{T},\theta ,Z,Z^{\prime
},C,D\right) \right\vert ^{2}} &=&\frac{\kappa }{N}\omega \left( J,\theta -%
\frac{\left\vert Z-Z_{1}\right\vert }{c},Z_{1},\Psi \right) \omega \left(
J,\theta ,Z,\left\vert \Psi \right\vert ^{2}\right) \left( \mathcal{G}%
_{0}+\left\vert \Psi \left( \theta -\frac{\left\vert Z-Z^{\prime
}\right\vert }{c},Z^{\prime }\right) \right\vert ^{2}\right) \\
&&G^{\prime }\left( G^{-1}\left( \omega \left( J,\theta ,Z,\left\vert \Psi
\right\vert ^{2}\right) \right) \right) \\
&\simeq &\frac{\kappa }{N}\omega ^{2}\left( J,\theta ,Z,\left\vert \Psi
\right\vert ^{2}\right) G^{\prime }\left( G^{-1}\left( \omega \left(
J,\theta ,Z,\left\vert \Psi \right\vert ^{2}\right) \right) \right) \left( 
\mathcal{G}_{0}+\left\vert \Psi \left( \theta -\frac{\left\vert Z-Z^{\prime
}\right\vert }{c},Z^{\prime }\right) \right\vert ^{2}\right)
\end{eqnarray*}%
and (\ref{KNM}) writes:%
\begin{eqnarray}
&&K\left( \theta ,Z,Z^{\prime },\left\Vert \Psi \right\Vert ^{2},\left\Vert
\Gamma \right\Vert ^{2}\right) \\
&\simeq &\frac{1}{a}\rho D\frac{\kappa }{N}\left\Vert \Gamma \right\Vert
^{2}\left\vert \Psi \left( \theta -\frac{\left\vert Z-Z^{\prime }\right\vert 
}{c},Z^{\prime }\right) \right\vert ^{2}\left\vert Z-Z^{\prime }\right\vert
^{2}  \notag \\
&&\times \nabla _{Z}^{2}\left( \omega \left( J,\theta ,Z,\left\vert \Psi
\right\vert ^{2}\right) G^{\prime }\left( G^{-1}\left( \omega \left(
J,\theta ,Z,\left\vert \Psi \right\vert ^{2}\right) \right) \right) \left( 
\mathcal{G}_{0}+\left\vert \Psi \left( \theta -\frac{\left\vert Z-Z^{\prime
}\right\vert }{c},Z^{\prime }\right) \right\vert ^{2}\right) \right)  \notag
\end{eqnarray}

\subsection*{Resolution of saddle point equations without approximation}

To study the background state for $T$ and $\hat{T}$, we start from the
effective action (\ref{Tcf}) 
\begin{eqnarray}
&&\Gamma ^{\dag }\left( T,\hat{T},\theta ,Z,Z^{\prime }\right) \left( \nabla
_{T}\left( \nabla _{T}-\left( -\frac{1}{\tau \omega }T+\frac{\lambda }{%
\omega }\hat{T}\right) \left\vert \Psi \left( \theta ,Z\right) \right\vert
^{2}\right) \right) \Gamma \left( T,\hat{T},\theta ,Z,Z^{\prime }\right) \\
&&+\Gamma ^{\dag }\left( T,\hat{T},\theta ,Z,Z^{\prime }\right) \left(
\nabla _{\hat{T}}\left( \nabla _{\hat{T}}-\frac{\rho }{\omega \left(
J,\theta ,Z,\left\vert \Psi \right\vert ^{2}\right) }\right. \right.  \notag
\\
&&\times \left. \left. \left( \left( h\left( Z,Z^{\prime }\right) -\hat{T}%
\right) C\left( \theta \right) \left\vert \Psi \left( \theta ,Z\right)
\right\vert ^{2}h_{C}-\eta H\left( \delta -T\right) -D\left( \theta \right) 
\hat{T}\left\vert \Psi \left( \theta -\frac{\left\vert Z-Z^{\prime
}\right\vert }{c},Z^{\prime }\right) \right\vert ^{2}h_{D}\right) \right)
\right) \Gamma \left( T,\hat{T},\theta ,Z,Z^{\prime }\right)  \notag
\end{eqnarray}

As in the text, we consider points such that:%
\begin{equation*}
h\left( Z,Z^{\prime }\right) \left\langle C_{Z,Z^{\prime }}\left( \theta
\right) h_{C}\left( \omega \left( \theta ,Z\right) \right) \left\vert \Psi
\left( \theta ,Z\right) \right\vert ^{2}\right\rangle -\eta >0
\end{equation*}%
for which the averages are:%
\begin{equation}
\left\langle T\left( Z,Z^{\prime }\right) \right\rangle =\lambda \tau
\left\langle \hat{T}\left( Z,Z^{\prime }\right) \right\rangle =\frac{\lambda
\tau h\left( Z,Z^{\prime }\right) \left\langle C_{Z,Z^{\prime }}\left(
\theta \right) h_{C}\left\vert \Psi \left( \theta ,Z\right) \right\vert
^{2}\right\rangle }{C_{Z,Z^{\prime }}\left( \theta \right) \left\vert \Psi
\left( \theta ,Z\right) \right\vert ^{2}h_{C}+D_{Z,Z^{\prime }}\left( \theta
\right) \left\vert \Psi \left( \theta -\frac{\left\vert Z-Z^{\prime
}\right\vert }{c},Z^{\prime }\right) \right\vert ^{2}h_{D}}
\end{equation}%
which allows to rewrite the effective action as:%
\begin{equation}
\Gamma ^{\dag }\left( T,\hat{T},\theta ,Z,Z^{\prime }\right) \left( \nabla
_{T}\left( \nabla _{T}+u\left( T-\left\langle T\right\rangle \right)
+s\left( \hat{T}-\left\langle \hat{T}\right\rangle \right) \right) +\nabla _{%
\hat{T}}\left( \nabla _{\hat{T}}+v\left( \hat{T}-\left\langle \hat{T}%
\right\rangle \right) \right) \right) \Gamma \left( T,\hat{T},\theta
,Z,Z^{\prime }\right)
\end{equation}%
with:%
\begin{eqnarray*}
u &=&\frac{\left\vert \Psi _{0}\left( Z\right) \right\vert ^{2}}{\tau \omega
_{0}\left( Z\right) } \\
v &=&\rho C\frac{\left\vert \Psi _{0}\left( Z\right) \right\vert
^{2}h_{C}\left( \omega _{0}\left( Z\right) \right) }{\omega _{0}\left(
Z\right) }+\rho D\frac{\left\vert \Psi _{0}\left( Z^{\prime }\right)
\right\vert ^{2}h_{D}\left( \omega _{0}\left( Z^{\prime }\right) \right) }{%
\omega _{0}\left( Z\right) } \\
s &=&-\frac{\lambda \left\vert \Psi _{0}\left( Z\right) \right\vert ^{2}}{%
\omega _{0}\left( Z\right) }
\end{eqnarray*}%
Imposing the constraint that the norm of $\Gamma \left( T,\hat{T},\theta
,Z,Z^{\prime }\right) $ is a given number determined by some average number
of connections all over the thread, the saddle point equations becomes:%
\begin{equation*}
\left( \nabla _{T}\left( \nabla _{T}+u\left( T-\left\langle T\right\rangle
\right) +s\left( \hat{T}-\left\langle \hat{T}\right\rangle \right) \right)
+\nabla _{\hat{T}}\left( \nabla _{\hat{T}}+v\left( \hat{T}-\left\langle \hat{%
T}\right\rangle \right) \right) +\alpha \right) \Gamma \left( T,\hat{T}%
,\theta ,Z,Z^{\prime }\right) =0
\end{equation*}%
with $\alpha $ the Lagrange multiplier.This also rewrite matricially:%
\begin{equation}
\left( \mathbf{\nabla }^{2}+\left( \mathbf{\nabla }\right) ^{t}\gamma 
\mathbf{x}+\alpha \right) \Gamma \left( T,\hat{T},\theta ,Z,Z^{\prime
}\right) =0  \label{ctm}
\end{equation}%
with:%
\begin{equation*}
\gamma =\left( 
\begin{array}{cc}
u & s \\ 
0 & v%
\end{array}%
\right)
\end{equation*}

\subsection*{Solution for Fourier transform}

This is solved by considering the Fourier transform of this equation:%
\begin{equation}
\left( -\mathbf{k}^{2}-\left( \mathbf{k}\right) ^{t}\gamma \mathbf{\nabla }_{%
\mathbf{k}}+\alpha \right) \Gamma \left( \mathbf{k},\theta ,Z,Z^{\prime
}\right) =0
\end{equation}%
We write the solution:%
\begin{equation*}
\Gamma \left( \mathbf{k},\theta ,Z,Z^{\prime }\right) =\exp \left( -\frac{1}{%
2}\mathbf{k}^{t}N\mathbf{k}\right) \hat{\Gamma}\left( \mathbf{k},\theta
,Z,Z^{\prime }\right)
\end{equation*}%
where the matrix $N$ satisifies:%
\begin{equation*}
-\mathbf{k}^{2}+\left( \mathbf{k}\right) ^{t}\gamma N\mathbf{k}=0
\end{equation*}%
and:%
\begin{equation*}
\left( -\left( \mathbf{k}\right) ^{t}\gamma \mathbf{\nabla }_{\mathbf{k}%
}+\alpha \right) \hat{\Gamma}\left( \mathbf{k},\theta ,Z,Z^{\prime }\right)
=0
\end{equation*}%
Equation for $N$ writes:%
\begin{equation*}
\frac{1}{2}\left( \gamma N+N\gamma ^{t}\right) =I
\end{equation*}%
where $I$ is the identity matrix. The solution is:%
\begin{equation*}
N=\left( 
\begin{array}{cc}
\frac{1}{u}\left( 1+\frac{s^{2}}{v\left( u+v\right) }\right) & -\frac{s}{%
v\left( u+v\right) } \\ 
-\frac{s}{v\left( u+v\right) } & \frac{1}{v}%
\end{array}%
\right)
\end{equation*}%
The equation for $\hat{\Gamma}\left( \mathbf{k},\theta ,Z,Z^{\prime }\right) 
$ becomes:%
\begin{equation}
\left( -\left( \mathbf{k}\right) ^{t}\gamma \mathbf{\nabla }_{\mathbf{k}%
}+\alpha \right) \hat{\Gamma}\left( \mathbf{k},\theta ,Z,Z^{\prime }\right)
=0  \label{Prl}
\end{equation}%
If we diagonalize $\gamma $:%
\begin{equation*}
\gamma =PDP^{-1}
\end{equation*}%
with:%
\begin{eqnarray*}
D &=&\left( 
\begin{array}{cc}
u & 0 \\ 
0 & v%
\end{array}%
\right) \\
P &=&\left( 
\begin{array}{cc}
1 & 1 \\ 
0 & \frac{v-u}{s}%
\end{array}%
\right)
\end{eqnarray*}%
and define:%
\begin{equation*}
\mathbf{\hat{k}}=P^{t}\mathbf{k}
\end{equation*}%
equation (\ref{Prl}) is:%
\begin{equation*}
\left( -\left( \mathbf{\hat{k}}\right) ^{t}D\mathbf{\nabla }_{\mathbf{\hat{k}%
}}+\alpha \right) \hat{\Gamma}\left( \mathbf{k},\theta ,Z,Z^{\prime }\right)
=0
\end{equation*}%
with solution:%
\begin{equation*}
\hat{\Gamma}\left( \mathbf{k},\theta ,Z,Z^{\prime }\right) =\hat{k}_{1}^{%
\frac{\alpha \delta }{u}}\hat{k}_{2}^{\frac{\left( 1-\delta \right) \alpha }{%
v}}
\end{equation*}%
where $\hat{k}_{1}$ and $\hat{k}_{2}$ are\ the component of $\mathbf{\hat{k}}
$:%
\begin{eqnarray*}
\hat{k}_{1} &=&k_{1} \\
\hat{k}_{1} &=&k_{1}+\frac{v-u}{s}k_{2}
\end{eqnarray*}%
and ultimately:%
\begin{equation*}
\Gamma _{\delta }\left( \mathbf{k},\theta ,Z,Z^{\prime }\right) =\exp \left(
-\frac{1}{2}\mathbf{k}^{t}N\mathbf{k}\right) k_{1}^{\frac{\alpha \delta }{u}%
}\left( k_{1}+\frac{v-u}{s}k_{2}\right) ^{\frac{\left( 1-\delta \right)
\alpha }{v}}
\end{equation*}%
The parameter $\delta $ implies the possibility of several local global
minima for each $\left( Z,Z^{\prime }\right) $. As explained in the text,
this parameter is not free if we aim at obtaining a minimum at all points $%
\left( Z,Z^{\prime }\right) $.

\subsubsection*{Background field}

To come back to the background field we compute the inverse Fourier
transform:%
\begin{equation*}
\Gamma _{\delta }\left( T,\hat{T},\theta ,Z,Z^{\prime }\right) =\int \exp
\left( -\frac{1}{2}\mathbf{k}^{t}N\mathbf{k}-i\mathbf{k\Delta T}\right)
k_{1}^{\frac{\alpha \delta }{u}}\left( k_{1}+\frac{v-u}{s}k_{2}\right) ^{%
\frac{\left( 1-\delta \right) \alpha }{v}}\frac{d\mathbf{k}}{2\pi }
\end{equation*}%
where:%
\begin{equation*}
\mathbf{\Delta T=}\left( 
\begin{array}{c}
T-\left\langle T\right\rangle \\ 
\hat{T}-\left\langle \hat{T}\right\rangle%
\end{array}%
\right)
\end{equation*}%
To estimate the integral, we diagonalize $N$:%
\begin{equation*}
N=\left( 
\begin{array}{cc}
\frac{1}{u}\left( 1+\frac{s^{2}}{v\left( u+v\right) }\right) & -\frac{s}{%
v\left( u+v\right) } \\ 
-\frac{s}{v\left( u+v\right) } & \frac{1}{v}%
\end{array}%
\right) =PDP^{-1}
\end{equation*}%
with:%
\begin{equation*}
D=\left( 
\begin{array}{cc}
\lambda _{+} & 0 \\ 
0 & \lambda _{-}%
\end{array}%
\right)
\end{equation*}%
and the eigenvalues:%
\begin{equation*}
\lambda _{\pm }=\frac{\frac{1}{u}\left( 1+\frac{s^{2}}{v\left( u+v\right) }%
\right) +\frac{1}{v}}{2}\pm \sqrt{\left( \frac{\frac{1}{u}\left( 1+\frac{%
s^{2}}{v\left( u+v\right) }\right) -\frac{1}{v}}{2}\right) ^{2}+\left( \frac{%
s}{v\left( u+v\right) }\right) ^{2}}
\end{equation*}%
The matrix $P$ is orthogonal:%
\begin{equation*}
P=\left( 
\begin{array}{cc}
\cos x & -\sin x \\ 
\sin x & \cos x%
\end{array}%
\right)
\end{equation*}%
\begin{equation*}
P^{-1}=P^{t}
\end{equation*}%
and $x$ satifies:%
\begin{eqnarray*}
\left( \frac{\lambda _{+}-\lambda _{-}}{2}\right) \sin 2x &=&-\frac{s}{%
v\left( u+v\right) } \\
\left( \frac{\lambda _{+}+\lambda _{-}}{2}\right) +\left( \frac{\lambda
_{+}-\lambda _{-}}{2}\right) \left( \cos 2x\right) &=&\frac{1}{u}\left( 1+%
\frac{s^{2}}{v\left( u+v\right) }\right)
\end{eqnarray*}%
with:%
\begin{equation*}
\frac{\lambda _{+}+\lambda _{-}}{2}=\frac{\frac{1}{u}\left( 1+\frac{s^{2}}{%
v\left( u+v\right) }\right) +\frac{1}{v}}{2}
\end{equation*}%
As a consequence, we obtain:%
\begin{equation*}
\tan 2x=\frac{-\frac{2s}{v\left( u+v\right) }}{\frac{\frac{1}{u}\left( 1+%
\frac{s^{2}}{v\left( u+v\right) }\right) -\frac{1}{v}}{2}}=-\frac{4su}{%
v^{2}-u^{2}+s^{2}}
\end{equation*}%
It thus implies that $\Gamma _{\delta }\left( T,\hat{T},\theta ,Z,Z^{\prime
}\right) $ is given by:

\begin{eqnarray*}
\Gamma _{\delta }\left( T,\hat{T},\theta ,Z,Z^{\prime }\right) &=&\int \exp
\left( -\frac{1}{2}\mathbf{k}^{t}D\mathbf{k}-i\mathbf{k\Delta T}^{\prime
}\right) \times \\
&&\times \left( k_{1}\cos x-k_{2}\sin x\right) ^{\frac{\alpha \delta }{u}%
}\left( k_{1}\left( \cos x+\frac{v-u}{s}\sin x\right) +\left( \frac{v-u}{s}%
\cos x-\sin x\right) k_{2}\right) ^{\frac{\left( 1-\delta \right) \alpha }{v}%
}\frac{d\mathbf{k}}{2\pi }
\end{eqnarray*}%
with:%
\begin{equation*}
\mathbf{\Delta T}^{\prime }=P^{t}\mathbf{\Delta T}
\end{equation*}%
In the approximation given in the text, we have $s<<1$ and:%
\begin{eqnarray*}
&&\Gamma _{\delta }\left( T,\hat{T},\theta ,Z,Z^{\prime }\right) \\
&\simeq &\int \exp \left( -\frac{1}{2}\mathbf{k}^{t}D\mathbf{k}-i\mathbf{%
k\Delta T}^{\prime }\right) \left( k_{1}-xk_{2}\right) ^{\frac{\alpha \delta 
}{u}}\left( \frac{v-u}{s}k_{2}+k_{1}\right) ^{\frac{\left( 1-\delta \right)
\alpha }{v}}\frac{d\mathbf{k}}{2\pi } \\
&\simeq &\int \exp \left( -\frac{1}{2}\mathbf{k}^{t}D\mathbf{k}-i\mathbf{%
k\Delta T}^{\prime }\right) \left( k_{1}\right) ^{\frac{\alpha \delta }{u}%
}\left( \frac{v-u}{s}k_{2}\right) ^{\frac{\left( 1-\delta \right) \alpha }{v}%
}\left( 1-x\frac{\alpha \delta }{u}\frac{k_{2}}{k_{1}}\right) \left( 1+\frac{%
s}{u-v}\frac{\left( 1-\delta \right) \alpha }{v}\frac{k_{1}}{k_{2}}\right) 
\frac{d\mathbf{k}}{2\pi } \\
&\simeq &\left( \frac{v-u}{s}\right) ^{\frac{\left( 1-\delta \right) \alpha 
}{v}}\int \exp \left( -\frac{1}{2}\mathbf{k}^{t}D\mathbf{k}-i\mathbf{k\Delta
T}^{\prime }\right) \left( k_{1}\right) ^{\frac{\alpha \delta }{u}}\left(
k_{2}\right) ^{\frac{\left( 1-\delta \right) \alpha }{v}}\left( 1-x\frac{%
\alpha \delta }{u}\frac{k_{2}}{k_{1}}+\frac{s\left( 1-\delta \right) \alpha 
}{v\left( u-v\right) }\frac{k_{1}}{k_{2}}\right) \frac{d\mathbf{k}}{2\pi }
\end{eqnarray*}%
These integrals are sums of products of parabolic cylinder functions:%
\begin{eqnarray*}
&&\Gamma _{\delta }\left( T,\hat{T},\theta ,Z,Z^{\prime }\right) \\
&\simeq &\left( \frac{v-u}{s}\right) ^{\frac{\left( 1-\delta \right) \alpha 
}{v}}2^{\frac{1}{\alpha u}+1}\prod\limits_{i=1}^{2}\exp \left( -\left(
\left( \frac{D^{-\frac{1}{2}}P^{t}\mathbf{\Delta T}}{2}\right) _{i}\right)
^{2}\right) \\
&&\times \left\{ \prod\limits_{i=1}^{2}D_{p_{i}}\left( \left( \frac{D^{-%
\frac{1}{2}}P^{t}\mathbf{\Delta T}}{4}\right) _{i}\right) -x\alpha \frac{%
\delta \prod\limits_{i=1}^{2}D_{p_{i}^{\left( 1\right) }}\left( \left( 
\frac{D^{-\frac{1}{2}}P^{t}\mathbf{\Delta T}}{4}\right) _{i}\right) }{u}+%
\frac{s\left( 1-\delta \right) \alpha
\prod\limits_{i=1}^{2}D_{p_{i}^{\left( 2\right) }}\left( \left( \frac{D^{-%
\frac{1}{2}}P^{t}\mathbf{\Delta T}}{4}\right) _{i}\right) }{v\left(
u-v\right) }\right\}
\end{eqnarray*}%
where:%
\begin{eqnarray*}
p_{1} &=&\frac{\alpha \delta }{u},p_{2}=\frac{\left( 1-\delta \right) \alpha 
}{v} \\
p_{1}^{\left( 1\right) } &=&\frac{\alpha \delta }{u}-1,p_{2}^{\left(
1\right) }=\frac{\left( 1-\delta \right) \alpha }{v}+1 \\
p_{1}^{\left( 1\right) } &=&\frac{\alpha \delta }{u}+1,p_{2}^{\left(
1\right) }=\frac{\left( 1-\delta \right) \alpha }{v}-1
\end{eqnarray*}%
The approximation made in the text, i.e. $s<<1$ corresponds to the first
terms:%
\begin{equation*}
\Gamma _{\delta }\left( T,\hat{T},\theta ,Z,Z^{\prime }\right) \simeq 
\mathcal{N}\prod\limits_{i=1}^{2}\exp \left( -\left( \left( \frac{D^{-\frac{%
1}{2}}P^{t}\mathbf{\Delta T}}{2}\right) _{i}\right) ^{2}\right)
\prod\limits_{i=1}^{2}D_{p_{i}}\left( \left( \frac{D^{-\frac{1}{2}}P^{t}%
\mathbf{\Delta T}}{4}\right) _{i}\right)
\end{equation*}%
where $\mathcal{N}$ is a normalization factor.

\subsubsection*{Condition for minima}

The conditions for finding a minimum with $\mathcal{N}\neq 0$ is similar to
those presented in the text. Action (\ref{Tcf}) for $\eta =0$:%
\begin{eqnarray}
&&\Gamma ^{\dag }\left( T,\hat{T},\theta ,Z,Z^{\prime }\right) \left( \nabla
_{T}\left( \nabla _{T}-\left( -\frac{1}{\tau \omega }T+\frac{\lambda }{%
\omega }\hat{T}\right) \left\vert \Psi \left( \theta ,Z\right) \right\vert
^{2}\right) \right) \Gamma \left( T,\hat{T},\theta ,Z,Z^{\prime }\right) \\
&&+\Gamma ^{\dag }\left( T,\hat{T},\theta ,Z,Z^{\prime }\right) \left(
\nabla _{\hat{T}}\left( \nabla _{\hat{T}}-\frac{\rho }{\omega \left(
J,\theta ,Z,\left\vert \Psi \right\vert ^{2}\right) }\right. \right.  \notag
\\
&&\times \left. \left. \left( \left( h\left( Z,Z^{\prime }\right) -\hat{T}%
\right) C\left( \theta \right) \left\vert \Psi \left( \theta ,Z\right)
\right\vert ^{2}h_{C}-D\left( \theta \right) \hat{T}\left\vert \Psi \left(
\theta -\frac{\left\vert Z-Z^{\prime }\right\vert }{c},Z^{\prime }\right)
\right\vert ^{2}h_{D}\right) \right) \right) \Gamma \left( T,\hat{T},\theta
,Z,Z^{\prime }\right)  \notag
\end{eqnarray}%
reduces, after using the saddle point equation (\ref{ctm}), to:%
\begin{equation}
\alpha \int \left\vert \Gamma \left( Z,Z^{\prime }\right) \right\vert ^{2}
\label{rfn}
\end{equation}%
where:%
\begin{equation*}
\left\vert \Gamma \left( Z,Z^{\prime }\right) \right\vert ^{2}=\int
\left\vert \Gamma \left( T,\hat{T},\theta ,Z,Z^{\prime }\right) \right\vert
^{2}d\left( T,\hat{T}\right)
\end{equation*}%
Using $p_{1}=\frac{\alpha \delta }{u},p_{2}=\frac{\left( 1-\delta \right)
\alpha }{v}$ allows to find $\alpha $ and $\delta $. \ As in the text, if we
ensures that the action has a minimum with $\left\vert \Gamma \left(
Z,Z^{\prime }\right) \right\vert ^{2}>0$ at each point $\left( Z,Z^{\prime
}\right) $ for $\hat{T}$, we find $p_{2}=\frac{1}{2}$ and thus $\delta =1-%
\frac{v}{2\alpha }$. The value of $p_{1}=\frac{\alpha -\frac{v}{2}}{u}$.

Then, if we include the potential for $\left\vert \Gamma \left( T,\hat{T}%
,\theta ,Z,Z^{\prime }\right) \right\vert ^{2}$, equation (\ref{ctm}) is
modified by shifting:%
\begin{equation*}
\alpha \rightarrow \alpha -U^{\prime }\left( \left\vert \Gamma
_{p_{1},p_{2}}\left( T,\hat{T},\theta ,Z,Z^{\prime }\right) \right\vert
^{2},Z,Z^{\prime }\right)
\end{equation*}%
so that the action (\ref{rfn}) becomes:%
\begin{equation*}
\alpha \int \left\vert \Gamma _{p_{1},p_{2}}\left( Z,Z^{\prime }\right)
\right\vert ^{2}+\hat{U}\left( \left\vert \Gamma _{p_{1},p_{2}}\left( T,\hat{%
T},\theta ,Z,Z^{\prime }\right) \right\vert ^{2},Z,Z^{\prime }\right)
\end{equation*}%
with:%
\begin{equation*}
\hat{U}\left( \left\vert \Gamma _{p_{1},p_{2}}\left( T,\hat{T},\theta
,Z,Z^{\prime }\right) \right\vert ^{2},Z,Z^{\prime }\right) -\left\vert
\Gamma _{p_{1},p_{2}}\left( T,\hat{T},\theta ,Z,Z^{\prime }\right)
\right\vert ^{2}\hat{U}^{\prime }\left( \left\vert \Gamma
_{p_{1},p_{2}}\left( T,\hat{T},\theta ,Z,Z^{\prime }\right) \right\vert
^{2},Z,Z^{\prime }\right)
\end{equation*}%
The action is minimal for:%
\begin{equation*}
\alpha +\hat{U}^{\prime }\left( \left\vert \Gamma _{p_{1},p_{2}}\left( T,%
\hat{T},\theta ,Z,Z^{\prime }\right) \right\vert ^{2}\right) =0
\end{equation*}%
and the norm of $\Gamma _{p_{1},p_{2}}\left( T,\hat{T},\theta ,Z,Z^{\prime
}\right) $ is given by:%
\begin{equation*}
\left\vert \Gamma _{p_{1},p_{2}}\left( T,\hat{T},\theta ,Z,Z^{\prime
}\right) \right\vert ^{2}=\hat{U}^{\prime }\left( -\alpha ,Z,Z^{\prime
}\right)
\end{equation*}%
If:%
\begin{equation}
\alpha \int U^{\prime }\left( -\alpha \right) +\hat{U}\left( U^{\prime
}\left( -\alpha \right) ,Z,Z^{\prime }\right) <0  \label{cdT}
\end{equation}%
there is a non trivial state at $\left( Z,Z^{\prime }\right) $. Otherwise $%
\left\vert \Gamma _{p_{1},p_{2}}\left( T,\hat{T},\theta ,Z,Z^{\prime
}\right) \right\vert ^{2}=0$. In the text we have assumed that (\ref{cdT})
is satisfied at each point. The value of $\alpha $ is determined by the
condition:%
\begin{equation*}
\int \hat{U}^{\prime }\left( -\alpha ,Z,Z^{\prime }\right) d\left(
Z,Z^{\prime }\right) =\left\Vert \Gamma _{p_{1},p_{2}}\right\Vert ^{2}
\end{equation*}%
where $\left\Vert \Gamma _{p_{1},p_{2}}\right\Vert ^{2}$ is the norm of $%
\Gamma _{p_{1},p_{2}}$.

\section*{Appendix 3 Static background state for the system}

We look for a static background state for the whole system. In the static
case, we assume that the static background field $\Psi _{0}\left( Z\right) $
is the minimum of $V\left( \Psi \right) $.

\subsection*{General equations}

An approximate static solution of (\ref{nqf}) can be found for the constant
background and a constant current, i.e. $J=\bar{J}$. We also set:%
\begin{equation*}
T\left( Z,Z_{1}\right) =\bar{T}\left( Z,\theta ,Z_{1},\theta -\frac{%
\left\vert Z-Z_{1}\right\vert }{c}\right)
\end{equation*}%
From now on, the quantity $T\left( Z,Z_{1}\right) $ refers to the average of
the connectivity function at points $\left( Z,Z_{1}\right) $, in the
background state defined above, i.e. $T\left( Z,Z_{1}\right) $ refers to $%
\left\langle T\left( Z,Z_{1}\right) \right\rangle $ defined as: 
\begin{equation*}
\left\langle T\left( Z,Z_{1}\right) \right\rangle =\int T\left\vert \Gamma
\left( T,\hat{T},\theta ,Z,Z_{1}\right) \right\vert ^{2}dT
\end{equation*}%
For points such that $T\left( Z,Z^{\prime }\right) \neq 0$, it is defined by
the set of equations (\ref{qf}) or (\ref{nqf}), (\ref{vrg}) (\ref{mnt}) if:%
\begin{equation*}
h\left( Z,Z^{\prime }\right) C_{Z,Z^{\prime }}\left( \theta \right)
h_{C}\left( \omega \left( \theta ,Z\right) \right) \left\vert \Psi \left(
\theta ,Z\right) \right\vert ^{2}>0
\end{equation*}%
\bigskip We choose $h_{C}\left( \omega \right) =\omega $ and $h_{C}\left(
\omega ^{\prime }\right) =\omega ^{\prime }$. As explained in the text, the
resolution is in three steps. We solve first for $\omega \left( Z\right) $.

\subsubsection*{Expression of $\protect\omega \left( Z\right) $}

We use (\ref{mnt}) to replace $\omega ^{\prime }\left\vert \Psi \left(
Z^{\prime }\right) \right\vert ^{2}$:

\begin{equation*}
\omega ^{\prime }\left\vert \Psi \left( Z^{\prime }\right) \right\vert ^{2}=%
\frac{\lambda \tau \exp \left( -\frac{\left\vert Z-Z^{\prime }\right\vert }{%
\nu c}\right) -T\left( Z,Z^{\prime }\right) }{T\left( Z,Z^{\prime }\right) }%
\left( \frac{1}{\alpha _{D}\tau _{D}}+\omega \left\vert \Psi \left( Z\right)
\right\vert ^{2}\right) -\frac{1}{\alpha _{C}\tau _{C}}
\end{equation*}%
This allows to rewrite (\ref{nqf}) as an equation for $\omega ^{-1}\left(
Z\right) $:%
\begin{eqnarray}
\omega \left( Z\right) &=&G\left( \int \frac{\kappa }{N}\frac{\omega \left(
J,Z_{1}\right) T\left\vert \Gamma \left( T,\hat{T},Z,Z_{1}\right)
\right\vert ^{2}}{\omega \left( Z\right) }\left( \mathcal{G}_{0}+\left\vert
\Psi \left( Z_{1}\right) \right\vert ^{2}\right) dZ_{1}\right)  \label{frnn}
\\
&\rightarrow &G\left( \int \frac{\kappa }{N}\left( \frac{\lambda \tau \exp
\left( -\frac{\left\vert Z-Z_{1}\right\vert }{\nu c}\right) -T\left(
Z,Z_{1}\right) }{T\left( Z,Z_{1}\right) }\left( \frac{1}{\alpha _{D}\tau _{D}%
}+\omega \left\vert \Psi \left( Z\right) \right\vert ^{2}\right) -\frac{1}{%
\alpha _{C}\tau _{C}}\right) \frac{T\left\vert \Gamma \left( T,\hat{T}%
,\theta ,Z,Z_{1}\right) \right\vert ^{2}}{\omega \left( J,\theta
,Z,\left\vert \Psi \right\vert ^{2}\right) }dTdZ_{1}\right)  \notag \\
&\simeq &G\left( \int \frac{\kappa }{N}\left( \left( \lambda \tau \exp
\left( -\frac{\left\vert Z-Z_{1}\right\vert }{\nu c}\right) -T\left(
Z,Z_{1}\right) \right) \left( \left( \frac{1}{\alpha _{D}\tau _{D}}-\frac{%
T\left( Z,Z_{1}\right) }{\alpha _{C}\tau _{C}}\right) \omega
^{-1}+\left\vert \Psi \left( Z\right) \right\vert ^{2}\right) \right)
dZ_{1}\right)  \notag
\end{eqnarray}

We can replace $T\left( Z,Z_{1}\right) $ in the integral by its average:%
\begin{equation*}
\frac{1}{V}T\left( Z\right) =\frac{1}{V}\int T\left( Z,Z_{1}\right) dZ_{1}
\end{equation*}%
so that:%
\begin{equation}
\omega \left( Z\right) =G\left( \frac{\kappa }{N}\left( \left( \lambda \tau
\nu c-VT\left( Z\right) \right) \left( \left( \frac{1}{\alpha _{D}\tau _{D}}-%
\frac{T\left( Z\right) }{V\alpha _{C}\tau _{C}}\right) \omega
^{-1}+\left\vert \Psi \left( Z\right) \right\vert ^{2}\right) \right) \right)
\label{BSN}
\end{equation}%
with solution defined by a function:%
\begin{equation*}
\omega \left( Z\right) =\hat{G}\left( T\left( Z\right) ,\left\vert \Psi
\left( Z\right) \right\vert ^{2}\right)
\end{equation*}

\subsubsection*{Finding $T\left( Z\right) $ and $T\left( Z,Z^{\prime
}\right) $}

In a second step, we can insert the solution for $\omega $ in the expression
for $T\left( Z,Z^{\prime }\right) $ that rewrites:%
\begin{eqnarray*}
T\left( Z,Z^{\prime }\right) &=&\frac{\lambda \tau \exp \left( -\frac{%
\left\vert Z-Z^{\prime }\right\vert }{\nu c}\right) }{1+\frac{\alpha _{D}}{%
\alpha _{C}}\frac{\frac{1}{\tau _{C}}+\alpha _{C}\omega ^{\prime }\left\vert
\Psi \left( Z^{\prime }\right) \right\vert ^{2}}{\frac{1}{\tau _{D}}+\alpha
_{D}\omega \left\vert \Psi \left( Z\right) \right\vert ^{2}}} \\
&=&\frac{\lambda \tau \exp \left( -\frac{\left\vert Z-Z^{\prime }\right\vert 
}{\nu c}\right) }{1+\frac{\alpha _{D}}{\alpha _{C}}\frac{\frac{1}{\tau _{C}}%
+\alpha _{C}\hat{G}\left( T\left( Z^{\prime }\right) ,\left\vert \Psi \left(
Z^{\prime }\right) \right\vert ^{2}\right) \left\vert \Psi \left( Z^{\prime
}\right) \right\vert ^{2}}{\frac{1}{\tau _{D}}+\alpha _{D}\hat{G}\left(
T\left( Z\right) ,\left\vert \Psi \left( Z\right) \right\vert ^{2}\right)
\left\vert \Psi \left( Z\right) \right\vert ^{2}}}
\end{eqnarray*}%
Then integrating over $Z^{\prime }$ and replacing:%
\begin{equation*}
\hat{G}\left( T\left( Z^{\prime }\right) ,\left\vert \Psi \left( Z^{\prime
}\right) \right\vert ^{2}\right) \left\vert \Psi \left( Z^{\prime }\right)
\right\vert ^{2}
\end{equation*}%
by its average over the volume $V$\ leads to:%
\begin{equation}
T\left( Z\right) =\frac{\lambda \tau \nu c}{1+\frac{\frac{1}{\tau _{C}\alpha
_{C}}+\Omega }{\frac{1}{\tau _{D}\alpha _{D}}+\hat{G}\left( T\left( Z\right)
,\left\vert \Psi \left( Z\right) \right\vert ^{2}\right) \left\vert \Psi
\left( Z\right) \right\vert ^{2}}}=\frac{\lambda \tau \nu c\left( \frac{1}{%
\tau _{D}\alpha _{D}}+\hat{G}\left( T\left( Z\right) ,\left\vert \Psi \left(
Z\right) \right\vert ^{2}\right) \left\vert \Psi \left( Z\right) \right\vert
^{2}\right) }{\frac{1}{\tau _{D}\alpha _{D}}+\frac{1}{\tau _{C}\alpha _{C}}%
+\Omega +\hat{G}\left( T\left( Z\right) ,\left\vert \Psi \left( Z\right)
\right\vert ^{2}\right) \left\vert \Psi \left( Z\right) \right\vert ^{2}}
\label{Tr}
\end{equation}%
where:%
\begin{equation}
\Omega =\frac{1}{V}\int \hat{G}\left( T\left( Z^{\prime }\right) ,\left\vert
\Psi \left( Z^{\prime }\right) \right\vert ^{2}\right) \left\vert \Psi
\left( Z^{\prime }\right) \right\vert ^{2}dZ^{\prime }  \label{MG}
\end{equation}%
and this leads to the following formula for $T\left( Z,Z^{\prime }\right) $: 
\begin{equation}
T\left( Z,Z^{\prime }\right) =\frac{\lambda \tau \exp \left( -\frac{%
\left\vert Z-Z^{\prime }\right\vert }{\nu c}\right) }{1+\frac{\frac{1}{\tau
_{C}\alpha _{C}}+\omega ^{\prime }\left\vert \Psi \left( Z^{\prime },\omega
^{\prime }\right) \right\vert ^{2}}{\frac{1}{\tau _{D}\alpha _{D}}+\omega
\left\vert \Psi \left( Z,\omega \right) \right\vert ^{2}}}=\frac{\lambda
\tau \exp \left( -\frac{\left\vert Z-Z^{\prime }\right\vert }{\nu c}\right) 
}{1+\frac{\frac{1}{\tau _{C}\alpha _{C}}+\hat{G}\left( T\left( Z^{\prime
}\right) ,\left\vert \Psi \left( Z^{\prime }\right) \right\vert ^{2}\right)
\left\vert \Psi \left( Z^{\prime },\omega ^{\prime }\right) \right\vert ^{2}%
}{\frac{1}{\tau _{D}\alpha _{D}}+\hat{G}\left( T\left( Z\right) ,\left\vert
\Psi \left( Z\right) \right\vert ^{2}\right) \left\vert \Psi \left( Z,\omega
\right) \right\vert ^{2}}}  \label{Ttp}
\end{equation}

\subsubsection*{Closing the system with minimization equations}

The system is then closed by minimizing the action for the field $\Psi
\left( \theta ,Z\right) $: 
\begin{eqnarray*}
&&\int \Psi ^{\dagger }\left( \theta ,Z\right) \left( -\nabla _{\theta
}\left( \frac{\sigma _{\theta }^{2}}{2}\nabla _{\theta }-\frac{1}{\hat{G}%
\left( \left( T\left( Z,Z_{1}\right) \right) _{Z_{1}},\left\vert \Psi \left(
Z\right) \right\vert ^{2}\right) }\right) \right) \Psi \left( \theta
,Z\right) \\
&&+V\left( \left\vert \Psi \left( \theta ,Z\right) \right\vert ^{2}-\int
T\left( Z^{\prime },Z_{1}\right) \left\vert \Psi _{0}\left( Z\right)
\right\vert ^{2}dZ_{1}\right)
\end{eqnarray*}%
We have assumed that the field is constrained by a potential limiting the
activit around some average $\left\vert \Psi _{0}\left( Z\right) \right\vert
^{2}$. We choose:%
\begin{equation*}
V=\frac{1}{2}\left( \left\vert \Psi \left( Z\right) \right\vert ^{2}-\int
T\left( Z^{\prime },Z_{1}\right) \left\vert \Psi _{0}\left( Z\right)
\right\vert ^{2}dZ_{1}\right)
\end{equation*}%
Performing the changement of variable:%
\begin{equation*}
\Psi \left( \theta ,Z\right) \rightarrow \exp \left( -\int \frac{1}{G\left(
\left\vert \Psi \left( \theta ,Z\right) \right\vert ^{2}\frac{\kappa }{N}%
\left( \lambda \tau \nu c-\int T\left( Z^{\prime },Z_{1}\right)
dZ_{1}\right) \right) }d\theta \right) \Psi \left( \theta ,Z\right)
\end{equation*}%
rewrites the action as:%
\begin{eqnarray*}
&&\int \Psi ^{\dagger }\left( \theta ,Z\right) \left( -\frac{\sigma _{\theta
}^{2}}{2}\nabla _{\theta }^{2}+\left( \frac{1}{\hat{G}\left( \left( T\left(
Z,Z_{1}\right) \right) _{Z_{1}},\left\vert \Psi \left( Z\right) \right\vert
^{2}\right) }\right) ^{2}\right) \Psi \left( \theta ,Z\right) \\
&&-\frac{1}{2}\int \nabla _{\theta }\left( \frac{1}{\hat{G}\left( \left(
T\left( Z,Z_{1}\right) \right) _{Z_{1}},\left\vert \Psi \left( Z\right)
\right\vert ^{2}\right) }\right) \left\vert \Psi \left( \theta ,Z\right)
\right\vert ^{2}+V\left( \left\vert \Psi \left( \theta ,Z\right) \right\vert
^{2}-\int T\left( Z,Z_{1}\right) \left\vert \Psi _{0}\left( Z\right)
\right\vert ^{2}dZ_{1}\right)
\end{eqnarray*}%
In the perspective of a static equilibrium, we aim thus at minimizing:%
\begin{equation*}
\int \left( \frac{1}{\hat{G}\left( \left( T\left( Z,Z_{1}\right) \right)
_{Z_{1}},\left\vert \Psi \left( Z\right) \right\vert ^{2}\right) }\right)
^{2}\left\vert \Psi \left( \theta ,Z\right) \right\vert ^{2}+V\left(
\left\vert \Psi \left( \theta ,Z\right) \right\vert ^{2}-\int T\left(
Z^{\prime },Z_{1}\right) \left\vert \Psi _{0}\left( Z\right) \right\vert
^{2}dZ_{1}\right)
\end{equation*}%
with saddle point equation:%
\begin{eqnarray*}
0 &=&\left( \frac{1}{\hat{G}\left( \left( T\left( Z,Z_{1}\right) \right)
_{Z_{1}},\left\vert \Psi \left( Z\right) \right\vert ^{2}\right) }\right)
^{2}\Psi \left( \theta ,Z\right) \\
&&-2\frac{G_{\left\vert \Psi \left( \theta ,Z\right) \right\vert
^{2}}^{\prime }\left( \left\vert \Psi \left( \theta ,Z\right) \right\vert
^{2}\frac{\kappa }{N}\left( \lambda \tau \nu c-T\left( Z\right) \right)
\right) }{\hat{G}^{3}\left( \left( T\left( Z,Z_{1}\right) \right)
_{Z_{1}},\left\vert \Psi \left( Z\right) \right\vert ^{2}\right) }\left\vert
\Psi \left( \theta ,Z\right) \right\vert ^{2}\Psi \left( \theta ,Z\right) \\
&&+\left( \left\vert \Psi \left( Z\right) \right\vert ^{2}-\int T\left(
Z^{\prime },Z_{1}\right) \left\vert \Psi _{0}\left( Z\right) \right\vert
^{2}dZ_{1}\right) \Psi \left( \theta ,Z\right) \\
&\simeq &\left( \left( \frac{1}{\hat{G}\left( \left( T\left( Z,Z_{1}\right)
\right) _{Z_{1}},\left\vert \Psi \left( Z\right) \right\vert ^{2}\right) }%
\right) ^{2}+\left( \left\vert \Psi \left( Z\right) \right\vert ^{2}-\int
T\left( Z^{\prime },Z_{1}\right) \left\vert \Psi _{0}\left( Z\right)
\right\vert ^{2}dZ_{1}\right) \right) \Psi \left( \theta ,Z\right)
\end{eqnarray*}%
with solutions:%
\begin{equation*}
\Psi \left( \theta ,Z\right) =0
\end{equation*}%
or $\left\vert \Psi \left( Z\right) \right\vert ^{2}$ satisfying:%
\begin{equation*}
\left( \frac{1}{\hat{G}\left( \left( T\left( Z,Z_{1}\right) \right)
_{Z_{1}},\left\vert \Psi \left( Z\right) \right\vert ^{2}\right) }\right)
^{2}+\left\vert \Psi \left( Z\right) \right\vert ^{2}\simeq \int T\left(
Z,Z^{\prime }\right) \left\vert \Psi _{0}\left( Z^{\prime }\right)
\right\vert ^{2}k\left( Z,Z^{\prime }\right) dZ^{\prime }
\end{equation*}%
This equation can be approximated by:%
\begin{eqnarray}
\left( \frac{1}{\hat{G}\left( \left( T\left( Z,Z_{1}\right) \right)
_{Z_{1}},\left\vert \Psi \left( Z\right) \right\vert ^{2}\right) }\right)
^{2}+\left\vert \Psi \left( Z\right) \right\vert ^{2} &\simeq &T\left(
Z\right) \frac{\int \left\vert \Psi _{0}\left( Z^{\prime }\right)
\right\vert ^{2}k\left( Z,Z^{\prime }\right) dZ^{\prime }}{V}  \label{Pl} \\
&\equiv &T\left( Z\right) \left\langle \left\vert \Psi _{0}\left( Z^{\prime
}\right) \right\vert ^{2}\right\rangle _{Z}  \notag
\end{eqnarray}%
As a consequence the system is reduced to two variables $T\left( Z\right) $
and $\left\vert \Psi \left( Z\right) \right\vert ^{2}$ together with (\ref%
{Tr}) and (\ref{Pl}). The average connectivity being then retrieved by (\ref%
{Ttp}).

\subsection*{Solving for $\left\vert \Psi \left( Z\right) \right\vert ^{2}$
and $T\left( Z\right) $}

To solve (\ref{Tr}) and (\ref{Pl}) for $T\left( Z\right) $ and $\left\vert
\Psi \left( Z\right) \right\vert ^{2}$, we use (\ref{Tr}):

\begin{equation*}
T\left( Z\right) =\frac{\lambda \tau \nu c\left( \frac{1}{\tau _{D}\alpha
_{D}}+\hat{G}\left( T\left( Z\right) ,\left\vert \Psi \left( Z\right)
\right\vert ^{2}\right) \left\vert \Psi \left( Z\right) \right\vert
^{2}\right) }{\frac{1}{\tau _{D}\alpha _{D}}+\frac{1}{\tau _{C}\alpha _{C}}%
+\Omega +\hat{G}\left( T\left( Z\right) ,\left\vert \Psi \left( Z\right)
\right\vert ^{2}\right) \left\vert \Psi \left( Z\right) \right\vert ^{2}}
\end{equation*}%
and express $\hat{G}\left( T\left( Z\right) ,\left\vert \Psi \left( Z\right)
\right\vert ^{2}\right) \left\vert \Psi \left( Z\right) \right\vert ^{2}$ as
a function of $T\left( Z\right) $: 
\begin{equation}
\hat{G}\left( T\left( Z\right) ,\left\vert \Psi \left( Z\right) \right\vert
^{2}\right) \left\vert \Psi \left( Z\right) \right\vert ^{2}=\frac{\left( 
\frac{1}{\tau _{D}\alpha _{D}}+\frac{1}{\tau _{C}\alpha _{C}}+\Omega \right)
T\left( Z\right) -\frac{1}{\tau _{D}\alpha _{D}}\lambda \tau \nu c}{\lambda
\tau \nu c-T\left( Z\right) }  \label{Tng}
\end{equation}%
Inserting this result in (\ref{Pl}) yields the equation for $\left\vert \Psi
\left( Z\right) \right\vert ^{2}$:

\begin{equation}
\left( \frac{\lambda \tau \nu c-T\left( Z\right) }{\left( \frac{1}{\tau
_{D}\alpha _{D}}+\frac{1}{\tau _{C}\alpha _{C}}+\Omega \right) T\left(
Z\right) -\frac{1}{\tau _{D}\alpha _{D}}\lambda \tau \nu c}\left\vert \Psi
\left( Z\right) \right\vert ^{2}\right) ^{2}+\left\vert \Psi \left( Z\right)
\right\vert ^{2}\simeq T\left( Z\right) \left\langle \left\vert \Psi
_{0}\left( Z^{\prime }\right) \right\vert ^{2}\right\rangle _{Z}
\end{equation}%
with solution:%
\begin{equation}
\left\vert \Psi \left( Z\right) \right\vert ^{2}=\frac{2T\left( Z\right)
\left\langle \left\vert \Psi _{0}\left( Z^{\prime }\right) \right\vert
^{2}\right\rangle _{Z}}{\left( 1+\sqrt{1+4\left( \frac{\lambda \tau \nu
c-T\left( Z\right) }{\left( \frac{1}{\tau _{D}\alpha _{D}}+\frac{1}{\tau
_{C}\alpha _{C}}+\Omega \right) T\left( Z\right) -\frac{1}{\tau _{D}\alpha
_{D}}\lambda \tau \nu c}\right) ^{2}T\left( Z\right) \left\langle \left\vert
\Psi _{0}\left( Z^{\prime }\right) \right\vert ^{2}\right\rangle _{Z}}%
\right) }  \label{Pss}
\end{equation}%
Ultimately, inserting this result in (\ref{Tng}) yields the following
equation for $T\left( Z\right) $:%
\begin{eqnarray}
\frac{\hat{\Omega}T\left( Z\right) -\frac{1}{\tau _{D}\alpha _{D}}\lambda
\tau \nu c}{\lambda \tau \nu c-T\left( Z\right) } &=&\hat{G}\left( T\left(
Z\right) ,\frac{2T\left( Z\right) \left\langle \left\vert \Psi _{0}\left(
Z^{\prime }\right) \right\vert ^{2}\right\rangle _{Z}}{\left( 1+\sqrt{%
1+4\left( \frac{\lambda \tau \nu c-T\left( Z\right) }{\hat{\Omega}T\left(
Z\right) -\frac{1}{\tau _{D}\alpha _{D}}\lambda \tau \nu c}\right)
^{2}T\left( Z\right) \left\langle \left\vert \Psi _{0}\left( Z^{\prime
}\right) \right\vert ^{2}\right\rangle _{Z}}\right) }\right)  \label{Tqnn} \\
&&\times \frac{2T\left( Z\right) \left\langle \left\vert \Psi _{0}\left(
Z^{\prime }\right) \right\vert ^{2}\right\rangle _{Z}}{\left( 1+\sqrt{%
1+4\left( \frac{\lambda \tau \nu c-T\left( Z\right) }{\hat{\Omega}T\left(
Z\right) -\frac{1}{\tau _{D}\alpha _{D}}\lambda \tau \nu c}\right)
^{2}T\left( Z\right) \left\langle \left\vert \Psi _{0}\left( Z^{\prime
}\right) \right\vert ^{2}\right\rangle _{Z}}\right) }  \notag
\end{eqnarray}

with:%
\begin{equation*}
\hat{\Omega}=\left( \frac{1}{\tau _{D}\alpha _{D}}+\frac{1}{\tau _{C}\alpha
_{C}}+\Omega \right)
\end{equation*}%
This equation has in general several solutions (see below) corresponding to
several regime of activity, depending on the point. Once $T\left( Z\right) $
is found, one can obtain $\left\vert \Psi \left( Z\right) \right\vert ^{2}$
using (\ref{Pss}). To obtain more precise formula for these solutions, we
will detail a particular case below. The system is ultimately determined by
finding $\Omega $.

\subsection*{Identification of $\Omega $}

The resolution is finalized by using (\ref{MG}) to identify the constant $%
\Omega $. Writing:

\begin{eqnarray*}
\Omega &=&\frac{1}{V}\int \hat{G}\left( T\left( Z\right) ,\left\vert \Psi
\left( Z\right) \right\vert ^{2}\right) \left\vert \Psi \left( Z\right)
\right\vert ^{2} \\
&=&\frac{1}{V}\int \frac{\left( \frac{1}{\tau _{D}\alpha _{D}}+\frac{1}{\tau
_{C}\alpha _{C}}+\Omega \right) T\left( Z\right) -\frac{1}{\tau _{D}\alpha
_{D}}\lambda \tau \nu c}{\lambda \tau \nu c-T\left( Z\right) } \\
&\simeq &\frac{\left( \frac{1}{\tau _{D}\alpha _{D}}+\frac{1}{\tau
_{C}\alpha _{C}}+\Omega \right) \bar{T}-\frac{1}{\tau _{D}\alpha _{D}}%
\lambda \tau \nu c}{\lambda \tau \nu c-\bar{T}}
\end{eqnarray*}%
where 
\begin{equation*}
\bar{T}=\frac{1}{V}\int T\left( Z\right) dZ
\end{equation*}%
is the average activity of the system, we can find $\Omega $ as a function
of $\bar{T}$:%
\begin{equation}
\Omega =\frac{\left( \frac{1}{\tau _{D}\alpha _{D}}+\frac{1}{\tau _{C}\alpha
_{C}}\right) \bar{T}-\frac{1}{\tau _{D}\alpha _{D}}\lambda \tau \nu c}{%
\lambda \tau \nu c-2\bar{T}}  \label{Mgt}
\end{equation}%
or $\bar{T}$ as a function of $\Omega $:%
\begin{equation}
\bar{T}=\frac{\lambda \tau \nu c\left( \Omega +\frac{1}{\tau _{D}\alpha _{D}}%
\right) }{\left( \frac{1}{\tau _{D}\alpha _{D}}+\frac{1}{\tau _{C}\alpha _{C}%
}\right) +2\Omega }  \label{mtg}
\end{equation}

Inserting this result inside (\ref{Tqnn}), integrating over $Z$, replacing $%
T\left( Z\right) $ by its average $\bar{T}$ inside the expression and using (%
\ref{Mgt}), yields: 
\begin{equation}
\Omega =\hat{G}\left( \bar{T},\frac{2\bar{T}\left\langle \left\vert \Psi
_{0}\right\vert ^{2}\right\rangle }{\left( 1+\sqrt{1+\frac{4\bar{T}}{\Omega
^{2}}\left\langle \left\vert \Psi _{0}\right\vert ^{2}\right\rangle }\right) 
}\right) \frac{2\bar{T}\left\langle \left\vert \Psi _{0}\right\vert
^{2}\right\rangle }{\left( 1+\sqrt{1+\frac{4\bar{T}}{\Omega ^{2}}%
\left\langle \left\vert \Psi _{0}\right\vert ^{2}\right\rangle }\right) }
\label{mgb}
\end{equation}%
with:%
\begin{equation*}
\left\langle \left\vert \Psi _{0}\right\vert ^{2}\right\rangle \equiv
\left\langle \left\langle \left\vert \Psi _{0}\left( Z^{\prime }\right)
\right\vert ^{2}\right\rangle _{Z}\right\rangle
\end{equation*}

The system (\ref{mgb}) and (\ref{mtg}) yields the possible values for $%
\Omega $ and $\bar{T}$. Once these constants derived, they can be replaced
in (\ref{Tqnn}) to find $T\left( Z\right) $ and finally $\left\vert \Psi
\left( Z\right) \right\vert ^{2}$ byusing (\ref{Ps}).

\subsection*{Particular case}

For $G$ an increasing function of the form $G\left( x\right) \simeq b_{0}x$
for $x<1$, we can solve the system. We start with the derivation of $\Omega $%
.

\subsubsection*{Derivation of $\Omega $}

In this particular case, we rewrite equation (\ref{BSN}) as:%
\begin{eqnarray*}
\omega \left( Z\right) &=&G\left( \frac{\kappa }{N}V\left( \left( \lambda
\tau \nu c-T\left( Z\right) \right) \left( \left( \frac{1}{\alpha _{D}\tau
_{D}}-\frac{T\left( Z\right) }{V\alpha _{C}\tau _{C}}\right) \omega
^{-1}+\left\vert \Psi \left( Z\right) \right\vert ^{2}\right) \right) \right)
\\
&\simeq &b_{0}\left( \frac{\kappa }{N}V\left( \left( \lambda \tau \nu
c-T\left( Z\right) \right) \left( \left( \frac{1}{\alpha _{D}\tau _{D}}-%
\frac{T\left( Z\right) }{V\alpha _{C}\tau _{C}}\right) \omega
^{-1}+\left\vert \Psi \left( Z\right) \right\vert ^{2}\right) \right) \right)
\end{eqnarray*}%
and for $\frac{1}{\alpha _{D}\tau _{D}}<<1$ and $\frac{1}{V\alpha _{C}\tau
_{C}}<<1$, this yields in first approximation:%
\begin{equation}
\omega \left( Z\right) \simeq b_{0}\left( \frac{\kappa }{N}\left( \left(
\lambda \tau \nu c-T\left( Z\right) \right) \left\vert \Psi \left( Z\right)
\right\vert ^{2}\right) \right)  \label{CT}
\end{equation}%
and equation (\ref{mgb}) writes:%
\begin{equation*}
\Omega =b\left( \left( \lambda \tau \nu c-\bar{T}\right) \frac{2\bar{T}%
\left\langle \left\vert \Psi _{0}\right\vert ^{2}\right\rangle }{\left( 1+%
\sqrt{1+\frac{4\bar{T}}{\Omega ^{2}}\left\langle \left\vert \Psi
_{0}\right\vert ^{2}\right\rangle }\right) }\right) \frac{2\bar{T}%
\left\langle \left\vert \Psi _{0}\right\vert ^{2}\right\rangle }{\left( 1+%
\sqrt{1+\frac{4\bar{T}}{\Omega ^{2}}\left\langle \left\vert \Psi
_{0}\right\vert ^{2}\right\rangle }\right) }
\end{equation*}%
We use (\ref{mtg}) under approximation $\frac{1}{\tau _{D}\alpha _{D}}\simeq 
\frac{1}{\tau _{C}\alpha _{C}}<<1$, so that:%
\begin{equation}
\bar{T}\simeq \frac{\lambda \tau \nu c}{2}  \label{RT}
\end{equation}%
and (\ref{MGG}) reduces to the formula quoted in the paper:

\begin{equation}
\Omega =b\bar{T}\left( \frac{2\bar{T}\left\langle \left\vert \Psi
_{0}\right\vert ^{2}\right\rangle }{\left( 1+\sqrt{1+\frac{4\bar{T}}{\Omega
^{2}}\left\langle \left\vert \Psi _{0}\right\vert ^{2}\right\rangle }\right) 
}\right) ^{2}  \label{MGG}
\end{equation}%
with $b=b_{0}\frac{\kappa }{N}V$. Equation (\ref{MGG}) can be transformed as:

\begin{equation*}
\left( 1+\sqrt{1+\frac{4\bar{T}}{\Omega ^{2}}\left\langle \left\vert \Psi
_{0}\right\vert ^{2}\right\rangle }\right) ^{2}=4\frac{b\bar{T}}{\Omega }%
\left( \bar{T}\left\langle \left\vert \Psi _{0}\right\vert ^{2}\right\rangle
\right) ^{2}
\end{equation*}%
and developping the square yields:%
\begin{equation*}
1+\frac{4\bar{T}}{\Omega ^{2}}\left\langle \left\vert \Psi _{0}\right\vert
^{2}\right\rangle =\left( 2\left( \frac{b\bar{T}}{\Omega }\left( \bar{T}%
\left\langle \left\vert \Psi _{0}\right\vert ^{2}\right\rangle \right) ^{2}-%
\frac{\bar{T}\left\langle \left\vert \Psi _{0}\right\vert ^{2}\right\rangle 
}{\Omega ^{2}}\right) -1\right) ^{2}
\end{equation*}%
for a final equation:%
\begin{equation*}
\frac{4\bar{T}}{\Omega ^{2}}\left\langle \left\vert \Psi _{0}\right\vert
^{2}\right\rangle =4\left( \frac{b\bar{T}}{\Omega }\left( \bar{T}%
\left\langle \left\vert \Psi _{0}\right\vert ^{2}\right\rangle \right) ^{2}-%
\frac{\bar{T}\left\langle \left\vert \Psi _{0}\right\vert ^{2}\right\rangle 
}{\Omega ^{2}}\right) ^{2}-4\left( \frac{b\bar{T}}{\Omega }\left( \bar{T}%
\left\langle \left\vert \Psi _{0}\right\vert ^{2}\right\rangle \right) ^{2}-%
\frac{\bar{T}\left\langle \left\vert \Psi _{0}\right\vert ^{2}\right\rangle 
}{\Omega ^{2}}\right)
\end{equation*}%
This leads to the equation for $\Omega $:%
\begin{equation}
1=\frac{b\bar{T}\Omega ^{3}}{\left( b\bar{T}^{2}\left\langle \left\vert \Psi
_{0}\right\vert ^{2}\right\rangle \Omega -1\right) ^{2}}  \label{FF}
\end{equation}%
Defining $b\bar{T}^{2}\left\langle \left\vert \Psi _{0}\right\vert
^{2}\right\rangle \Omega =X$, this equation becomes:%
\begin{equation}
\left( b\bar{T}\right) ^{2}\left( \bar{T}\left\langle \left\vert \Psi
_{0}\right\vert ^{2}\right\rangle \right) ^{3}=\frac{X^{3}}{\left(
X-1\right) ^{2}}  \label{RDCQT}
\end{equation}%
For $d=\left( b\bar{T}\right) ^{2}\left( \bar{T}\left\langle \left\vert \Psi
_{0}\right\vert ^{2}\right\rangle \right) ^{3}<\frac{27}{4}$ there is one
solution:%
\begin{equation*}
X\simeq \sqrt[3]{\left( b\bar{T}\right) ^{2}}\bar{T}\left\langle \left\vert
\Psi _{0}\right\vert ^{2}\right\rangle
\end{equation*}%
with:%
\begin{equation*}
\Omega \simeq \left( b\bar{T}\right) ^{-\frac{1}{3}}<<1
\end{equation*}

For $d=\left( b\bar{T}\right) ^{2}\left\langle \left\vert \Psi
_{0}\right\vert ^{2}\right\rangle ^{3}>\frac{27}{4}$ there are three
solutions. The first one is:%
\begin{equation*}
X\simeq \left( b\bar{T}\right) ^{2}\left( \bar{T}\left\langle \left\vert
\Psi _{0}\right\vert ^{2}\right\rangle \right) ^{3}
\end{equation*}%
with:%
\begin{equation*}
\Omega \simeq b\bar{T}\left( \bar{T}\left\langle \left\vert \Psi
_{0}\right\vert ^{2}\right\rangle \right) ^{2}
\end{equation*}%
The two other solutions are centered around $1$. We set $X=1\pm \delta $ and
(\ref{RCDTQT}) becomes: 
\begin{equation}
\left( b\bar{T}\right) ^{2}\left( \bar{T}\left\langle \left\vert \Psi
_{0}\right\vert ^{2}\right\rangle \right) ^{3}\simeq \frac{1}{\delta ^{2}}
\end{equation}%
so that:%
\begin{equation*}
X=1\pm \sqrt{\frac{1}{\left( b\bar{T}\right) ^{2}\left( \bar{T}\left\langle
\left\vert \Psi _{0}\right\vert ^{2}\right\rangle \right) ^{3}}}
\end{equation*}%
with:%
\begin{equation*}
\Omega =\frac{1\pm \sqrt{\frac{1}{\left( b\bar{T}\right) ^{2}\left( \bar{T}%
\left\langle \left\vert \Psi _{0}\right\vert ^{2}\right\rangle \right) ^{3}}}%
}{b\bar{T}^{2}\left\langle \left\vert \Psi _{0}\right\vert ^{2}\right\rangle 
}
\end{equation*}

\subsubsection*{Computation of $T\left( Z\right) $}

In this paragraph we compute also $Y$ defined by:%
\begin{equation}
Y=\frac{\left( \frac{1}{\tau _{D}\alpha _{D}}+\frac{1}{\tau _{C}\alpha _{C}}%
+\Omega \right) T\left( Z\right) -\frac{1}{\tau _{D}\alpha _{D}}\lambda \tau
\nu c}{\lambda \tau \nu c-T\left( Z\right) }  \label{md}
\end{equation}%
that will be used to derive the connectivity function $T\left( Z,Z^{\prime
}\right) $. Equation (\ref{Tqnn}) rewrites in this particular case:

\begin{equation}
Y=b\left( 2\bar{T}-T\left( Z\right) \right) \left( \frac{2T\left( Z\right)
\left\langle \left\vert \Psi _{0}\left( Z^{\prime }\right) \right\vert
^{2}\right\rangle _{Z}}{\left( 1+\sqrt{1+\frac{4T\left( Z\right) }{Y^{2}}%
\left\langle \left\vert \Psi _{0}\left( Z^{\prime }\right) \right\vert
^{2}\right\rangle _{Z}}\right) }\right) ^{2}  \label{dfn}
\end{equation}%
Similar computations as for $\Omega $ reduce equation (\ref{dfn}) to:%
\begin{equation}
1=\frac{b\left( 2\bar{T}-T\left( Z\right) \right) Y^{3}}{\left( b\left( 2%
\bar{T}-T\left( Z\right) \right) T\left( Z\right) \left\langle \left\vert
\Psi _{0}\left( Z^{\prime }\right) \right\vert ^{2}\right\rangle
_{Z}Y-1\right) ^{2}}  \label{tm}
\end{equation}%
Equations (\ref{md}) and (\ref{tm}) form a system allowing to find $T\left(
Z\right) $. To do so, we solve (\ref{tm}) for $T\left( Z\right) $: 
\begin{equation}
T\left( Z\right) =\frac{\lambda \tau \nu c\left( \frac{1}{\tau _{D}\alpha
_{D}}+Y\right) }{\frac{1}{\tau _{D}\alpha _{D}}+\frac{1}{\tau _{C}\alpha _{C}%
}+\Omega +Y}\simeq \frac{2\bar{T}Y}{\Omega +Y}  \label{nvd}
\end{equation}%
Along with (\ref{tm}), equation (\ref{nvd}) leads to the equation defining $%
Y $:%
\begin{equation}
1=\frac{b\frac{2\bar{T}\Omega }{\Omega +Y}Y^{3}}{\left( b\frac{2\bar{T}%
\Omega }{\Omega +Y}\frac{2\bar{T}Y}{\Omega +Y}\left\langle \left\vert \Psi
_{0}\left( Z^{\prime }\right) \right\vert ^{2}\right\rangle _{Z}Y-1\right)
^{2}}  \label{FY}
\end{equation}

\paragraph*{Lowest order approximation}

If we assume that the fluctations of $\left\langle \left\vert \Psi
_{0}\left( Z^{\prime }\right) \right\vert ^{2}\right\rangle _{Z}$ around $%
\left\vert \Psi _{0}\right\vert ^{2}$ are relatively low, we can assume in
first approximation that:%
\begin{equation*}
\Omega \simeq Y
\end{equation*}%
and thus;%
\begin{equation*}
T\left( Z\right) \simeq \frac{\lambda \tau \nu c\left( \frac{1}{\tau
_{D}\alpha _{D}}+\Omega \right) }{\frac{1}{\tau _{D}\alpha _{D}}+\frac{1}{%
\tau _{C}\alpha _{C}}+2\Omega }\simeq \frac{\lambda \tau \nu c}{2}
\end{equation*}%
Then the resolution of (\ref{FY}) is similar to the previous section.

For $d=\left( b\bar{T}\right) ^{2}\left( \bar{T}\left\langle \left\vert \Psi
_{0}\right\vert ^{2}\right\rangle \right) ^{3}<\frac{27}{4}$ there is one
solution:%
\begin{equation}
Y\simeq \left( b\bar{T}\right) ^{-\frac{1}{3}}<<1  \label{sltnn}
\end{equation}

For $d=\left( b\bar{T}\right) ^{2}\left( \bar{T}\left\langle \left\vert \Psi
_{0}\right\vert ^{2}\right\rangle \right) ^{3}>\frac{27}{4}$ there are three
solutions. The first one is:%
\begin{equation}
Y\simeq \left( b\bar{T}\right) ^{2}\left( \bar{T}\left\langle \left\vert
\Psi _{0}\left( Z^{\prime }\right) \right\vert ^{2}\right\rangle _{Z}\right)
^{3}  \label{sltdnn}
\end{equation}

The two other solutions are centered around $\frac{1}{b\bar{T}%
^{2}\left\langle \left\vert \Psi _{0}\right\vert ^{2}\right\rangle }$. We
obtain:%
\begin{equation}
Y=\frac{1\pm \sqrt{\frac{1}{\left( b\bar{T}\right) ^{2}\left( \bar{T}%
\left\langle \left\vert \Psi _{0}\left( Z^{\prime }\right) \right\vert
^{2}\right\rangle _{Z}\right) ^{3}}}}{b\bar{T}^{2}\left\langle \left\vert
\Psi _{0}\left( Z^{\prime }\right) \right\vert ^{2}\right\rangle _{Z}}
\label{sltddd}
\end{equation}%
We write $Y_{+}$ the solution (\ref{sltdnn}) and $Y_{-_{\left( \pm \right)
}} $ the solutions (\ref{sltddd}). In the sequel, we will consider only
these three solutions, written $Y_{+,-\left( _{\pm }\right) }$, and neglect (%
\ref{sltnn}) which corresponds to $\left\langle \left\vert \Psi _{0}\left(
Z^{\prime }\right) \right\vert ^{2}\right\rangle _{Z}^{3}<<1$, a point with
low activity.

\paragraph*{First order corrections}

Note also that the average connectivity at $Z$, i.e. $T\left( Z\right) $ can
be computed including the first order correction with respect to $\frac{%
\lambda \tau \nu c}{2}$, by using:%
\begin{equation}
T\left( Z\right) =\frac{\lambda \tau \nu cY+\lambda \tau \nu c\frac{1}{\tau
_{D}\alpha _{D}}}{Y+\hat{\Omega}}  \label{TFT}
\end{equation}%
with:%
\begin{equation*}
\hat{\Omega}=\frac{1}{\tau _{C}\alpha _{C}}+\frac{1}{\tau _{D}\alpha _{D}}%
+\Omega
\end{equation*}%
We compute these corrections, in the approximation $\frac{1}{\tau _{C}\alpha
_{C}}<<1$ and $\frac{1}{\tau _{D}\alpha _{D}}<<1$. \ We write equation (\ref%
{FY}) as:%
\begin{equation}
1=\frac{b\frac{2\bar{T}\left( \delta +Y\right) }{\delta +x+2Y}\left(
Y+x\right) ^{3}}{\left( b\frac{2\bar{T}\left( Y+\delta \right) }{\delta +x+2Y%
}\frac{2\bar{T}\left( Y+x\right) }{\delta +x+2Y}\left\langle \left\vert \Psi
_{0}\left( Z^{\prime }\right) \right\vert ^{2}\right\rangle _{Z}\left(
Y+x\right) -1\right) ^{2}}  \label{YF}
\end{equation}%
wth $x$\ is the deviation of $Y$ from its zeroth rdr value and where $\delta 
$ is the difference between $\Omega $ and $Y$ at the first order: 
\begin{equation*}
\delta =\Omega -Y\simeq 3\left( b\bar{T}\right) ^{2}\left( \bar{T}%
\left\langle \left\vert \Psi _{0}\left( Z^{\prime }\right) \right\vert
^{2}\right\rangle _{Z}\right) ^{2}\bar{T}\left( \left\langle \left\vert \Psi
_{0}\right\vert ^{2}\right\rangle ^{2}-\left\langle \left\vert \Psi
_{0}\left( Z^{\prime }\right) \right\vert ^{2}\right\rangle _{Z}^{2}\right)
\end{equation*}%
if (\ref{sltdnn}) is used, or:%
\begin{equation*}
\delta =\Omega -Y\simeq \mp \frac{1}{T^{3}b\left\langle \left\vert \Psi
_{0}\left( Z^{\prime }\right) \right\vert ^{2}\right\rangle _{Z}^{2}}\left( 
\frac{5}{2}\sqrt{\frac{1}{T^{5}b^{2}\left\langle \left\vert \Psi _{0}\left(
Z^{\prime }\right) \right\vert ^{2}\right\rangle _{Z}^{3}}}\pm 1\right) \bar{%
T}\left( \left\langle \left\vert \Psi _{0}\right\vert ^{2}\right\rangle
^{2}-\left\langle \left\vert \Psi _{0}\left( Z^{\prime }\right) \right\vert
^{2}\right\rangle _{Z}^{2}\right)
\end{equation*}%
if (\ref{sltddd}) is considered. Expanding equation (\ref{YF}) to the first
order yields the correction $\Delta Y$ to the lowest order solution,
whatever its form:%
\begin{equation*}
\Delta Y=-\frac{\delta }{\left( 5-4\frac{bT^{2}Y\left\langle \left\vert \Psi
_{0}\left( Z^{\prime }\right) \right\vert ^{2}\right\rangle _{Z}}{\left(
bT^{2}Y\left\langle \left\vert \Psi _{0}\left( Z^{\prime }\right)
\right\vert ^{2}\right\rangle _{Z}-1\right) }\right) }
\end{equation*}%
which corrects (\ref{sltdnn}):

\begin{equation}
Y_{+}\simeq \left( b\bar{T}\right) ^{2}\left( \bar{T}\left\langle \left\vert
\Psi _{0}\left( Z^{\prime }\right) \right\vert ^{2}\right\rangle _{Z}\right)
^{3}-\frac{3\left( b\bar{T}\right) ^{2}\left( \bar{T}\left\langle \left\vert
\Psi _{0}\left( Z^{\prime }\right) \right\vert ^{2}\right\rangle _{Z}\right)
^{2}}{5-4\frac{bT^{2}Y\left\langle \left\vert \Psi _{0}\left( Z^{\prime
}\right) \right\vert ^{2}\right\rangle _{Z}}{\left( bT^{2}Y\left\langle
\left\vert \Psi _{0}\left( Z^{\prime }\right) \right\vert ^{2}\right\rangle
_{Z}-1\right) }}\bar{T}\left( \left\langle \left\vert \Psi _{0}\right\vert
^{2}\right\rangle ^{2}-\left\langle \left\vert \Psi _{0}\left( Z^{\prime
}\right) \right\vert ^{2}\right\rangle _{Z}^{2}\right)
\end{equation}%
with:%
\begin{eqnarray*}
T_{+}\left( Z\right) &\simeq &\frac{2\bar{T}Y}{\Omega +Y}=\frac{2\bar{T}Y}{%
2Y+\delta }\simeq \bar{T}-\frac{\bar{T}}{2Y}\delta \\
&\simeq &\bar{T}-\frac{3\bar{T}\left( \left\langle \left\vert \Psi
_{0}\right\vert ^{2}\right\rangle ^{2}-\left\langle \left\vert \Psi
_{0}\left( Z^{\prime }\right) \right\vert ^{2}\right\rangle _{Z}^{2}\right) 
}{2\left\langle \left\vert \Psi _{0}\left( Z^{\prime }\right) \right\vert
^{2}\right\rangle _{Z}}
\end{eqnarray*}%
\bigskip and (\ref{sltddd}): 
\begin{eqnarray*}
Y_{-\pm } &\simeq &Y=\frac{1\pm \sqrt{\frac{1}{\left( b\bar{T}\right)
^{2}\left( \bar{T}\left\langle \left\vert \Psi _{0}\left( Z^{\prime }\right)
\right\vert ^{2}\right\rangle _{Z}\right) ^{3}}}}{b\bar{T}^{2}\left\langle
\left\vert \Psi _{0}\left( Z^{\prime }\right) \right\vert ^{2}\right\rangle
_{Z}} \\
&&\pm \frac{\frac{1}{T^{3}b\left\langle \left\vert \Psi _{0}\left( Z^{\prime
}\right) \right\vert ^{2}\right\rangle _{Z}^{2}}\left( \frac{5}{2}\sqrt{%
\frac{1}{T^{5}b^{2}\left\langle \left\vert \Psi _{0}\left( Z^{\prime
}\right) \right\vert ^{2}\right\rangle _{Z}^{3}}}\pm 1\right) }{2\left( 5-4%
\frac{bT^{2}Y\left\langle \left\vert \Psi _{0}\left( Z^{\prime }\right)
\right\vert ^{2}\right\rangle _{Z}}{\left( bT^{2}Y\left\langle \left\vert
\Psi _{0}\left( Z^{\prime }\right) \right\vert ^{2}\right\rangle
_{Z}-1\right) }\right) }\bar{T}\left( \left\langle \left\vert \Psi
_{0}\right\vert ^{2}\right\rangle ^{2}-\left\langle \left\vert \Psi
_{0}\left( Z^{\prime }\right) \right\vert ^{2}\right\rangle _{Z}^{2}\right)
\end{eqnarray*}%
with:%
\begin{eqnarray*}
T_{-\pm }\left( Z\right) &\simeq &\bar{T}-\frac{\bar{T}}{2Y}\delta \\
&\simeq &\bar{T}\pm \frac{b\bar{T}\left( \frac{5}{2}\sqrt{\frac{1}{%
T^{5}b^{2}\left\langle \left\vert \Psi _{0}\left( Z^{\prime }\right)
\right\vert ^{2}\right\rangle _{Z}^{3}}}\pm 1\right) }{\left( b\bar{T}\left( 
\bar{T}\left\langle \left\vert \Psi _{0}\left( Z^{\prime }\right)
\right\vert ^{2}\right\rangle _{Z}\right) \pm \sqrt{\frac{1}{\bar{T}%
\left\langle \left\vert \Psi _{0}\left( Z^{\prime }\right) \right\vert
^{2}\right\rangle _{Z}}}\right) }\bar{T}\left( \left\langle \left\vert \Psi
_{0}\right\vert ^{2}\right\rangle ^{2}-\left\langle \left\vert \Psi
_{0}\left( Z^{\prime }\right) \right\vert ^{2}\right\rangle _{Z}^{2}\right)
\end{eqnarray*}%
we define also the average of $T_{-+}\left( Z\right) $ and $T_{--}\left(
Z\right) $:%
\begin{equation*}
T_{-}\left( Z\right) \simeq \bar{T}+\frac{5}{2}\sqrt{\frac{1}{\bar{T}%
^{2}b^{2}\left\langle \bar{T}\left\vert \Psi _{0}\left( Z^{\prime }\right)
\right\vert ^{2}\right\rangle _{Z}^{5}}}\bar{T}\left( \left\langle
\left\vert \Psi _{0}\right\vert ^{2}\right\rangle ^{2}-\left\langle
\left\vert \Psi _{0}\left( Z^{\prime }\right) \right\vert ^{2}\right\rangle
_{Z}^{2}\right)
\end{equation*}

\subsubsection*{Connectivity functions}

We ultimately write the connectivity functions:%
\begin{eqnarray*}
T\left( Z,Z^{\prime }\right) &=&\frac{\lambda \tau \exp \left( -\frac{%
\left\vert Z-Z^{\prime }\right\vert }{\nu c}\right) }{1+\frac{\alpha _{D}}{%
\alpha _{C}}\frac{\frac{1}{\tau _{C}}+\alpha _{C}\omega ^{\prime }\left\vert
\Psi \left( Z^{\prime }\right) \right\vert ^{2}}{\frac{1}{\tau _{D}}+\alpha
_{D}\omega \left\vert \Psi \left( Z\right) \right\vert ^{2}}} \\
&=&\frac{\lambda \tau \exp \left( -\frac{\left\vert Z-Z^{\prime }\right\vert 
}{\nu c}\right) }{1+\frac{\frac{1}{\alpha _{C}\tau _{C}}+\hat{G}\left(
T\left( Z^{\prime }\right) ,\left\vert \Psi \left( Z^{\prime }\right)
\right\vert ^{2}\right) \left\vert \Psi \left( Z^{\prime }\right)
\right\vert ^{2}}{\frac{1}{\tau _{D}\alpha _{D}}+\hat{G}\left( T\left(
Z\right) ,\left\vert \Psi \left( Z\right) \right\vert ^{2}\right) \left\vert
\Psi \left( Z\right) \right\vert ^{2}}}
\end{eqnarray*}%
Since:%
\begin{equation*}
Y=\hat{G}\left( T\left( Z\right) ,\left\vert \Psi \left( Z\right)
\right\vert ^{2}\right) \left\vert \Psi \left( Z\right) \right\vert ^{2}
\end{equation*}%
we write $T\left( Z,Z^{\prime }\right) $:%
\begin{equation*}
T\left( Z,Z^{\prime }\right) =\frac{\lambda \tau \exp \left( -\frac{%
\left\vert Z-Z^{\prime }\right\vert }{\nu c}\right) \left( \frac{1}{\tau
_{D}\alpha _{D}}+Y\left( Z\right) \right) }{\frac{1}{\tau _{D}\alpha _{D}}%
+Y\left( Z\right) +\frac{1}{\alpha _{C}\tau _{C}}+Y\left( Z^{\prime }\right) 
}
\end{equation*}%
with $Y\left( Z\right) $ given by (\ref{sltdnn}) and (\ref{sltddd}). There
are nine possibilities for the connectivity function:%
\begin{equation*}
T\left( Z_{+,-\left( _{\pm }\right) },Z_{+,-\left( _{\pm }\right) }^{\prime
}\right) =\frac{\lambda \tau \exp \left( -\frac{\left\vert Z-Z^{\prime
}\right\vert }{\nu c}\right) \left( \frac{1}{\tau _{D}\alpha _{D}}%
+Y_{+,-\left( _{\pm }\right) }\left( Z\right) \right) }{\frac{1}{\tau
_{D}\alpha _{D}}+Y_{+,-\left( _{\pm }\right) }\left( Z\right) +\frac{1}{%
\alpha _{C}\tau _{C}}+Y_{+,-\left( _{\pm }\right) }\left( Z^{\prime }\right) 
}
\end{equation*}%
Given that the solution (\ref{sltddd}) are both centered around $\frac{1}{%
b\left\langle \left\vert \Psi _{0}\left( Z^{\prime }\right) \right\vert
^{2}\right\rangle _{Z}}$, and thus relatively close from each other, we can
gather them in one approximative solution $\frac{1}{b\left\langle \left\vert
\Psi _{0}\left( Z^{\prime }\right) \right\vert ^{2}\right\rangle _{Z}}$ and
replace $Y_{+,-\left( _{\pm }\right) }$ by $Y_{\pm }\left( Z\right) $ given
by: 
\begin{equation*}
T\left( Z_{\pm },Z_{\pm }^{\prime }\right) =\frac{\lambda \tau \exp \left( -%
\frac{\left\vert Z-Z^{\prime }\right\vert }{\nu c}\right) \left( \frac{1}{%
\tau _{D}\alpha _{D}}+Y_{\pm }\left( Z\right) \right) }{\frac{1}{\tau
_{D}\alpha _{D}}+Y_{\pm }\left( Z\right) +\frac{1}{\alpha _{C}\tau _{C}}%
+Y_{\pm }\left( Z^{\prime }\right) }
\end{equation*}%
and this yields four possibilities as detailed in the text.

Note that (\ref{TFT}) can also be written as:%
\begin{equation*}
T\left( Z_{\pm }\right) =\frac{\lambda \tau \nu cY_{\pm }\left( Z\right)
+\lambda \tau \nu c\frac{1}{\tau _{D}\alpha _{D}}}{Y_{\pm }\left( Z\right) +%
\frac{1}{\tau _{D}\alpha _{D}}+\frac{1}{\alpha _{C}\tau _{C}}+\Omega _{\pm }}
\end{equation*}%
where:%
\begin{equation*}
\Omega _{\pm }=\left( \Omega _{+},\Omega _{-}\right)
\end{equation*}%
with:%
\begin{equation*}
\Omega _{+}=b\bar{T}\left( \bar{T}\left\langle \left\vert \Psi
_{0}\right\vert ^{2}\right\rangle \right) ^{2},\Omega _{-}=\frac{1}{b\bar{T}%
^{2}\left\langle \left\vert \Psi _{0}\right\vert ^{2}\right\rangle }
\end{equation*}

\subsubsection*{Activities}

We can rewrite the activity (\ref{CT}) as:%
\begin{eqnarray*}
\omega _{+}\left( Z\right) &\simeq &b_{0}\left( \frac{\kappa }{N}\left(
\lambda \tau \nu c-T_{+}\left( Z\right) \right) \left\vert \Psi \left(
Z\right) \right\vert ^{2}\right) \\
&=&b_{0}\frac{\kappa }{N}\bar{T}\left( 1-\frac{3\left( \left\langle
\left\vert \Psi _{0}\left( Z^{\prime }\right) \right\vert ^{2}\right\rangle
_{Z}^{2}-\left\langle \left\vert \Psi _{0}\right\vert ^{2}\right\rangle
^{2}\right) }{2\left\langle \left\vert \Psi _{0}\left( Z^{\prime }\right)
\right\vert ^{2}\right\rangle _{Z}^{2}}\right) \left\vert \Psi \left(
Z\right) \right\vert ^{2}
\end{eqnarray*}%
and:%
\begin{eqnarray*}
\omega _{-}\left( Z\right) b_{0} &\simeq &\left( \frac{\kappa }{N}\left(
\lambda \tau \nu c-T_{+}\left( Z\right) \right) \left\vert \Psi \left(
Z\right) \right\vert ^{2}\right) \\
&\simeq &b_{0}\frac{\kappa }{N}\bar{T}\left( \left( 1+\frac{5}{2}\sqrt{\frac{%
1}{\bar{T}^{2}b^{2}\left\langle \bar{T}\left\vert \Psi _{0}\left( Z^{\prime
}\right) \right\vert ^{2}\right\rangle _{Z}^{5}}}\left( \left\langle
\left\vert \Psi _{0}\left( Z^{\prime }\right) \right\vert ^{2}\right\rangle
_{Z}^{2}-\left\langle \left\vert \Psi _{0}\right\vert ^{2}\right\rangle
^{2}\right) \right) \left\vert \Psi \left( Z\right) \right\vert ^{2}\right)
\end{eqnarray*}%
where $\omega _{-}\left( Z\right) $ gathers the formula for $\omega
_{-}\left( Z\right) $

The square $\left\vert \Psi \left( Z\right) \right\vert ^{2}$ depends on $%
\Omega _{\pm }$. Thus we write:%
\begin{equation*}
\left\vert \Psi _{\pm }\left( Z\right) \right\vert ^{2}=\frac{2T_{\pm
}\left( Z\right) \left\langle \left\vert \Psi _{0}\left( Z^{\prime }\right)
\right\vert ^{2}\right\rangle _{Z}^{2}}{\left( 1+\sqrt{1+4\left( \frac{%
\lambda \tau \nu c-T_{\pm }\left( Z\right) }{\left( \frac{1}{\tau _{D}\alpha
_{D}}+\frac{1}{\tau _{C}\alpha _{C}}+\Omega _{\pm }\right) T_{\pm }\left(
Z\right) -\frac{1}{\tau _{D}\alpha _{D}}\lambda \tau \nu c}\right)
^{2}T_{\pm }\left( Z\right) \left\langle \left\vert \Psi _{0}\left(
Z^{\prime }\right) \right\vert ^{2}\right\rangle _{Z}}\right) }
\end{equation*}%
Given that:%
\begin{equation*}
\sqrt{1+4\left( \frac{\lambda \tau \nu c-T_{\pm }\left( Z\right) }{\left(
\Omega _{\pm }\right) T_{\pm }\left( Z\right) }\right) ^{2}T_{\pm }\left(
Z\right) \left\langle \left\vert \Psi _{0}\left( Z^{\prime }\right)
\right\vert ^{2}\right\rangle _{Z}}\simeq \sqrt{1+\left( \frac{1}{\Omega
_{\pm }}\right) ^{2}T_{\pm }\left( Z\right) \left\langle \left\vert \Psi
_{0}\left( Z^{\prime }\right) \right\vert ^{2}\right\rangle _{Z}}
\end{equation*}%
we find:%
\begin{equation*}
\left\vert \Psi _{+}\left( Z\right) \right\vert ^{2}=\frac{2T_{+}\left(
Z\right) \left\langle \left\vert \Psi _{0}\left( Z^{\prime }\right)
\right\vert ^{2}\right\rangle _{Z}^{2}}{\left( 1+\sqrt{1+\frac{T_{+}\left(
Z\right) \left\langle \left\vert \Psi _{0}\left( Z^{\prime }\right)
\right\vert ^{2}\right\rangle _{Z}}{\left( b\bar{T}\left( \bar{T}%
\left\langle \left\vert \Psi _{0}\right\vert ^{2}\right\rangle \right)
^{2}\right) ^{2}}}\right) }\simeq 2\bar{T}\left\langle \left\vert \Psi
_{0}\left( Z^{\prime }\right) \right\vert ^{2}\right\rangle _{Z}^{2}
\end{equation*}%
and:%
\begin{equation*}
\left\vert \Psi _{-}\left( Z\right) \right\vert ^{2}=\frac{2T_{-}\left(
Z\right) \left\langle \left\vert \Psi _{0}\left( Z^{\prime }\right)
\right\vert ^{2}\right\rangle _{Z}^{2}}{\left( 1+\sqrt{1+4\left( b\bar{T}%
^{2}\left\langle \left\vert \Psi _{0}\right\vert ^{2}\right\rangle \right)
^{2}T_{-}\left( Z\right) \left\langle \left\vert \Psi _{0}\left( Z^{\prime
}\right) \right\vert ^{2}\right\rangle _{Z}}\right) }<<1
\end{equation*}%
for $\bar{T}>>1$.

Thus, given our assumptions:%
\begin{equation*}
\omega _{-}\left( Z\right) <<\omega _{+}\left( Z\right)
\end{equation*}

\subsection*{Appendix 4 Solution of classical action's first order condition}

To solve equations (\ref{psG})\ and (\ref{pSG}), the dependency of $\omega
^{-1}\left( J\left( \theta \right) ,\theta ,Z,\mathcal{G}_{0}+\left\vert
\Psi \right\vert ^{2}\right) $ in $\left\vert \Psi \right\vert ^{2}$ has to
be explicited. Note that in first approximation, the solution of (\ref{pSG})
is:%
\begin{eqnarray*}
\delta \Psi \left( \theta ,Z\right) &\simeq &-\frac{\nabla _{\theta }\omega
^{-1}\left( J\left( \theta \right) ,\theta ,Z,\mathcal{G}_{0}+\left\vert
\Psi \right\vert ^{2}\right) }{U^{\prime \prime }\left( X_{0}\right) }\Psi
_{0}\left( \theta ,Z\right) \\
&=&\frac{\nabla _{\theta }\omega \left( J\left( \theta \right) ,\theta ,Z,%
\mathcal{G}_{0}+\left\vert \Psi _{0}\right\vert ^{2}\right) }{U^{\prime
\prime }\left( X_{0}\right) \omega ^{2}\left( J\left( \theta \right) ,\theta
,Z,\mathcal{G}_{0}+\left\vert \Psi _{0}\right\vert ^{2}\right) }\Psi
_{0}\left( \theta ,Z\right)
\end{eqnarray*}%
and this approximation is sufficient as a first approximation.

However, to find a more precise expression for $\delta \Psi \left( \theta
,Z\right) $, we use (\ref{fqt}) that defines $\omega ^{-1}\left( J\left(
\theta \right) ,\theta ,Z,\mathcal{G}_{0}+\left\vert \Psi \right\vert
^{2}\right) $ at the classical order:%
\begin{eqnarray}
&&\omega ^{-1}\left( J,\theta ,Z,\left\vert \Psi \right\vert ^{2}\right)
\label{ftq} \\
&=&G\left( J\left( \theta ,Z\right) +\int \frac{\kappa }{N}\frac{\omega
\left( J,\theta -\frac{\left\vert Z-Z_{1}\right\vert }{c},Z_{1},\Psi \right)
T\left( Z,\theta ,Z_{1},\theta -\frac{\left\vert Z-Z_{1}\right\vert }{c}%
\right) }{\omega \left( J,\theta ,Z,\left\vert \Psi \right\vert ^{2}\right) }%
\right.  \notag \\
&&\times \left. \left( \mathcal{G}_{0}\left( Z_{1}\right) +\left\vert \Psi
\left( \theta -\frac{\left\vert Z-Z_{1}\right\vert }{c},Z_{1}\right)
\right\vert ^{2}\right) dZ_{1}\right)  \notag
\end{eqnarray}%
Using (\ref{ftq}), the defining equation (\ref{psG}) for $\delta \Psi \left(
\theta ,Z\right) $ becomes:%
\begin{eqnarray*}
&&G^{-1}\left( -\frac{U^{\prime \prime }\left( X_{0}\right) }{X_{0}}%
\int^{\theta }\delta \Psi \left( \theta ,Z\right) \right) \\
&=&\int \frac{\kappa }{N}\frac{\omega \left( J,\theta -\frac{\left\vert
Z-Z_{1}\right\vert }{c},Z_{1},\Psi \right) }{\omega \left( J,\theta
,Z,\left\vert \Psi \right\vert ^{2}\right) }T\left( Z,\theta ,Z_{1},\theta -%
\frac{\left\vert Z-Z_{1}\right\vert }{c}\right) \left( \mathcal{G}_{0}\left(
Z_{1}\right) +\left\vert \Psi \left( \theta -\frac{\left\vert
Z-Z_{1}\right\vert }{c},Z_{1}\right) \right\vert ^{2}\right) dZ_{1}
\end{eqnarray*}%
This equation can be rewritten in the local approximation:%
\begin{equation}
G^{-1}\left( -\frac{U^{\prime \prime }\left( X_{0}\right) }{X_{0}}%
\int^{\theta }\delta \Psi \left( \theta ,Z\right) \right) \simeq J\left(
\theta ,Z\right) +\frac{\left( -\Gamma \nabla _{\theta }+\Gamma ^{\prime
}\nabla _{Z}^{2}\right) \left( \omega \left( J,\theta ,Z\right) \left( 
\mathcal{G}_{0}\left( Z\right) +\left\vert \Psi \left( \theta ,Z\right)
\right\vert ^{2}\right) \right) }{\omega \left( J,\theta ,Z,\left\vert \Psi
\right\vert ^{2}\right) }  \label{pSM}
\end{equation}%
where $\Gamma $ and $\Gamma ^{\prime }$ are defined by: 
\begin{eqnarray*}
\Gamma &=&\int \frac{\kappa }{N}\frac{\left\vert Z-Z_{1}\right\vert }{c}%
T\left( Z,\theta ,Z_{1},\theta -\frac{\left\vert Z-Z_{1}\right\vert }{c}%
\right) dZ_{1} \\
\Gamma ^{\prime } &=&\int \frac{\kappa }{N}\left\vert Z-Z_{1}\right\vert
^{2}T\left( Z,\theta ,Z_{1},\theta -\frac{\left\vert Z-Z_{1}\right\vert }{c}%
\right) dZ_{1}
\end{eqnarray*}%
At the lowest order in derivatives, equation (\ref{pSM}) becomes:%
\begin{eqnarray}
G^{-1}\left( -\frac{U^{\prime \prime }\left( X_{0}\right) }{X_{0}}%
\int^{\theta }\delta \Psi \left( \theta ,Z\right) \right) &\simeq &J\left(
\theta ,Z\right) -\frac{\Gamma \nabla _{\theta }\omega \left( J,\theta
,Z\right) \left( \mathcal{G}_{0}\left( Z_{1}\right) +\left\vert \Psi \left(
\theta ,Z\right) \right\vert ^{2}\right) }{\omega \left( J,\theta
,Z,\left\vert \Psi \right\vert ^{2}\right) }  \label{rtm} \\
&=&J\left( \theta ,Z\right) -\Gamma \nabla _{\theta }\left\vert \Psi \left(
\theta ,Z\right) \right\vert ^{2}+\Gamma \frac{\delta \Psi \left( \theta
,Z\right) }{\int^{\theta }\delta \Psi \left( \theta ,Z\right) }\left( 
\mathcal{G}_{0}\left( Z\right) +\left\vert \Psi \left( \theta ,Z\right)
\right\vert ^{2}\right)  \notag \\
&\simeq &J\left( \theta ,Z\right) -\Gamma \sqrt{X_{0}}\nabla _{\theta
}\delta \Psi \left( \theta ,Z\right) +\Gamma \frac{\mathcal{G}_{0}\left(
Z\right) +X_{0}+\sqrt{X_{0}}\delta \Psi \left( \theta ,Z\right) }{%
\int^{\theta }\delta \Psi \left( \theta ,Z\right) }\delta \Psi \left( \theta
,Z\right)  \notag \\
&\simeq &J\left( \theta ,Z\right) +\Gamma \frac{\mathcal{G}_{0}\left(
Z_{1}\right) +X_{0}}{\int^{\theta }\delta \Psi \left( \theta ,Z\right) }%
\delta \Psi \left( \theta ,Z\right)  \notag
\end{eqnarray}%
We set:%
\begin{equation*}
Y=\ln \left( \int \delta \Psi \left( \theta ,Z\right) \right)
\end{equation*}%
and (\ref{rtm}) writes: 
\begin{eqnarray}
G^{-1}\left( -\frac{U^{\prime \prime }\left( X_{0}\right) }{X_{0}}\exp
Y\right) &=&J\left( \theta ,Z\right) +\Gamma \left( \mathcal{G}_{0}\left(
Z_{1}\right) +X_{0}\right) \nabla _{\theta }Y  \label{qrt} \\
&\simeq &\left\langle J\right\rangle \left( Z\right) +\Gamma \left( \mathcal{%
G}_{0}\left( Z_{1}\right) +X_{0}\right) \nabla _{\theta }Y  \notag
\end{eqnarray}%
where $\left\langle J\right\rangle \left( Z\right) $ is the current averaged
over time. The solution of (\ref{qrt}) is: 
\begin{equation*}
\int \delta \Psi \left( \theta ,Z\right) =\exp \left( Y\right) =\exp \left(
H^{-1}\left( \frac{\theta }{\Gamma \left( \mathcal{G}_{0}\left( Z_{1}\right)
+X_{0}\right) }+d\right) \right)
\end{equation*}%
with: 
\begin{equation*}
H\left( Y\right) =\int \frac{dY}{G^{-1}\left( -\frac{U^{\prime \prime
}\left( X_{0}\right) }{X_{0}}\exp Y\right) -\left\langle J\right\rangle
\left( Z\right) }
\end{equation*}%
and:%
\begin{eqnarray}
\delta \Psi \left( \theta ,Z\right) &=&\left( G^{-1}\left( -\frac{U^{\prime
\prime }\left( X_{0}\right) }{X_{0}}\exp \left( H^{-1}\left( \frac{\theta }{%
\Gamma \left( \mathcal{G}_{0}\left( Z_{1}\right) +\sqrt{X_{0}}\right) }%
+d\right) \right) \right) -\left\langle J\right\rangle \left( Z\right)
\right)  \label{nls} \\
&&\times \exp \left( H^{-1}\left( \frac{\theta }{\Gamma \left( \mathcal{G}%
_{0}\left( Z_{1}\right) +\sqrt{X_{0}}\right) }+d\right) \right)  \notag
\end{eqnarray}%
The constant $d$ is chosen so that $\lim_{\theta \rightarrow \infty }\delta
\Psi \left( \theta ,Z\right) =0$. For slowly varyng currents, $\left\langle
J\right\rangle \left( Z\right) $ can replaced by $J\left( \theta ,Z\right) $
in the formula.

\end{document}